\documentclass[twocolumn]{aastex701}

\newcommand{\farsec}{\mbox{\ensuremath{.\!\!^{\prime\prime}}}} 
\newcommand{\hi}{H{\sc i}}

\newcommand{\hdue}{H$_2$}
\newcommand{\msun}{M$_{\odot}$}

\usepackage{caption}
\usepackage{newtxtext,newtxmath}
\usepackage[normalem]{ulem}
\usepackage{upgreek}
\usepackage{amsmath}

\newcommand{\uu}{\`{}$\bar{\rm U}$\`{}$\bar{\rm u}$}

\begin{document}

\title{The MaNGA Low-mass disks HUnt for CO (MaLHUCO) Survey}

\author[0000-0003-4675-3246]{M. Grossi}
\affiliation{Observatório do Valongo, Universidade Federal do Rio de Janeiro, Ladeira Pedro Antônio 43, Saúde, Rio de Janeiro, RJ, 20080-090, Brazil} 
\email{grossi@astro.ufrj.br}

\author[0000-0002-5788-2628]{E. Corbelli}
\affiliation{INAF - Osservatorio Astrofisico di Arcetri, L. E. Fermi 5, 50125, Firenze, IT}
\email{edvige.corbelli@inaf.it}

\author[0000-0001-9388-7146]{D. R. Gonçalves}
\affiliation{Observatório do Valongo, Universidade Federal do Rio de Janeiro, Ladeira Pedro Antônio 43, Saúde, Rio de Janeiro, RJ, 20080-090, Brazil} 
\email{}

\author[0000-0003-2690-5345]{E. D. Paspaliaris}
\affiliation{INAF - Osservatorio Astrofisico di Arcetri, L. E. Fermi 5, 50125, Firenze, IT}
\email{}


\begin{abstract}

We present James Clerk Maxwell Telescope (JCMT) observations of the $^{12}$CO(J = 2-1) emission of 42 low-mass, star-forming disk galaxies of morphological type Scd or later from the Mapping Nearby Galaxies at Apache Point Observatory (MaNGA) survey.
The sample,  which probes the M33-like stellar-mass regime, is complemented with metallicities, star formation rates, and \hi\ masses used to investigate the star formation process and to test scaling relations involving molecular gas mass 
in low-mass systems. We detect CO emission in 55\% of the sample and derive H$_2$ masses using both a constant Galactic and a metallicity-dependent CO-to-H$_2$ conversion factor. 
The 12 $\upmu$m  luminosity, which includes polycyclic aromatic hydrocarbon features, exhibits a tight linear correlation with the CO line emission, making it a robust tracer of  global molecular gas content. The molecular gas mass - star formation rate relation, i.e. the Kennicutt-Schmidt law, is the most fundamental  one and it is found to remain linear down to 
low stellar masses. 
We also find that the mean molecular gas depletion time is slightly shorter in low-mass late-type galaxies than in more massive systems, consistent with their higher specific star formation rates.
Finally, while the specific molecular gas mass ($M_{\rm H_2}/M_*$) shows no significant dependence on stellar mass and a large intrinsic scatter, the H{\sc i}-to-stellar mass ratio ($M_{\rm HI}/M_*$) decreases with increasing stellar mass and molecular fraction ($M_{\rm H_2}/M_{\rm gas}$), highlighting the progressive transition from atomic- to molecular-dominated interstellar medium along the galaxy population.

\end{abstract}




\section{Introduction}

Stars condense out of molecular hydrogen which forms on dust grains from dense atomic gas \citep{1971ApJ...163..155H,2009ApJ...693..216K}. The molecular gas content  and the star formation rate (SFR) of galaxies show an extremely tight correlation across a wide range of morphological types as it has been shown  by the Kennicutt-Schimdt (KS) relation \citep{1959ApJ...129..243S,1998ApJ...498..541K} in the local Universe. Different scaling relations have been  investigated to establish the molecular gas efficiency and the role of atomic and molecular gas, dust and stellar masses in driving the global and local star formation rate (SFR) of present-day galaxies \citep{1993ApJ...411..170E,2012A&A...542A..32C,2013AJ....146...19L,2016A&A...590A..27G,2017ApJS..233...22S,2022ARA&A..60..319S}. 
The main sequence (MS) of star-forming galaxies, for example, defines
an approximately linear relation between the stellar mass ($M_*$) and the SFR of star-forming galaxies
\citep{2004MNRAS.351.1151B,2015ApJ...808L..49G,2015ApJ...801L..29R,2016ApJ...821L..26C}. Galaxies above the MS are experiencing a phase of enhanced star formation activity, while those in the process of quenching evolve downward in the $M_*$ -- SFR plane, forming the so-called passive cloud \citep{2015ApJ...801L..29R}.
\citet{2021ApJ...907..114D}  claim that the more fundamental scaling relations are those between specific molecular gas mass (or molecular gas-to-stellar mass ratio, $M_{\rm H_2}$/$M_*$), specific star formation rate (sSFR=SFR/$M_*$), and star formation efficiency (SFE = SFR/$M_{\rm H_2}$), and they show that galaxies form a single sequence in the sSFR --  $M_{\rm H_2}$/$M_*$  -- SFE plane  from quenched to starbursts. 
Star formation law studies at high redshift have become feasible  \citep{2010Natur.463..781T,2016ApJ...833...70D,2020ApJ...902..110D,2025A&A...701A.260L}
and complement local surveys to understand galaxy formation and evolution through cosmic times.

In a hierarchical universe, as it is described by the  $\Lambda$ Cold Dark Matter paradigm, it is extremely important to investigate star formation laws and their scatter also in low-mass disks and dwarf galaxies, since they are most abundant at any redshift. The interstellar medium (ISM) of low-mass star-forming galaxies
is metal poor,  making them close analogs of galaxies in the early Universe. At the same time the low metal abundances make the molecular gas difficult to detect through the emission of its widely used tracer, the carbon monoxide \citep[CO;][]{2012AJ....143..138S,2015A&A...583A.114H,2016A&A...590A..27G,2020A&A...643A.180H}. 

Recent studies have aimed to extend observations of
cold molecular gas to lower-mass systems, assembling statistically significant samples \citep{2017A&A...604A..53C,2017ApJS..233...22S,2020A&A...643A.180H,2024A&A...687A.244H}. 
The extended CO Legacy Database for GASS \citep[xCOLDGASS,][]{2017ApJS..233...22S} included 532 local galaxies  at redshift 0.01 $< z <$ 0.05, randomly selected from the SDSS, enabling to investigate scaling relations down to $M_* \simeq 10^9$ M$_{\odot}$.
The Apex Low-redshift Legacy Survey of Molecular Gas (ALLSMOG) survey \citep{2017A&A...604A..53C} 
carried out an extended CO survey of star-forming galaxies at intermediate and low stellar masses -- $8.5 <$ log ($M_*$/M$_{\odot}$) $< 10$ -- extracted from the
MPA-JHU catalogue of SDSS DR7 \citep[][]{2004MNRAS.351.1151B}.
\citet{2017A&A...604A..53C} found that star-forming galaxies across more than two orders of magnitude in $M_*$ obey similar scaling relations between CO luminosity and $M_*$, SFR, and gas-phase metallicity. However,  few molecular gas mass estimates are available in ALLSMOG for star-forming galaxies with sub-solar metallicities
or stellar masses below 3$\times$ 10$^9$~M$_\odot$ \citep{2017A&A...604A..53C}. 
Lastly, the Metallicity and Gas for Mass
Assembly (MAGMA) sample \citep{2020A&A...643A.180H}, analyzed molecular and atomic gas properties of 392 galaxies with available CO observations, spanning a broad range of stellar masses,
($M_* \sim 10^7 - 10^{11}$ M$_{\odot}$). MAGMA also includes galaxies from xCOLDGASS and ALLSMOG, with approximately 30\%  being dwarfs ($M_* < 3 \times 10^9$ M$_{\odot}$). Nonetheless, the coverage of CO-detected systems at low $M_*$ remains limited.

Due to the difficulty to use CO as H$_2$ tracer it is not yet clear whether the molecular gas reservoir of low-mass galaxies
is consistent with a simple mass-scaling  (stellar, baryonic or total dynamical mass) or additional factors such as disk dynamics, dust-to-gas ratio, and gas inflows or outflows  play a major role. In such galaxies  cooling is slow, gravity and spiral arms are weak, feedback is strong: these effects can reduce the abundance and size distribution of cold gas clumps and affect their efficiency to form stars  \citep{2005ApJ...625..763L,2012AJ....143..138S}. 

To further investigate how the molecular gas content of low-mass disks is linked to their global properties -- such as $M_*$, morphology, star formation activity, metallicity and atomic gas mass  ($M_{\rm HI}$) -- we selected 42 star-forming low-mass disk galaxies from the MaNGA survey \citep{2015ApJ...798....7B} and assembled the MaNGA Low-mass disks HUnt for CO (MaLHUCO) sample. 
 We considered galaxies with morphologies equal or later than Scd types, spanning more than one order of magnitude in stellar mass and covering a wide range of SFRs. 
We present and discuss here results relative to $^{12}$CO(J=2-1) observations of these galaxies.
By using the MaNGA survey we have access to a large set of ancillary data such as SFR, metallicity, and stellar surface density maps. 
Previous and ongoing CO surveys of MaNGA galaxies are mostly focused on more massive systems and do not cover very late morphological types, e.g. ALMA-MaNGA QUEnching and STar formation Survey  \citep[ALMaQUEST,][]{2019ApJ...884L..33L}, MaNGA-Arizona Radio Observatory Survey of CO Targets \citep[MASCOT,][]{2022MNRAS.510.3119W}, and Kiloparsec Investigations of Local Objects’ Gas And Star-formation (KILOGAS)\footnote{\url{https://kilogas.space/index.html}}. Our sample allows us to explore a different range of galaxy properties, as it includes 
later morphological types and it extends toward lower stellar masses.
Our closest Scd galaxy, M33, fully classifies as a typical MaLHUCO galaxy and, given its well known properties inferred from high resolution observations, it can be used as a reference for MaLHUCO galaxies \citep{2014A&A...572A..23C,2025A&A...700A..57C}. At the same time, this work can also help understanding if M33 is forming stars and molecules at a rate comparable to its siblings. 

The paper is organized as follows: in Section 2 we present the sample selection and the list of the ancillary data used in this work; in Section 3 we describe the CO observations; in Section 4 we compare the properties of CO-detected and non-detected galaxies and  derive the \hdue\ gas masses in Secion 5. Section 6 is dedicated to the study of the main scaling relations for our sample. The main results are discussed in Section 7 and summarized in Section 8. CO spectra are displayed in the Appendix.

\begin{figure*}
 \begin{center}
\includegraphics[bb=170 270 440 520,width=3.69cm, clip]{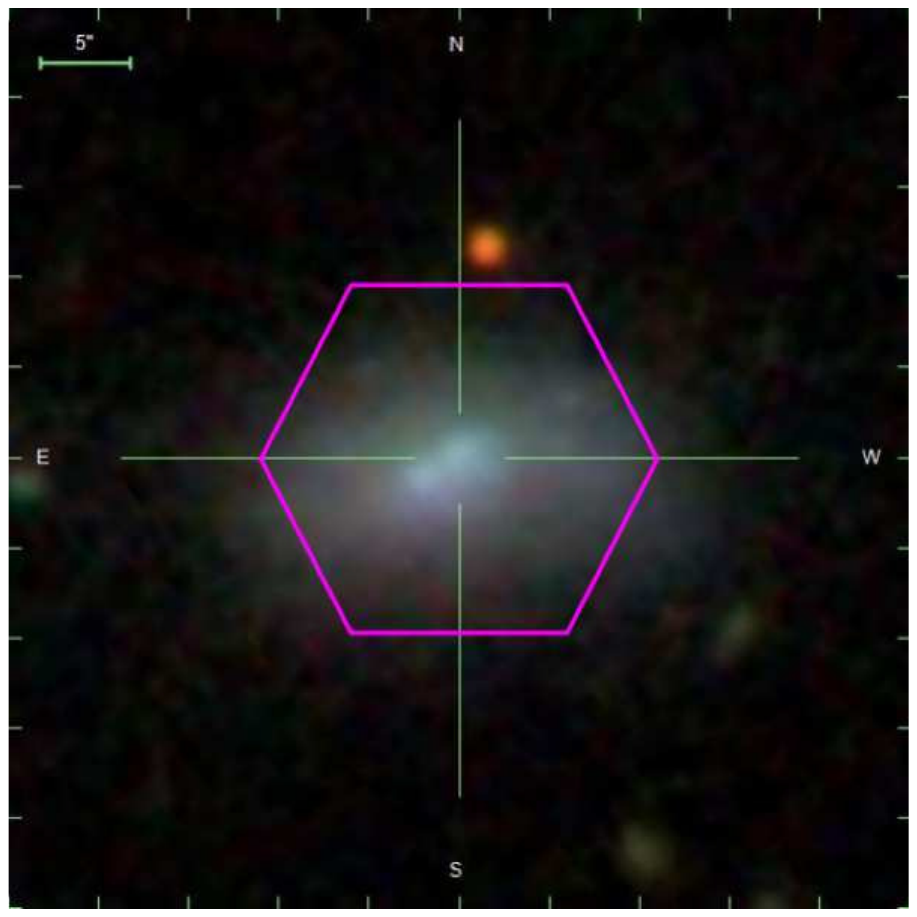}
\includegraphics[bb=140 240 470 550,width=3.64cm, clip]{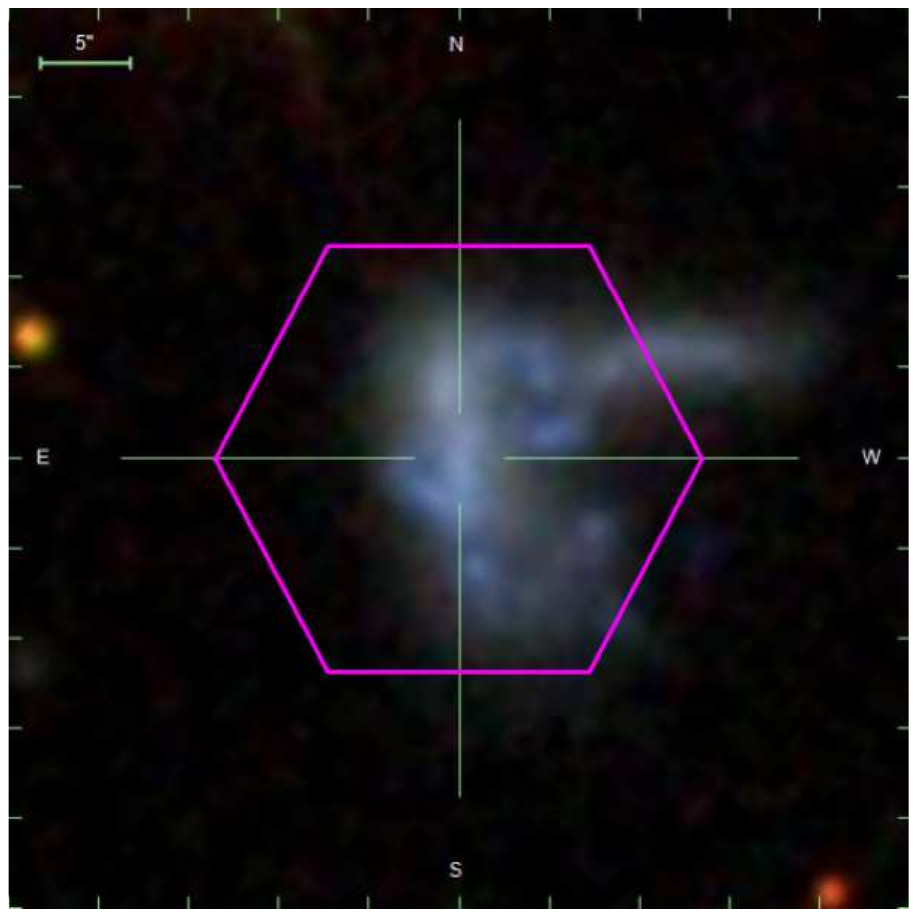}
\includegraphics[bb=70 60 370 380, width=3.22cm, clip]{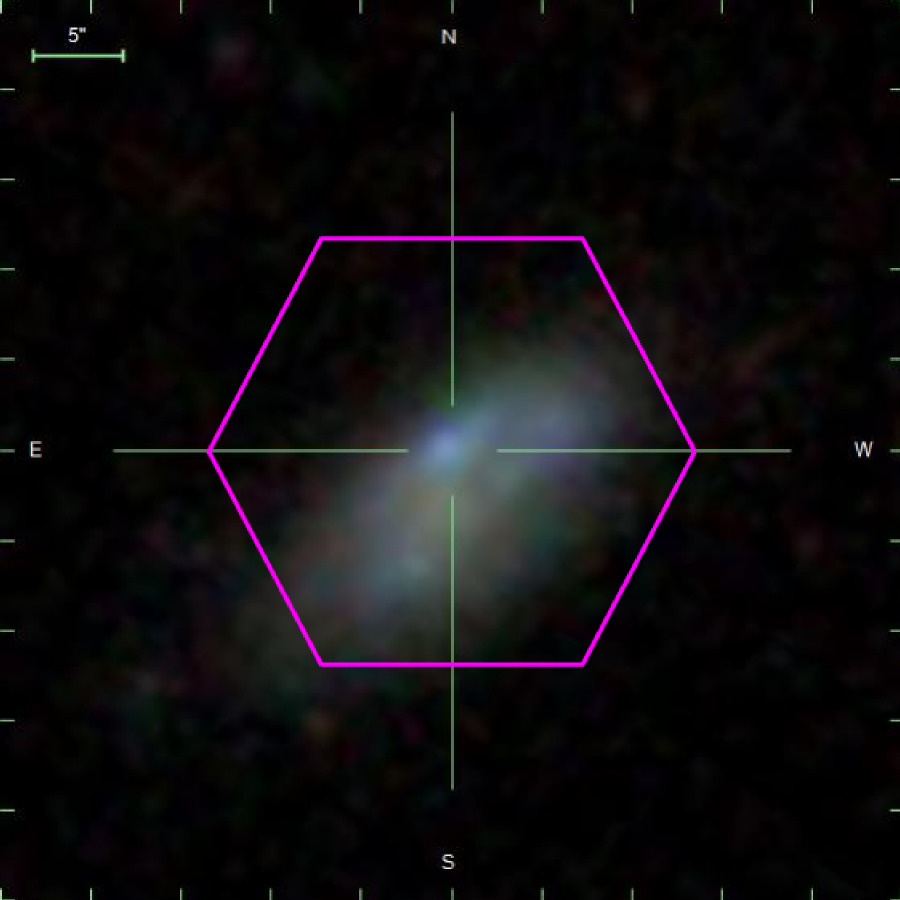}
\includegraphics[bb=50 40 390 400, width=3.25cm, clip]{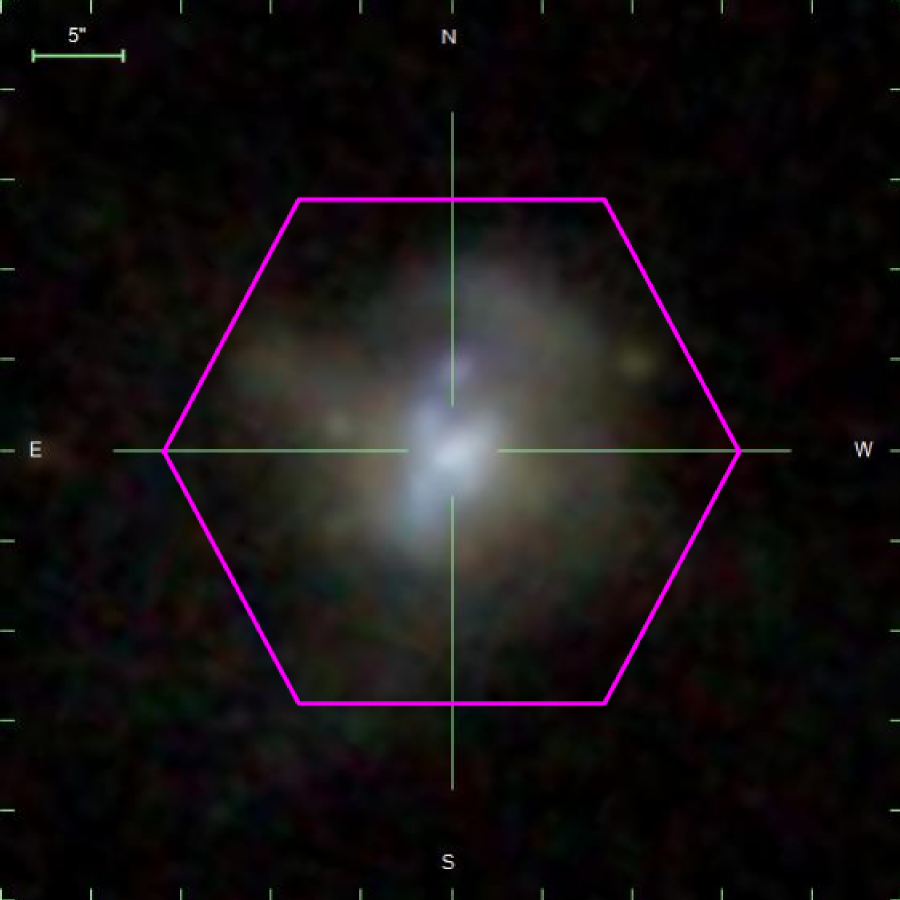}\\
\includegraphics[bb=125 230 420 530,width=3.65cm, clip]{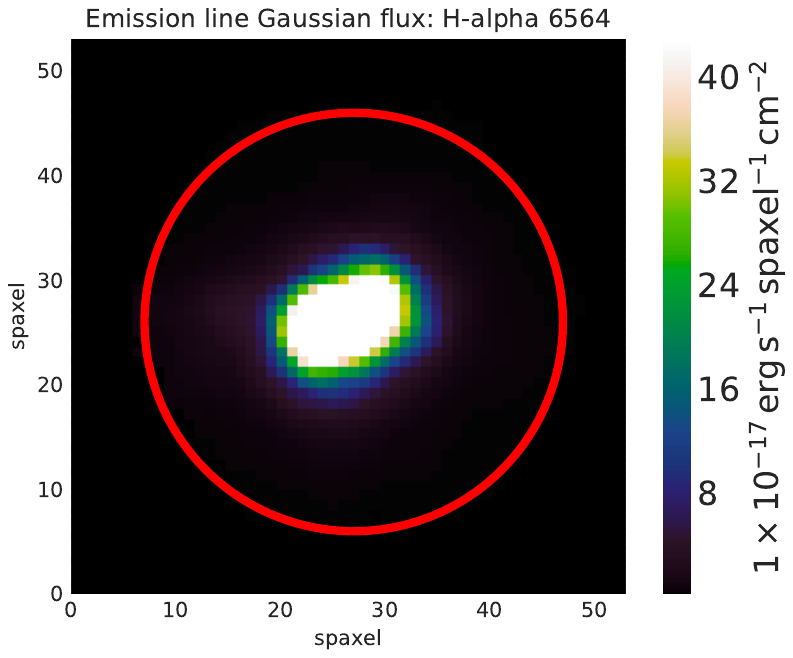}
\includegraphics[bb=133 230 420 530,width=3.55cm, clip]{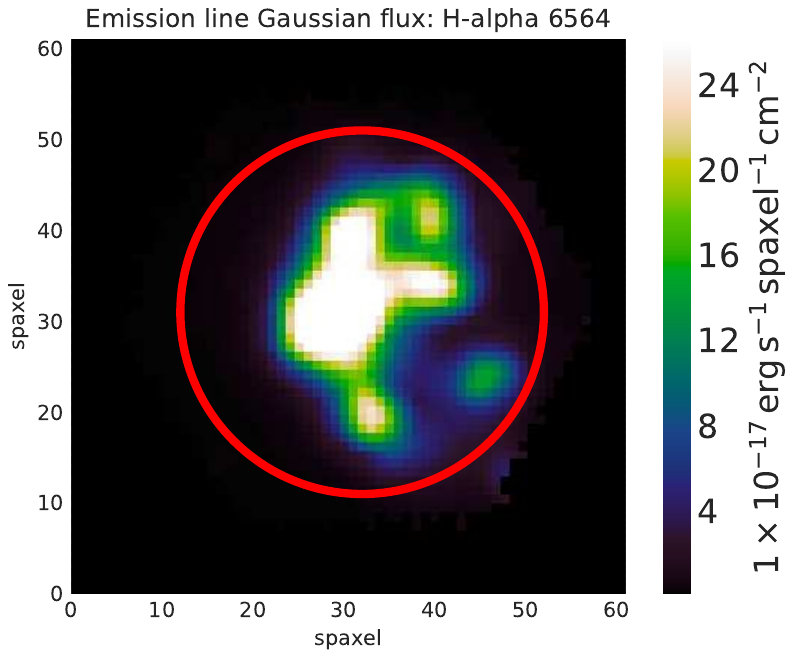}
\includegraphics[bb=10 0 360 375,width=3.45cm, clip]{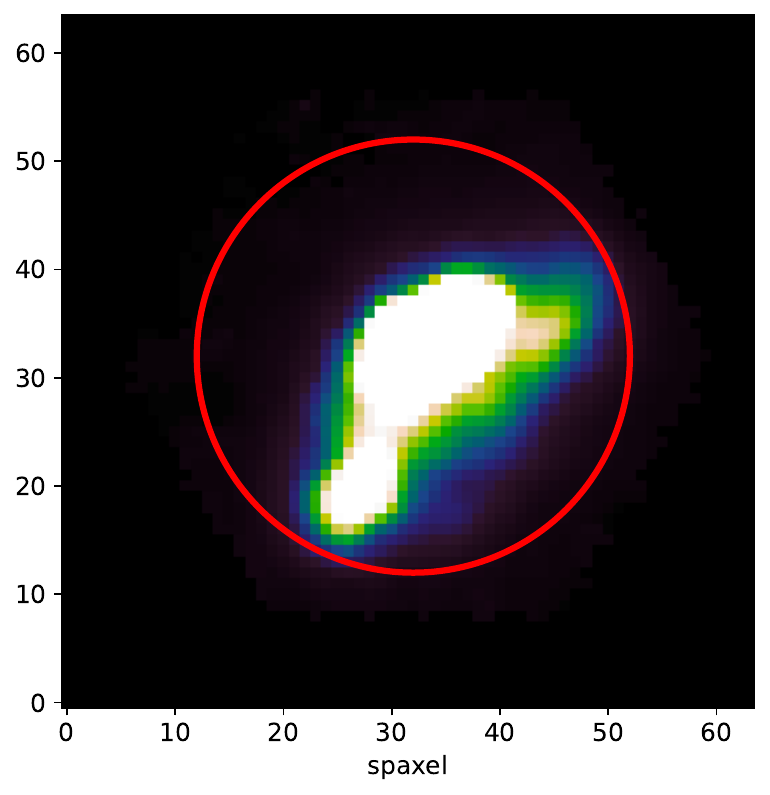}
\includegraphics[bb=10 0 360 375,width=3.45cm, clip]{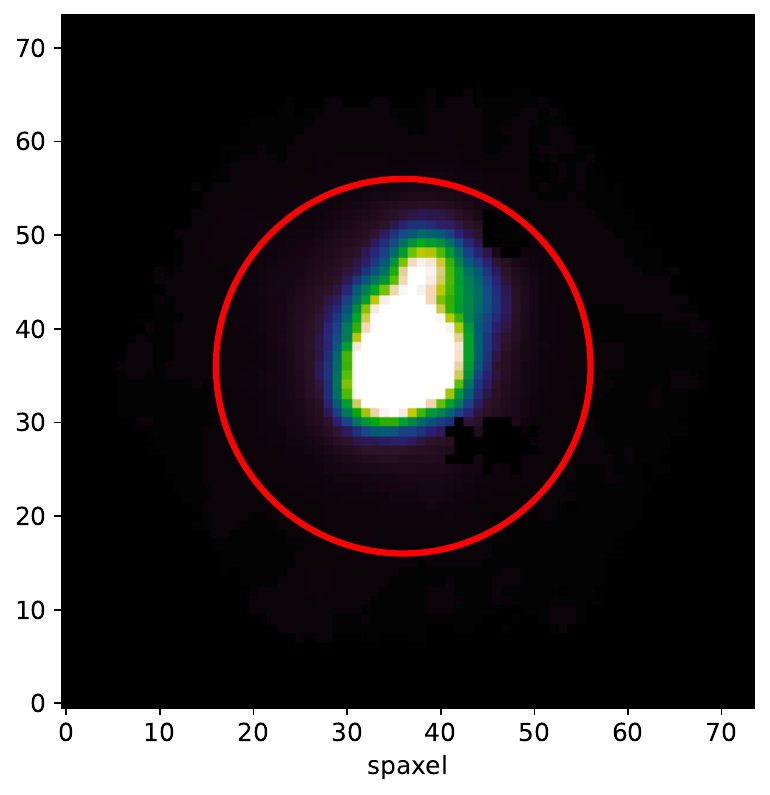}
\caption{Examples of MaLHUCO galaxies. Combined $gri$ SDSS images with the MaNGA hexagonal field of view overlaid (32$^{\prime\prime}$ diameter, top row), and the corresponding H$\upalpha$ 
emission with the JCMT beam at 230 GHz overlaid (20$^{\prime\prime}$ diameter, bottom row).
}
\label{fig:Ha}
\end{center}
\end{figure*}

\section{Sample selection and ancillary data}

Galaxies in our sample have been selected from the MANGA survey.
MaNGA is an integral field unit (IFU) spectroscopic survey of $\sim$ 10000 galaxies at $z < 0.15$ that uses the Baryon Oscillation Spectroscopic Survey spectrograph \citep{2006AJ....131.2332G,2013AJ....146...32S} on the Sloan Foundation 2.5 m telescope at Apache Point Observatory \citep{2015ApJ...798....7B}. The spectrograph has a resolution R $\sim$ 2000 and covers the wavelength range between
3600 and 10000 \AA. The fibers are arranged into hexagonal bundles with a field of view between 12$^{\prime\prime}$ and 32$^{\prime\prime}$.  The effective point-spread function of the survey is 2\farsec5, corresponding to $\sim$2 kpc at $z \sim$ 0.04, the highest redshift in our sample. 
We use the final
data release of the survey  \citep[DR17;][]{2022ApJS..259...35A}.

We first  considered all MaNGA galaxies of morphological type equal or later than Scd (6 $\leq$ T-type $\leq$ 11) using the MaNGA Visual  Morphology Catalog \citep[][]{2022MNRAS.512.2222V}. We shall refer to this sample as {\em MaNGA-late}. Then, we selected galaxies with stellar masses lower than the  Local Group galaxy M33, adopted as a reference. M33 is a Scd-type galaxy with $M_* = 5.5 \times 10^9$ M$_{\odot}$, $M_{\rm H_2} = 3 \times 10^8$~M$_{\odot}$, SFR $\sim$ 0.45~M$_\odot$~yr$^{-1}$, half-solar metallicity, and a well established CO-to-H$_2$ conversion factor \citep{2009A&A...493..453V,2010A&A...512A..63M,2014A&A...567A.118D, 2014A&A...572A..23C}. 
Stellar masses were taken from \citet{2013MNRAS.435.2764M} for SDSS DR8\footnote{\url{https://www.sdss.org/dr19/data\_access/value-added-catalogs/?vac\_id=-10}} 
($M_*^{\rm SED}$). 

We considered only galaxies with SFR as traced by H$\upalpha$ line, SFR$_{\rm H\upalpha}$, around or larger than  0.1~M$_{\odot}$ yr$^{-1}$ and atomic gas mass $M_{\rm HI} > 3\times 10^8$~M$_\odot$, as provided by the \hi-MaNGA VAC\footnote{\url{https://www.sdss4.org/dr17/data\_access/value-added-catalogs/?vac\_id=hi-manga-dr3}} \citep{2019MNRAS.488.3396M,2020RNAAS...4....3G,2021MNRAS.503.1345S}. 
Then we visually inspected the morphology of the H$\upalpha$ emission
and included only galaxies whose emission is less extended than the size of the JCMT telescope beam at 230~GHz ($\sim$ 20$^{\prime\prime}$), as shown in Fig.~\ref{fig:Ha}. This led to a final sample of 42 objects. 
Half of them are of morphological type Scd or Sd ($T$ = 6,7), which are the most abundant morphologies in {\em MaNGA-late}, while 17 additional galaxies are classified as Sdm or Sm ($T$ = 8,9). MaLHUCO includes only 4 systems that are of irregular or peculiar morphology, as the MANGA survey lacks a wide selection of this type of galaxies. 

\begin{table*}[ht]
\begin{center}
\caption{The MaLHUCO sample}
\begin{tabular}{rllllrrlcrrr}
\hline\hline
  \multicolumn{1}{c}{ID} &
  \multicolumn{1}{c}{R.A.} &
  \multicolumn{1}{c}{Decl.} &
  \multicolumn{1}{c}{$z$} &
  \multicolumn{1}{c}{$V_{\rm opt}$} &
  \multicolumn{1}{c}{$V_{\rm HI}$} &
  \multicolumn{1}{c}{Type} &
  \multicolumn{1}{c}{T} &
  \multicolumn{1}{c}{Bar$^a$} &
  \multicolumn{1}{c}{$D_{\rm L}$} &
  \multicolumn{1}{c}{t} \\
& $\:\;\,$[J2000] & $\:\;\,$[J2000]  & & [km s$^{-1}$] & [km s$^{-1}$] & & & &[Mpc] & [h]\\
\hline \hline
  1-119871 & 00:00:02.1 & +15:52:54 & 0.0200 & 6001   & 6007    & SABdm    & 8  & 0.5  &  86.0 & 0.23\\
  1-120743 & 00:23:14.9 & +14:48:34 & 0.0177 & 5301   & 5308    & SBcd     & 6  & 1.0  &  75.8 & 0.60\\
  1-39107  & 00:32:34.9 & +15:02:09 & 0.0179 & 5380   & 5401    & Sdm      & 8  & 0.0  &  76.8 & 1.92\\
  1-37167  & 02:56:40.7 & -00:14:44 & 0.0290 & 8706   & 8706    & SBdm     & 8  & 1.0  & 125.6 & 0.68\\
  1-51008  & 03:12:06.7 & -08:13:57 & 0.0136 & 4091   & 4107    & Sm       & 9  & 0.0  &  58.3 & 0.62\\
  1-50963  & 03:15:10.8 & -07:29:55 & 0.0317 & 9515   & 9486    & SBd      & 7  & 1.0  & 137.8 & 0.65\\
  1-71167  & 07:36:45.6 & +33:07:21 & 0.0162 & 4855   & 4859    & SBd      & 7  & 1.0  &  69.3 &0.80\\
  1-152563 & 07:42:36.1 & +28:15:41 & 0.0157 & 4711   & 4743    & Sm       & 9  & 0.0  &  67.7 &0.58\\
  1-201183 & 07:46:23.9 & +27:20:18 & 0.0160 & 4799   & 4800    & Scd      & 6  & 0.0  &  68.2 &0.60\\
  1-152828 & 07:51:43.0 & +30:26:55 & 0.0142 & 4269   & 4269    & SBm      & 9  & 1.0  &  60.9 &0.85\\
  1-382306 & 07:52:51.3 & +17:33:36 & 0.0159 & 4759   & 4745    & Sbcd     & 6  & 1.0  &  68.0 &0.83\\
  1-604885 & 07:55:22.5 & +15:05:38 & 0.0156 & 4670   & 4652    & SBdm     & 8  & 1.0  &  66.7 &0.24\\
  1-121773 & 07:55:51.7 & +36:18:55 & 0.0199 & 5991   & 6021    & Sd       & 7  & 0.0  &  86.4 &0.54\\
  1-584678 & 07:56:14.6 & +17:01:13 & 0.0169 & 5056   & 5084    & SABdm    & 8  & 0.5  &  72.4 &2.00\\
  1-44745  & 08:00:47.8 & +46:41:25 & 0.0194 & 5810   & 5789    & SBm      & 9  & 1.0  &  83.2 &0.14\\
  1-45112  & 08:08:55.2 & +45:41:15 & 0.0404 & 12115  & 12090   & SBdm     & 8  & 1.0  & 176.2 &0.03\\
  1-72402  & 08:12:24.4 & +40:44:35 & 0.0235 & 7037   & 7089    & SABd     & 7  & 0.5  & 101.7 &0.35\\
  1-218435 & 08:17:32.4 & +28:21:28 & 0.0199 & 5954   & 5934    & SBdm     & 8  & 1.0  &  85.1 &1.21\\
  43-47    & 08:27:18.0 & +46:01:57 & 0.0073 & 2206   & 2181    & dSph     & 10 & 0.0  &  31.5 &0.69\\
  1-216951 & 09:05:06.7 & +41:13:51 & 0.0252 & 7550   & 7578    & SBm      & 9  & 1.0  & 108.4 &0.84\\
  1-605833 & 09:23:48.0 & +02:06:45 & 0.0240 & 7207   & 7200    & SBd      & 7  & 0.75 & 103.6 &0.71\\
  1-156037 & 09:35:00.9 & +48:58:36 & 0.0250 & 7482   & 7503    & SBd      & 7  & 1.0  & 107.8 &1.04\\
  1-78143  & 09:43:08.6 & +03:56:25 & 0.0196 & 5879   & 5875    & SABcd    & 6  & 0.5  &  84.4 &1.50\\
  1-80572  & 10:50:17.2 & +03:30:45 & 0.0220 & 6602   & 6605    & Sdm      & 8  & 0.0  &  95.4 &0.40\\
  1-62169  & 11:05:14.9 & +00:59:21 & 0.0213 & 6394   & 6395    & SABdm   & 8  & 0.5   &  92.0 &1.40\\
  1-187394 & 11:32:00.2& +53:42:50 & 0.0268 & 8037   & 8033    & SBd      & 7  & 1.0  & 115.7 &0.82\\
  1-401878 & 12:21:16.5 & +36:20:03 & 0.0315 & 9450   & 9466    & Sdm      & 8  & 0.0  & 136.9 &0.65\\
  1-194947 & 12:48:25.4 & +54:12:05 & 0.0164 & 4927   & 4930    & SBcd     & 6  & 1.0  &  70.5 &0.54\\
  1-624292 & 13:06:19.8 & +32:58:25 & 0.0261 & 7813   & 7867    & Sd       & 7  & 0.0  & 113.2 &0.22\\
  1-625180 & 13:16:41.9 & +31:22:03 & 0.0193 & 5771 
  & 5760 
  & SABcd    & 6  & 0.5  & 82.3&1.20\\
  1-418224 & 13:41:10.4 & +37:01:07 & 0.0190 & 5701   &  5621   & SBcd     & 6  & 0.75 &  81.5 &0.80\\
  1-629221 & 14:10:59.5 & +38:45:55 & 0.0187 & 5604   &  5596   & SABcd    & 6  & 0.5  &  80.2&0.31\\
  1-631727 & 15:00:17.9 & +11:20:36 & 0.0201 & 6011   & 6001    & SABm     & 9  & 0.5  &  86.0 &0.34\\
  1-199432 & 15:38:28.9 & +43:44:00 & 0.0184 & 5509   & 5519    & SBcd     & 6  & 1.0  &  79.0 &0.34\\
  1-440301 & 15:39:09.6 & +24:49:51 & 0.0162 & 4858   & 4866    & Sm       & 9  & 0.0  &  69.5 &1.44\\
  1-265069 & 15:59:21.2 & +28:23:29 & 0.0215 & 6433   & 6421    & SABcd    & 6  & 0.5  &  92.1 &0.72\\
  1-210358 & 16:24:39.5 & +41:01:11 & 0.0271 & 8130   & 8174    & Irr      & 10 & 0.0  & 117.9 &1.05\\
  1-295155 & 16:30:07.4 & +23:55:33 & 0.0149 & 4454   &  4457   & SBd      & 7  & 1.0  &  63.5 &0.74\\
  1-325250 & 16:50:47.8 & +28:50:44 & 0.0326 & 9794   & 9821    & S        & 11 & 0.0  & 142.5 &0.07\\
  1-294140 & 16:51:42.7 & +19:25:43 & 0.0220 & 6603   & 6641    & SABd     & 7  & 0.5  &  95.6 &0.56\\
  1-177270 & 17:07:16.2 & +34:49:21 & 0.0367 & 10998  & 11010   & S-merger & 11 & 0.0  & 159.6 &0.44\\
  1-635629 & 21:30:25.8 & -00:28:28 & 0.0199 & 5964   & 6015    & Sd       & 7  & -0.5 &  85.5 &0.98\\\hline\hline
\end{tabular}
\label{table:sample}
\end{center}
\footnotesize{$^a$  1=conspicuous straight bar, 0.75=clear conspicuous bar, 0.5=clear bar in the inner regions of the galaxy, 0.25=typically a roundish structure, 0=no trace of bar, -0.5=difficult to distinguish.}
\end{table*}

Coordinates 
and redshifts of the MaLHUCO sample are listed in Table~\ref{table:sample}. 
The median galaxy redshift is $z = 0.024$.
We also display the optical and \hi\ recession velocities, $V_{\rm opt}$ and $V_{\rm HI}$ (both defined following the optical convention), the morphological type ($T$), the bar parameter (Bar), the luminosity distance ($D_{\rm L}$), and the total (ON+OFF) JCMT integration time ($t$). The bar parameter is taken from the MaNGA Visual Morphologies 
Catalog \citep{2022MNRAS.512.2222V} and it represents the probability that a galaxy hosts a bar:  0 indicates no bar, 1 a very likely barred galaxy, and negative values are assigned when the presence of a bar is difficult to determine.  The luminosity distance $D_{\rm L}$ has been computed for $H_0$ = 73~km~s$^{-1}$~Mpc$^{-1}$ according to a Flat $\Lambda$CDM with $\Omega_{\rm M}=0.3$. 

\subsection{The Pipe3D Value Added Catalog}

\begin{figure}
 \begin{center}
\includegraphics[trim=0 20 30 85, clip, scale=0.64]{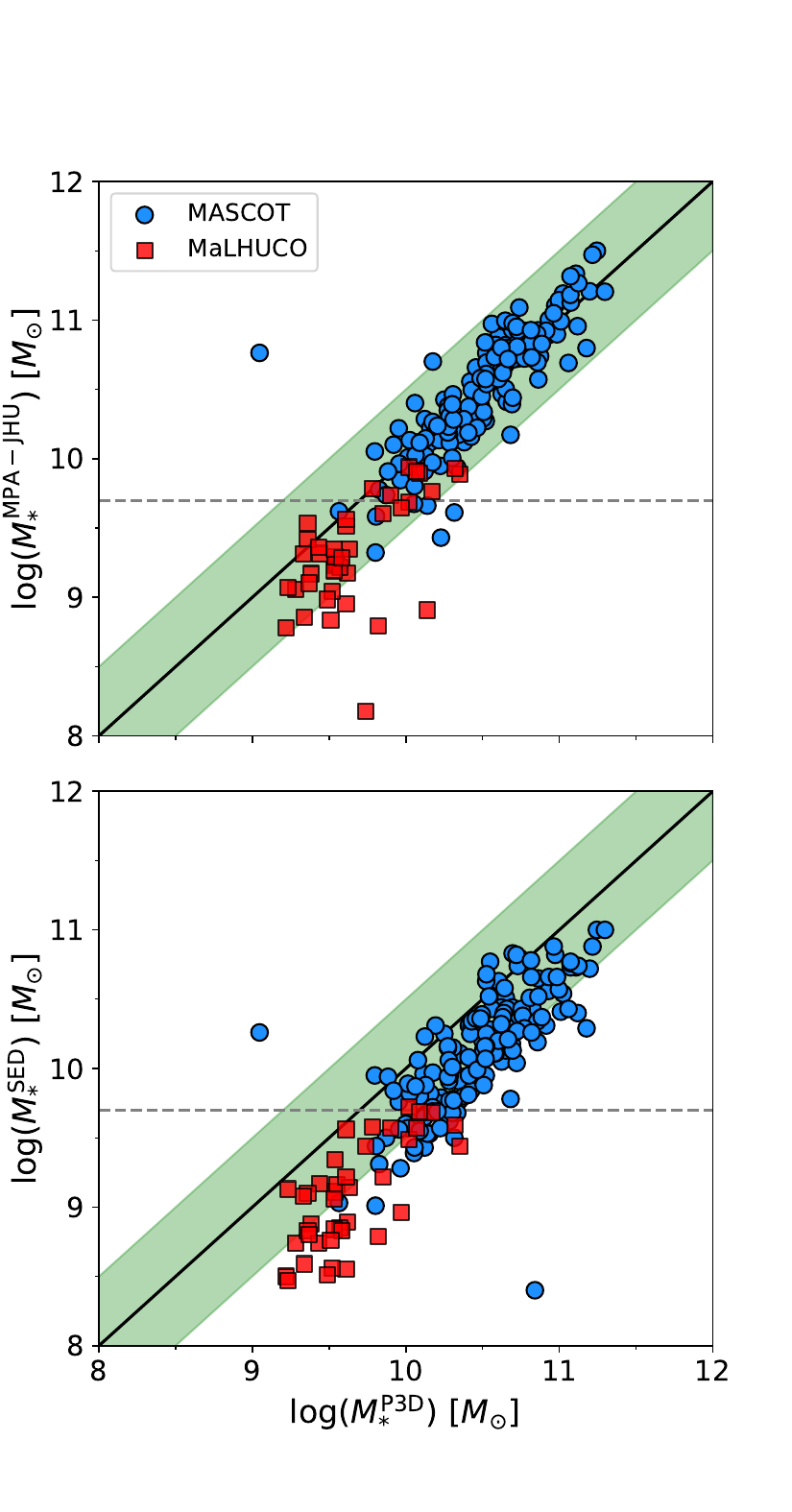}
\caption{Comparison between different stellar mass estimates for the MaLHUCO (red squares) and MASCOT (blue circles) samples. $M^{\rm P3D}_*$ are compared with stellar masses from the MPA-JHU catalog (top) and from \citet[][bottom]{2013MNRAS.435.2764M}. The solid line indicates to the one-to-one relation, while the green shaded region marks a $\pm$ 0.5 dex dispersion. The dashed horizontal line at log($M_*$/M$_{\odot}$) = 9.7 highlights the threshold mass adopted to define the MaLHUCO galaxies.}
\label{fig:compare_mstar}
\end{center}
\end{figure}

\begin{figure}
 \begin{center}
\includegraphics[scale=0.6]{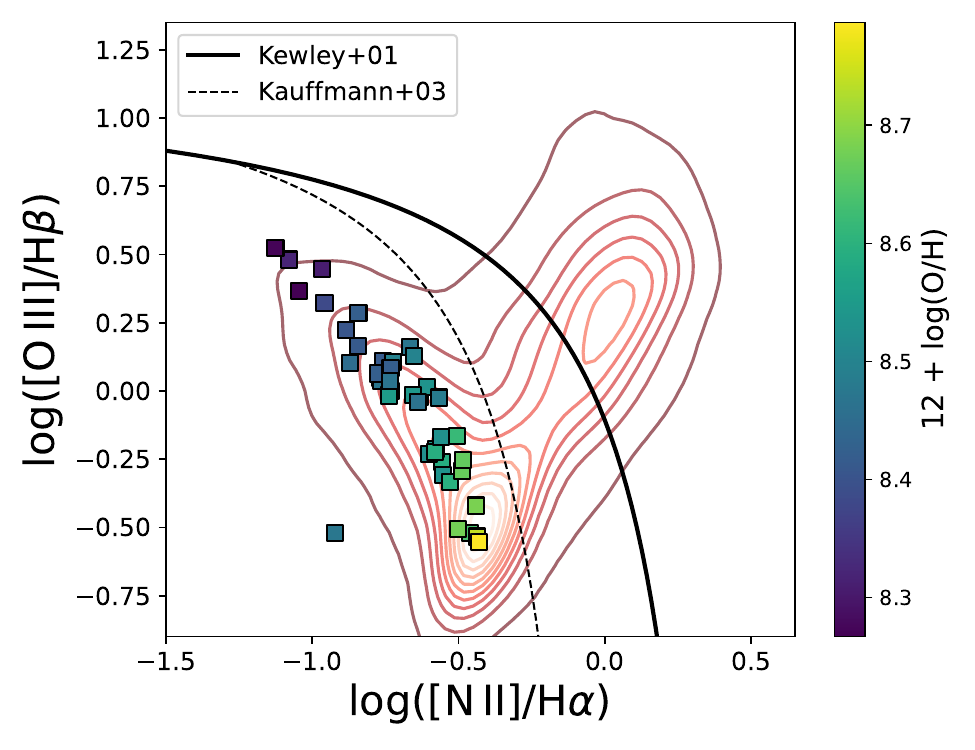}
\caption{Central emission line ratios [O~{\sc iii}]/H$\upbeta$ versus [N~{\sc ii}]/H$\upalpha$ of the MaLHUCO sample. Galaxies below the dashed line from \citet{2003MNRAS.346.1055K} are unlikely to host an AGN. Targets are color-coded by their oxygen abundance. 
Galaxy 43-47 is not included in this plot because its properties are not available in the MaNGA database.}
\label{fig:bpt}
\end{center}
\end{figure}

The main properties of the galaxies are obtained from the MaNGA Pipe3D VAC\footnote{\url{https://www.sdss4.org/dr17/data\_access/value-added-catalogs/?vac\_id=manga-pipe3d-value-added-catalog:-spatially-resolved-and-integrated-properties-of-galaxies-for-dr17}} \citep{2022ApJS..262...36S}, which is based on the Pipe3D pipeline \citep{2016RMxAA..52..171S,2018RMxAA..54..217S}. We use this catalog because it provides a uniform database of the main properties of MaNGA galaxies, enabling a comparison between our sample and other CO surveys based on MaNGA targets (see Sect. \ref{sec:MASCOT}). As the Pipe3D VAC was not available at the time of our first observing run, we instead adopted the \citet{2013MNRAS.435.2764M} catalog for the initial sample definition ($M_*^{\rm SED}$, see column 2 of Table \ref{table:prop}).
The pipeline determines stellar masses by fitting the continuum of the IFU data with stellar population models adopting the \citet{1955ApJ...121..161S}
initial mass function (IMF). Uncertainties in log$M_*^{\rm P3D}$ are in the range 0.06 and 0.1 dex.  Subtracting  the best-fitting stellar population model from the original cube allows the analysis of the ionized gas emission lines \citep{2016RMxAA..52..171S,2018RMxAA..54..217S}. The SFRs are 
estimated in two ways: a) by integrating the  total H$\upalpha$ flux within the MaNGA field of view and applying the conversion given by \citet{1998ApJ...498..541K} (SFR$_{\rm H\upalpha}^{\rm P3D}$); b) by integrating the SFR derived from simple stellar population (SSP) modeling over a time scale $t <$ 32 Myr (SFR$_{\rm SSP}^{\rm P3D}$). 
Stellar masses and SFRs from the Pipe3D catalog are used in this work after correcting the original values by $-$0.215 dex and -0.20 respectively to recover the Chabrier IMF \citep{2014ARA&A..52..415M}. 
The values of $M_*^{\rm P3D}$ are consistent with those used for the galaxy selection, $M_*^{\rm SED}$, derived by \citet{2013MNRAS.435.2764M}, in terms of the adopted IMF. 
However, $M_*^{\rm P3D}$ are on average 0.6 dex higher than $M_*^{\rm SED}$, as we show in the bottom panel of Fig. \ref{fig:compare_mstar}. We also display in the top panel the comparison between $M_*^{\rm P3D}$ and stellar masses from the MPA-JHU catalog \citep{2004MNRAS.351.1151B}, adopted to select galaxies in the xCOLDGASS and ALLSMOG surveys. In this case, the difference between the two estimates is more significant at low stellar masses. This comparison indicates that Pipe3D systematically overestimate the stellar masses of our galaxies. A similar positive difference between $M_*^{\rm P3D}$ and stellar masses derived with different methods has been reported in previous works \citep[e.g.][]{2021MNRAS.505.3135A} for late-type, low-mass star forming galaxies.  
Nevertheless, we adopt the Pipe3D mass estimates throughout this work for convenience, and our sample indeed traces the overall low-mass distribution of $M_*^{\rm P3D}$ values, as shown in Section~5. 

The catalog also provides 
the oxygen abundance at one effective radius ($R_e$) 
using the O3N2 = $\log$[([O$\,${\sc iii}]/H$\upbeta$)/([N$\,${\sc ii}]/H$\upalpha$)] indicator according to the calibration of  \citet{2004MNRAS.348L..59P}.
This indicator is used to estimate the CO-to-H$_2$ conversion, following the prescription of \citet{2017MNRAS.470.4750A}, that will be adopted in this work. 
The metallicities of our sample are in the range 8.1 $<$ 12 + log(O/H) $<$ 8.8 dex with a peak of the distribution around 8.4 dex.
The availability of metallicity measurements allowed us to select a sample that is not extremely metal poor, which would otherwise require very high CO-to-H$_2$ conversion factors and long integration times to secure detection. The log values of $M_*^{\rm P3D}$, SFR$_{\rm H\upalpha}^{\rm P3D}$, SFR$_{\rm SSP}^{\rm P3D}$, and oxygen abundances from the Pipe3D catalog are listed in Table~\ref{table:prop}.

Analysis of the Baldwin, Phillips, Terlevich (BPT) diagram, \citet[][see Fig.~{\ref{fig:bpt}}]{1981PASP...93....5B} indicates that the emission-line ratios in the central region of our galaxies are characteristic of star forming galaxies, with no evidence of AGN activity \citep{2003MNRAS.346.1055K}, 
and that they form a continuous sequence ranging from lower-metallicity objects (high [O {\sc iii}]/H$\upbeta$ and low [N {\sc ii}]/H$\upalpha$ flux ratios) to higher-metallicity objects (low [O {\sc iii}]/H$\upbeta$ and high [N {\sc ii}]/H$\upalpha$ ratios).

\begin{table*}
\begin{center}
\caption{Properties of the MaLHuCO sample}
\begin{tabular}{lcrcccccc}
\hline\hline
  \multicolumn{1}{c}{MaNGA} &
  \multicolumn{1}{c}{log $M_*^{SED}$} &
  \multicolumn{1}{c}{log $M_*^{{\rm P3D},a}$} &
  \multicolumn{1}{c}{log SFR$_{\rm H\upalpha}^{{\rm P3D},a}$} &
  \multicolumn{1}{c}{log SFR$_{\rm SSP}^{{\rm P3D},a}$} &
  \multicolumn{1}{c}{log $M_{\rm HI}$} &
    \multicolumn{1}{c}{log $L_{12}$} &
    \multicolumn{1}{c}{log $L_{22}$} &
  \multicolumn{1}{c}{12 + log O/H$^{b}$} \\
      ID     &[M$_{\odot}$] &  [M$_{\odot}$] $\:\:\:\:\:$ &  [M$_{\odot}$ yr$^{-1}$] & [M$_{\odot}$ yr$^{-1}$]  & [M$_{\odot}$] &   erg~s$^{-1}$ & erg~s$^{-1}$ & dex \\
  \hline\hline
1-119871 & 9.57 &  9.90     $\pm$ 0.07 &  0.01 &  0.01 & 9.29     &  42.47 &  42.19 &  8.68 $\pm$  0.01   \\
1-120743 & 9.17 &  9.44     $\pm$ 0.08 & -0.70 & -0.73 & 9.47     &  41.08 &  41.35 &  8.44 $\pm$  0.03   \\
1-39107  & 9.10 &  9.36     $\pm$ 0.08 & -0.58 & -0.81 & 8.92     &  41.72 &  41.46 &  8.55 $\pm$  0.01   \\
1-37167  & 9.49 & 10.02     $\pm$ 0.08 &  0.02 &  0.19 & 9.71     &  42.56 &  42.24 &  8.67 $\pm$  0.01   \\
1-51008  & 8.74 &  9.28     $\pm$ 0.08 & -0.67 & -0.72 & 8.84     &  41.46 &  41.28 &  8.41 $\pm$  0.01   \\
1-50963  & 9.68 & 10.17     $\pm$ 0.08 &  0.00 &  0.42 & 9.88     &  42.35 &  41.97 &  8.47 $\pm$  0.01   \\
1-71167  & 9.11 &  9.54     $\pm$ 0.07 & -0.51 & -0.27 & 9.44     &  41.51 &  41.38 &  8.40 $\pm$  0.03   \\
1-152563 & 8.50 &  9.22     $\pm$ 0.07 & -0.92 & -0.97 & 8.50     &  41.09 &  41.30 &  8.40 $\pm$  0.01   \\
1-201183 & 8.88 &  9.38     $\pm$ 0.07 & -0.74 & -0.68 & 9.01     &  41.68 &  41.33 &  8.59 $\pm$  0.01   \\
1-152828 & 8.47 &  9.23     $\pm$ 0.08 & -0.60 & -0.48 & 9.34     &  41.17 &  41.39 &  8.32 $\pm$  0.01   \\
1-382306 & 9.08 &  9.33     $\pm$ 0.09 & -1.01 & -0.98 & 8.87     &  41.39 &  41.29 &  8.66 $\pm$  0.02   \\
1-604885 & 8.55 &  9.61     $\pm$ 0.07 & -0.16 & -0.29 & 9.48     &  42.01 &  42.04 &  8.42 $\pm$  0.01   \\
1-121773 & 9.16 &  9.55     $\pm$ 0.08 & -0.90 & -0.23 & 9.57     &  41.09 &  41.52 &  8.45 $\pm$  0.05   \\
1-584678 & 8.83 &  9.36     $\pm$ 0.08 & -0.49 & -0.37 & 9.19     &  41.53 &  41.42 &  8.41 $\pm$  0.02   \\
1-44745  & 8.96 &  9.97     $\pm$ 0.07 &  0.03 & -0.12 & 9.52     &  42.23 &  42.13 &  8.48 $\pm$  0.01   \\
1-45112  & 9.22 &  9.85     $\pm$ 0.08 & -0.18 & -0.16 & 9.39     &  42.29 &  42.22 &  8.54 $\pm$  0.01   \\
1-72402  & 9.44 &  9.74     $\pm$ 0.08 &  0.07 & -0.28 & 9.44     &  42.42 &  42.17 &  8.59 $\pm$  0.01   \\
1-218435 & 8.59 &  9.34     $\pm$ 0.07 & -0.38 & -0.51 & 9.46     &  41.51 &  41.63 &  8.27 $\pm$  0.01   \\
  43-47  & 7.44 &8.60$^{c}$ $\pm$ 0.15 & -1.05 & -1.05 & 8.69     &  40.80 &  40.91 &  $\:\,$8.14 $\pm$ 0.01$^d$ \\
1-216951 & 9.14 &  9.63     $\pm$ 0.07 & -0.15 & -0.36 & 9.25     &  42.19 &  42.05 &  8.53 $\pm$  0.01   \\
1-605833 & 9.72 & 10.02     $\pm$ 0.08 & -0.15 & -0.27 & 9.96     &  42.34 &  41.95 &  8.64 $\pm$  0.01   \\
1-156037 & 8.89 &  9.62     $\pm$ 0.08 & -0.24 & -0.34 & 9.44     &  41.74 &  41.64 &  8.38 $\pm$  0.01   \\
1-78143  & 8.56 &  9.52     $\pm$ 0.08 & -0.85 & -0.57 & 9.36     &  41.23 &  41.20 &  8.46 $\pm$  0.03   \\
1-80572  & 9.58 &  9.78     $\pm$ 0.09 &  0.10 & -0.01 & 9.57     &  42.48 &  42.23 &  8.47 $\pm$  0.01   \\
1-62169  & 8.74 &  9.43     $\pm$ 0.08 & -0.69 & -0.59 & 9.25     &  41.67 &  41.29 &  8.52 $\pm$  0.02   \\
1-187394 & 9.22 &  9.61     $\pm$ 0.08 & -0.07 & -0.28 & 9.59     &  42.09 &  41.96 &  8.46 $\pm$  0.02   \\
1-401878 & 7.67 & 10.14     $\pm$ 0.08 &  0.28 &  0.39 & 9.53     &  42.76 &  42.40 &  8.68 $\pm$  0.01   \\
1-194947 & 8.85 &  9.57     $\pm$ 0.07 & -1.01 & -0.48 & 9.22     &  41.30 &  41.21 &  8.53 $\pm$  0.01   \\
1-624292 & 9.69 & 10.09     $\pm$ 0.07 &  0.37 &  0.25 & 9.46     &  42.84 &  42.60 &  8.74 $\pm$  0.01   \\
1-625180 & 9.06$^e$&9.53    $\pm$ 0.07 & -0.59 & -0.65 & 9.47     &  41.56 &  41.25 &  8.47 $\pm$  0.01   \\
1-418224 & 9.34 &  9.54     $\pm$ 0.07 & -0.70 & -0.71 & 9.69     &  41.40 &  41.39 &  8.42 $\pm$  0.02   \\
1-629221 & 8.84 &  9.53     $\pm$ 0.08 & -0.91 & -0.62 & 9.62     &  41.44 &  41.36 &  8.53 $\pm$  0.01   \\
1-631727 & 9.56 &  9.61     $\pm$ 0.08 & -0.39 & -0.51 & 9.13     &  41.86 &  41.42 &  8.60 $\pm$  0.01   \\
1-199432 & 8.51 &  9.49     $\pm$ 0.08 & -0.97 & -0.63 & 9.09     &  41.29 &  41.03 &  8.55 $\pm$  0.02   \\
1-440301 & 9.13 &  9.23     $\pm$ 0.08 & -0.70 & -0.78 & 9.16     &  41.35 &  41.16 &  8.47 $\pm$  0.01   \\
1-265069 & 8.83 &  9.58     $\pm$ 0.09 & -0.48 & -0.37 & 9.68     &  41.60 &  41.40 &  8.42 $\pm$  0.02   \\
1-210358 & 8.79 &  9.82     $\pm$ 0.07 &  0.06 & -0.05 & 9.61     &  41.78 &  41.87 &  8.27 $\pm$  0.01   \\
1-295155 & 8.80 &  9.37     $\pm$ 0.08 & -0.69 & -0.57 & 9.05     &  41.69 &  41.46 &  8.50 $\pm$  0.00   \\
1-325250 & 9.44 & 10.35     $\pm$ 0.07 &  0.82 &  0.27 & 9.81     &  43.05 &  43.05 &  8.50 $\pm$  0.00   \\
1-294140 & 8.76 &  9.51     $\pm$ 0.07 &  0.07 & -0.38 & 9.99$^f$ &  42.09 &  42.10 &  8.31 $\pm$  0.00   \\
1-177270 & 9.59 & 10.32     $\pm$ 0.08 &  0.43 &  0.52 & 9.43     &  43.32 &  43.21 &  8.62 $\pm$  0.00   \\
1-635629 & 9.57 & 10.07     $\pm$ 0.08 &  0.02 &  0.15 & 9.78$^f$ &  42.73 &  42.36 &  8.79 $\pm$  0.00   \\
\hline\hline
\end{tabular}
\label{table:prop}
\end{center}
\footnotesize{$^a$Corrected for a Chabrier IMF. }
\footnotesize{$^b$ Oxygen abundance at 1 $R_e$ based on the O3N2 indicator \citep{2004MNRAS.348L..59P}, used by Accurso et al. (2017) to calculate M$_{\rm H_2}$. $^c$ Stellar mass estimated from the WISE W1 magnitude \citep{2023ApJ...946...95J}. $^d$ Metallicity obtained from the SDSS spectra using the the O3N2 indicator \citep{2004MNRAS.348L..59P}. $^e$ This is the photometric stellar mass because the \citet{2013MNRAS.435.2764M} estimate is unavailable. $^f$\texttt{CONFPROB} $>$ 0.6.}
\end{table*}

\subsection{H{\sc i}-MaNGA}

The \hi-MaNGA survey provides a catalog of 21-cm detections (or upper limits) 
for MaNGA galaxies \citep{2019MNRAS.488.3396M}.
It uses Green Bank Telescope (GBT) observations to obtain global
\hi\ measurements 
at a depth comparable to the  Arecibo Legacy Fast Arecibo L-band Feed Array (ALFALFA) survey \citep{2018ApJ...861...49H}. ALFALFA data are used in the regions of the sky that overlap with the MaNGA survey.
The DR3 of \hi-MaNGA \citep{2021MNRAS.503.1345S}
contains \hi\ measurements of 6632 
galaxies. 

The \hi\ masses of the MaLHUCO sample are displayed in Table \ref{table:prop}.
However, as the beams of the radio telescopes
at 21 cm are significantly larger than the size of a typical MaNGA galaxy (3$^{\prime}$ for Arecibo, and 9$^{\prime}$ for GBT, compared to $< 1^{\prime}$ for most MaNGA galaxies), caution is needed about the possibility of the \hi\ observations being contaminated by the emission of
neighboring galaxies \citep{2022RNAAS...6....1S}.  
In the analysis of the \hi\ properties of our sample, 
we excluded \hi\ detections with a probability larger than 0.6 that more than 20$\%$  of the \hi\ comes from galaxies other than primary target (i.e. \texttt{CONFPROB} $>$ 0.6 in \hi-MaNGA VAC). Only two galaxies do not meet this criterion, and they are highlighted in Table \ref{table:prop}.

\subsection{WISE data}

The Wide-field Infrared Survey Explorer \citep[WISE;][]{2010AJ....140.1868W} is a NASA space telescope that  scanned the whole sky in four mid-infrared (MIR) passbands: W1, W2, W3, and W4, centred at at 3.4, 4.6, 12, and 22 $\upmu$m, respectively. The angular resolution is 6\farcs1, 6\farcs4, 6\farcs5, and 12$^{\prime\prime}$, in bandpasses W1–W4, respectively. The W3 band is dominated by emission from the 11.3~$\upmu$m feature from polycyclic aromatic hydrocarbons (PAH) excited by ultraviolet (UV) radiation from young stars, while the W4 band traces warm dust continuum heated by the same star-forming population \citep{2014ApJ...782...90C}.
We cross-matched the AllWISE Source Catalog \citep{2013wise.rept....1C} with the MaLHUCO galaxies to obtain W3 and W4 photometry. The catalog provides multiple aperture photometry measurements for extended sources. We adopted the photometry obtained within an aperture radius of 11$^{\prime\prime}$ (w3mag\_3, w4mag\_1), comparable to the size of the JCMT beam at 230 GHz. 
In Table~\ref{table:prop} we report the luminosities at 12~$\upmu$m, $L_{12}$, and 22~$\upmu$m $L_{22}$. Uncertainties 
are obtained from the AllWISE catalog, with mean values of 0.02 and 0.06 dex for $L_{12}$ and $L_{22}$, respectively. 

\subsection{Comparison sample: the MASCOT survey}
\label{sec:MASCOT}

MASCOT is a survey for molecular gas in galaxies 
selected from MaNGA \citep{2022MNRAS.510.3119W}. The main goal of the  survey  is to probe the molecular gas content of star-forming galaxies with  M$_*^{P3D}> 10^{9.5}$ M$_\odot$. The targets span a range of 3 orders of magnitude in sSFR, including MS galaxies, as well as a number of green valley \citep{2004ApJ...600..681B,2007ApJS..173..342M} and starburst systems. MASCOT has only 13 galaxies which are in {\em MANGA-late}. 
Observations of the $^{12}$CO(J=1-0) line have been conducted at the Arizona Radio Observatory. The first data release includes CO observations of 187 galaxies.  
In this work, stellar masses and SFRs of the MASCOT sample are obtained from the SDSS DR17 Pipe3D (v3.1) catalog, whereas \citet{2022MNRAS.510.3119W} used an earlier release of the catalog. We exclude from the comparison sample 9 galaxies for which the metallicity at 1 $R_e$, derived from the O3N2 
indicator \citep{2004MNRAS.348L..59P}, was not available in the Pipe3D catalog.

\section{Observations and data reduction}

We used the \uu\ (230 GHz) receiver \citep{2020SPIE11453E..3TM} of the JCMT to observe the $^{12}$CO(J=2-1) emission line and trace the molecular gas content  of the selected sample of low-mass disk galaxies.

The $^{12}$CO(J=2-1) observations were carried out in raster map scanning mode in two semesters, 2023B  and 2024B (Program ID: M23BP020, E24XP014, M24BP043) with the auto-correlation spectrometer imaging system \citep[ACSIS,][]{2022SPIE12190E..31L} as backend. 
We used a 1 GHz bandwidth configuration with a resolution of
0.488 MHz -- i.e. 0.43 km s$^{-1}$ for the studied $^{12}$CO(J=2-1) transition.
The JCMT beam at this wavelength has a full width at half maximum (FWHM) of 20$^{\prime\prime}$. Given this spatial scale, we do not resolve the CO emission.
The observed area encompasses most of the MaNGA IFU field of view, containing the main H$\upalpha$-emitting regions (see Fig. \ref{fig:Ha}).
Each galaxy was sampled until the target noise (at a resolution of 10 km s$^{-1}$) was reached.
The total allocated time to observe our sample of 42 sources was 41.5 hours, obtained with an atmosphere opacity $0.12 < \tau_{225} < 0.15$ (weather Band 4). The total integration time for each target is shown in Table~\ref{table:sample}.

Observations were reduced using standard procedures of the Starlink/ORAC-DP pipeline \citep{2015A&C.....9...40J}.
The output spectra are in units of the observed source antenna
temperature. 
The main beam temperature $T_{\rm mb}$  was derived from the antenna temperature according to $T_{\rm mb} = T^*_A /\eta_{\rm mb}$, where $\eta_{\rm mb} = 0.66$\footnote{\url{https://www.eaobservatory.org/jcmt/instrumentation/heterodyne/calibration/}} is the main beam efficiency. 
A first- to third-order polynomial baseline was fitted and subtracted from the spectra. The root mean square  noise ($\sigma_{\rm rms}$) was then measured at a velocity resolution of 10 km s$^{-1}$. The results of the observations of the MaLHUCO sample including $\sigma_{\rm rms}$, line widths, and integrated intensities, are listed in Table \ref{table:tab1}.
$I_{\rm tot}$ (column~2) is the line brightness obtained by summing the flux density over all channels within the velocity range $\Delta V$ (column~4).  For undetected sources we computed the upper limits as
\begin{equation}
    I_{\rm tot} < {3 \times \, \sigma_{\rm rms}} \, \times  dv  \times \sqrt{\Delta V/dv}
\end{equation}

\noindent
using a nominal line width of $\Delta V=150$~km~s$^{-1}$, where $dv =$ 10 km~s$^{-1}$ is the spectral channel width and $N = \Delta V/dv$ is the number of channels corresponding to the adopted velocity width. The uncertainties in $I_{\rm tot}$ have been estimated as $\sigma_{\rm rms} \times  dv \times \sqrt{N}$. 
In column~5 we display the mean recession velocity of the detected signal ($V_{\rm mean}$). Columns 6 and 7 report the integrated line brightness ($I_{\rm g}$) and the full width at half maximum derived from a Gaussian fit to the line (FWHM$_{\rm g}$), using the Continuum and Line Analysis Single-dish Software (\textsc{CLASS}) of the Grenoble Image and Line Data Analysis Software (\textsc{GILDAS})\footnote{\url{http://iram.fr/IRAMFR/GILDAS/}} package.

A galaxy is considered detected if the following conditions are satisfied: i) $I_{\rm tot} > 4\times \sigma_{\rm rms}$, ii) $\bigr|V_{\rm mean} -V_{\rm opt}\bigl|<$ 100~km~s$^{-1}$, iii) $I_{\rm g}$ is larger than 3 times its quoted  uncertainty. In the last column of Table~\ref{table:tab1} we include a flag providing the quality of the detection: 1 for detected targets and 0 for non-detections. In two cases, 1-152828 and 1-187394, a flag value of 2 denotes  tentative detections. As these objects do not meet one of the previous criteria, they have not been included in the detected sample presented in the rest of this work. Therefore, out of the 42 observed galaxies, CO has been detected in 23. The $^{12}$CO(J = 2-1) spectra of the MaLHUCO sample are displayed in Fig. \ref{fig:CO_spectra} in Appendix~\ref{app:B}.

The CO luminosity in K km s$^{-1}$ pc$^2$ units has been computed using the beam diluted velocity integrated brightness temperature $I_{\rm tot}$ (in main beam temperature units) according to Eq. (3) of \citet{2005ARA&A..43..677S}:

\begin{equation}
L_{\rm CO}^{2-1}=3.25 \times 10^{7} \times {18.06} \, I_{\rm tot} \, \nu^{-2}_{\rm rest} (1+z)^{-1} D^2_{\rm L}
\end{equation}

\noindent which is valid for a source of any size, where $\nu_{\rm rest} = 230.538$ GHz is the $^{12}$CO(J = 2-1) rest frequency, $z$ is the galaxy redshift, $D_{\rm L}$ the luminosity distance in Mpc. We have used the relation between the flux density in Jy and the antenna temperature in K  $S_{\nu} = 15.6\ T^*_A/\eta_a  = 18.06\ T_{mb} \eta_{mb} /\eta_a$ where $\eta_a = 0.57$ is the aperture efficiency\footnote{\url{https://www.eaobservatory.org/jcmt/instrumentation/heterodyne/calibration/}}.

\begin{figure*}
 \begin{center}
\includegraphics[scale=0.60]{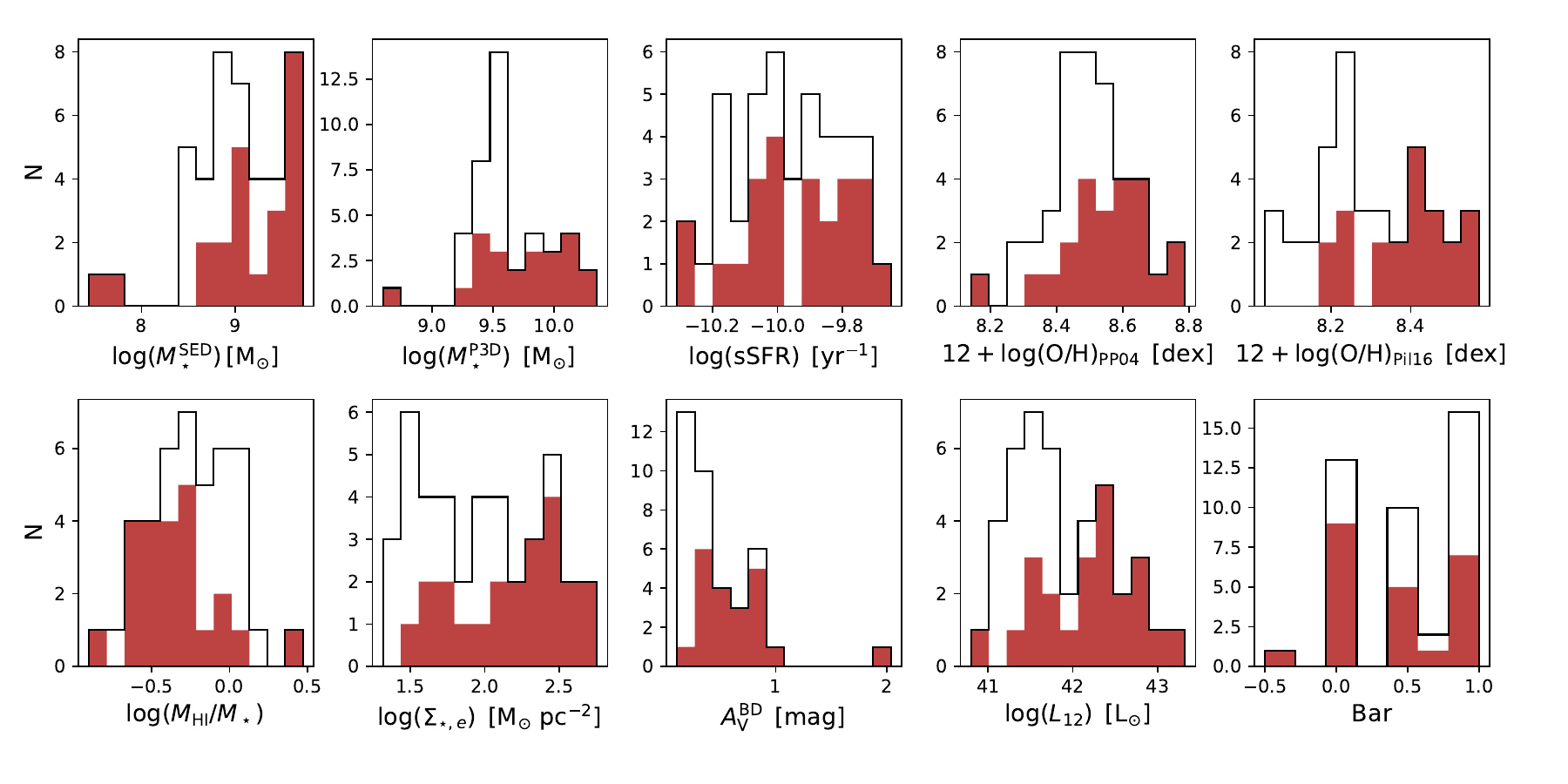}
\caption{Distribution of CO detections (filled histogram) and non-detections (open histrogram) as a function of $M_*$, sSFR, oxygen abundance at 1 $R_e$, $M_{\rm HI}/M_*$, stellar surface density at 1 $R_e$, $\Sigma_{*, e}$, visual attenuation, $A_{\rm V}^{\rm BD}$, 12 $\upmu$m luminosity, $L_{12}$, and bar parameter.
}
\label{fig:02a}
\end{center}
\end{figure*}

\section{Linking CO detections to galaxy properties} 

Figure \ref{fig:02a} displays detected  and non-detected galaxies in the $^{12}$CO(J=2-1) line as
a function of $M_*$, sSFR, oxygen abundance at $R_e$, atomic-to-stellar-mass ratio ($M_{\rm HI}/M_*$), stellar surface density at $R_e$ ($\Sigma_{*, e}$), dust attenuation in the $V$-band derived from the Balmer decrement, $A_{\rm V}^{\rm BD}$, 
$L_{12}$, and the bar parameter.
The histograms show that the majority of the non-detections, as expected, have, on average, lower stellar masses and metallicities. The threshold value below which the detection rate (i.e. the ratio of the number of detections to the number of sources per bin) drops below $\sim$100\% depends on the adopted stellar mass calibration: $\sim 10^{9.4}$ M$_{\odot}$ when using the \citet{2013MNRAS.435.2764M} stellar masses, and $\sim 10^{9.7}$ M$_{\odot}$ for the Pipe3D catalog.
Similarly, the oxygen abundance threshold 
depends on the adopted metallicity calibration.  In the figure we show two examples, one for the calibrations of \citet{2004MNRAS.348L..59P}, with a detection rate starting to decrease below 12 + log(O/H) $\sim$ 8.57, and another for the calibration proposed by \citet{2016MNRAS.457.3678P},  12 + log(O/H) $\sim$ 8.4.
The majority of the non-detected galaxies appear to have a high $M_{\rm HI}/M_*$, as 78\% 
of non-detections have log($M_{\rm HI}/M_*$) $> -0.2$. The dependence of  the detection rate on sSFR  
is less strong, as it appears approximately constant in most of the histogram bins.  

The detection of molecular gas in galaxies is strongly correlated with the local stellar surface density, $\Sigma_*$, reflecting the role of the stellar potential in regulating the physical conditions of the ISM. Regions with higher stellar surface density provide deeper gravitational wells and higher mid-plane pressures, which favor the formation and survival of molecular clouds by enhancing gas compression \citep{1993ApJ...411..170E,2002ApJ...569..157W,2006ApJ...650..933B,2025A&A...700A..57C,2025A&A...702A.264P}.  Almost all galaxies with 
$ \Sigma_* \gtrsim$ 100 M$_{\odot}$ pc$^{-2}$, measured at 1 $R_e$, exhibit a CO detection.

Dust grains efficiently shield \hdue\ and other molecules from UV radiation, and this is especially important for CO, which is 
more easily photo-dissociated by ambient UV fields \citep{2012A&A...542A..32C,2013ARA&A..51..207B,2018ARA&A..56..673G,2019A&A...626A..63D,2025A&A...699A.346S}.
The visual attenuation of a galaxy depends on the amount 
of dust along the line of sight and can be used as an indicator of its dust content \citep{2019MNRAS.486L..91C,2020MNRAS.492.2651B,2025A&A...702A.264P}. Figure~\ref{fig:02a} shows that 
nearly all galaxies with $A_{\rm V}^{\rm BD} > 0.4$ are detected at 230 GHz.

The luminosity at 12 $\upmu$m, $L_{12}$, traces the emission from PAH. Fig. \ref{fig:02a} highlights that the CO detection rate is high for $L_{12} > 10^{42}$~L$_{\odot}$ and decreases below this threshold. 
The relation between $L_{\rm CO}$ and $L_{12}$ will be discussed in more detail later. 

Bars drive gas inflows toward galaxy centers, triggering and fueling star formation activity \citep{1991ApJ...370L..65B,1994mtia.conf..143A,2013MNRAS.429.1949A}.
CO observations have shown that high central molecular gas concentrations are more common in barred spiral galaxies \citep{1999ApJ...525..691S,2005ApJ...632..217S,2007PASJ...59..117K,2019MNRAS.484.5192C,2022A&A...666A.175Y}. 
As a result, the enhanced molecular gas concentrations in barred low-mass disks may increase the likelihood of a CO detection.
Therefore, in the last panel of the figure,  we display the number of detections and non-detections as a function of bar presence. The bar parameter ranges from 0 to 1, with 0 indicating no bar and 1 indicating a clear central bar. Negative values correspond to galaxies for which the bar classification is uncertain or ambiguous. We do not find a clear correlation between CO detections and the bar parameter, as the detection rate appears slightly higher in non-barred galaxies ($\sim$~60\%) compared to barred ones ($\lesssim$~50\%).

\begin{table*}
\begin{center}
\caption{$^{12}$CO(J=2-1) emission line properties of MaLHUCO galaxies.}
\begin{tabular}{lccccccccccc}
 \hline \hline
   MaNGA      &  $I_{\rm tot}$        &   rms  &$\Delta$V&V$_{\rm mean}$&    $I_{\rm g}$            & FWHM$_{\rm g}$ & log$\,M_{\rm H_2}^{{\rm MW},a}$ & 
    log$\,M_{\rm H_2}^{{\rm A17},b}$  &  log$\,M_{\rm H_2, R_{21}}^{{\rm A17},c}$ & $R_{21}$ & Det$^d$ \\ 
    ID &      [K km s$^{-1}$]     &  [mK]  &[km s$^{-1}$]&[km s$^{-1}$]& [K km s$^{-1}$]&[km s$^{-1}$] & [M$_{\odot}$] &  [M$_{\odot}$] &   [M$_{\odot}$] & & \\
 \hline \hline
 1-119871 & 1.98  $\pm$ 0.19  & 4.0 & 215 &  5991 &  1.95 $\pm$  0.22   & 123 &  8.99 $\pm$ 0.14   &  9.02 $\pm$ 0.17 &    9.05 $\pm$ 0.18 &  0.65  &  1 \\  
 1-120743 & $<0.29$           & 2.5 & 150 &    -- &                  -- & --  &  $<$8.05           &       $<$ 8.45   &    $<$ 8.57        &  0.53  &  0 \\
 1-39107  & 0.31 $\pm$ 0.04   & 1.3 & 105 &  5379 & 0.36 $\pm$ 0.06     &  91 &  8.09 $\pm$ 0.15   &  8.30 $\pm$ 0.17 &    8.42 $\pm$ 0.18 &  0.52  &  1 \\
 1-37167  & 0.98  $\pm$ 0.09  & 2.6 & 119 &  8705 &  0.97 $\pm$  0.14   &  70 &  9.01 $\pm$ 0.15   &  9.07 $\pm$ 0.17 &    9.08 $\pm$ 0.18 &  0.68  &  1 \\
 1-51008  & 0.38 $\pm$ 0.08   & 2.8 &  88 &  4124 & 0.32 $\pm$0.10      &  31 &  7.94 $\pm$ 0.17   &  8.39 $\pm$ 0.19 &    8.51 $\pm$ 0.19 &  0.53  &  1 \\
 1-50963  & 0.68  $\pm$ 0.10  & 2.5 & 181 &  9490 &  0.72 $\pm$  0.12   & 112 &  8.93 $\pm$ 0.16   &  9.31 $\pm$ 0.18 &    9.29 $\pm$ 0.19 &  0.73  &  1 \\
 1-71167  & 0.53  $\pm$ 0.09  & 2.2 & 163 &  4825 &  0.54 $\pm$  0.11   & 102 &  8.24 $\pm$ 0.16   &  8.71 $\pm$ 0.18 &    8.78 $\pm$ 0.19 &  0.60  &  1 \\
 1-152563 & $<0.29$           & 2.5 & 150 &    -- &                  -- & --  &  $<$7.96           &       $<$ 8.39   &    $<$8.54         &  0.50  &  0 \\
 1-201183 & 0.60 $\pm$ 0.08   & 2.0 & 176 &  4790 & 0.61$\pm$ 0.11      & 123 &  8.28 $\pm$ 0.15   &  8.42 $\pm$ 0.17 &    8.54 $\pm$ 0.18 &  0.54  &  1 \\
 1-152828 & $<0.24$           & 2.1 & 150 &   --  &                 --  &  -- &  $<$7.78           &       $<$ 8.39   &    $<$8.48         &  0.57  &  2 \\
 1-382306 &  0.23$\pm$0.02    & 2.3 &  38 &  4768 & 0.25$\pm$0.05       &  24 &  7.86 $\pm$ 0.14   &  7.88 $\pm$ 0.17 &    8.03 $\pm$ 0.17 &  0.50  &  1 \\
 1-604885 & $<0.34$           & 2.9 & 150 &   --  &   --                & --  &  $<$8.01           &       $<$ 8.45   &    $<$8.52         &  0.60  &  0 \\
 1-121773 & $<0.29$           & 2.5 & 150 &   --  & --                  & --  &  $<$8.17           &       $<$ 8.56   &    $<$8.62         &  0.61  &  0 \\
 1-584678 & 0.13$\pm 0.01$    & 1.0 &  98 &  5096 & 0.16$\pm$ 0.04      &  55 &  7.67 $\pm$ 0.14   &  8.12 $\pm$ 0.17 &    8.20 $\pm$ 0.17 &  0.59  &  1 \\
 1-44745  & 0.55  $\pm$ 0.10  & 3.8 &  76 &  5756 &  0.64 $\pm$  0.15   &  68 &  8.41 $\pm$ 0.16   &  8.74 $\pm$ 0.18 &    8.79 $\pm$ 0.19 &  0.63  &  1 \\
 1-45112  & 0.33  $\pm$ 0.05  & 1.8 & 135 & 12124 &  0.28 $\pm$  0.05   &  49 &  8.82 $\pm$ 0.15   &  9.07 $\pm$ 0.18 &    9.12 $\pm$ 0.18 &  0.62  &  1 \\
 1-72402  & 0.82  $\pm$ 0.10  & 2.2 & 182 &  7064 &  0.80 $\pm$  0.11   & 110 &  8.75 $\pm$ 0.15   &  8.92 $\pm$ 0.17 &    8.99 $\pm$ 0.18 &  0.60  &  1 \\
 1-218435 & $<0.23$           & 2.0 & 150 &   --  &   --                &  -- &  $<$8.05           &       $<$ 8.73   &    $<$8.83         &  0.57  &  0 \\
 43-47    & 0.22 $\pm$ 0.05   & 1.6 &  90 &  2241 &  0.21 $\pm$0.06     &  73 &  7.18 $\pm$ 0.17   &  8.07 $\pm$ 0.19 &    8.22 $\pm$ 0.20 &  0.49  &  1 \\
 1-216951 & 0.34  $\pm$ 0.04  & 1.3 &  85 &  7573 &  0.32 $\pm$  0.05   &  58 &  8.43 $\pm$ 0.15   &  8.68 $\pm$ 0.17 &    8.76 $\pm$ 0.18 &  0.59  &  1 \\
 1-605833 & 1.16  $\pm$ 0.09  & 2.1 & 198 &  7213 &  1.14 $\pm$  0.09   &  92 &  8.92 $\pm$ 0.14   &  9.00 $\pm$ 0.17 &    9.06 $\pm$ 0.17 &  0.60  &  1 \\
 1-156037 &   $<0.23$         & 2.0 & 150 &   --  & --                  &  -- &  $<$8.25           &       $<$ 8.75   &    $<$8.83         &  0.59  &  0 \\
 1-78143  & $<$0.15           & 1.3 & 150 &   --  & --                  &  -- &  $<$7.86           &       $<$ 8.22   &    $<$8.32         &  0.56  &  0 \\
 1-80572  & 1.30  $\pm$ 0.13  & 2.5 & 263 &  6653 &  1.32 $\pm$  0.15   & 168 &  8.89 $\pm$ 0.15   &  9.27 $\pm$ 0.17 &    9.31 $\pm$ 0.18 &  0.65  &  1 \\
 1-62169  & $<$ 0.36          & 3.1 & 150 &   --  & --                  &  -- &  $<$8.31           &       $<$ 8.57   &    $<$8.68         &  0.55  &  0 \\
 1-187394 & $<0.28$           & 2.4 & 150 &   --  & --                  &  -- &  $<$8.40           &       $<$ 8.77   &    $<$8.84         &  0.60  &  2 \\
 1-401878 & 0.95  $\pm$ 0.11  & 2.4 & 217 &  9487 &  0.93 $\pm$  0.12   & 108 &  9.06 $\pm$ 0.15   &  9.12 $\pm$ 0.17 &    9.11 $\pm$ 0.18 &  0.72  &  1 \\
 1-194947 & $<$0.33           & 2.8 & 150 &   --  & --                  &  -- &  $<$8.05           &       $<$ 8.29   &    $<$8.38         &  0.57  &  0 \\
 1-624292 & 2.23  $\pm$ 0.15  & 3.0 & 223 &  7866 &  2.32 $\pm$  0.18   & 177 &  9.28 $\pm$ 0.14   &  9.22 $\pm$ 0.17 &    9.22 $\pm$ 0.18 &  0.70  &  1 \\
 1-625180 & $<0.20$           & 1.7 & 150 &   --  & --                  & --  &  $<$7.96           &       $<$ 8.30   &    $<$8.41         &  0.54  &  0 \\ 
 1-418224 & $<0.26$           & 2.2 & 150 &   --  & --                  & --  &  $<$8.07           &       $<$ 8.48   &    $<$8.60         &  0.54  &  0 \\
 1-629221 & $<0.36$           & 3.1 & 150 &   --  & --                  & --  &  $<$8.20           &  $<$8.45         &    $<$8.55         &  0.55  &  0 \\
 1-631727 & 0.55 $\pm$ 0.04   & 4.0 &  92 &  6038 &  0.60$\pm$ 0.16     &  77 &  8.44 $\pm$ 0.14   &  8.58 $\pm$ 0.17 &    8.67 $\pm$ 0.17 &  0.57  &  1 \\
 1-199432 & $<$0.43           & 3.7 & 150 &   --  & --                  & --  &  $<$8.26           &       $<$ 8.47   &    $<$8.58         &  0.55  &  0 \\
 1-440301 & $<0.16$           & 1.4 & 150 &   --  & --                  & --  &  $<$7.72           &       $<$ 8.07   &    $<$8.20         &  0.53  &  0 \\
 1-265069 & $<0.34$           & 2.9 & 150 &   --  & --                  & --  &  $<$8.29           &       $<$ 8.73   &    $<$8.81         &  0.59  &  0 \\
 1-210358 &   $<0.26$         & 2.2 & 150 &   --  &   --                & --  &  $<$8.37           &       $<$ 9.08   &    $<$9.12         &  0.64  &  0 \\
 1-295155 & $<0.28$           & 2.4 & 150 &   --  & --                  & --  &  $<$7.89           &       $<$ 8.20   &    $<$8.30         &  0.56  &  0 \\
 1-325250 & 1.36  $\pm$ 0.19  & 6.4 &  90 &  9805 &  1.39 $\pm$  0.25   &  70 &  9.25 $\pm$ 0.15   &  9.58 $\pm$ 0.18 &    9.58 $\pm$ 0.18 &  0.70  &  1 \\
 1-294140 & 0.26  $\pm$ 0.06  & 2.3 &  68 &  6618 &  0.28 $\pm$  0.08   &  51 &  8.19 $\pm$ 0.17   &  8.83 $\pm$ 0.19 &    8.90 $\pm$ 0.20 &  0.59  &  1 \\
 1-177270 & 0.89  $\pm$ 0.12  & 3.6 & 111 & 11016 &  0.89 $\pm$  0.14   &  58 &  9.17 $\pm$ 0.16   &  9.32 $\pm$ 0.18 &    9.29 $\pm$ 0.19 &  0.75  &  1 \\
 1-635629 & 4.16  $\pm$ 0.10  & 1.7 & 323 &  5957 &  4.26 $\pm$  0.23   & 222 &  9.31 $\pm$ 0.14   &  9.17 $\pm$ 0.17 &    9.18 $\pm$ 0.17 &  0.68  &  1 \\
\hline \hline
\end{tabular}
\label{table:tab1}
\end{center}
\footnotesize{$^a$ Using $\alpha_{\rm CO}^{\rm MW} = 4.36$ M$_{\odot}$ pc$^{-2}$ (K km s$^{-1}$)$^{-1}$ and including correction for helium.
$^b$ Based on the method of \citet{2017MNRAS.470.4750A} using $M_*^{\rm P3D}$ and a fixed $R_{21} = 0.7$.
{$^c$ Based on the method of \citet{2017MNRAS.470.4750A} adopting $M_*^{\rm P3D}$ and a variable $R_{21}$ (as given in the next column).}
$^d$ (0=no, 1=yes, 2=tentative but not considered). Masses  include helium, upper limits have been computed using the 3 integrated sigma value for a nominal line width as indicated. For two gaussian fits the fitted velocity is the flux weighted mean, the width is the sum of the two widths.}
\end{table*}

\section{Molecular gas masses}

Molecular gas mass estimates rely on the value of CO-to-H$_2$ conversion factor, $\alpha_{\rm CO}$. This is expected to vary with the physical properties of the ISM, such as metal content, ionizing radiation field strength, and
gas surface density \citep[see][ and references therein]{2013ARA&A..51..207B}. Metallicity is the dominant parameter affecting $\alpha_{\rm CO}$. Indeed, it is found that it sharply increases 
in systems with metallicities below 12 + log(O/H) $\approx$ 8.4, or half the solar value \citep{1997A&A...328..471I,1998AJ....116.2746T,2001PASJ...53L..45M,2002Ap&SS.281..127B,2012AJ....143..138S}. 
We use  the following expression to convert $L_{\rm CO}^{2-1}$ into $M_{\rm H_2}$ in units of M$_\odot$:
\begin{equation}
M_{\rm H_2}= \alpha_{\rm CO} {L_{\rm CO}^{1-0}} \qquad {\hbox{with}}\ {L_{\rm CO}^{1-0}}={L_{\rm CO}^{2-1}\over R_{21}},
\end{equation}

\noindent where L$_{\rm CO}^{2-1}$ and L$_{\rm CO}^{1-0}$ are the luminosities of the $^{12}$CO(J=2-1) and $^{12}$CO(J=1-0) lines respectively, and R$_{21}$ is their  ratio. In the rest of the paper we use the symbol $L_{\rm CO}$ to indicate ${L_{\rm CO}^{1-0}}$.

The use of $^{12}$CO(2-1) line as a tracer of molecular gas requires  knowledge of $R_{21}$. Usually, a constant $R_{21} = 0.7$ is assumed \citep{2011ApJ...730L..13B}. Recently \citet{2025ApJ...988..162D} suggest that, for studies of main-sequence galaxy samples, a fixed $R_{21} = 0.64^{+0.16}_{-0.14}$ does not significantly bias molecular gas mass estimates from CO(J=2–1) either on galaxy-wide average scales or on kiloparsec scales.
However, it has also recently been shown that $R_{21}$  correlates with the SFR and properties related with it, such as sSFR, SFR surface density, and SFE, while it appears to be insensitive to $M_*$ and galaxy size \citep[]{2025ApJ...979..228K}. The dispersion is however very high at low stellar masses and there are no conclusions on possible dependencies on metallicity, given the low number of objects  with 12 + log(O/H) $< 8.5$ investigated in the same work. In M33, where the metallicity is half solar, \citet{2014A&A...567A.118D} have found a constant value of $R_{21}$=0.8 throughout the whole star-forming disk. They conclude that the lower metallicity and the lower optical depth of CO lines in the outer disk likely balance the expected decrease of $R_{21}$. 
Given the similarities between the properties of our sample and M33, and the above mentioned uncertainties on the decrease of $R_{21}$ with SFR, in this work we use a constant $R_{21} = 0.7$.
However, we also show in Table~\ref{table:tab1} the values of $M_{\rm H_2}$, assuming a dependence of $R_{21}$ on SFR, following the prescription of \citet{2025ApJ...979..228K}. 
We find that $R_{21}$ varies  between 0.5 and 0.75 in the MaLHUCO sample.

We do not apply aperture corrections to the molecular mass estimates,  
as the targets are selected  
such that the extent of the star forming regions,  traced by the H$\upalpha$ emission, is smaller than the JCMT beam size. 
Moreover, aperture corrections often rely on assumptions about the radial scaling of the molecular gas surface brightness. As shown by \citet{2025A&A...699A.346S}, the CO scale radius varies significantly, and radial profiles of the molecular gas surface density can deviate considerably from the median trend. This is especially true for late-type spiral galaxies such as Sd-Sm systems, 
whose gas surface density profiles typically have very small scale lengths.

The \hdue\ masses are obtained using two different  CO-to-H$_2$ conversion factors: 1) the Milky Way conversion factor, $\alpha^{\rm MW}_{\rm CO}$ = 4.36 M$_{\odot}$ pc$^{-2}$ (K km s$^{-1}$)$^{-1}$, with a typical uncertainty of $\pm$ 30\% \citep[and references therein]{2001ApJ...547..792D,2006A&A...454..781L,2013ARA&A..51..207B}; 2) a metallicity dependent $\alpha_{\rm CO}$ adopting the calibration of \citet{2017MNRAS.470.4750A}, with an error
in log$\alpha_{\rm CO}$ of $\sim$ 0.165 dex. In this case the calculation of $\alpha_{\rm CO}$ depends on galaxy metallicity and on the sSFR offset  from the expected MS at a given redshift $z$: 

\begin{equation}
\log \alpha_{\rm CO} = 
14.752 - 1.623[12 + \log({\rm O/H})] + \\
0.062 \log \Delta({\rm MS}),
\label{eq:varco}
\end{equation}

\noindent
where

\begin{equation}
\Delta({\rm MS}) = \frac{{\rm sSFR_{\rm measured}}}{\rm sSFR_{\rm MS} ({\mathit{z}}, M_{*})}
\end{equation}
  
\noindent
is 
the offset from the MS at  redshift $z$ and 

\begin{equation}
\begin{aligned}
\log({\rm sSFR}_{\rm MS} (z, M_{*})) = -1.12 + 1.14z - 0.19z^2 - \\ (0.3 + 0.13z)
\times(\log M_{*} - 10.5) \, [{\rm Gyr}^{-1} ],
\end{aligned}
\end{equation}

\begin{figure}
 \begin{center}
\includegraphics[scale=0.37]{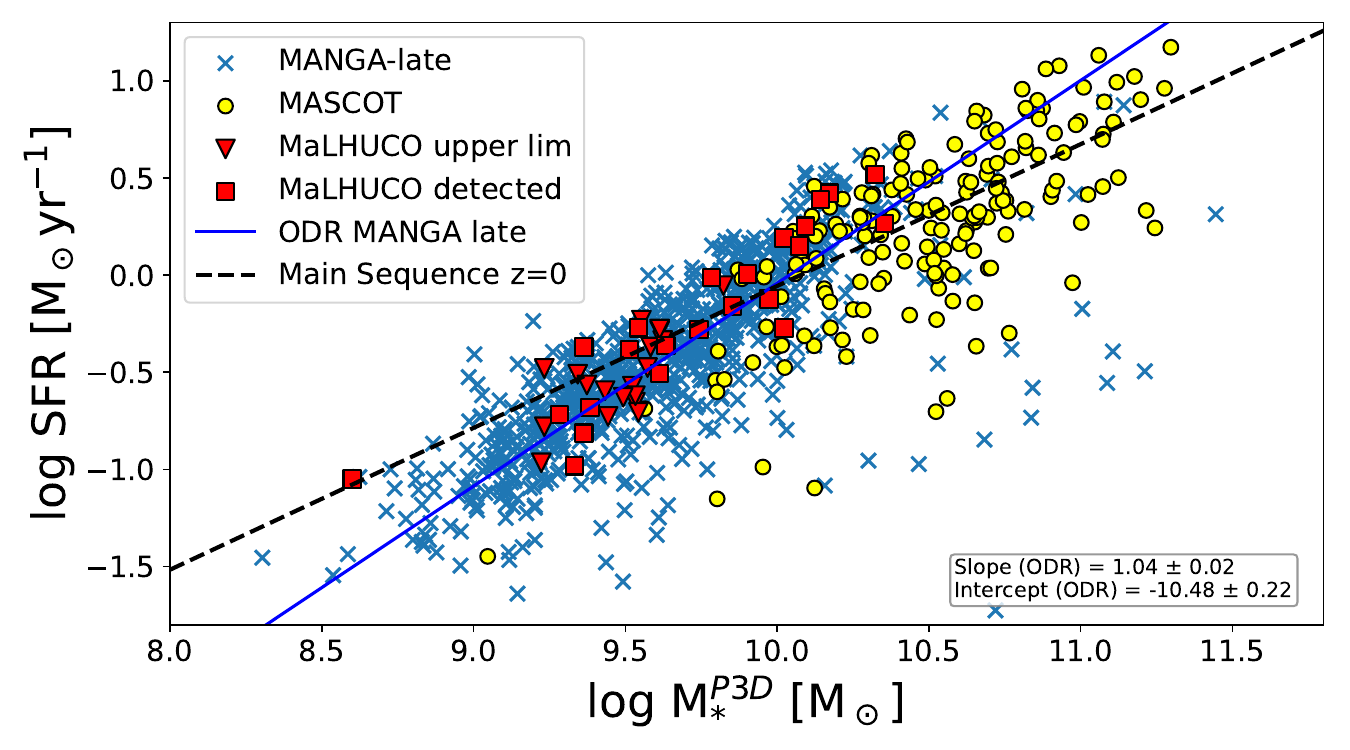}
\includegraphics[scale=0.42]{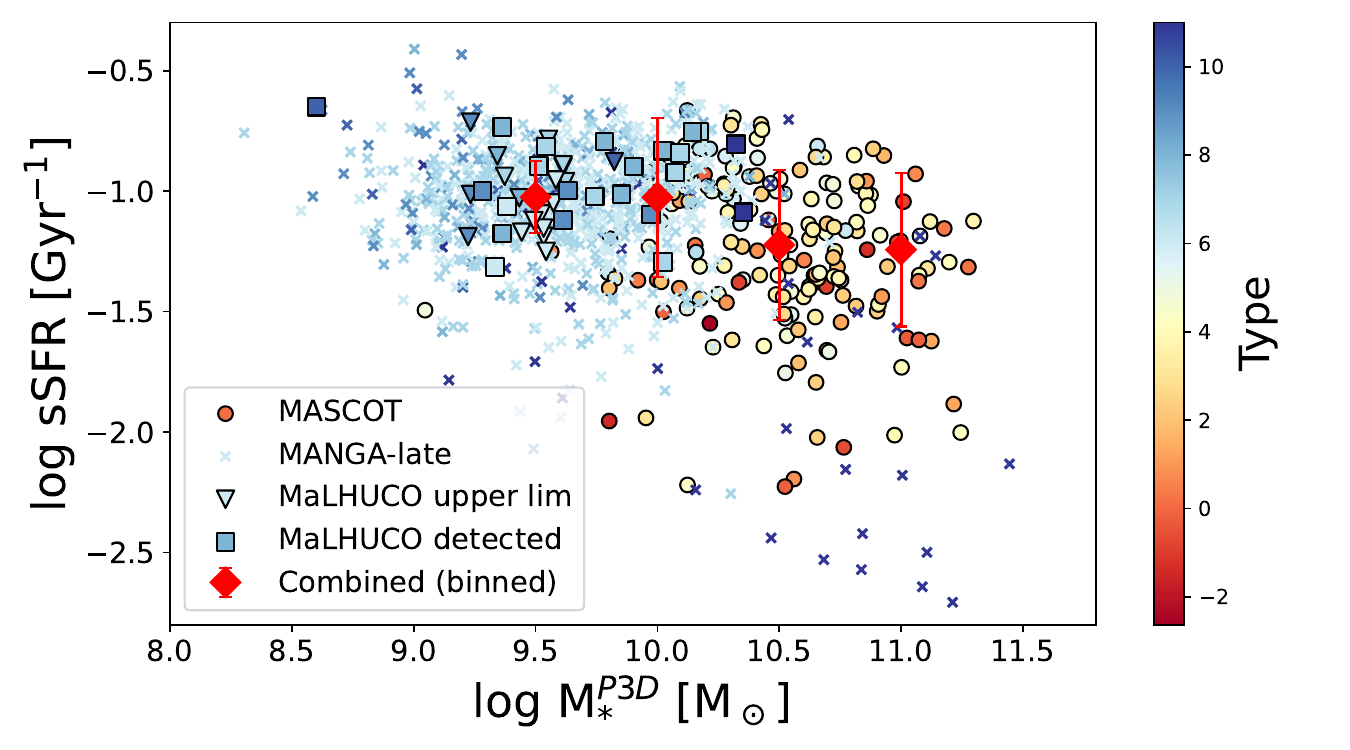}
\caption{Top: Relation between ${\rm SFR}_{\rm SSP}^{\rm P3D}$ and $M_*^{\rm P3D}$ for {\em MaNGA-late} (crosses), MASCOT (circles) and  MaLHUCO galaxies (squares: CO detections; downward-pointing
triangles: CO non-detections).
The solid line shows the ODR fit to the {\em MaNGA-late} sample.
The dashed line corresponds to the main sequence at $z =$ 0  \citep{2018MNRAS.477.3014B}.
{Bottom:  The specific SFR (sSFR=SFR$_{\rm SSP}^{\rm P3D}$/$M_*^{\rm P3D}$} as a function of $M_*^{\rm P3D}$. Symbols are the same as in the upper panel.
MaLHUCO and MASCOT samples have been color-coded by the morphological type.}
\label{fig:02}
\end{center}
\end{figure}

\noindent
is the analytical definition of the MS according to \citet{2012ApJ...754L..29W}\footnote{We note that the offset from the star-forming MS is computed relative to the relation of \citet{2012ApJ...754L..29W}, where SFRs and stellar masses are derived using different methodologies from those adopted in this work and in  \citet{2017MNRAS.470.4750A}. However, because the dependence of $\alpha_{\rm CO}$
on $\Delta$(MS) in Eq. \ref{eq:varco} is weak (the coefficient of the $\Delta$(MS) term is $\sim$0.06), systematic differences between SFR and $M_*$ calibrations have only a minor impact on the inferred molecular gas masses.}. Both conversion factors include the correction for helium.
Columns~8, 9 and 10 of Table~\ref{table:tab1} display the computed molecular masses, or their upper limits, for the galactic ($M^{\rm MW}_{\rm H_2}$) and metallicity dependent $\alpha_{\rm CO}$  ($M^{\rm A17}_{\rm H_2}$), respectively. For the latter estimate, we show molecular masses for a fixed $R_{21} = 0.7$, $M^{\rm A17}_{\rm H_2}$, and for a variable $R_{21}$, calculated using the relation between $R_{21}$ and SFR given in \citet{2025ApJ...979..228K}, $M^{\rm A17}_{\rm H_2,R_{21}}$. 
Both estimates are obtained using 
$M_*^{\rm P3D}$. 
Errors in $M_{\rm H_2}$ are calculated by summing in quadrature the uncertainties in the spectral line integrated emission and  the CO-to-H$_2$ conversion factor. 

Molecular masses in our sample range between $\sim 10^{7.2}$ M$_{\odot}$ and $\sim 10^{9.3}$ M$_{\odot}$ ($\sim 10^{7.9}$ M$_{\odot}$ and $\sim 10^{9.6}$ M$_{\odot}$) for the galactic (metallicity-dependent) conversion factor, and  a fixed $R_{21}$.
The inclusion of a variable line ratio, does not significantly alter molecular gas mass estimates: the mean value of $M^{\rm A17}_{\rm H_2,R_{21}}$  is 0.07~dex higher than $M^{\rm A17}_{\rm H_2}$. 
Unless stated otherwise, throughout this work we use $M^{\rm A17}_{\rm H_2}$ with $\alpha_{\rm CO}$ as defined in \citet{2017MNRAS.470.4750A}, and  a constant line ratio $R_{21}=0.7$ and we shall refer to this as $M_{\rm H_2}$.

\begin{table*}
\begin{center}
\caption{Results of ODR and Bayesian linear regression fits using the model $y = ax + b$ for the samples analyzed in this work: MaNGA-late, MaLHUCO, and MASCOT.}
\begin{tabular}{ccrrccccc}
\hline \hline
\hline
$x$  &  $y$     &  $a_{\rm ODR} \:\:\:\:\:$        &   $b_{\rm ODR}  \:\:\:\:\:$  & $a_{\rm Bay}$&    $b_{\rm Bay}$            & $a_{\rm ODR}$        &   $b_{\rm ODR}$ & Figure \\ 
\hline \hline
& & \multicolumn{2}{c}{MaNGA-late} \\
\cline{3-4}
\hline
log$M_*^{\rm P3D}$ & logSFR   & 1.04$\pm$0.02 & -10.48$\pm$0.22 & -- & -- & -- & -- & Fig. \ref{fig:02} \\
\hline
& & \multicolumn{2}{c}{MaLHUCO + MASCOT} & \multicolumn{2}{c}{MaLHUCO} &  \multicolumn{2}{c}{MaLHUCO} &\\
\cline{1-4}\cline{5-9}
\hline
log$L_{12}$ & log$L_{\rm CO}$ & 0.97 $\pm$ 0.04 & -33.00 $\pm$ 1.60 & 0.99 $\pm$ 0.08 & -34.09 $\pm$ 3.35 & -- & -- & Fig. \ref{fig:03}\\
log$L_{22}$ & log$L_{\rm CO}$ & 0.93 $\pm$ 0.04 & -30.74 $\pm$ 1.86 & 1.05 $\pm$ 0.13 & -36.45 $\pm$ 5.66 & -- & -- & Fig. \ref{fig:03} \\
log$M_{*}^{\rm P3D}$ & logM$_{\rm H_2}^{\rm MW}$ & 1.26 $\pm$ 0.04 & -3.73 $\pm$ 0.37 & 1.61 $\pm$ 0.16 & -7.23 $\pm$ 1.55 & 1.41 $\pm$ 0.11 & -5.20 $\pm$ 1.08 & Fig. \ref{fig:04} \\
log$M_{*}^{\rm P3D}$ & logM$_{\rm H_2}^{\rm A17}$ & 1.04 $\pm$ 0.04 & -1.41 $\pm$ 0.45 & 1.24 $\pm$ 0.15 & -3.51 $\pm$ 1.47& 1.10 $\pm$ 0.12  & -1.93 $\pm$ 1.17 & Fig. \ref{fig:04} \\
log$M_{\rm H_2}^{\rm A17}$ & logSFR & 1.04 $\pm$ 0.05 & -9.58 $\pm$ 0.51 & -- & -- &  1.00 $\pm$ 0.10 & -8.98 $\pm$ 0.87 & Fig. \ref{fig:05} \\
logsSFR & log$\tau_{\rm H_2}$ & -1.00 $\pm$ 0.06 & -0.99 $\pm$ 0.08 & -- & -- &  -- & -- & Fig. \ref{fig:06} \\
$\log(M_{\rm HI}/M_{*}^{\rm P3D})$ & $(M_{\rm H_2}^{\rm A17}/M_{\rm gas})$ & -0.34 $\pm$ 0.03 & 0.11 $\pm$ 0.02 & -- & -- &  -- & -- & Fig. \ref{fig:08} \\
\hline \hline
\label{table:tabfit}
\end{tabular}
\end{center}
\end{table*}

\section{Scaling relations}

The aim of this section is to examine scaling relations and assess whether they are consistent from high- to low-mass galaxies.
When investigating the scaling relations of the MaLHUCO sample, which has reliable uncertainty estimates, we use the LinMix package \citep{2007ApJ...665.1489K} which applies a hierarchical Bayesian model (BAY) to perform linear fits in the $x-y$ plane. LinMix also accounts for upper limits in the dependent variable. For the analysis of larger samples, i.e. MaLHUCOS+MASCOT, we use orthogonal data regression (ODR)\footnote{Specifically, we use the python package scipy.odr} without accounting for uncertainties, as errors in the CO line fluxes are not available in the MASCOT catalog.
We use ODR instead of the traditional least-squares
method, because ODR minimizes the orthogonal distances to the fitted lines taking into account variances on both the x- and y-axes. The results of the linear regression analysis presented in this section 
are listed in Table \ref{table:tabfit}.

In the upper panel of Fig.~\ref{fig:02}, we show the MS of MaLHUCO galaxies 
($M_*^{\rm P3D}$ vs. SFR$_{\rm SSP}^{\rm P3D}$), where square and triangles indicate CO-detections  and non-detections, respectively. 
Our galaxies are compared with the {\em MaNGA-late} (crosses), and MASCOT (circles) samples. 
The figure illustrates that our survey is  an extension of MASCOT at lower stellar masses and
MaLHUCO galaxies closely follows the distribution of $M_*$ and SFR of galaxies in {\em MANGA-late}. The figure also shows that the MS of {\em MaNGA-late} galaxies is steeper than the MS traced at $z = 0$ by \citet{2018MNRAS.477.3014B}, showing a scaling of the SFR with $M_*$ close to linear. This change in the scaling law is also evident in the bottom panel of Fig.~\ref{fig:02}, where we plot the sSFR as a function of stellar mass: {\em MaNGA-late} and MaLHUCO galaxies have, on average, higher sSFRs than MASCOT. 
Lastly, one can see that the adopted morphological criterion also implies a selection of galaxies in terms of lower $M_*$.  
Figure \ref{fig:02} shows that there are very few galaxies in {\em MaNGA-late} with  log($M_*^{\rm P3D}/$M$_{\sun} > 10.35$, and that the most massive systems are already approaching the quenching phase.

In the rest of this  section we investigate relations between CO luminosity or molecular gas mass and
global galaxy properties. 

\begin{figure}
\begin{center}
\includegraphics[scale=0.58]{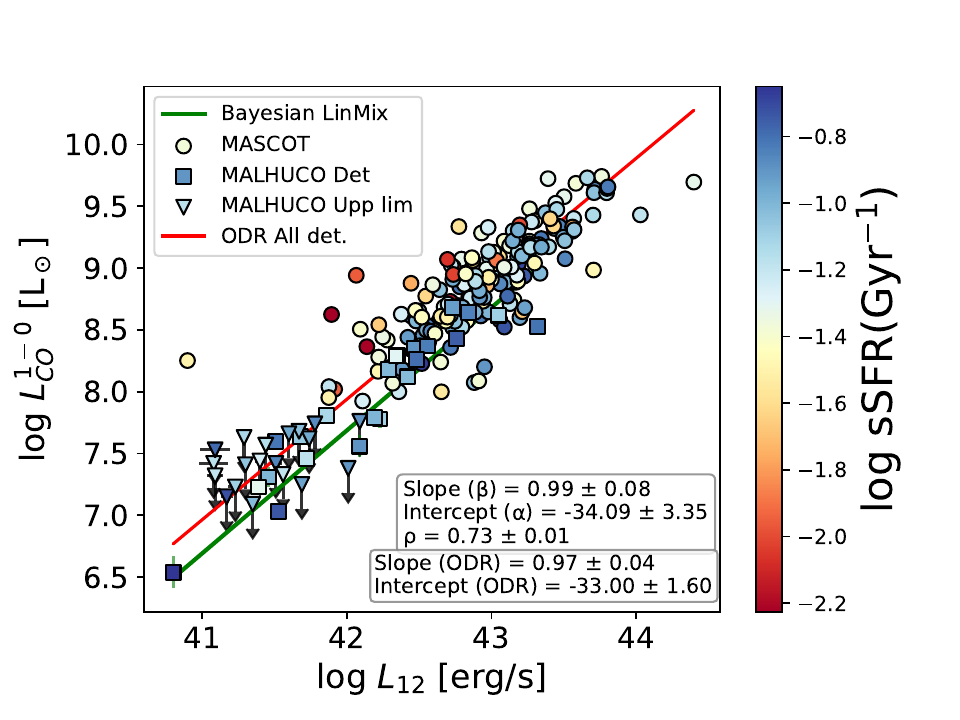}
\includegraphics[scale=0.58]{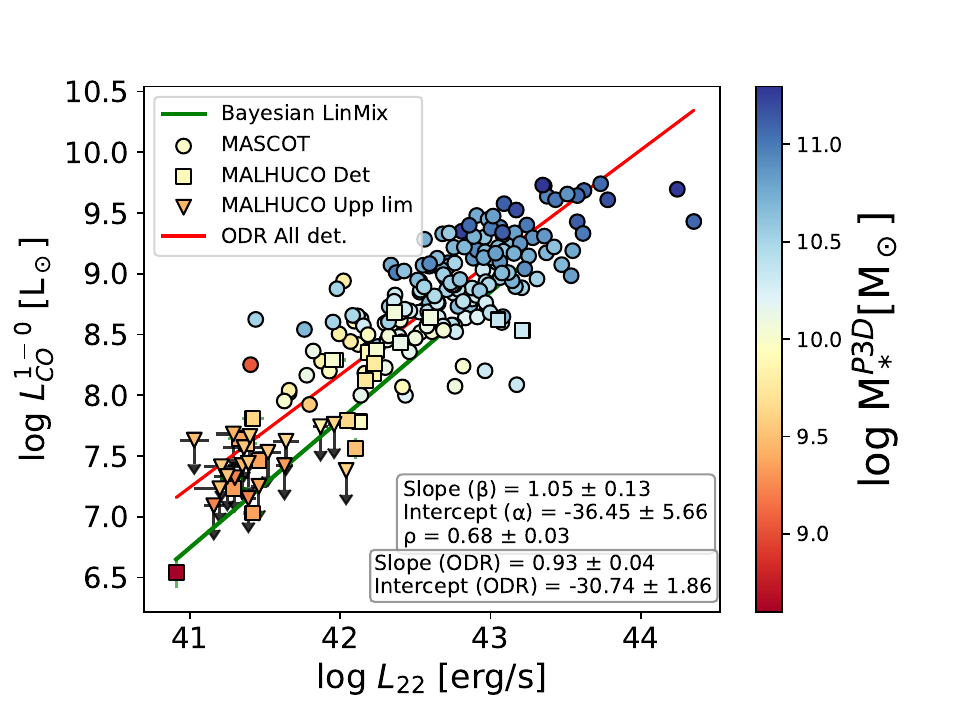}
\caption{$L_{\rm CO}$ versus $L_{12}$ (top panel) and $L_{\rm CO}$ versus $L_{22}$ (bottom panel) for MaLHUCO (squares: detections; downward-pointing
triangles: non-detections) and MASCOT (circles) samples. 
In both panels, the green lines show fitted relations  inferred using Bayesian methods for MALHUCO targets only. 
The red lines indicate the ODR best fit to MASCOT and MALHUCO CO-detected galaxies. Data have been color-coded by sSFR (left panel) and $M_*^{\rm P3D}$ (right panel).}
\label{fig:03}
\end{center}
\end{figure}

\subsection{The relationship between CO and PAH or mid-infrared emission}

In Fig. \ref{fig:03} we display the comparison between 
$L_{\rm CO}$ and the two WISE luminosities, $L_{12}$ and $L_{22}$, 
for both MaLHUCO 
and MASCOT 
samples.  In the top panel, galaxies are color-coded by sSFR, while in the bottom panel, they are color-coded by $M_*^{\rm P3D}$. 
The figure suggests that for a given $L_{\rm CO}$,  galaxies  with a lower sSFR also have a lower $L_{12}$, and that 
more massive galaxies are brighter both in MIR and in CO line emission.
We also show in the figure ODR (green lines) and BAY linear regression (red lines) fits for the MaLHUCO+MASCOT and MaLHUCO-only galaxies, respectively. In both cases we we find an approximately linear relationship between CO and MIR emissions in both MIR bands (see Table \ref{table:tabfit}) 
 The correlation with $L_{12}$ is tighter than with $L_{22}$. 
This behavior supports the use of MIR emission, in particular the 12 $\upmu$m luminosity, as one of the most reliable estimator of $L_{\rm CO}$ 
in galaxies, consistent with previous studies \citep{2015ApJ...799...92J,2019ApJ...887..172G,2022ApJ...940..133G,2023ApJ...944L..10L}. These studies underline that both CO and MIR emission are related to dust, as well to heating sources such as the interstellar radiation field and the SFR.

PAH molecules and CO are spatially mixed in molecular clouds and are excited/destroyed under similar conditions. Their relation is very tight because chemical abundance variations affect the luminosities of both gas tracers. Dense gas is found deeper inside the molecular clouds, while PAHs are found primarily on the cloud surfaces, leading to a weaker association between 12~$\upmu$m emission and dense gas tracers.
This may suggest that the intensity of PAH emission provides a more direct tracer of the
bulk of the molecular gas in the ISM, possibly including
non-CO-emitting gas \citep{2023ApJ...944L..10L}. The upper limits in Fig.~\ref{fig:03} suggest that the relation may break down at MIR luminosities $L_{12/22} \lesssim 10^{42}$ erg s$^{-1}$, where CO luminosities fall below 
the expected trend. However, due to the large number of CO-undetected galaxies below this luminosity threshold, deeper observations are required to confirm this behavior.


\begin{figure}
 \begin{center}
\includegraphics[scale=0.58]{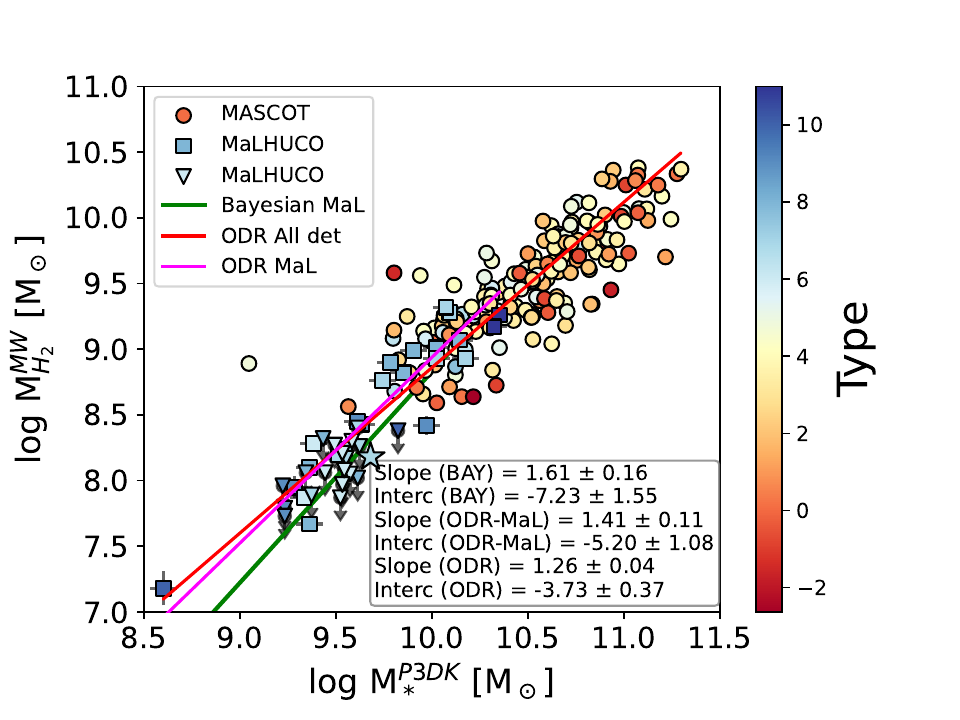}
\includegraphics[scale=0.58]{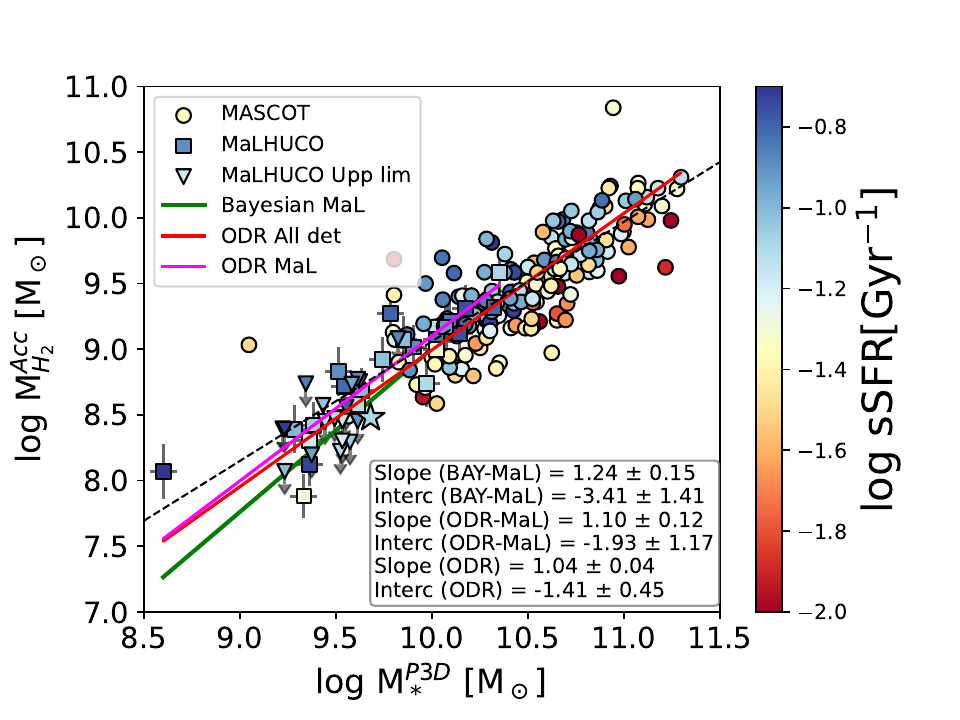}
\caption{The molecular gas main sequence ($M_{\rm H_2}$ vs. $M_*$). The top panel shows the relation for $M_{\rm H_2}^{\rm MW}$, while the bottom panel displays the relation for $M_{\rm H_2}^{\rm A17}$. In both panels, the red line corresponds to the ODR fit to the MaLHUCO+MASCOT sample, the magenta line is the ODR fit to the MaLHUCO galaxies with a CO detection (squares), the green line highlights the Bayesian fit to all MaLHUCO galaxies (squares and downward-pointing triangles). Galaxies are color-coded by morphological type (top panel) and  sSFR (bottom panel).   
The star symbol marks the location of M33. In the bottom panel, the dashed line indicates the Bayesian fit to ALLSMOG and xCOLDGASS samples by \citet[][see text for details]{2024A&A...687A.244H}.}
\label{fig:04}
\end{center}
\end{figure}

\subsection{The molecular gas main sequence }

In Fig. \ref{fig:04} we show the molecular gas main sequence \citep[MGMS;][]{2019ApJ...884L..33L} obtained for the combined MaLHUCO+MASCOT sample. Galaxies are color-coded by morphological type (top panel) and sSFR (bottom panel), respectively. 
The MGMS shows a tight correlation 
 close to linear for the MASCOT+MaLHUCO sample (1.04 $\pm$ 0.04) when using $M^{\rm A17}_{\rm H_2}$ 
(bottom panel of Fig. \ref{fig:04}). The slope is slightly steeper, but still consistent with linear scaling, if we consider only the MaLHUCO sample, according to either ODR (1.10 $\pm$ 0.12) or Bayesian (1.24 $\pm$ 0.15) fits (see also Table \ref{table:tabfit}). 
Both  results are very similar within the errors, implying that the ODR fitting 
is not significantly biased by the exclusion of upper limits.
Overall, the best-fit relations are steeper when we consider the molecular gas masses obtained with a galactic conversion factor (top panel of Fig. \ref{fig:04}, Table \ref{table:tabfit}). 

In the bottom panel of Fig. \ref{fig:04}, the dashed line indicates the scaling relation for ALLSMOG and xCOLDGASS joined sample through a Bayesian linear regression by \citet{2024A&A...687A.244H}. This has a slope of 0.91 considering the calibration of \citet{2017MNRAS.470.4750A} 
and using stellar masses from the SDSS DR7
MPA-JHU catalog\footnote{http://home.strw.leidenuniv.nl/~jarle/SDSS/}. The fit is only slightly shallower than what we infer for the joined MASCOT and MaLHUCO sample, likely because stellar masses from Pipe3D are higher than those from the MPA-JHU catalog at the low stellar mass end \citep[e.g.][]{2021MNRAS.505.3135A}.
The MGMS highlights that the amount of molecular hydrogen is strongly regulated by $M_*$, which traces the gravitational potential of a galaxy, favoring the conversion of \hi\ to \hdue\ as the stellar mass increases \citep{2019ApJ...884L..33L,2021ApJ...907..114D,2023MNRAS.518.4767B}. 
MASCOT early-type galaxies as well as galaxies with a low sSFR tend to lie below the fitted relations.

\begin{figure*}
\begin{center}
\includegraphics[scale=0.47]{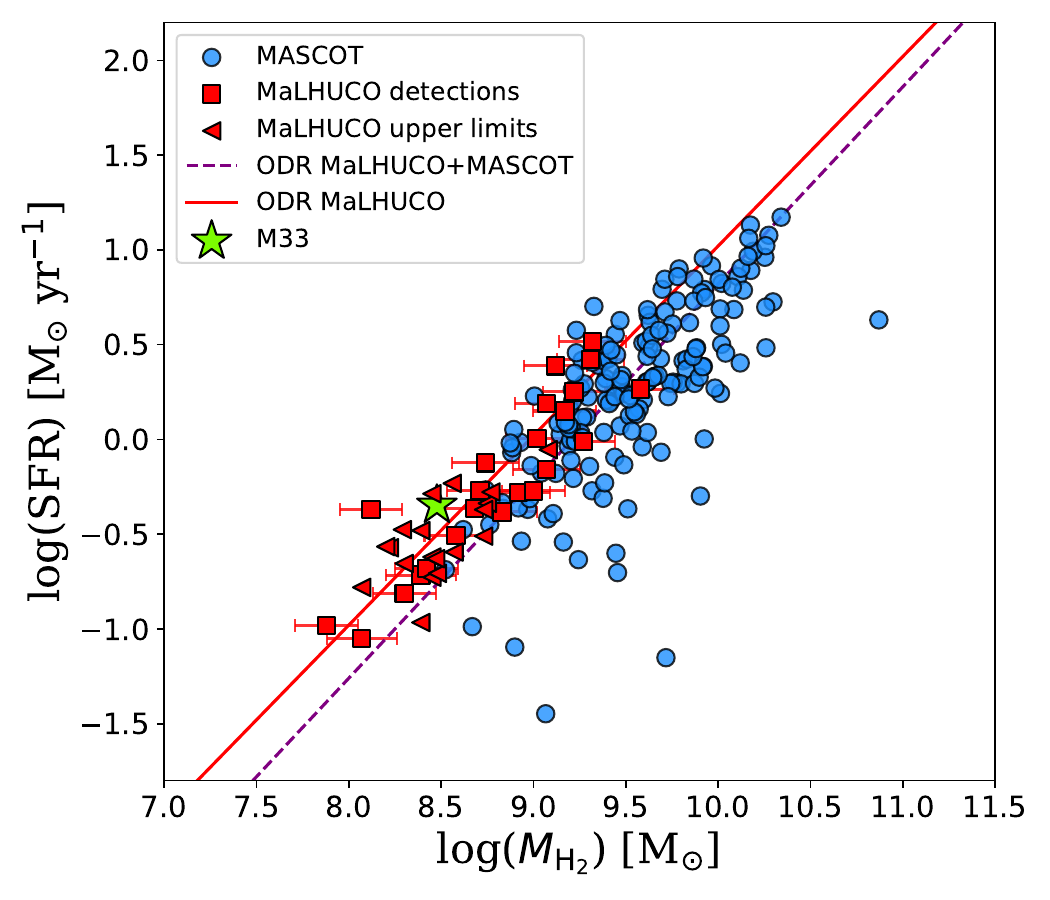}
\includegraphics[scale=0.54]{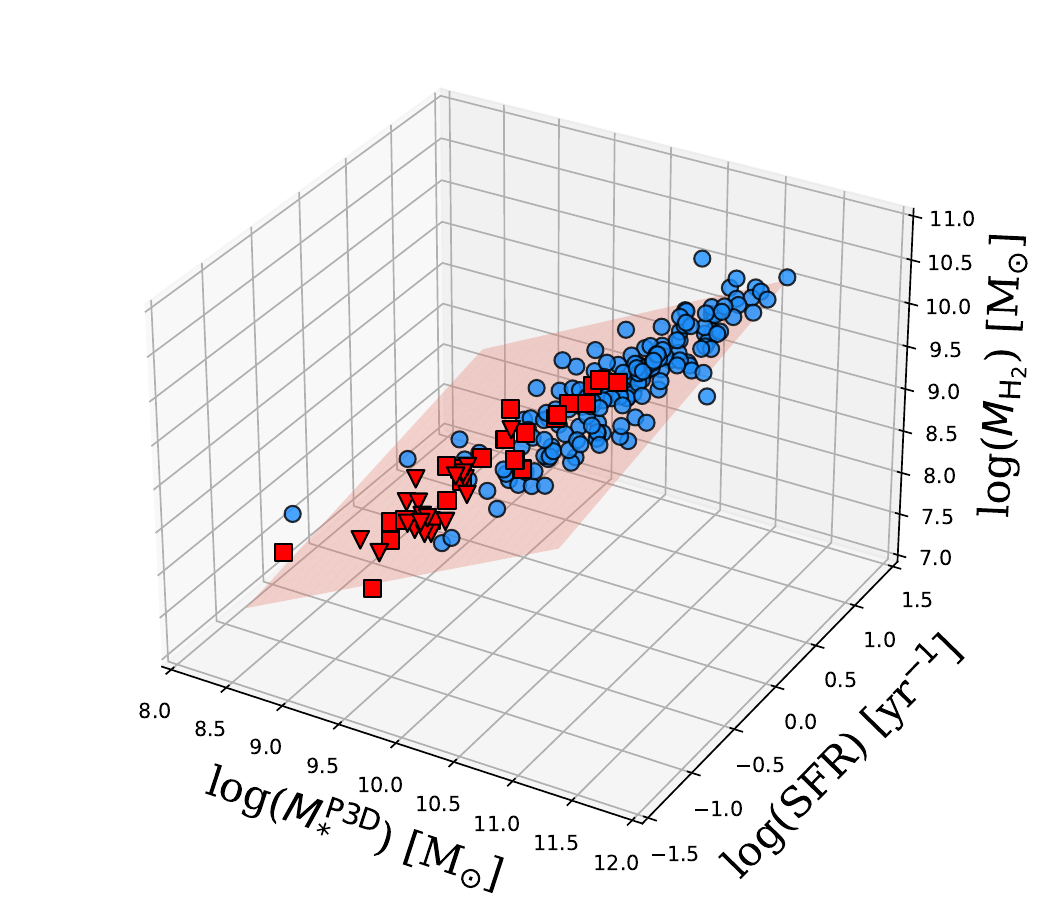}
\caption{Left: The global KS relation for the MaLHUCO (squares: detections;
left-pointing triangles: non-detections) and MASCOT (circles) sample. The dashed purple line is the result of the ODR fit to the CO detections of the full sample (MaLHUCO+MASCOT). The red solid line shows the fit for the MaLHUCO CO-detected galaxies only. 
The parameters of the two best-fit relations are given in Table \ref{table:tabfit}. Right: The 3D distribution of $M_{\rm H_2}$, SFR,  and $M_*$ including both samples, with the best-fit plane overlaid. Symbols are the same as in the left panel.}
\label{fig:05}
\end{center}
\end{figure*}

\subsection{The Kennicutt-Schmidt law and gas depletion-time scaling relations}

 The relation between the global molecular mass  and SFR is usually defined as the integrated KS law and its slope
found in previous studies ranges between 0.8 - 1.1 \citep{2014ApJ...793...19S,2016A&A...590A..27G,2021ApJ...907..114D,2025A&A...699A.346S}. 
The resolved KS law on kiloparsec or sub-kiloparsec scales also exhibits a close-to-linear relation \citep[$N \sim 1$;][]{2008AJ....136.2846B,2013AJ....146...19L,2019ApJ...884L..33L}.
In Fig. \ref{fig:05} (left panel) we display the KS relation for the samples examined in this work. The relation is linear when considering either the MaLHUCO sample or 
the combined MASCOT+MaLHUCO samples (Table \ref{table:tabfit}). 

The  three-dimensional relation deﬁned by
$M_*$, SFR, and $M_{\rm H_2}$ can be used to better 
constrain the molecular gas mass \citep{2019ApJ...884L..33L,2021ApJ...907..114D,2025A&A...699A.346S}. In the right panel of Fig. \ref{fig:05} we
show the 3D  plane with $M_*$, SFR, and $M_{\rm H_2}$ ($x$-, $y$-, and $z$-axes,
respectively). 
We fit the relation with the ODR method (without including upper limits), and the best-fit is given by 

\begin{equation}
\begin{aligned}
\log(M_{\rm H_2}/{\rm M}_{\odot}) = 0.66 \log(M_*/{\rm M}_{\odot}) \: + \\
0.28 \log({\rm SFR}/{\rm M}_{\odot} \, {\rm yr}^{-1}) + 2.47
\end{aligned}
\end{equation}

\noindent
with a dispersion of 0.19 dex.

\begin{figure}
 \begin{center}
\includegraphics[scale=0.5]{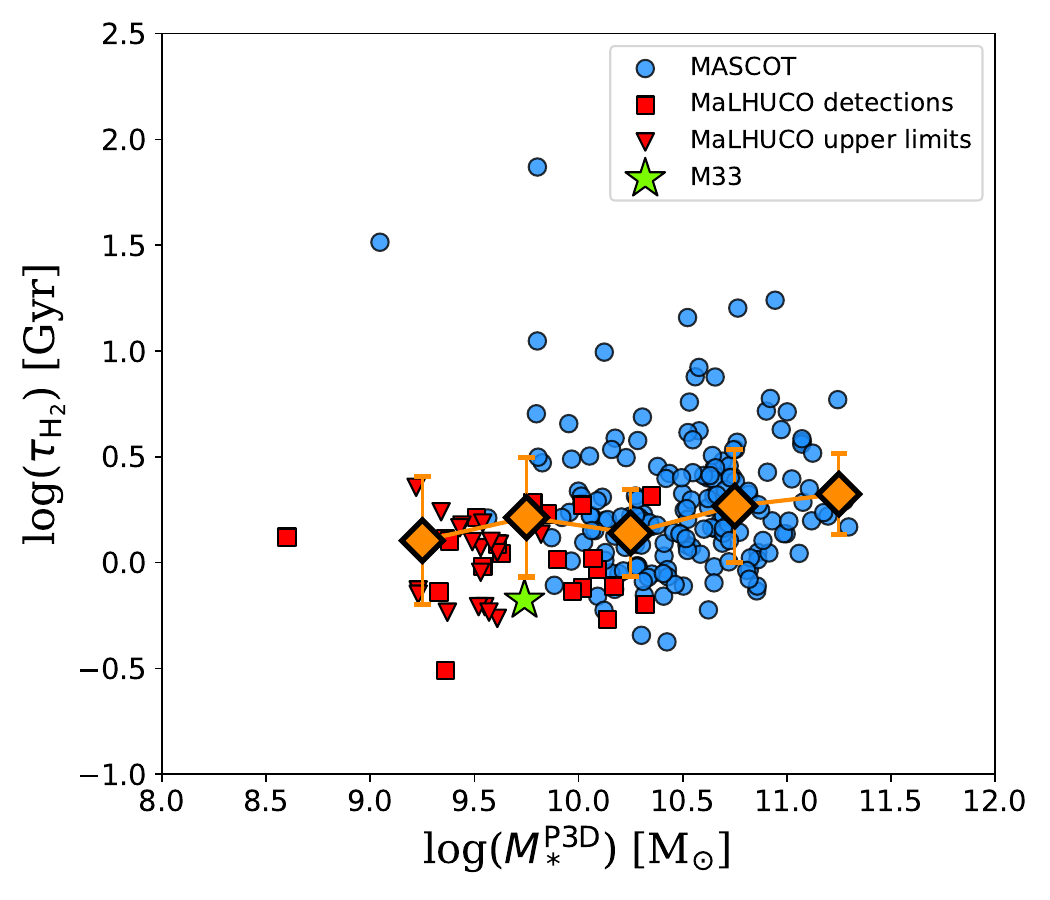}
\includegraphics[scale=0.48]{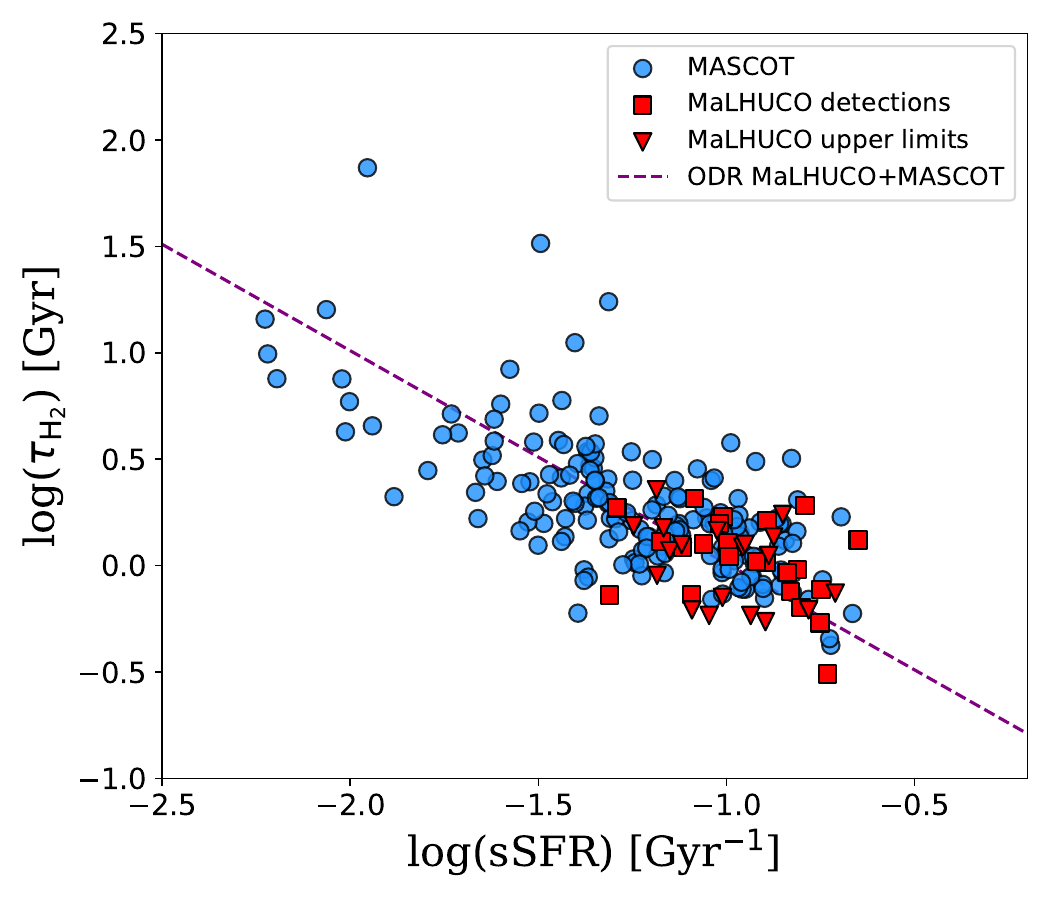}
\caption{Top: $\tau_{\rm H_2}$ as a function of the stellar mass for both MASCOT (circles) and MaLHUCO (squares: detections; downward-pointing triangles: non-detections) samples. 
Binned values (diamonds) are calculated as medians in stellar-mass intervals of 0.5 dex and restricted to detected galaxies with $\log(M_*^{\mathrm{P3D}}/\mathrm{M}_\odot) > 9$. 
The star marks the location of M33. 
Bottom: correlation between $\tau_{H_2}$ and sSFR for MASCOT and MALHUCO samples. Symbols are the same as in the left panel. The parameters of the best-fit relation (dashed line) are given in Table \ref{table:tabfit}.}
\label{fig:06}
\end{center}
\end{figure}

The inverse of the SFE, the depletion time ($\tau_{\rm H_2}$), provides the timescale over which star formation could be maintained at the current rate, given the available amount of molecular gas.  We display  $\tau_{\rm H_2}$ versus $M_{*}^{\rm P3D}$ for MaLHUCO and MASCOT samples in the top panel of Fig. \ref{fig:06}. The depletion time is  weakly dependent on $M_*$: its mean is 1.8~Gyr for the MASCOT sample while for MaLHUCO galaxies is 1.1~Gyr. Further molecular-gas observations of very low-mass
galaxies are needed to confirm this trend. On the other hand, $\tau_{\rm H_2}$  is clearly anti-correlated with sSFR, i.e. molecular hydrogen is consumed more rapidly as the SFR per unit stellar mass increases, as shown in the bottom panel of Fig. \ref{fig:06}. This trend is consistent with previous works, which reported similar behavior \citep{2014A&A...564A..66B,2017A&A...604A..53C,2020A&A...643A.180H, 2020ARA&A..58..157T, 2022ARA&A..60..319S}.

\begin{figure*}
\begin{center}
\includegraphics[trim=12 0 65 38, clip, scale=0.45]{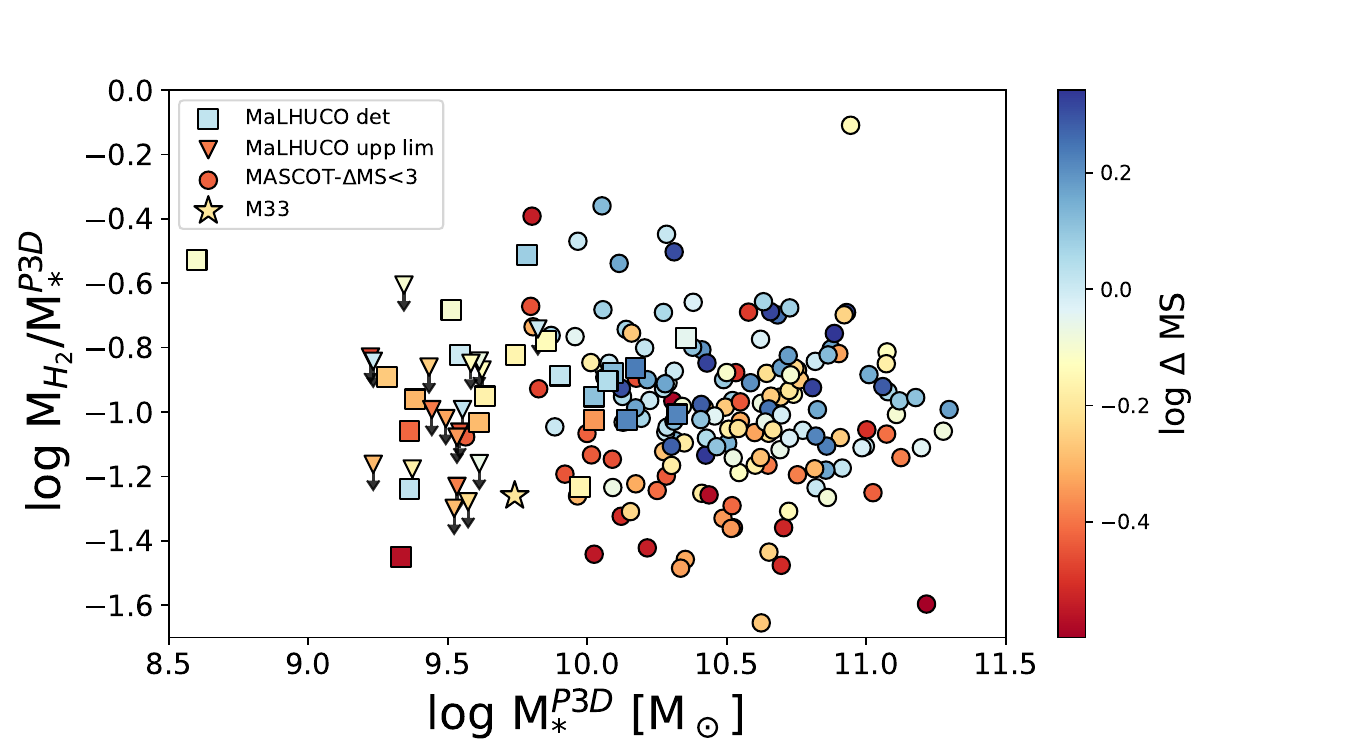}
\includegraphics[trim=0 0 30 38, clip, scale=0.45]{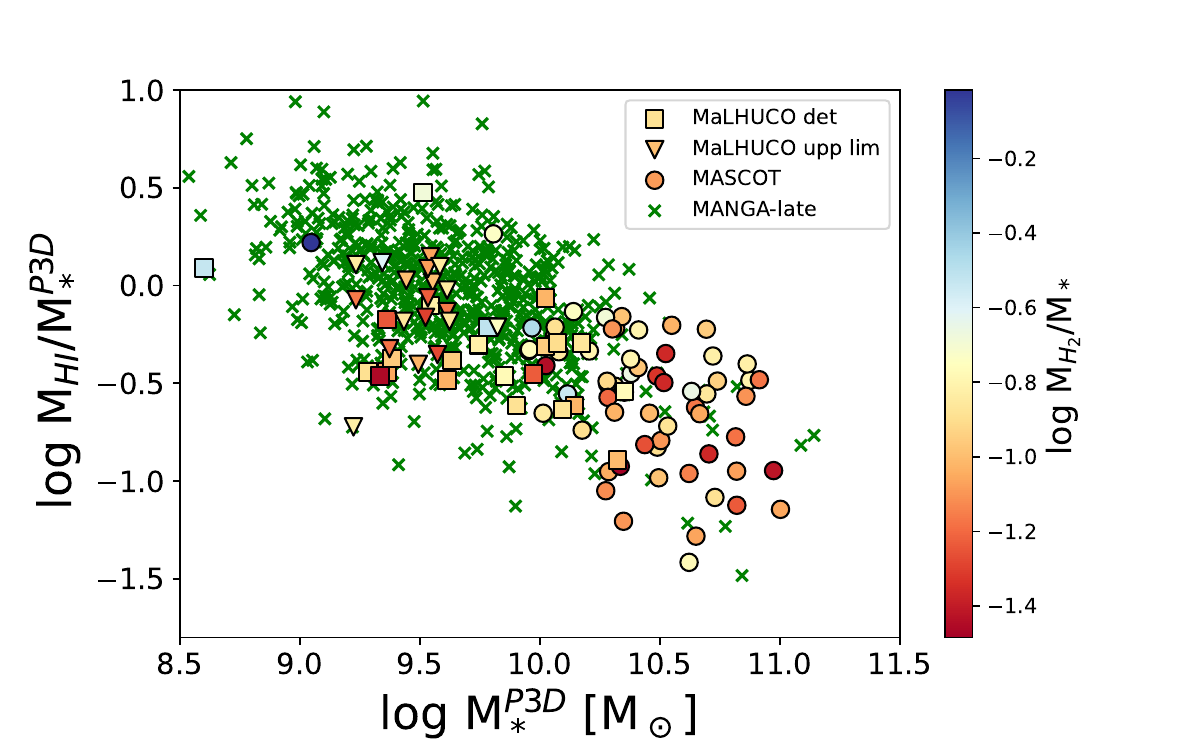}
\caption{Left: $M_{\rm H_2}$/$M_*^{\rm P3D}$ as a function of  $M_*^{\rm P3D}$ for MaLHUCO and MASCOT galaxies, color-coded by the offset from the main sequence, $\log \Delta$(MS). Right: $M_{\rm HI}/M_*^{\rm P3D}$ as a function of $M_*^{\rm P3D}$ for {\em MANGA-late},  MaLHUCO, and MASCOT galaxies
 with detections in the \hi-MaNGA catalog. 
Targets from the last two samples are color-coded by specific molecular mass ($M_{\rm H_2}/M_*$). Symbols are the same as in the previous figures.} 
\label{fig:07}
\end{center}
\end{figure*}

\subsection{Gas Fraction Scaling Relations}

The left panel of Fig. \ref{fig:07} shows the distribution of the specific molecular mass,  $M_{\rm H_2}/M_*$, as a function of $M_*$ with galaxies color-coded by their offset from the main sequence, $\Delta$(MS). $M_{\rm H_2}/M_*$ shows a large scatter: at a given stellar mass, this ratio can have up to one order of magnitude variation, showing no correlation with either $M_*$,  SFR,  or $M_{\rm HI}$.  This suggests that molecular formation is regulated by other factors, including environmental influences and local properties such as the local disk pressure \citep{2025A&A...700A..57C}. Although, on average, galaxies with a negative offset from the main sequence have a smaller specific molecular mass, there are exceptions. The $M_{\rm H_2}/M_*$ correlation with $\Sigma_{*, e}$ 
is also very weak (not shown here). M33 exhibits a very low specific molecular mass compared to the MaLHUCO sample.

In the right panel of Fig. \ref{fig:07}, we show the specific atomic-gas mass, or the \hi-to-stellar mass ratio, $M_{\rm HI}/M_*$, as a function of $M_*$, including the {\em MaNGA-late} sample with an \hi\ detection. MaLHUCO and MASCOT galaxies are color-coded by the specific molecular gas mass. Contrary to the $M_{\rm H_2}$/$M_*$ ratio, the trend here is clear: galaxies with a low stellar mass show a high $M_{\rm HI}/M_*$ and the ratio, on average, decreases as stellar and molecular mass increase \citep{2012ApJ...756..113H,2022ARA&A..60..319S}. 
Nonetheless, the color coding indicates that  $M_{\rm HI}/M_*$ shows no clear trend with $M_{\rm H_2}/M_*$.  

\begin{figure}
 \begin{center}
\includegraphics[scale=0.48]{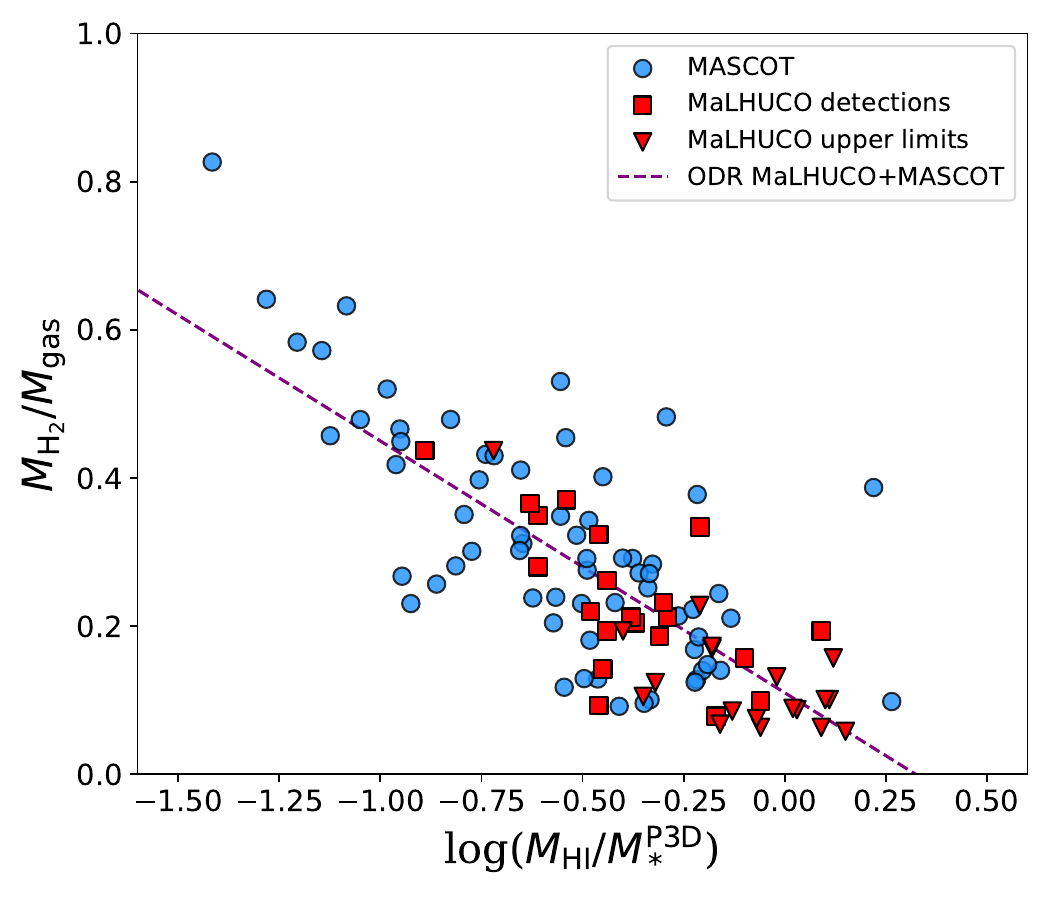}
\caption{$M_{\rm H_2}$/$M_{\rm gas}$ versus $M_{\rm HI/M_*^{\rm P3D}}$ for MASCOT and MaLHUCO galaxies
 with detections in the \hi-MaNGA catalog}. Symbols are the same as in the previous figures. The parameters of the best-fit relation (dashed line) are given in Table \ref{table:tabfit}.
\label{fig:08}
\end{center}
\end{figure}

The molecular fraction, defined as the ratio between the molecular gas mass and the total gas mass (\hi\ and H$_2$), $M_{\rm H_2}$/$M_{\rm gas}$, increases with both $M_*$ and SFR, but shows no clear trend with $M_{\rm HI}$. 
A way to normalize this relation is to consider how the molecular fraction varies with $M_{\rm HI}$/$M_*$, as shown in Fig. \ref{fig:08}. The figure indicates an increase in the molecular fraction with decreasing \hi-to-stellar-mass ratio. 
This underlines the role of the stellar gravitational potential in enhancing the gas density, favoring molecule formation. Galaxies with large \hi\ masses tipically host most of the excess atomic gas in the outer disk, where H$_2$ formation is inefficient. In the inner disk, \hi\ provides the shielding column density for molecule formation, particularly in metal-poor galaxies, 
due to their reduced dust content \citep{2025A&A...702A.264P}.  Gas-rich systems also exhibit lower molecular fractions because their low heavy-element content limits cooling and dust formation, both of which are closely linked to molecule formation. Such relation might be useful for estimating $M_{\rm H_2}$/$M_{\rm gas}$  in case  no CO measurements are available (Table \ref{table:tabfit}).

\subsection{Molecular mass scaling with gas attenuation}

\begin{figure*}
\begin{center}
\includegraphics[scale=0.55]{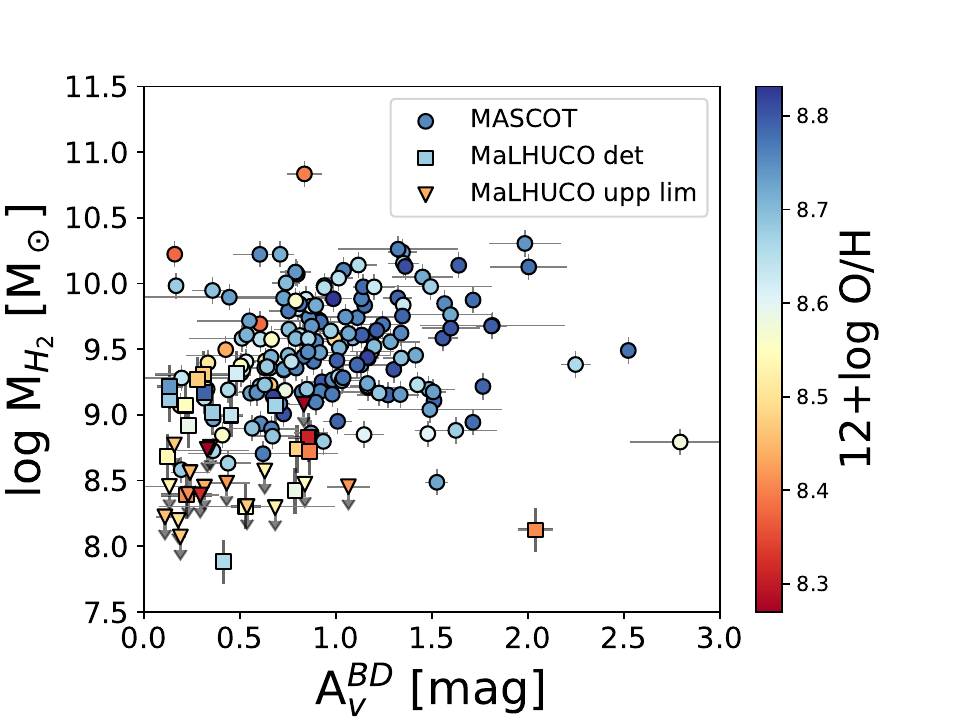}
\includegraphics[scale=0.55]{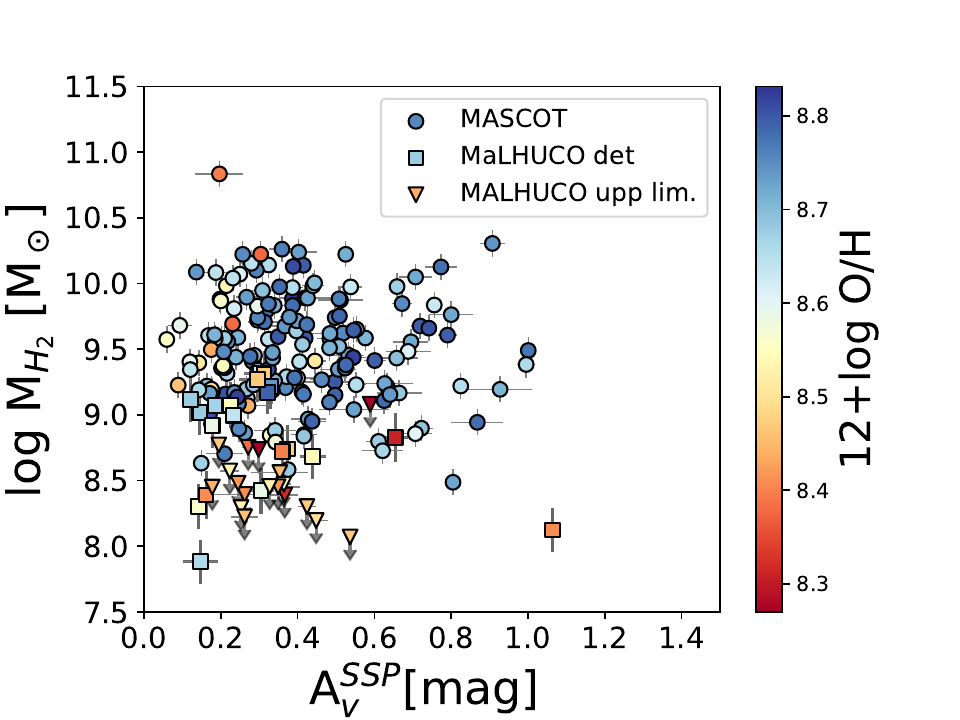}
\caption{$M_{\rm H_2}$ of MASCOT  and MaLHUCO samples as a function of the attenuation from the Balmer decrement and from SSP models, shown in the left and right panel, respectively. Data have been color-coded by the oxygen abundance. Symbols are the same as in the previous figures.}
\label{fig:09}
\end{center}
\end{figure*}

Dust acts as catalyst for molecule formation in the ISM of present-day galaxies and provides shielding against UV dissociating radiation. 
Scaling relations between dust and molecular gas masses are observed both globally and in spatially resolved regions of galaxies \citep{2012A&A...540A..52C,2020A&A...633A.100C,2021A&A...649A..18G,2025A&A...699A.346S,2025A&A...702A.264P}. Dust masses for MANGA galaxies  are not available, but we can use the visual attenuation $A_{\rm V}$ as a tracer of dust, and check whether it correlates with molecular masses. The MaNGA Pipe3D catalog provides both the attenuation from population synthesis model, $A_{\rm V}^{\rm SSP}$, and the gas attenuation from the Balmer decrement, $A_{\rm V}^{\rm BD}$, and here we use their mean values within $R_e$. Figure~\ref{fig:09} shows that only $A_{\rm V}^{\rm BD}$ establishes a mild correlation with $M_{\rm H_2}$ with a latent variable correlation $\rho=0.37$, as from Bayesian analysis. The color-coding of Fig.~\ref{fig:09} implies that galaxies with higher metal abundances also have higher extinction and  higher molecular masses. The different behavior in the two plots can be explained as following. Radiation from  H{\sc ii} regions is partially absorbed by dust associated with nearby molecular gas. On the other hand, the attenuation affecting the stellar population, traced by $A_{\rm V}^{\rm SSP}$, shows a much weaker correlation with the molecular gas content of the ISM. As shown by \citet{2025A&A...702A.264P}, only the attenuation of light from the young stellar population correlates with the Balmer decrement. $A_{\rm V}^{\rm BD}$ correlates more strongly with dust-related quantities such as $L_{12}$ ($\rho$=0.67) and $L_{22} \,$ ($\rho$=0.52) and with oxygen abundances ($\rho$=0.55) for both the MASCOT and the MaLHUCO sample.

\section{Discussion}

\subsection{Star formation drivers}


The SFR of today's galaxies  is related to the amount of cold gas present in the ISM, and to its capacity to cool and fragment. Molecular clouds are an intermediate phase between the cold atomic gas in the ISM and the gravitational bound gas fragments which collapse to make stars. 
Disk gravity is balanced by the ISM pressure, such that the fraction of the ISM in molecular form depends on the local stellar and gas surface densities
\citep[e.g.][]{1993ApJ...411..170E,2025A&A...700A..57C,Elmegreen26}. The efficiency of molecules to form stars,  is small and it does not vary much from galaxy to galaxy. Hence, the star formation rate is linked to the amount of molecular gas in a galaxy through the KS law \citep{1998ApJ...498..541K}. As massive stars are born, they heat and ionize the ISM creating turbulence and shocks which trigger new events of star formation and support the ISM against gravitational collapse. A link between the cold molecules and other warm/hot ISM components is then expected because both phases are connected with the star formation cycle. Hot dust as well as PAH, for example, correlate locally with the presence of young stars and molecules \citep{2004ApJ...613..986P,2006ApJ...646..192J,2017ApJ...850...68C,2019MNRAS.482.1618C,2024A&A...684A..71U}. We have seen in this paper the strong correlation between WISE band emission at 12~$\mu$m (which has PAH as well as hot dust contributions) and the CO luminosity.

\begin{figure}
 \begin{center}
\includegraphics[scale=0.43]{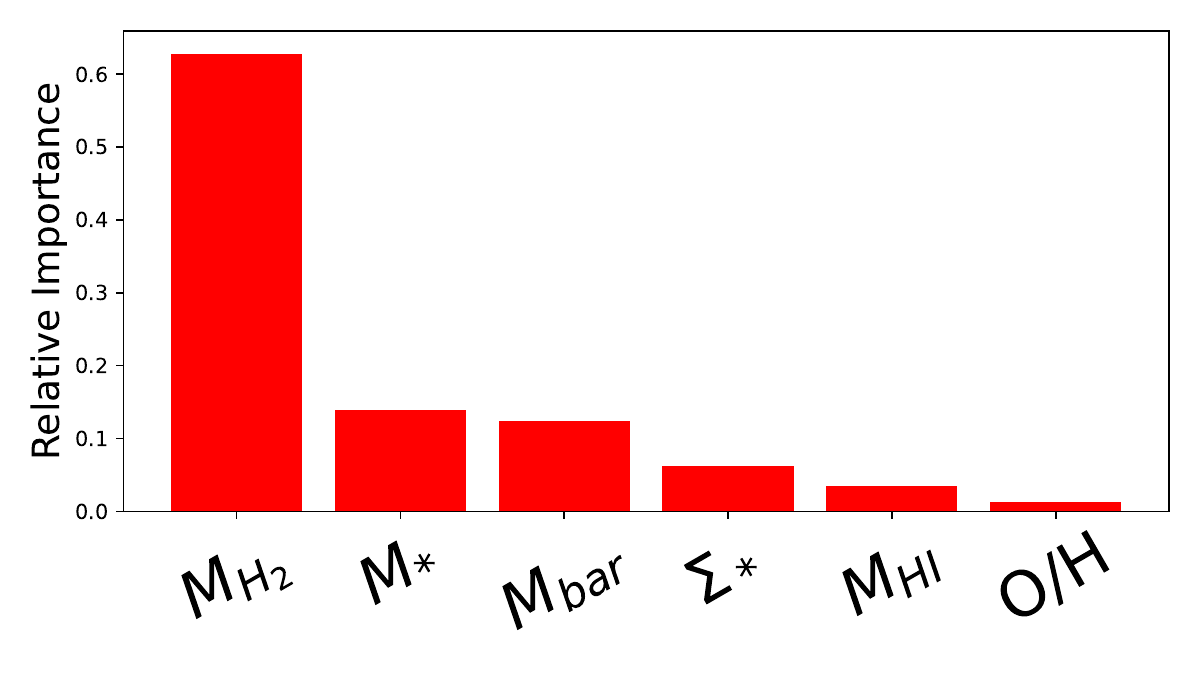}
\caption{Relative importance of galaxy parameters derived from the RF regression used to predict the global SFR in the combined MASCOT and MaLHUCO sample.}
\label{fig:10}
\end{center}
\end{figure}

Given the star formation cycle, one may ask which is the key ingredient which drives the rate of star formation for our low-mass disk sample. 
Scaling relations are often used to understand the trigger at the heart of a physical process. However, disentangling fundamental correlations from those that arise as by-products requires statistical methods which take into account the relative importance of variables, scatter, correlation ranking, etc. Three-dimensional correlations, like the one shown in Fig.~\ref{fig:05}, are also often used to infer if one relation results from a combination of other more fundamental relations, e.g. the MS in relation to the MGMS and KS relation \citep{2019ApJ...884L..33L}. In addition, non-linear trends, which might set up as the galaxy masses extend beyond a given mass interval, can further limit conclusions based on the scatter around a liner fit.

For these reasons, we use a random forest analysis (RF) to assess the relative importance of input variables in driving the SFR, some of which might be combinations of others.
Taking into account all galaxies from the combined sample (MaLHUCO+MASCOT), for which we have an estimate of the molecular and atomic masses, we run an RF regression considering global quantities such as $M_{\rm H_2}$, $M_{*}$, $M_{\rm HI}$, $M_{\rm bar}$ = $M_{\rm HI}$ + $M_{\rm H_2}$ + $M_{*}$, but also average quantities across the disk such as O/H, $\Sigma_*$.  We run the analysis for the full sample and find that the relative importance of molecular gas mass is as high as 60$\%$ in driving star formation, as shown in Fig. \ref{fig:10}. This implies that the KS relation is globally the most fundamental. 

Similarly, if we run a RF regression to assess whether the atomic gas mass or the stellar mass drives the amount of molecular hydrogen in a galaxy, we find that for the combined sample analyzed here, the relative importance of the stellar mass is as high as 80$\%$. Our results are therefore consistent with those presented by \citet{2023MNRAS.518.4767B} for larger samples and  underline the role of the stellar potential in driving spiral arms and ISM compression in galaxy disks, thereby enhancing molecule formation. Since we do not spatially resolve the MaLHUCO galaxies, we cannot consider additional variables such as the local disk hydrostatic pressure as potential drivers of star formation. Spatially resolved analyses have in fact shown that the combination of gas and stellar potential is more strongly linked to the SFR density than the local molecular gas surface density \citep[e.g.][]{2024MNRAS.52710201E, 2025A&A...700A..57C}, underlining the presence of diffuse, non-star forming molecular gas in the ISM.

\subsection{Variations in Scaling Relations: A Comparative Analysis}

We have seen in this work that the MS for MANGA-late galaxies is steeper than that found by \citet{2018MNRAS.477.3014B} for star forming MANGA galaxies, with a slope close to unity. Our targets are drawn from the MANGA-late sample, although most of the CO-detected galaxies lie at low end of the $M_{\rm HI}/M_*$ distribution. In fact, we have shown that the molecular gas fraction increases as $M_{\rm HI}/M_*$ decreases. The lack of resolved H{\sc i} maps does not allow us to measure the H{\sc i} mass and surface densities within the optical disk but, as shown by \citet{2025PASA...42...46L}, the mean H{\sc i} surface density within the optical disk increases as the stellar mass decreases (although the relation shows a large scatter). Less evolved galaxies with a high gas-to-stellar mass ratio might therefore still have sufficient gravitational potential from the gas and stellar disk to cool the gas and form molecules. Their lower molecular fraction is then likely due to reduced cooling efficiency associated with lower dust and metal content. In addition, these galaxies lack the large overdensities associated with prominent spiral arms, which are typically found in earlier morphological types, and at higher stellar surface density. 

The scaling between the molecular gas mass and the stellar mass, the MGMS, for the MaLHUCO sample is compatible with that for MASCOT galaxies when the appropriate CO-to-H$_2$ conversion factor is adopted. The  mean sSFR, of the order of 0.1~Gyr$^{-1}$, is higher for MaLHUCO than for MASCOT galaxies. Given the correlation between sSFR and $\tau_{\rm H_2}$, the molecular gas depletion time for MaLHUCO galaxies is on average slightly shorter than that of more massive galaxies. The typical depletion time scale for molecular gas in star-forming spiral galaxies in the local Universe is around 1 - 2  Gyr \citep{2011ApJ...730L..13B,2013AJ....146...19L,2017ApJ...849...26U}, consistent with what we find in this work. 
We found a mild positive correlation between $\tau_{\rm H_2}$ and $M_*$, although characterized by a large scatter, a trend that has been reported in previous works   
\citep{2011MNRAS.415...61S,2014MNRAS.443.1329H,2020A&A...643A.180H, 2022ARA&A..60..319S,2024A&A...687A.244H}, especially for MS galaxies.
However, some authors also report no statistically significant correlations between these two parameters 
\citep{2014A&A...564A..66B,2017MNRAS.470.4750A}. The absence of a relation between $\tau_{\rm H_2}$ and $M_*$ would indicate that star formation proceeds at a universal rate in galaxies, while a correlation might  be a consequence of global galaxy properties shaping the properties of molecular clouds, altering the fraction of dense gas available for star formation  \citep{2020ApJ...901L...8S,2022ARA&A..60..319S}. Given the uncertainties in the CO line ratio $R_{21}$, and in the molecular mass estimates due for example to displacements of the bulk of molecular gas from the beam center,  and given the small difference in the mean $\tau_{\rm H_2}$ between the  MaLHUCO and MASCOT samples, we cannot strongly favor either interpretation. 

This small difference in $\tau_{\rm H_2}$ implies that our galaxies lie above the global KS scaling relation of the combined sample (MaLHUCO+MASCOT; Fig. \ref{fig:05}, left panel). Nonetheless, they follow the same linear trend observed for the global KS relation of the combined sample.
A linear KS law has also been found for the combined ALLSMOG and xCOLDGASS samples by \citet{2024A&A...687A.244H} considering SFR along the x-axis and $M_{\rm H_2}$ along the y-axis.

The lack of a correlation between M$_{H_2}$/M$_*$ and M$_*$ and its large scatter, is in agreement with the weak or absent correlation reported in other studies of low mass galaxies  \citep[e.g.][]{2017ApJS..233...22S,2018RMxAA..54..443C,2020A&A...643A.180H}. Possible selection biases at low stellar masses leave the situation still uncertain in the dwarf regime.

Lastly, an additional source of uncertainty in the scaling relations is the adopted stellar-mass estimates. As shown in Fig. \ref{fig:compare_mstar}, $M_*^{\rm P3D}$ are systematically higher than $M_*^{\rm SED}$ and, to a lesser extent,  than the MPA-JHU estimates, with the offset increasing towards the low-mass end. As a consequence, the trends of $\tau_{\rm H_2}$ with sSFR (Fig. \ref{fig:06}, bottom panel) or of the gas fraction with $M_{\rm HI}/M_*$ (Fig. \ref{fig:08}) appear steeper when based on $M_*^{\rm P3D}$, since low-mass systems are displaced towards intermediate masses.
On the other hand, the dependence of $\tau_{\rm H_2}$ on $M_*$ (Fig. \ref{fig:06}, top panel) becomes shallower when using $M_*^{\rm P3D}$. By contrast, the KS relation remains essentially unchanged, as $M_{\rm H_2}$ estimates (Eq. \ref{eq:varco}) depend only weakly on the adopted stellar mass. Overall, although the choice of the stellar-mass catalog modifies the slope and normalization of the correlations involving $M_*$, the main trends discussed in this work are preserved.

\section{Summary and Conclusions}

We selected a sample of late-type galaxies from the MaNGA survey to search for $^{12}$CO(J=2-1) line emission. Galaxies were selected with morphological type later or equal than Scd (6 $\leq$ $T$ $\leq$ 11), stellar masses  in the range $M_*^{\rm SED} \lesssim 10^{9.7}$ M$_{\odot}$ (or $M_*^{\rm P3D} \lesssim 10^{10.3}$ M$_{\odot}$ if stellar masses are taken from the MaNGA Pipe3D catalog). Galaxies in the MaLHUCO sample span a range of properties typical of low-mass disk galaxies such as M33 in the Local Group. This work extends the analysis of molecular gas scaling relations to a lower stellar-mass regime.

CO emission was detected in 23 out of 42 targets (55\%). Two additional sources show tentative CO detections, which were not considered in the analysis presented here. 
$M_{\rm H_2}$ ranges between $10^{7.2}$ \msun\ and $10^{9.3}$ \msun\  or $ 10^{7.9}$ \msun\ and  $10^{9.6}$ \msun, whether we use the galactic or the CO-to-\hdue\ metallicity-dependent conversion factor \citep{2017MNRAS.470.4750A}. Because the metallicity range of our sample does not extend to very low values, with an average of 12 + log(O/H) $\sim$ 8.4, 
the impact of metallicity  variations in $\alpha_{\rm CO}$ is still not significant for the determination of the \hdue\ masses.
As found in previous studies, the CO detection rate increases with stellar mass, metallicity, stellar surface density, and PAH luminosity. 

We analyzed scaling relations by complementing our sample with data from the MASCOT survey, which targets more massive and earlier-type galaxies from the MaNGA survey.
We find a strong correlation between CO luminosity and WISE 12 $\upmu$m and 22 $\upmu$m luminosities. For $L_{12}$, the relation is tighter and approximately linear over 3 orders of magnitude, when only detected galaxies are considered. 
This implies that PAHs are good tracers of molecular gas in galaxies \citep{2023ApJ...944L..11C,2024A&A...691L...2S}, confirming the potential of $L_{12}$ for predicting CO brightness in star-forming systems. However, because some galaxies undetected in CO have $L_{12} \lesssim  10^{42}$ L$_{\odot}$, it is not possible to asses whether this scaling effectively applies at the very low-luminosity regime. Here deviations at the faint end are possible driven by PAH destruction and CO dissociation at low metallicities and stellar masses. More sensitive observations are required to address this issue in a more definitive way. On the other hand, the relation between attenuation  derived from the Balmer decrement and $M_{\rm H_2}$ shows  substantially larger scatter than the $L_{\rm CO} - L_{12}$ correlation.

The MGMS  and the KS law show a scaling close to linear for the combined low-mass (MaLHUCO) and high-mass (MASCOT) samples. On global scales, the KS relation is the most fundamental, as suggested by the RF regression analysis. The molecular gas depletion time scale exhibits a weak dependence on $M_*$ and shows a modest decrease at lower stellar masses and later morphological types.   
This is consistent with the higher sSFR observed in our sample.
Moreover, $\tau_{\rm H_2}$ is strongly anti-correlated with sSFR, 
in agreement with previous CO surveys \citep{2013AJ....146...19L,2014A&A...564A..66B,2014MNRAS.443.1329H,2016MNRAS.462.1749S,2020A&A...643A.180H}.  
This confirms that the level of star-formation activity in galaxies is determined by both the
amount of molecular gas available and the efficiency of the conversion of gas into stars \citep{2022ARA&A..60..319S}.

The specific molecular gas mass has a large dispersion and shows no clear correlation  with $M_*$,  
contrary to $M_{\rm HI}/M_*$. 
Similar trends for $M_{\rm H_2}/M_*$ and $M_{\rm HI}/M_*$  are found in the literature \citep[][and the xCOLDGASS sample]{2024A&A...687A.244H}  and they are also predicted by semi-analytical models  \citep{2012MNRAS.424.2701F} for galaxies with $M_* < 10^{10.5}$ \msun. At higher stellar masses a drop-off in the specific molecular mass is expected by simulations and is sometimes observed, depending on sample selection \citep[e.g.][]{2018RMxAA..54..443C}.

The molecular fraction, $M_{\rm H_2}/M_{\rm gas}$ is tightly anti-correlated with $M_{\rm HI}/M_*$. %
This scaling relation may be used as a proxy for the molecular content of galaxies, especially given the availability of wide-area 21-cm surveys. It is likely a reflection of the tendency of the molecular-to-atomic gas ratio to correlate with $M_*$, as $M_{\rm HI}/M_*$ decreases with stellar mass \citep{2018MNRAS.476..875C,2020A&A...643A.180H,2022ARA&A..60..319S,2024SCPMA..6799811Y}. Moreover, galaxies with low stellar masses have lower metal content and therefore need larger \hi\ surface densities to effectively shield molecules, leading to a further decrease in the molecular fraction with increasing $M_{\rm HI}/M_*$.

Future high-resolution and high-sensitivity CO observations with ALMA will spatially resolve the molecular gas distribution  in the MaLHUCO sample, enabling the study of the KS and $\tau_{\rm H_2}$ scaling relations for low mass systems on kpc/sub-kpc scales. Such observations, combined with MaNGA data sets, will provide a critical benchmark to assess whether there are genuine changes in local star formation physics in low-mass disk galaxies and to identify what drives the well-known properties of our closest low mass disk galaxy, M33. 

\begin{acknowledgements} 
We thank the anonymous referee for constructive comments and suggestions that helped to improve the paper.

MG acknowledges support from FAPERJ grant E-26/211.370/2021. DRG acknowledges FAPERJ (E-26/211.370/2021; E-26/211.527/2023) and CNPq (403011/2022-1; 315307/2023-4) grants. EC acknowledges financial support from INAF-Mini Grant RF-2023-SHAPES. EDP acknowledges  support from INAF-Mini Grant RF-2024 "Dust emission and optical extinction as gas tracers in star forming galaxies”.

These observations were obtained by the James Clerk Maxwell Telescope, operated by the East Asian Observatory on behalf of The National Astronomical Observatory of Japan; Academia Sinica Institute of Astronomy and Astrophysics; the Korea Astronomy and Space Science Institute; the National Astronomical Research Institute of Thailand; Center for Astronomical Mega-Science (as well as the National Key R\&D Program of China with No. 2017YFA0402700). Additional funding support is provided by the Science and Technology Facilities Council of the United Kingdom and participating universities and organizations in the United Kingdom and Canada.

The authors wish to recognize and acknowledge the very significant cultural role and reverence that the summit of Maunakea has always had within the indigenous Hawaiian community.  We are most fortunate to have the opportunity to conduct observations from this mountain.

\end{acknowledgements}

\bibliography{Manga_rev_dwarfs}{}
\bibliographystyle{aasjournalv7}


\begin{appendix}

\section{$^{12}$CO(J=2-1) spectra of target galaxies} \label{app:B}

In this section we present the $^{12}$CO(J=2-1) spectra of the MaLHUCo galaxies, obtained at the JCMT telescope at a spectral resolution of 10~km s$^{-1}$. In Fig.~\ref{fig:CO_spectra} the detected signals are highlighted in yellow, while marginal detections are displayed in light blue and were not included in our analysis. The vertical red line indicates the optical velocity, $V_{\rm opt}$, for each target. To the left of each spectrum, we show the SDSS $gri$ image with the FWHM of the JCMT beam and the MaNGA hexagonal field of view overlaid.

\begin{figure*}[ht]
\begin{center}
\includegraphics[scale=0.19 ]{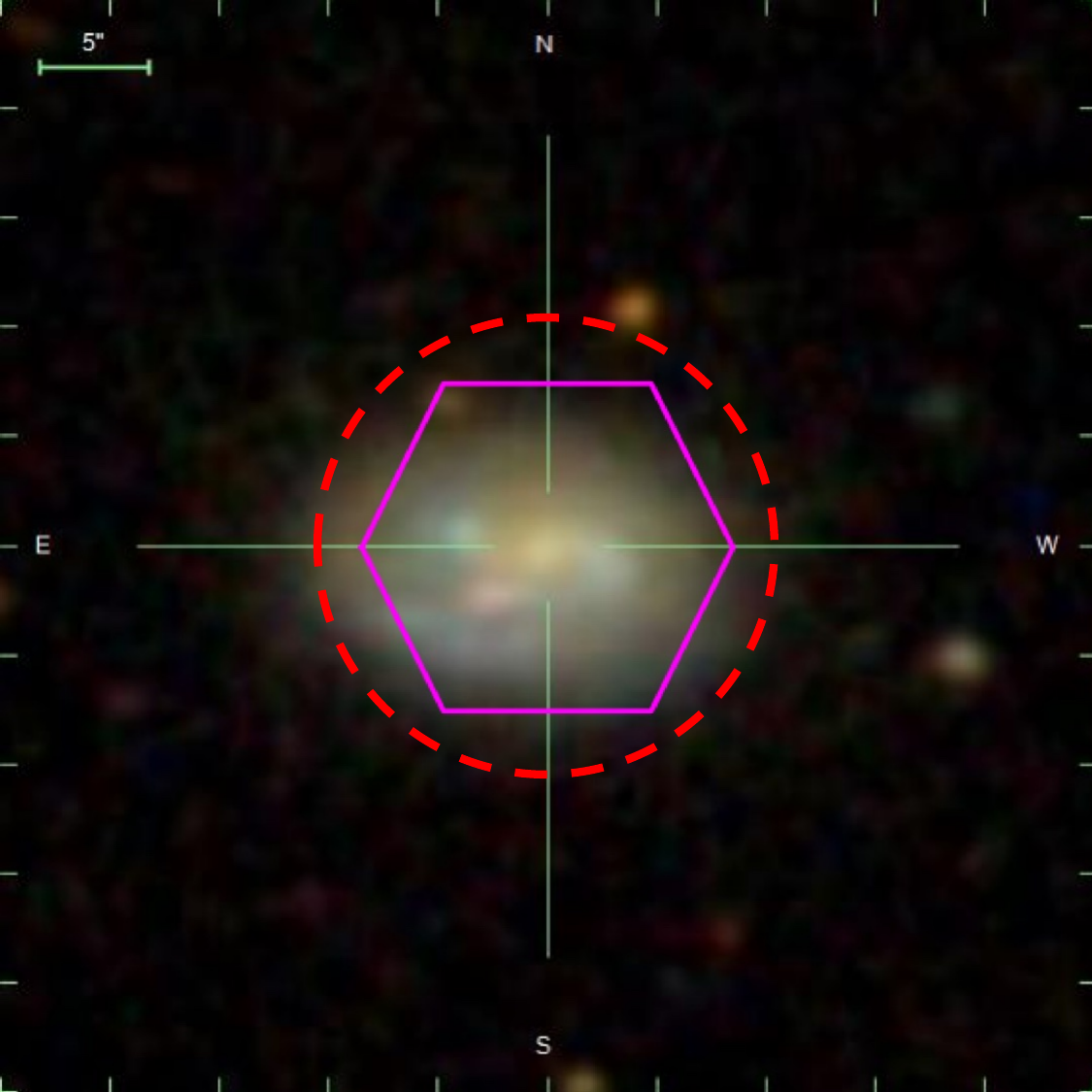}
\includegraphics[bb=5 5 495 405, scale=0.27]{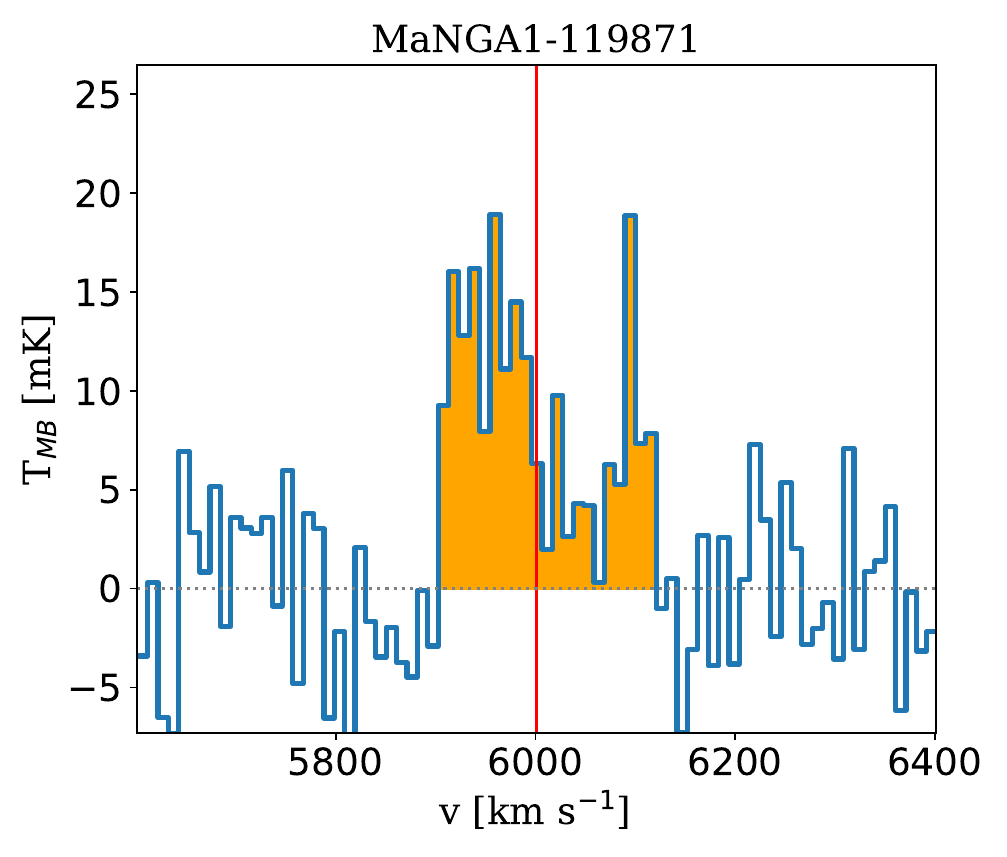}
\includegraphics[scale=0.19 ]{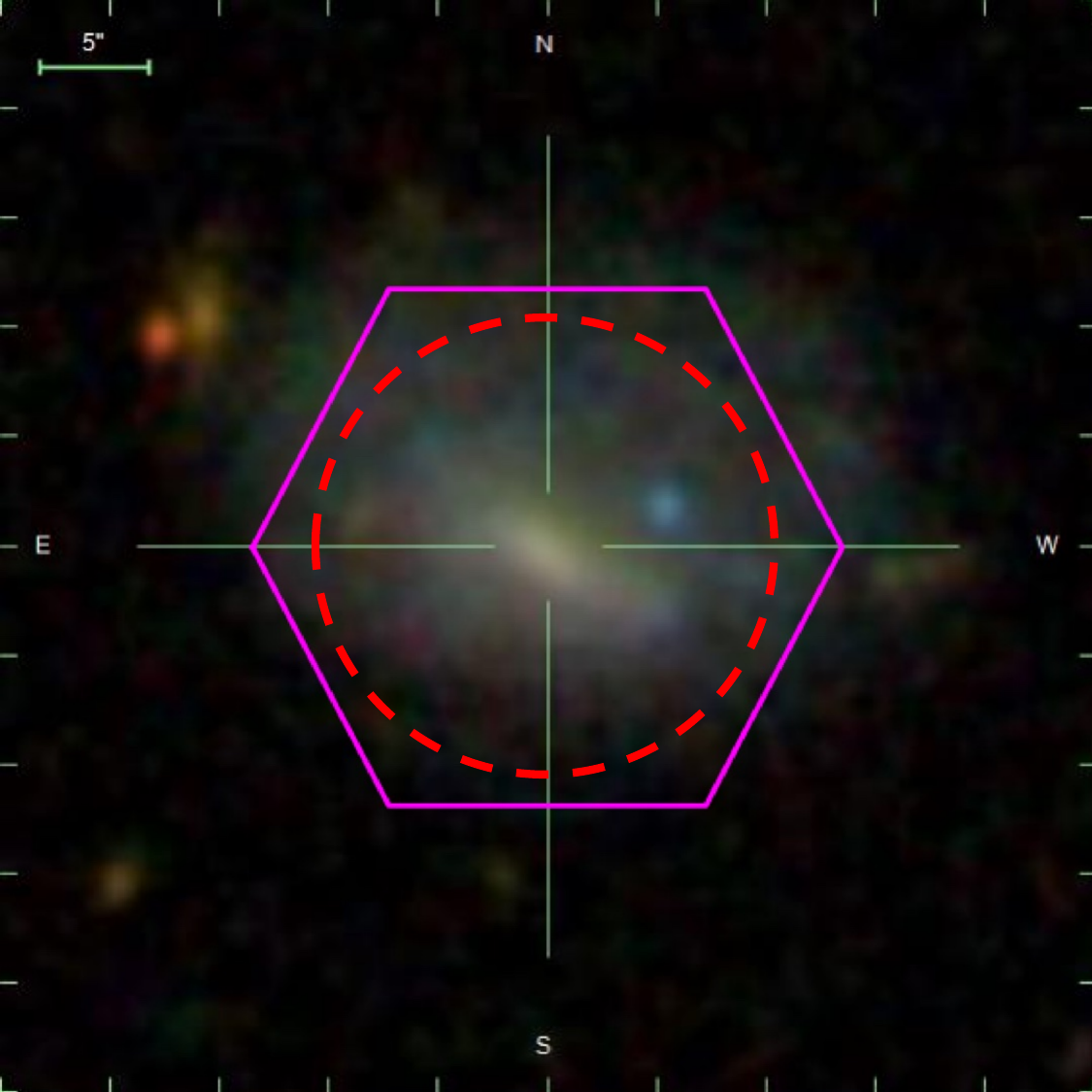}
\includegraphics[bb=5 5 495 405, scale=0.27]{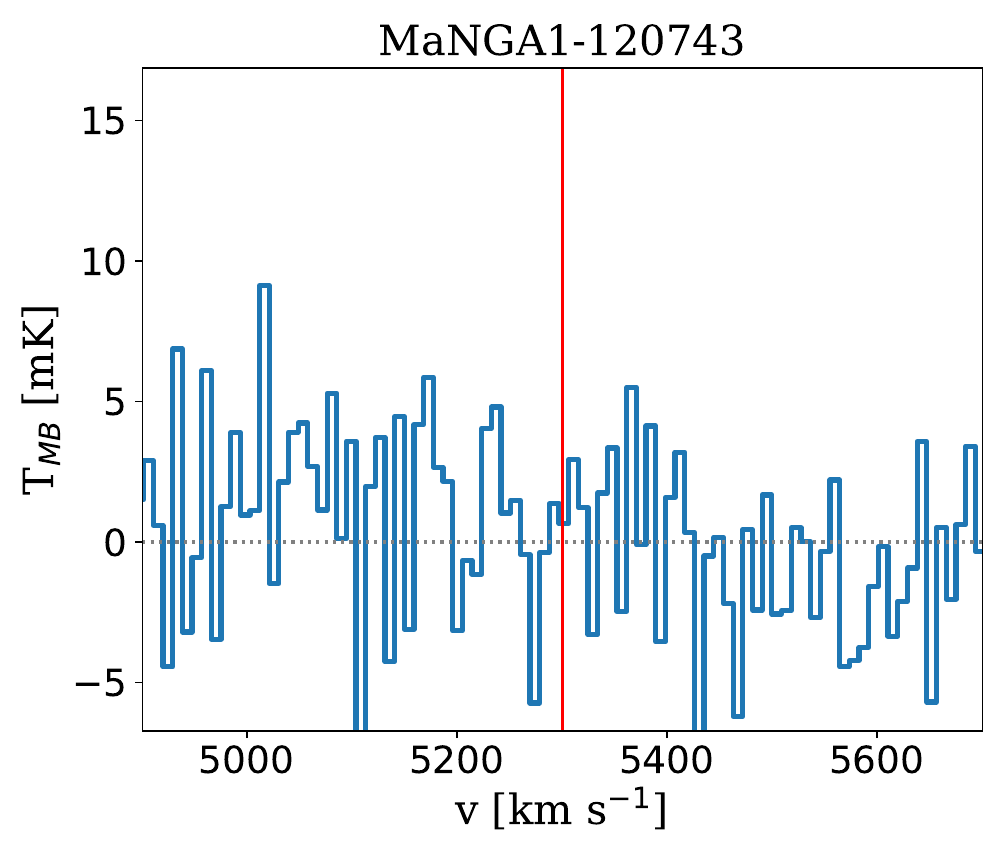}
\includegraphics[scale=0.19]{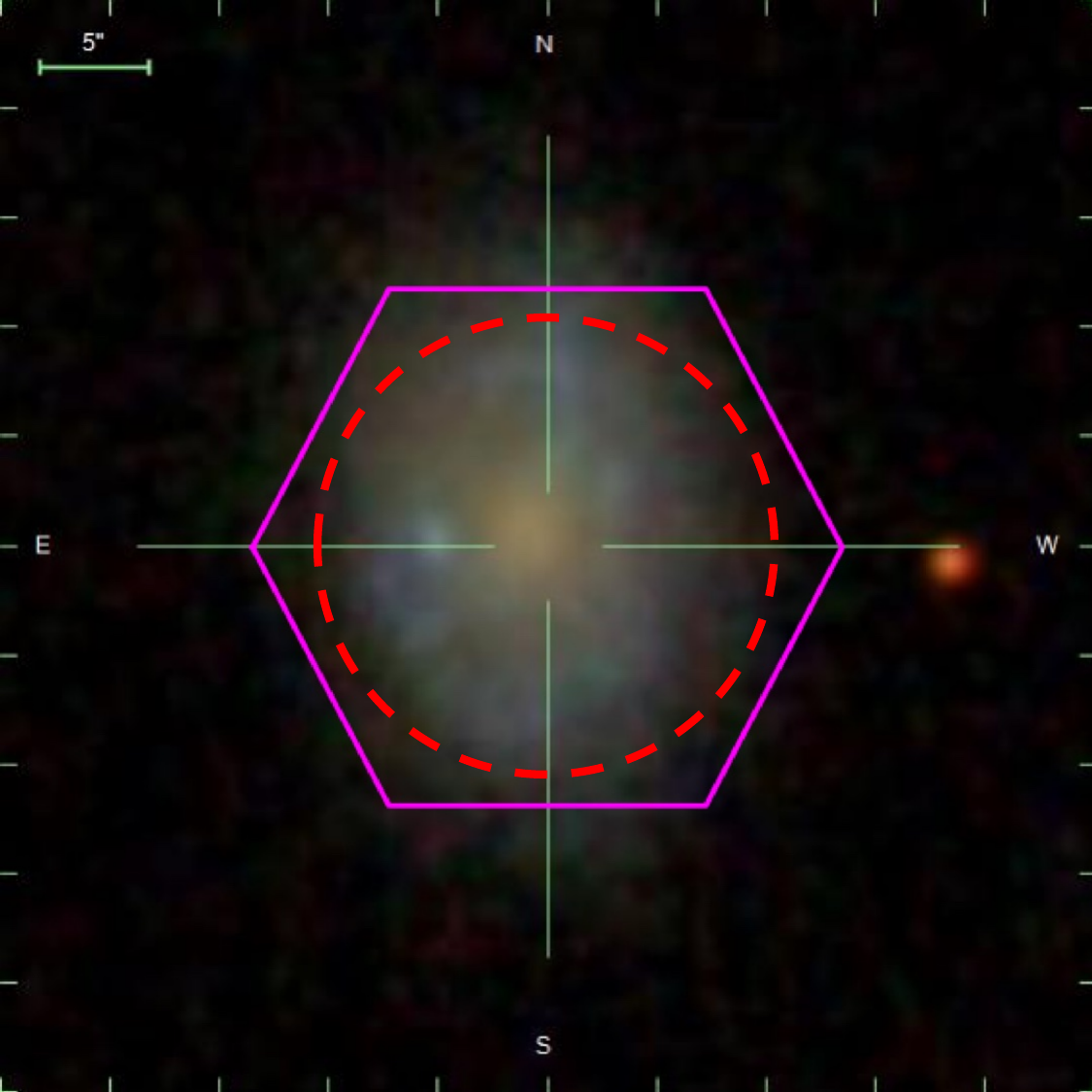}
\includegraphics[bb=5 5 495 405, scale=0.27]{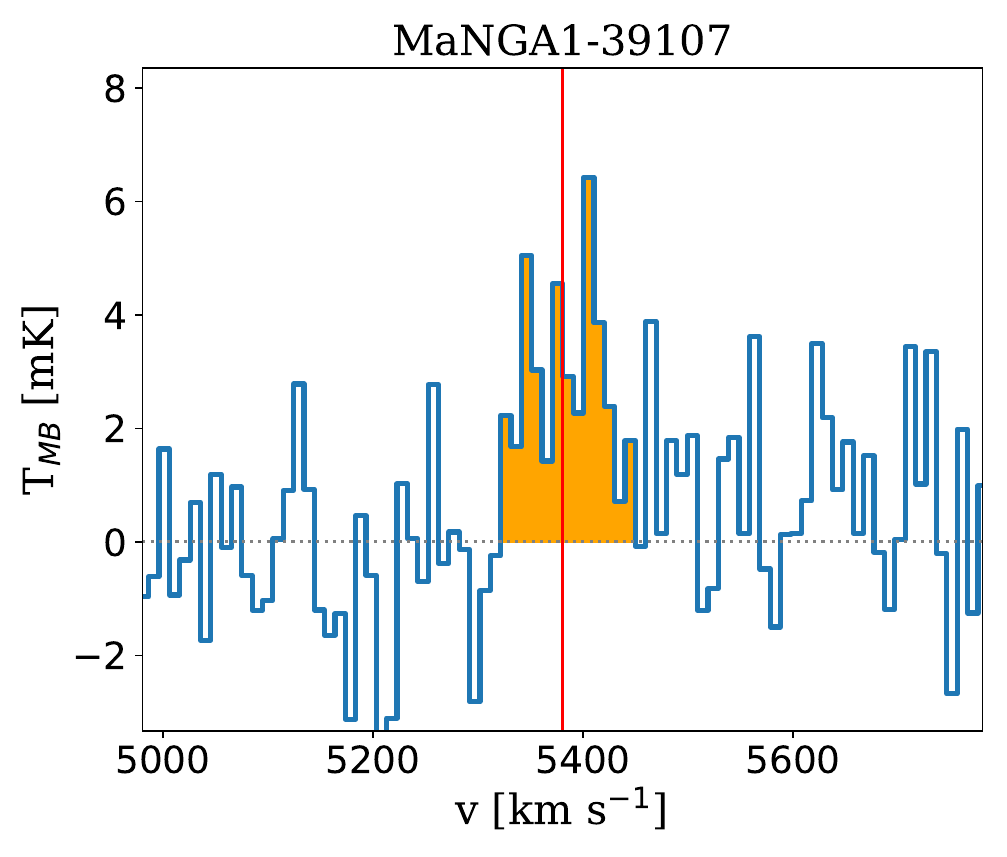}
\includegraphics[scale=0.19]{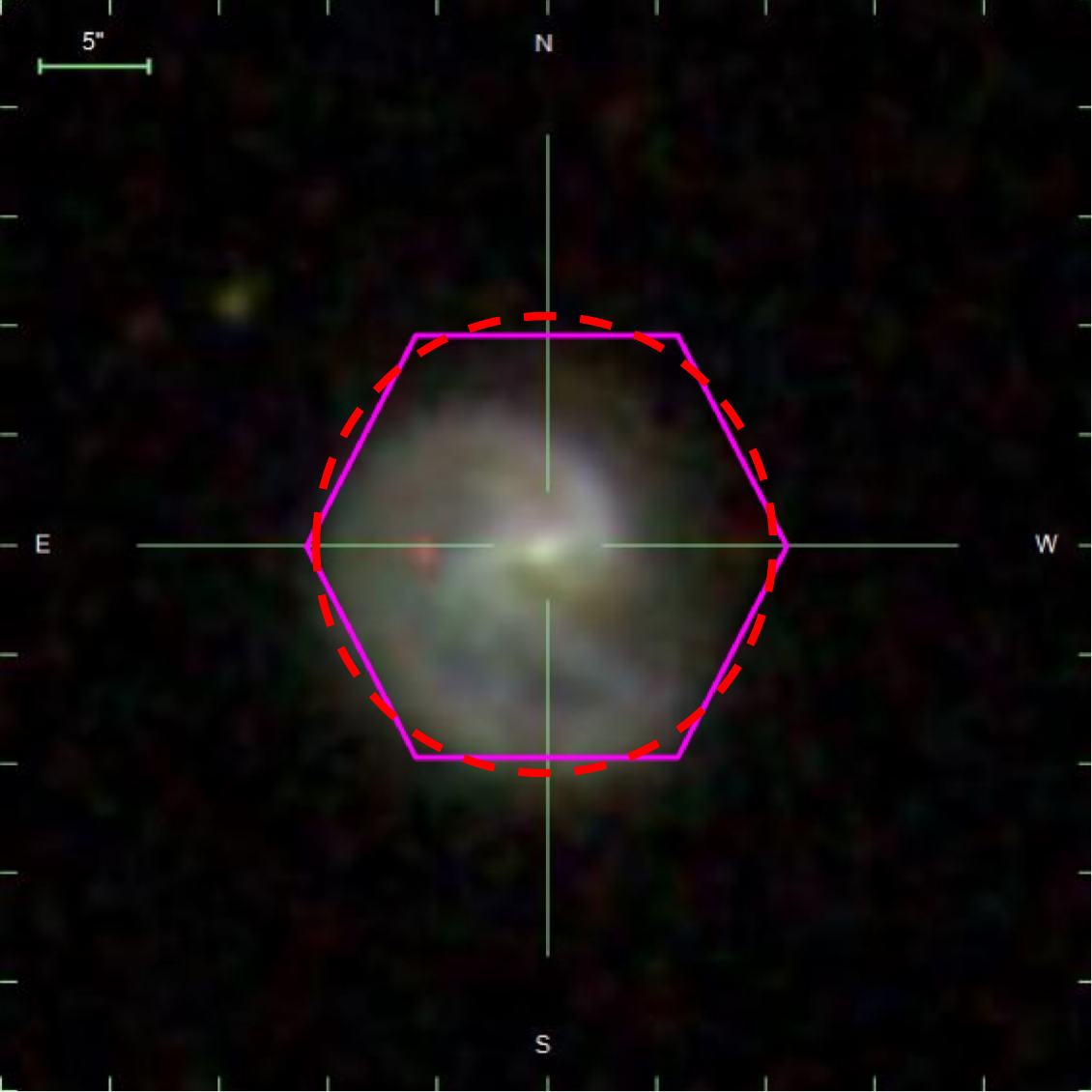}
\includegraphics[bb=5 5 495 405, scale=0.27]{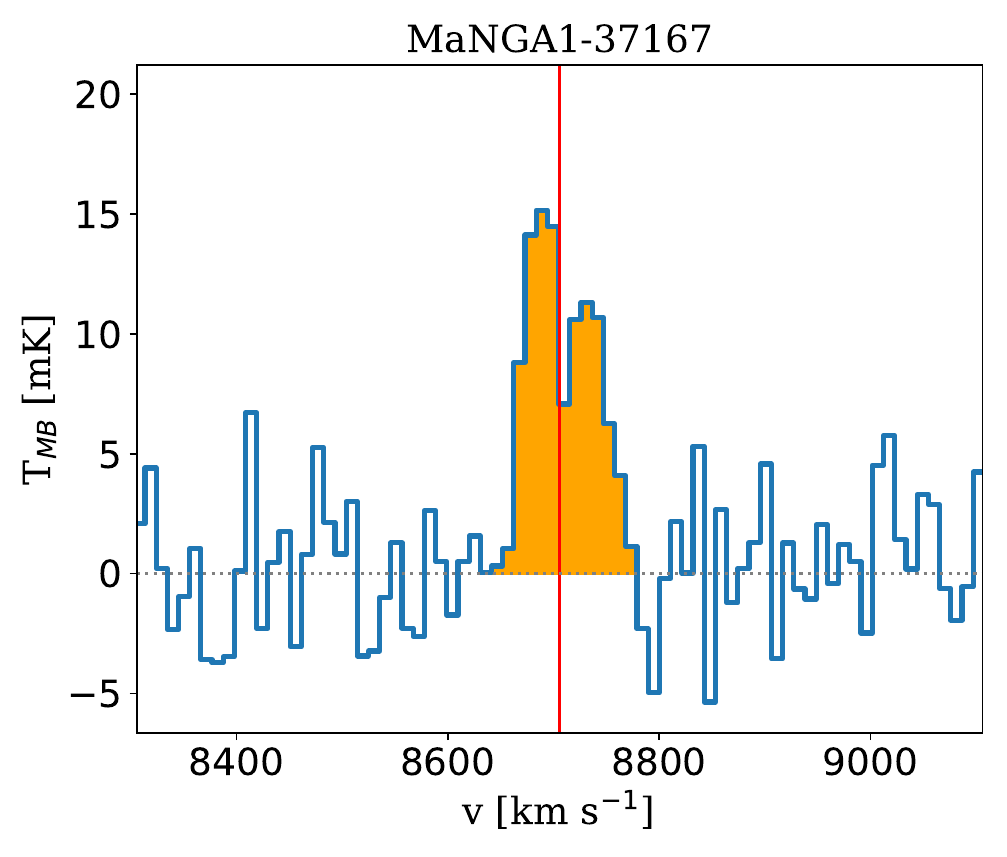}
\includegraphics[scale=0.19]{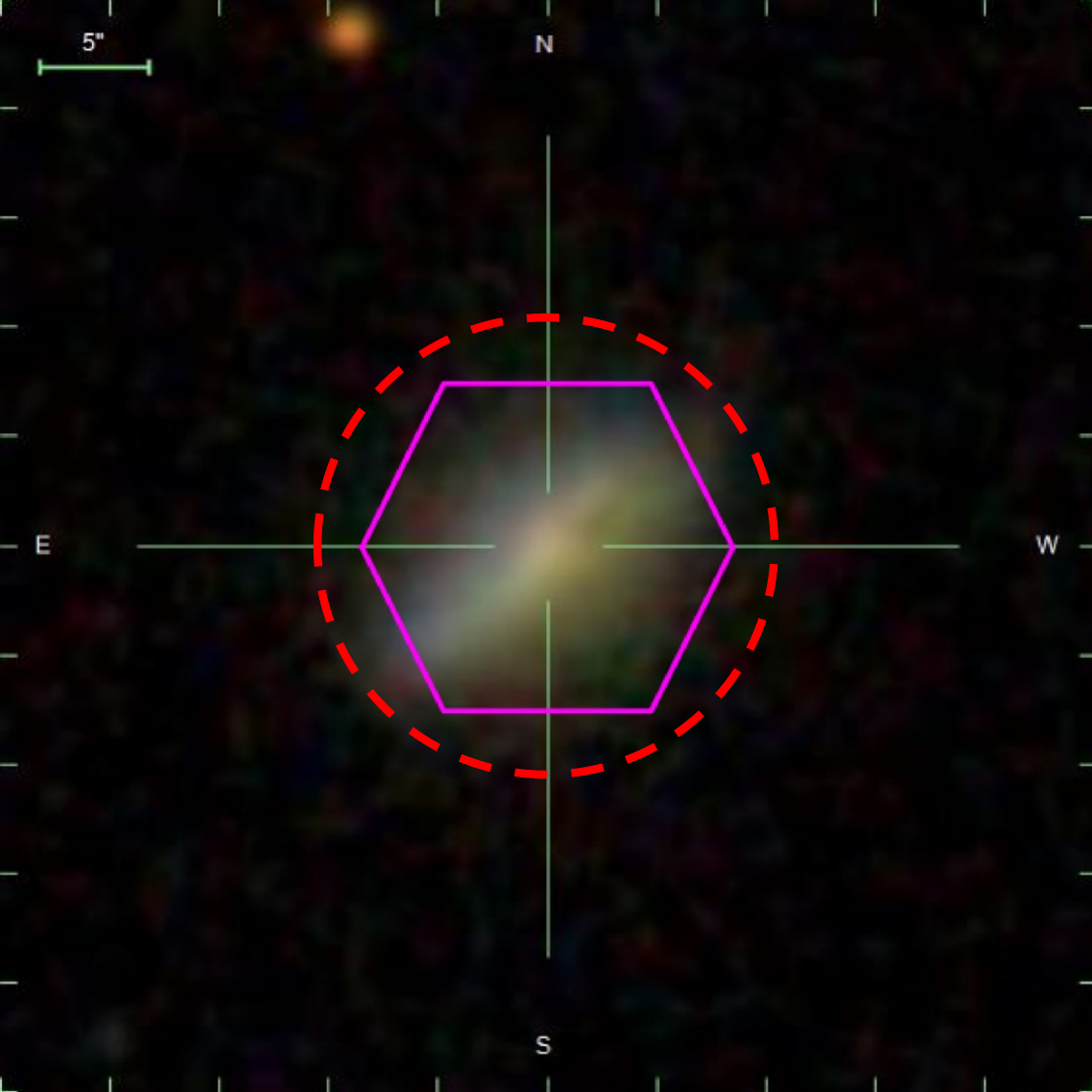}
\includegraphics[bb=5 5 495 405, scale=0.27]{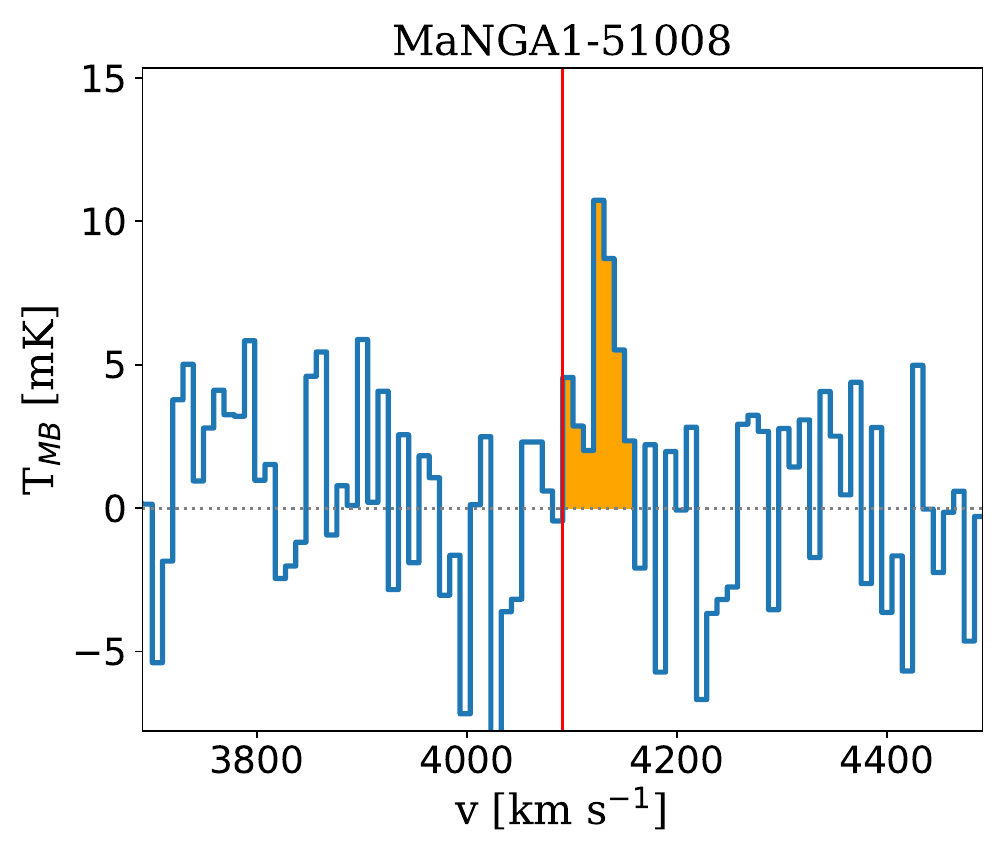}
\includegraphics[scale=0.19]{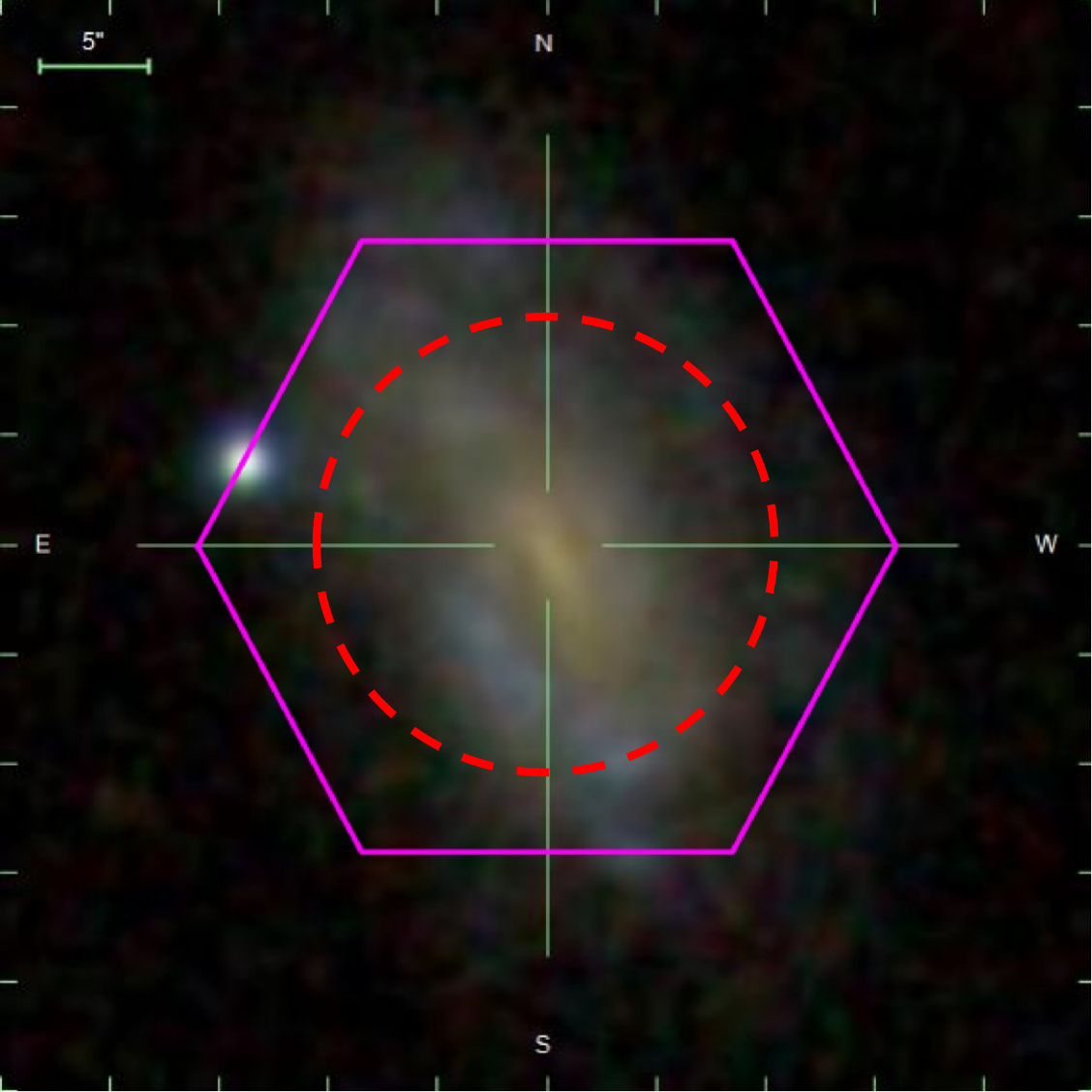}
\includegraphics[bb=5 5 495 405, scale=0.27]{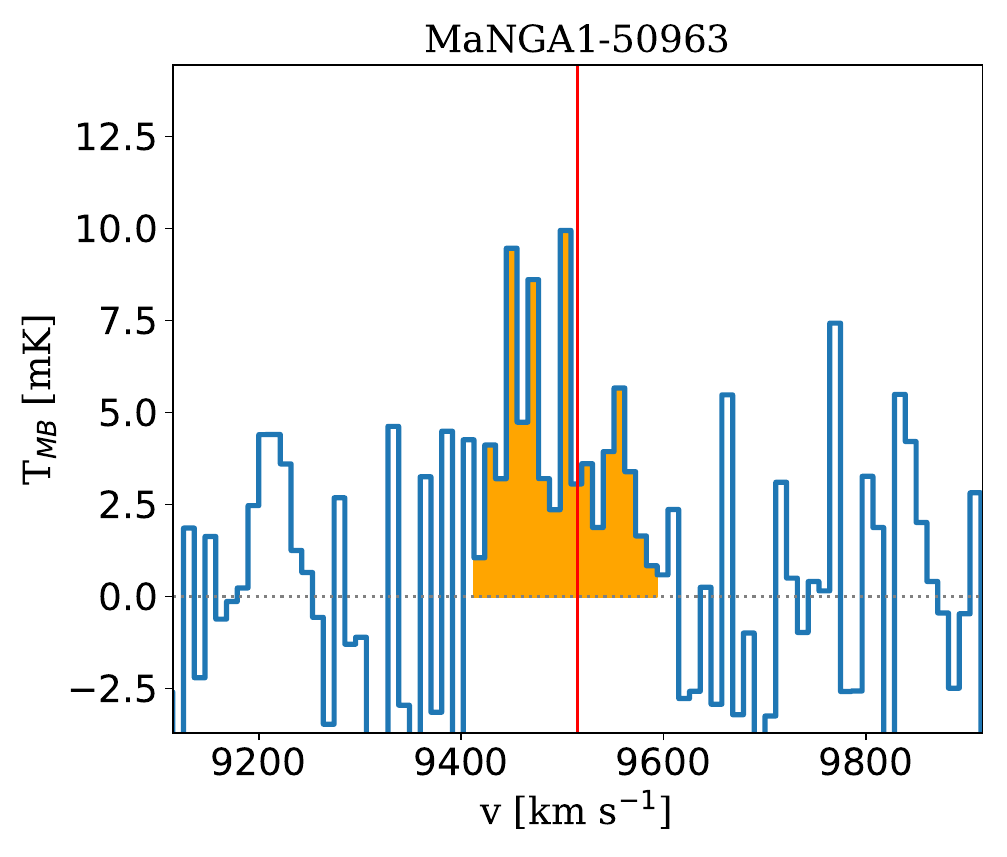}
\end{center}
\caption{Left: SDSS $gri$ image of a MaLHUCO galaxy, with the hexagonal MaNGA field of view and the FWHM of the JCMT beam at 230 GHz overlaid in magenta and red, respectively.
Right: the corresponding observed $^{12}$CO(J=2-1) spectrum. Detected signals are highlighted in yellow, while marginal detections, not included in our analysis are shown in light blue. The vertical red line indicates V$_{opt}$ for each target. This layout is repeated for two galaxy–spectrum pairs in each row.}
\end{figure*}

\begin{figure*}[ht]
  \ContinuedFloat
  \begin{center}
\includegraphics[scale=0.19]{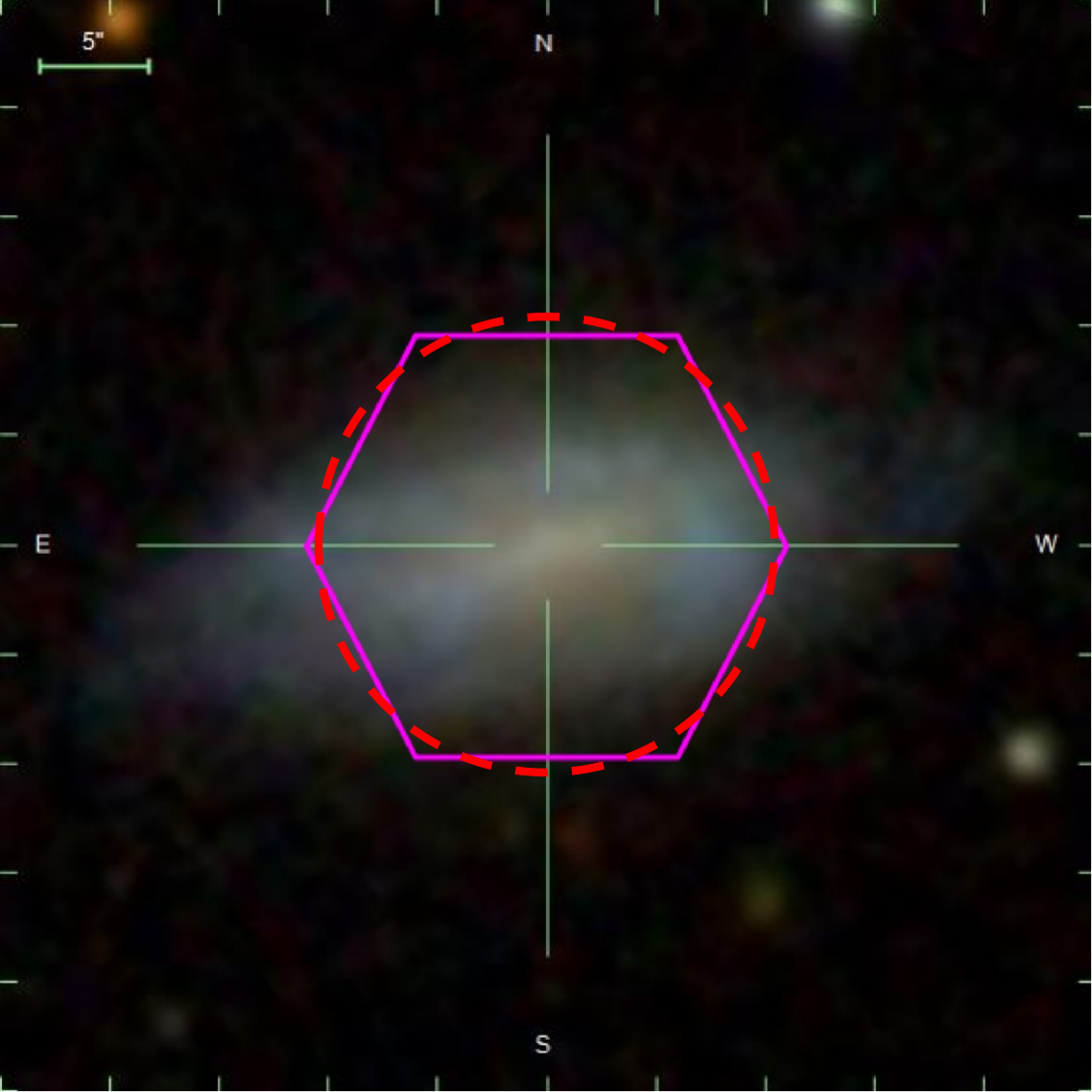}
\includegraphics[bb=5 5 495 405, scale=0.27]{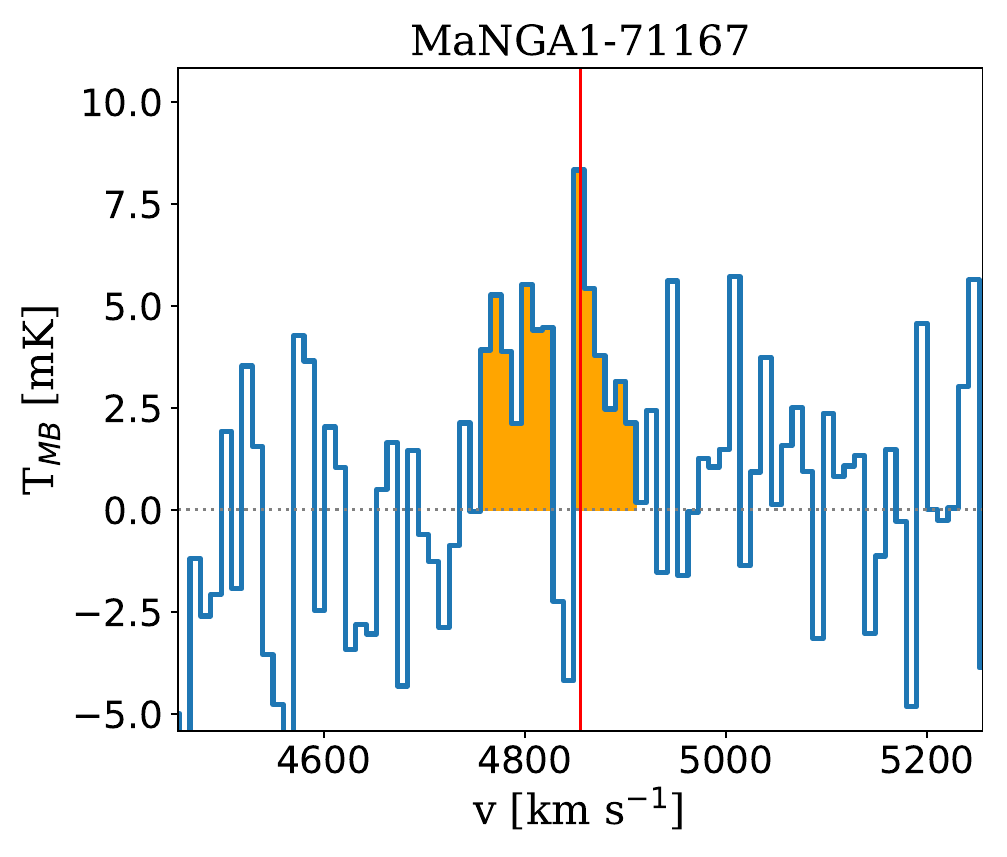}
\includegraphics[scale=0.19]{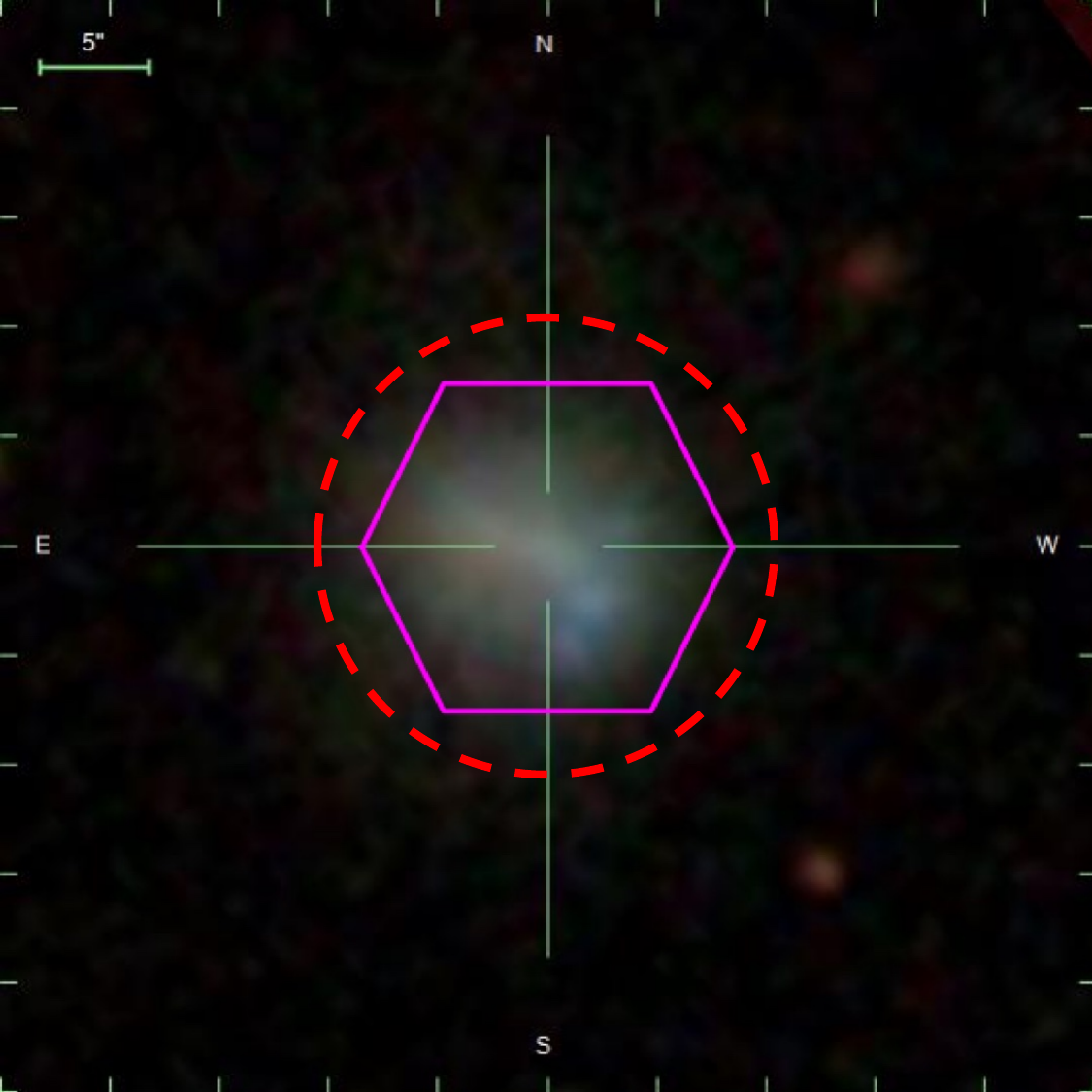}
\includegraphics[bb=5 5 495 405, scale=0.27]{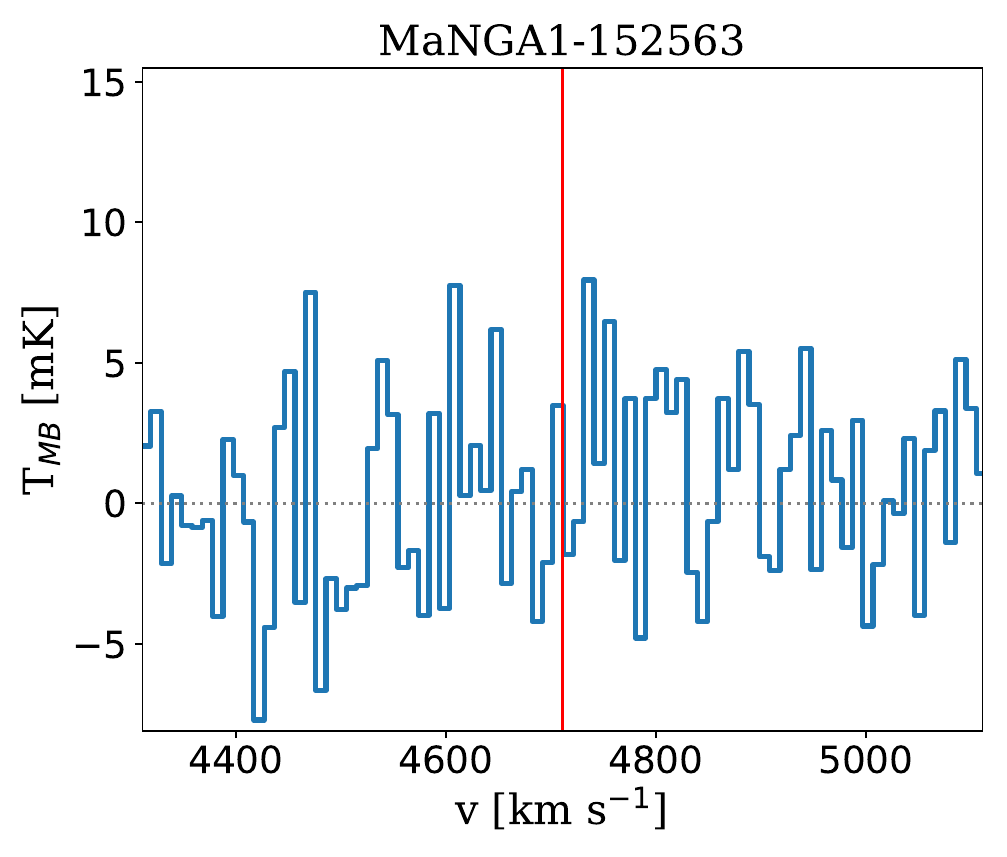}
\includegraphics[scale=0.19]{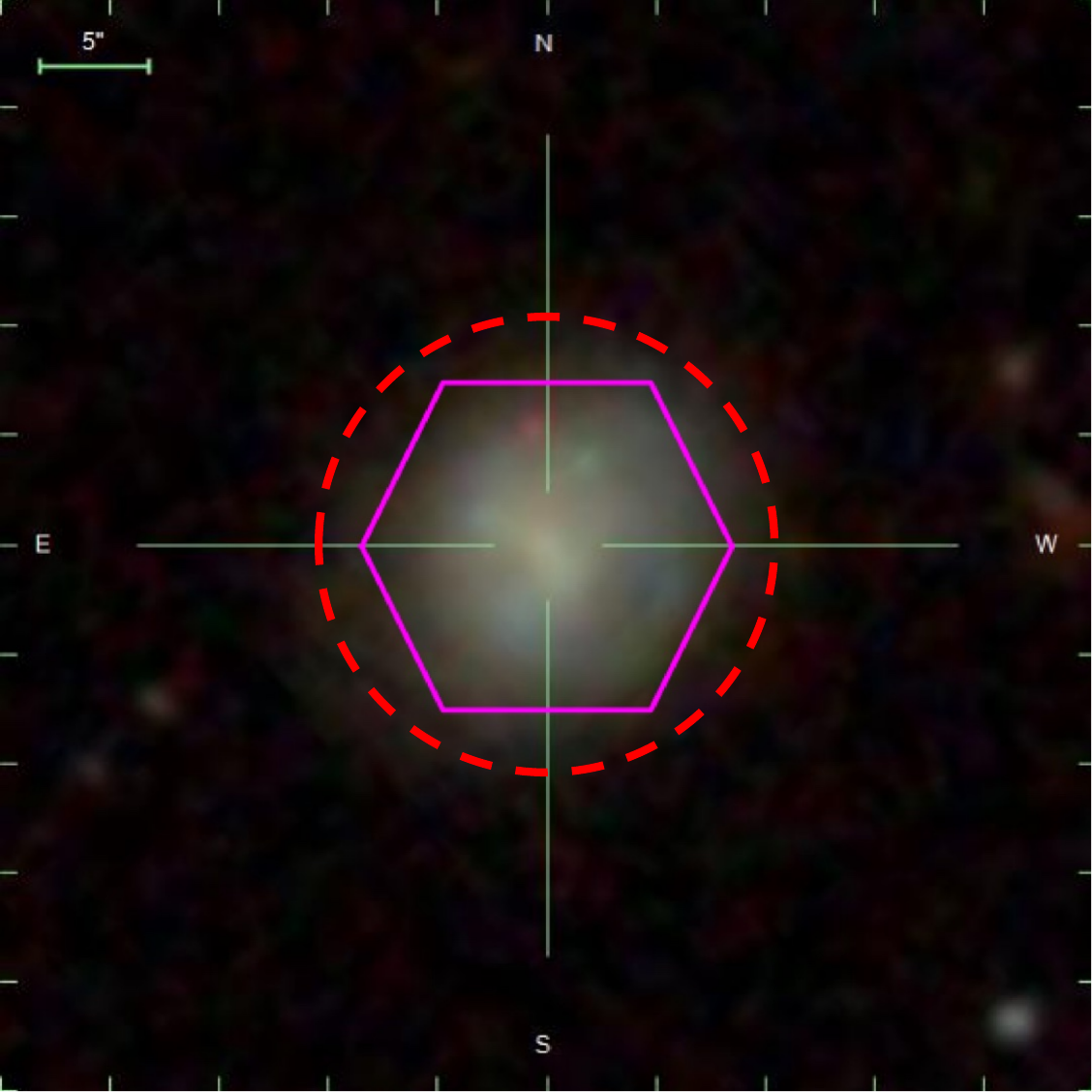}
\includegraphics[bb=5 5 495 405, scale=0.27]{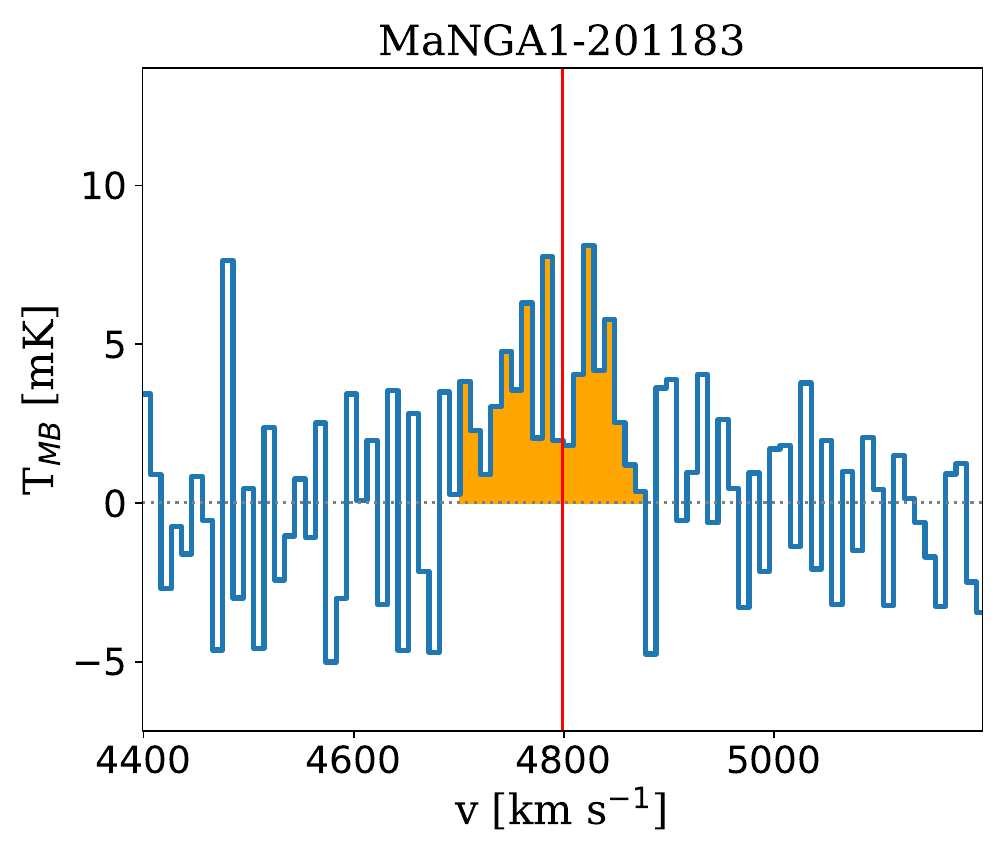}
\includegraphics[scale=0.19]{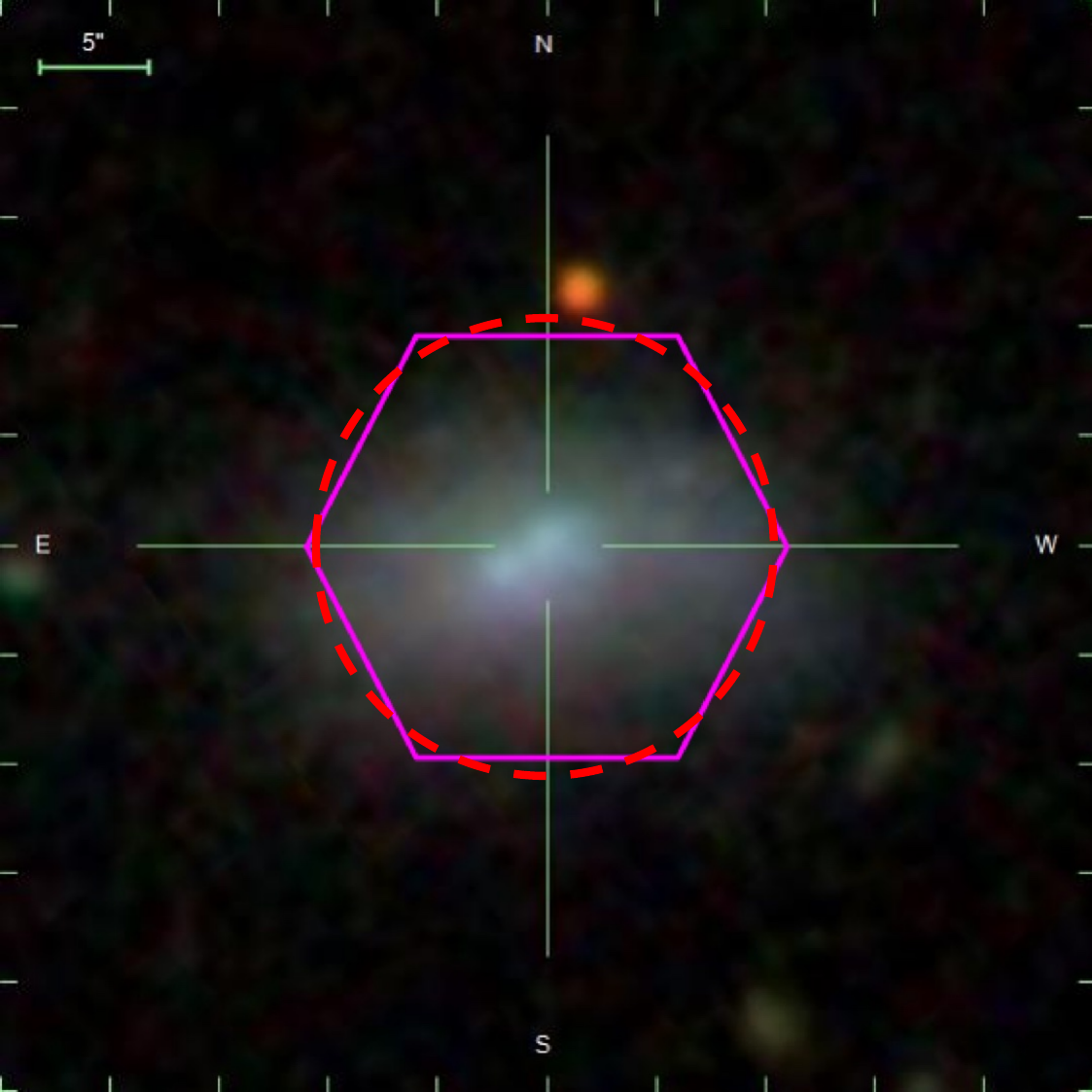}
\includegraphics[bb=5 5 495 405, scale=0.27]{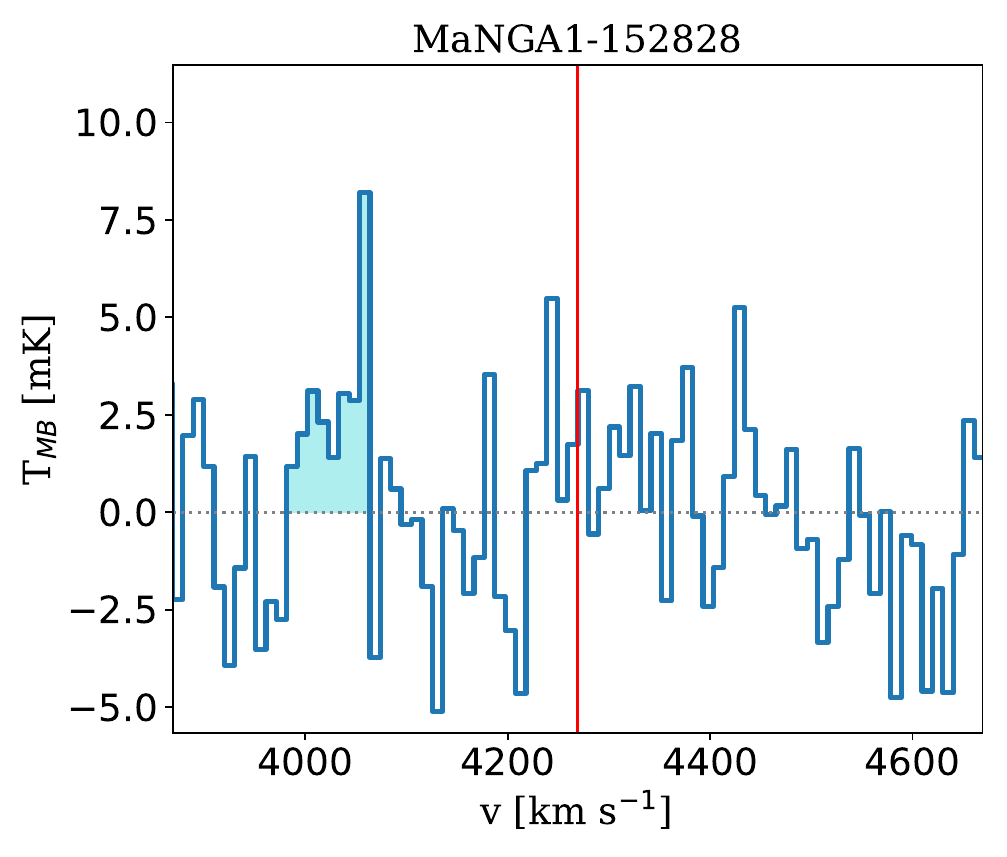}
\includegraphics[scale=0.19]{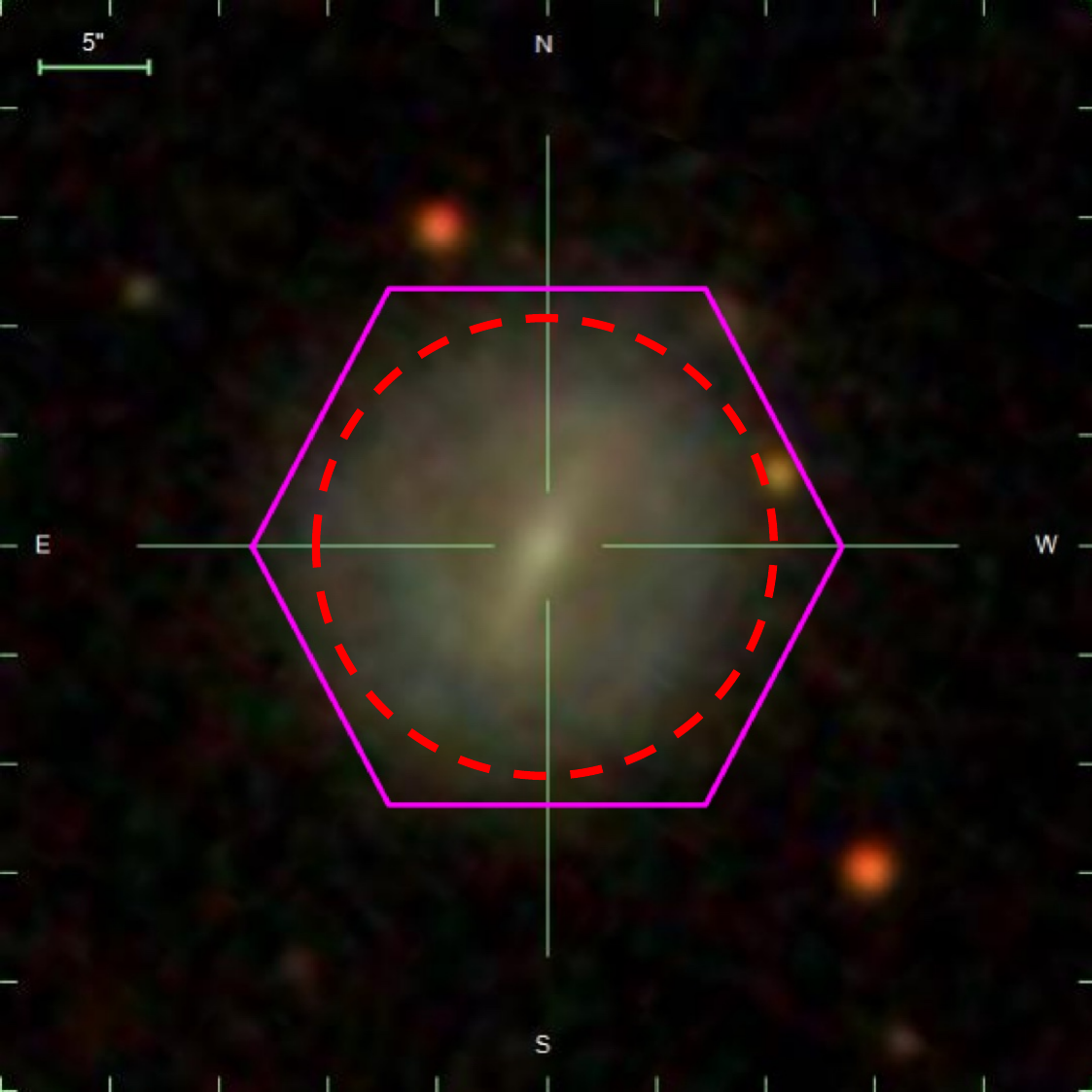}
\includegraphics[bb=5 5 495 405, scale=0.27]{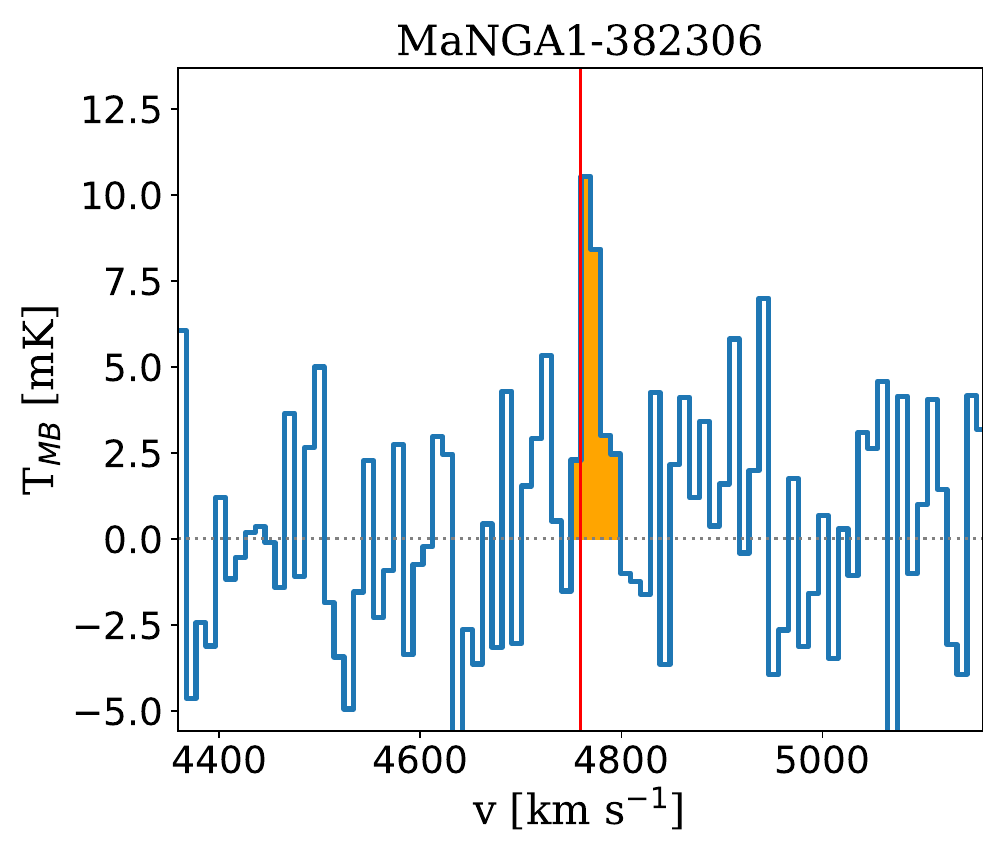}
\includegraphics[scale=0.19]{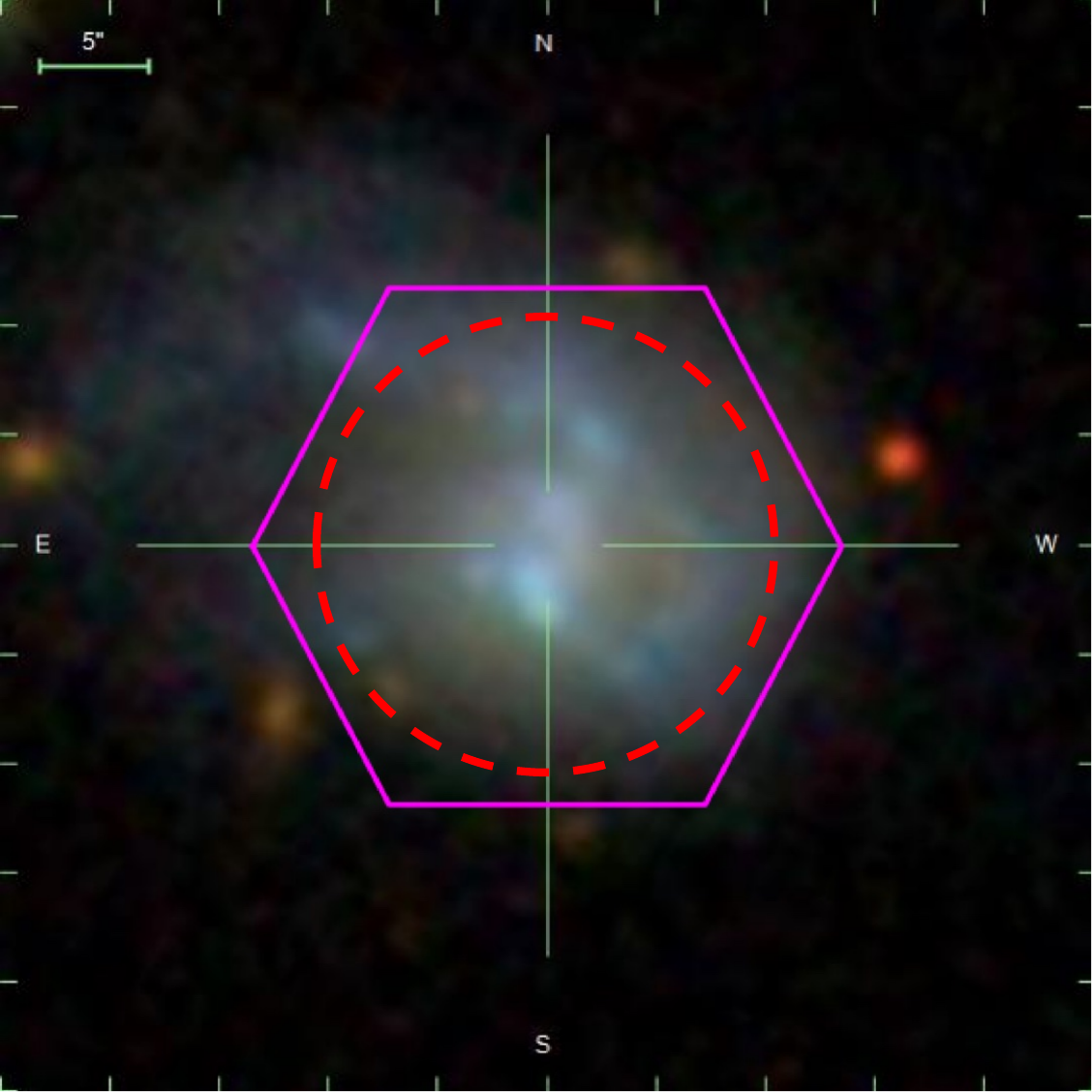}
\includegraphics[bb=5 5 495 405, scale=0.27]{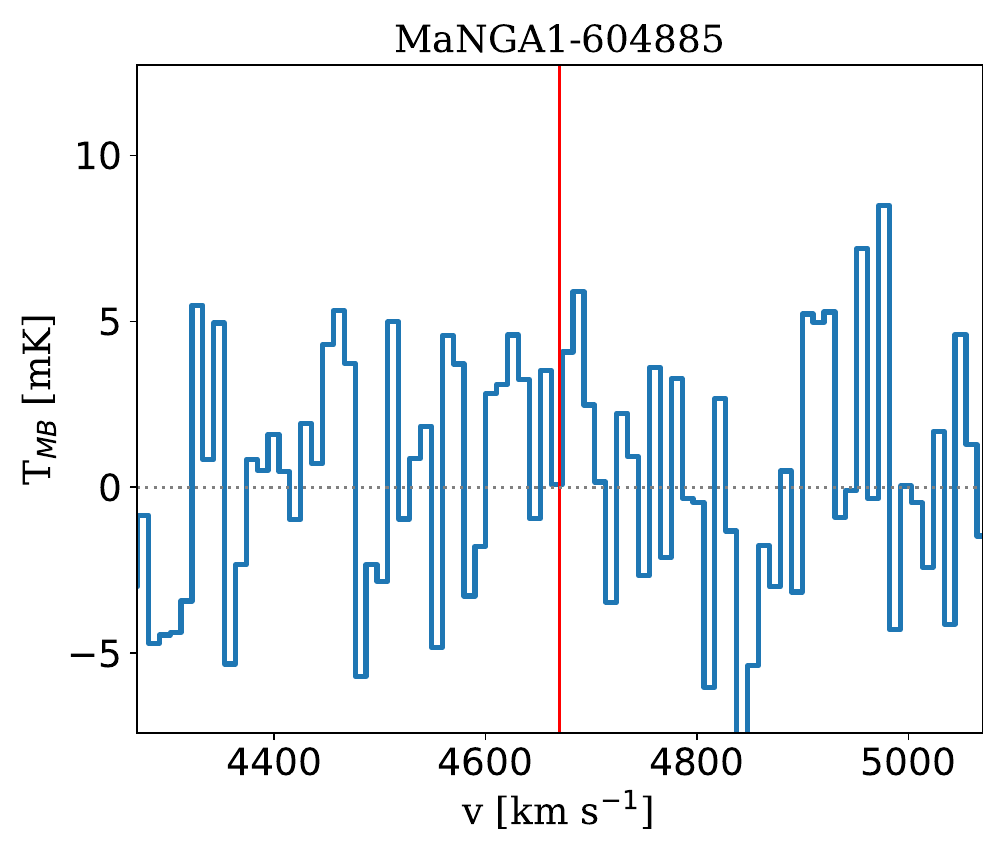}
\includegraphics[scale=0.19]{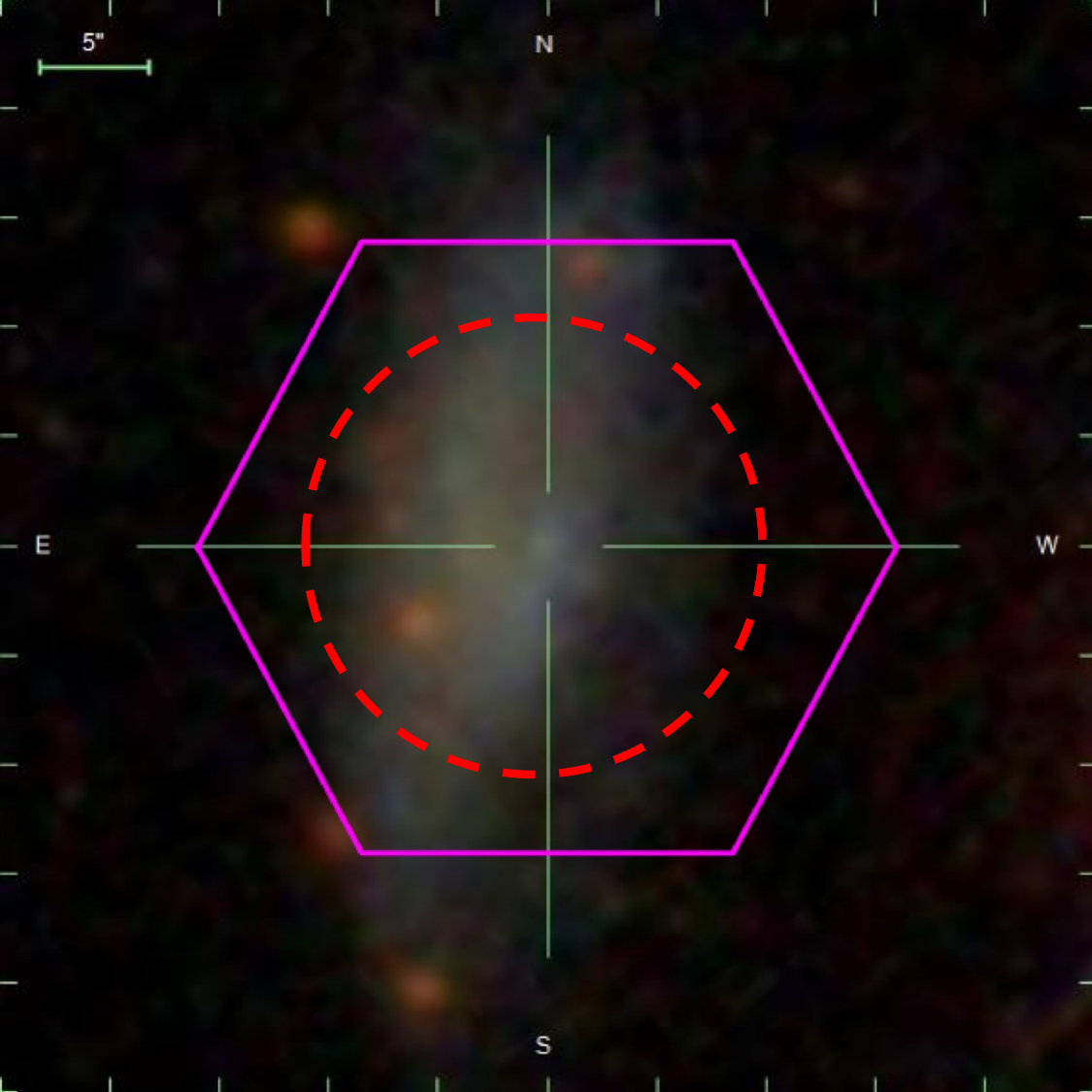}
\includegraphics[bb=5 5 495 405, scale=0.27]{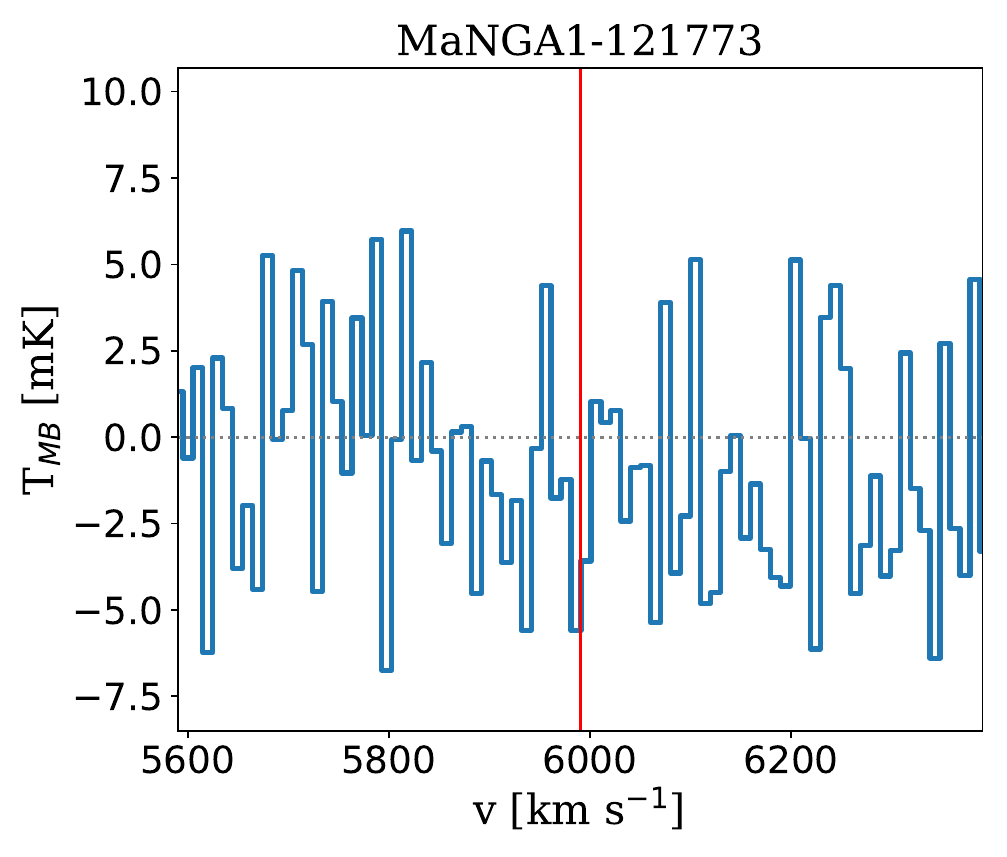}
\includegraphics[scale=0.19]{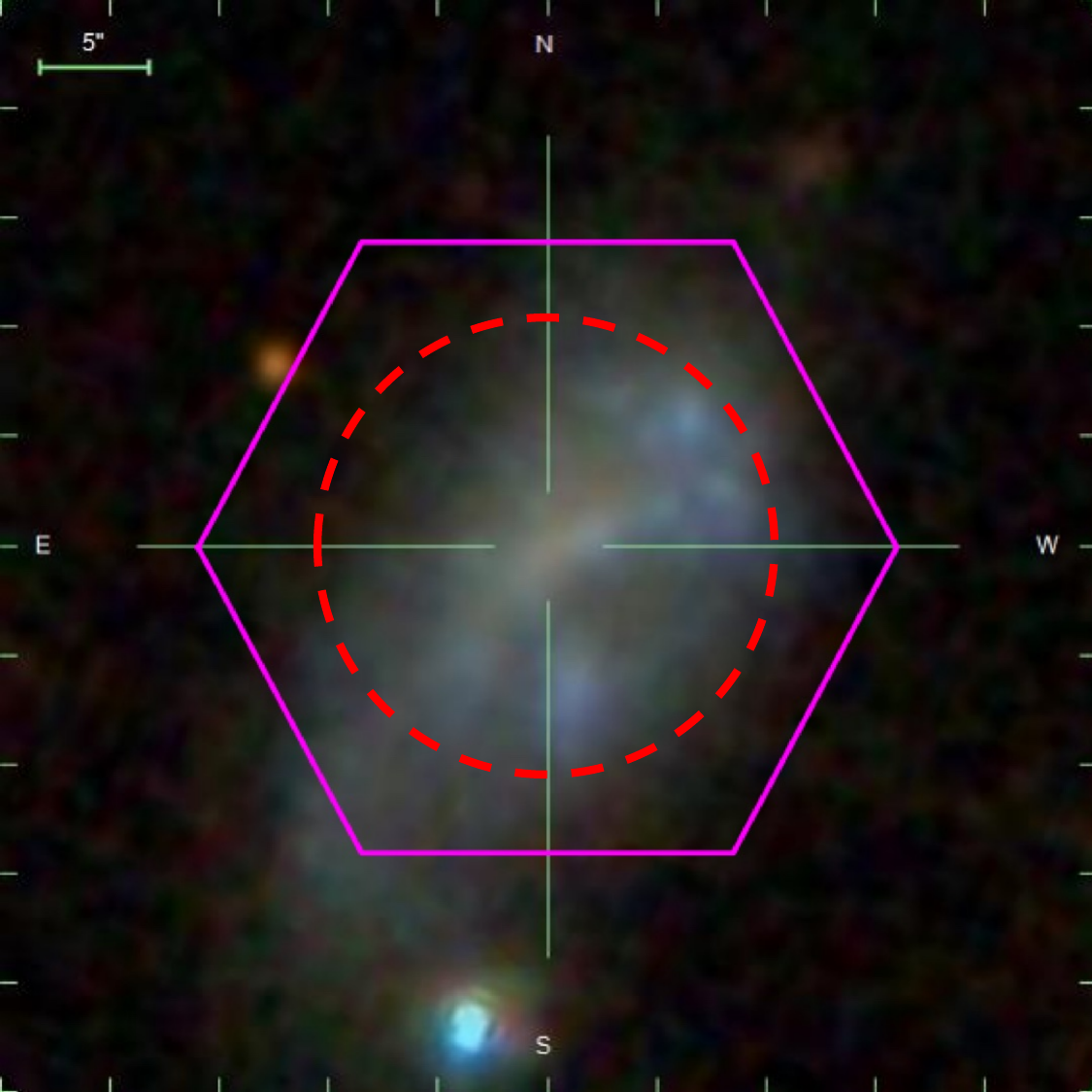}
\includegraphics[bb=5 5 495 405, scale=0.27]{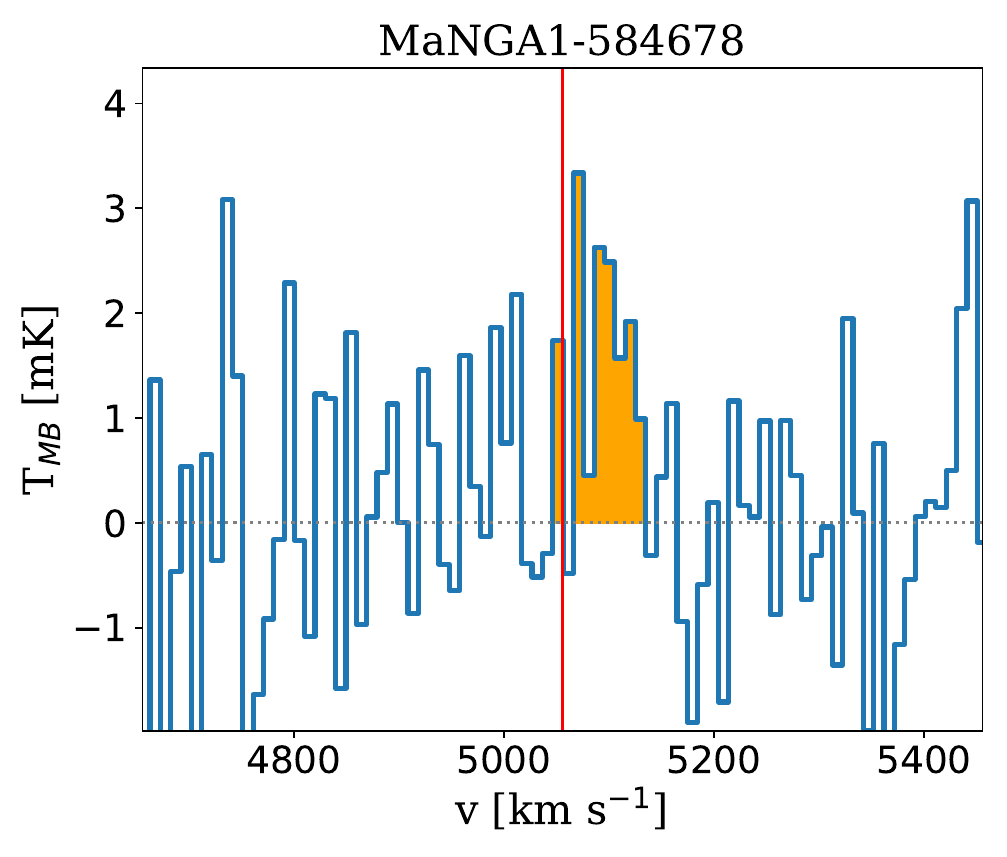}
\includegraphics[scale=0.19]{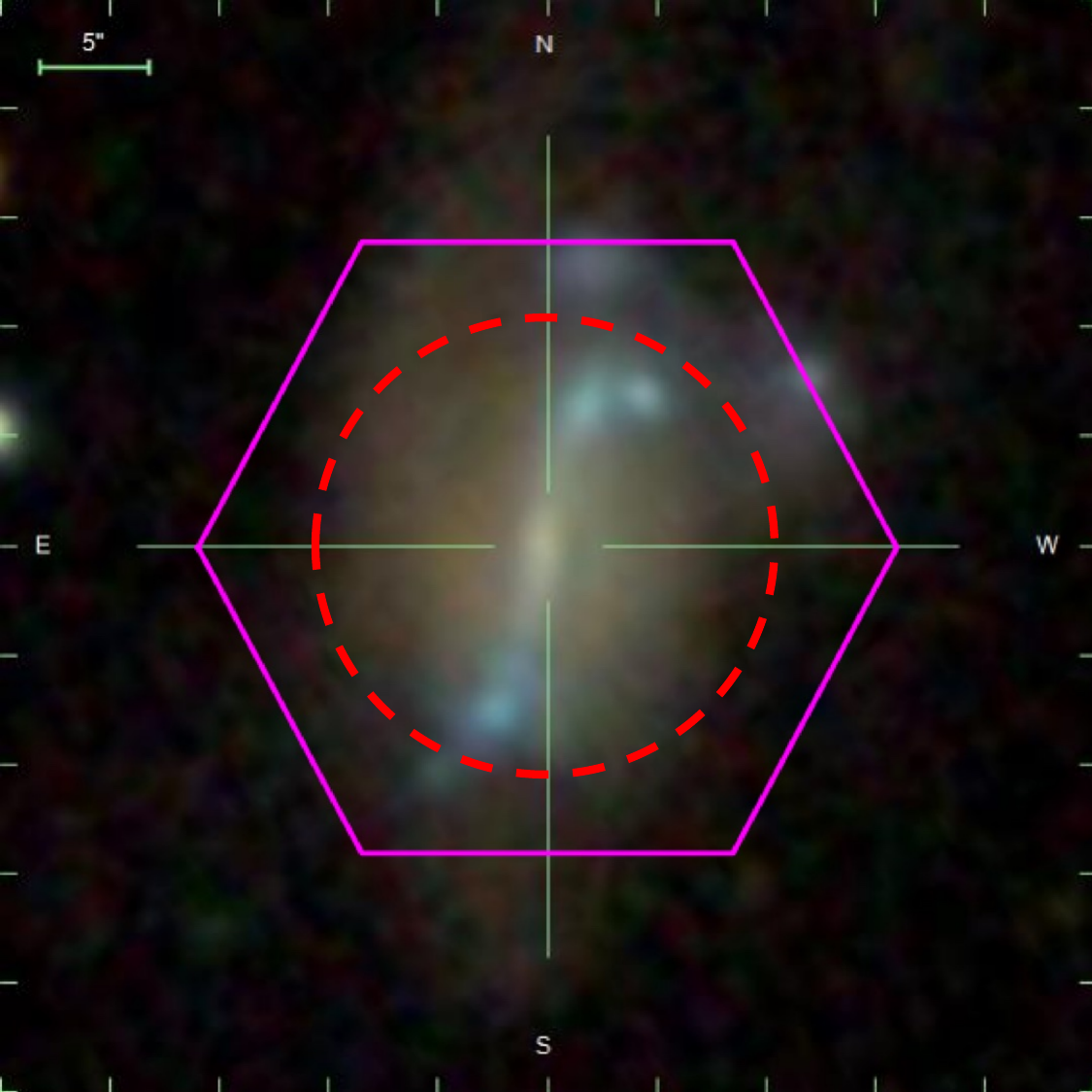}
\includegraphics[bb=5 5 495 405, scale=0.27]{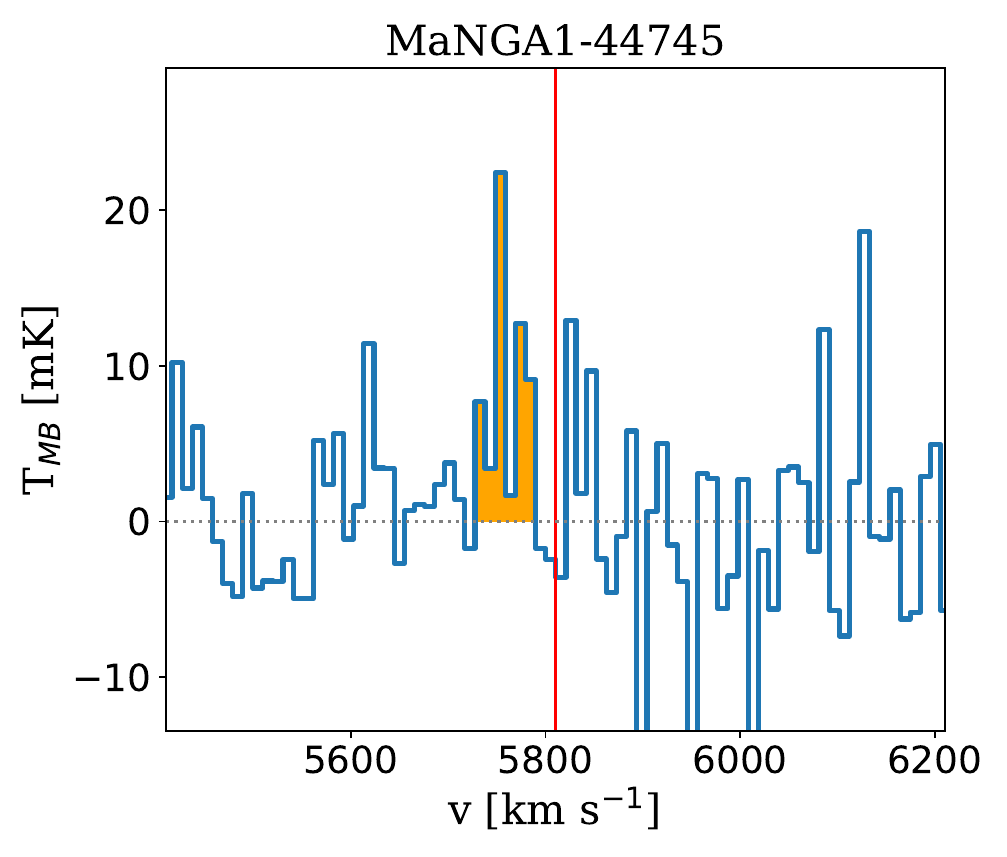}
\includegraphics[scale=0.19]{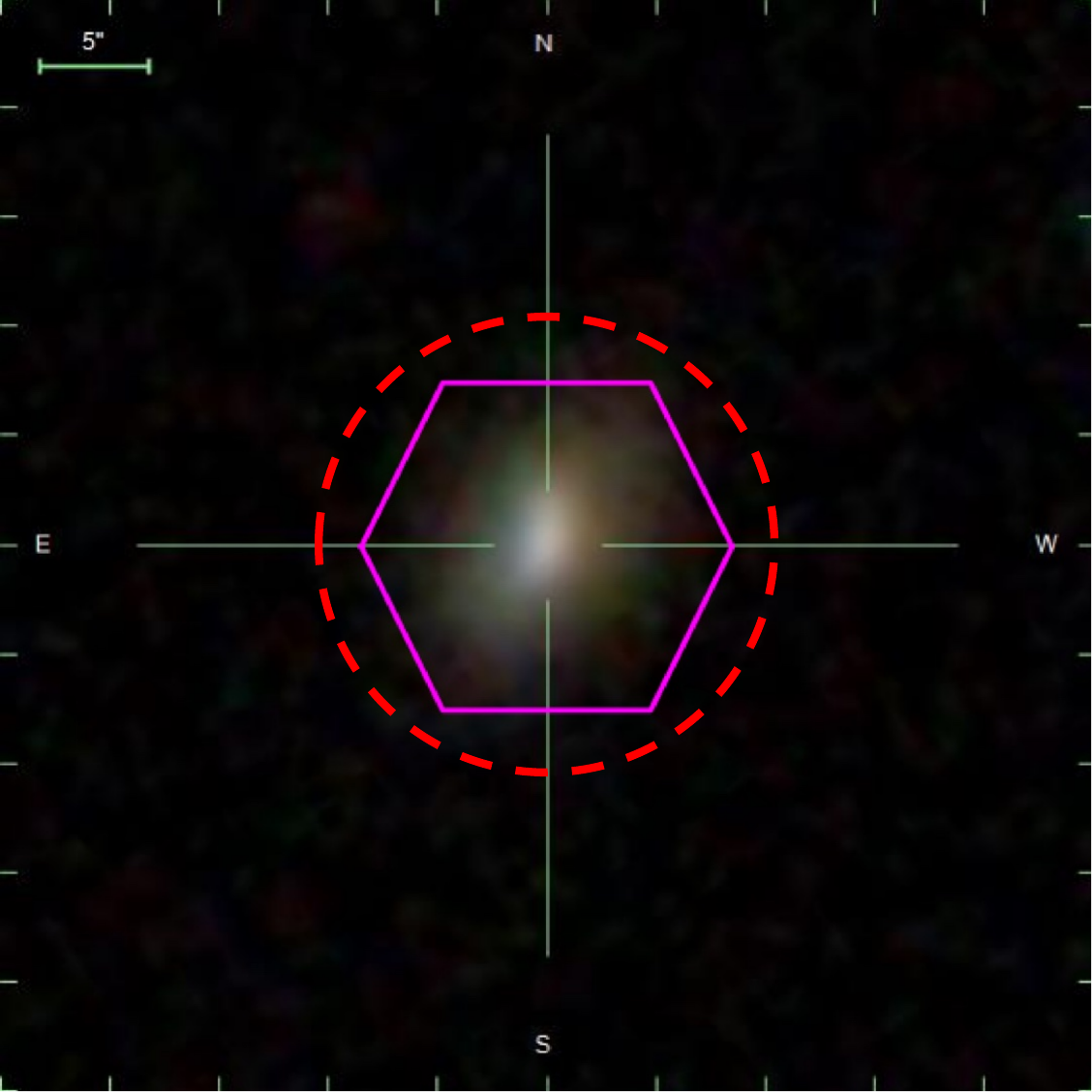}
\includegraphics[bb=5 5 495 405, scale=0.27]{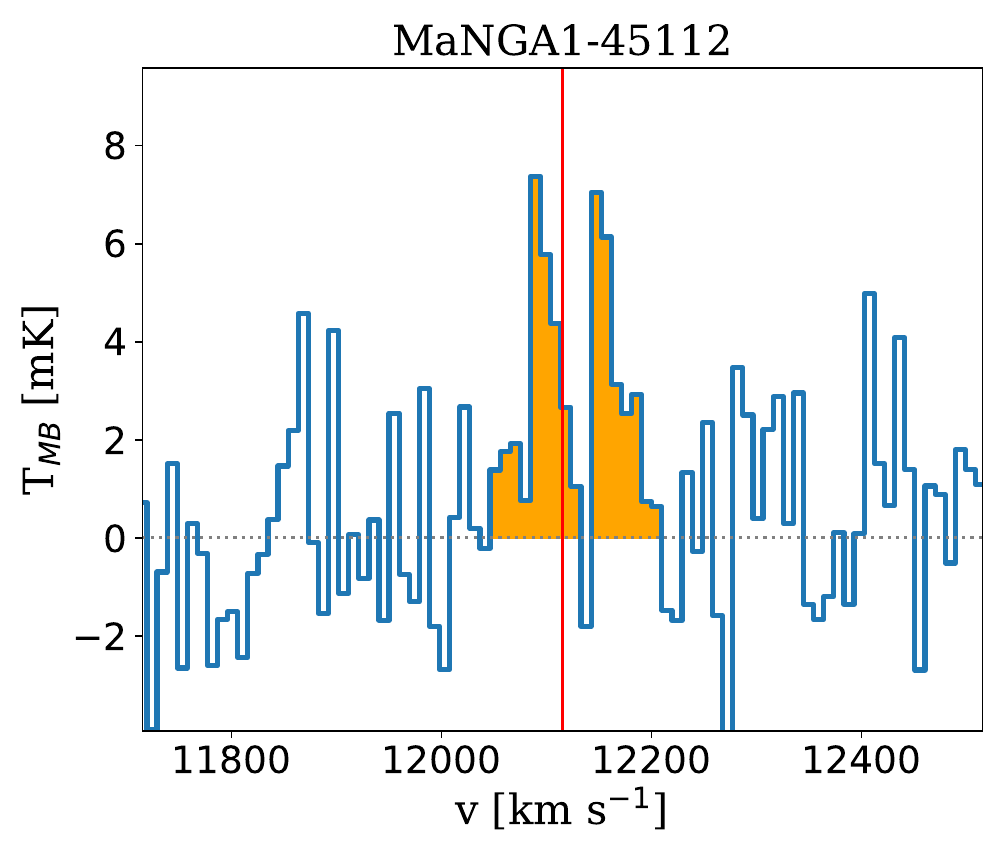}
\includegraphics[scale=0.19]{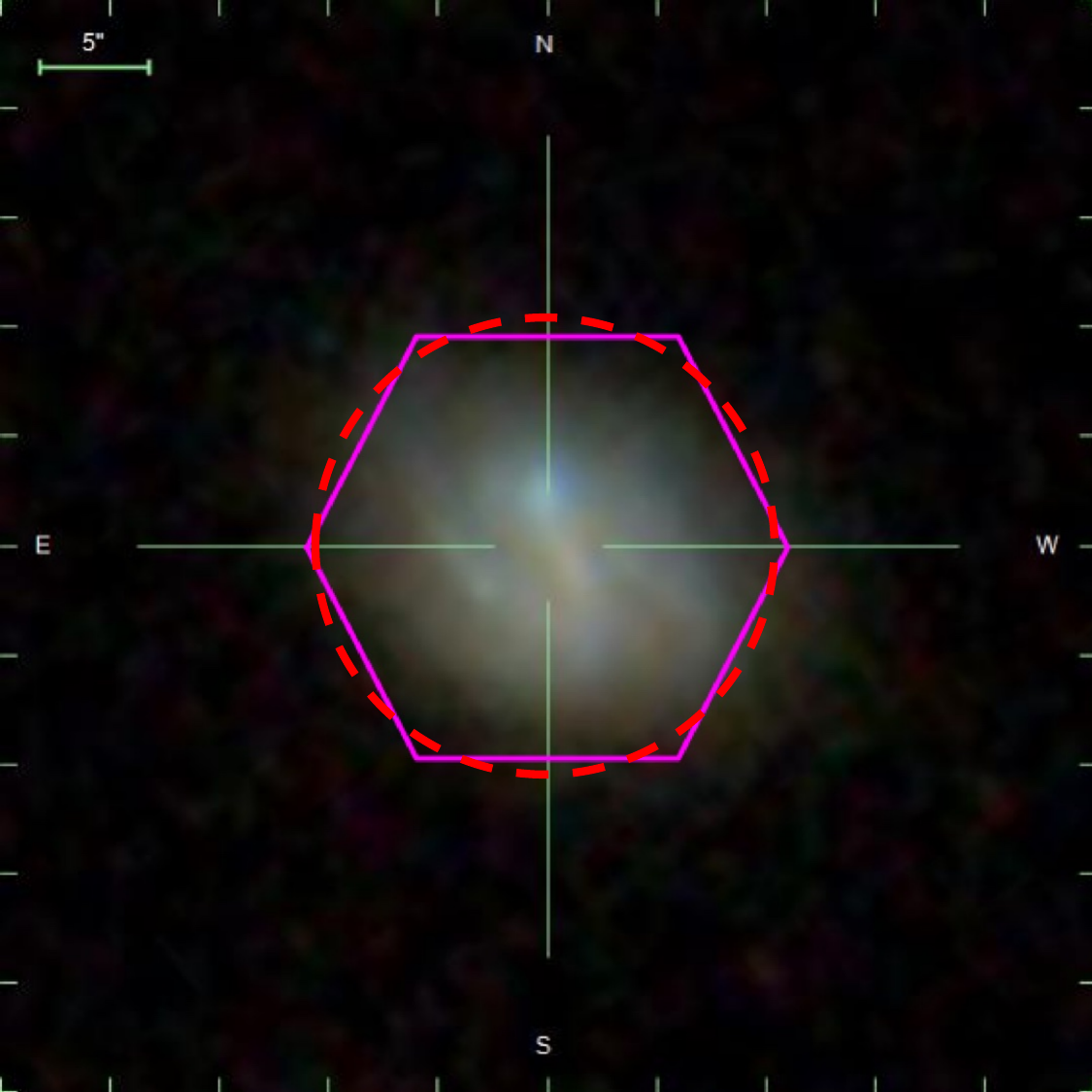}
\includegraphics[bb=5 5 495 405, scale=0.27]{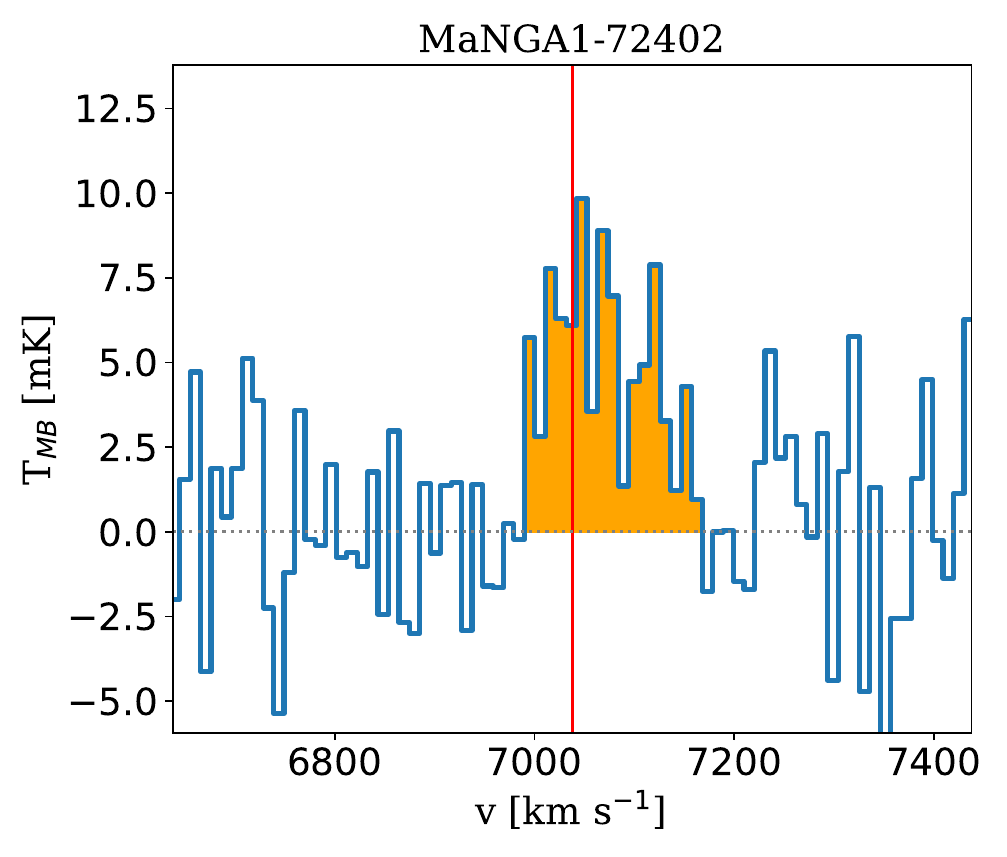}
\includegraphics[scale=0.19]{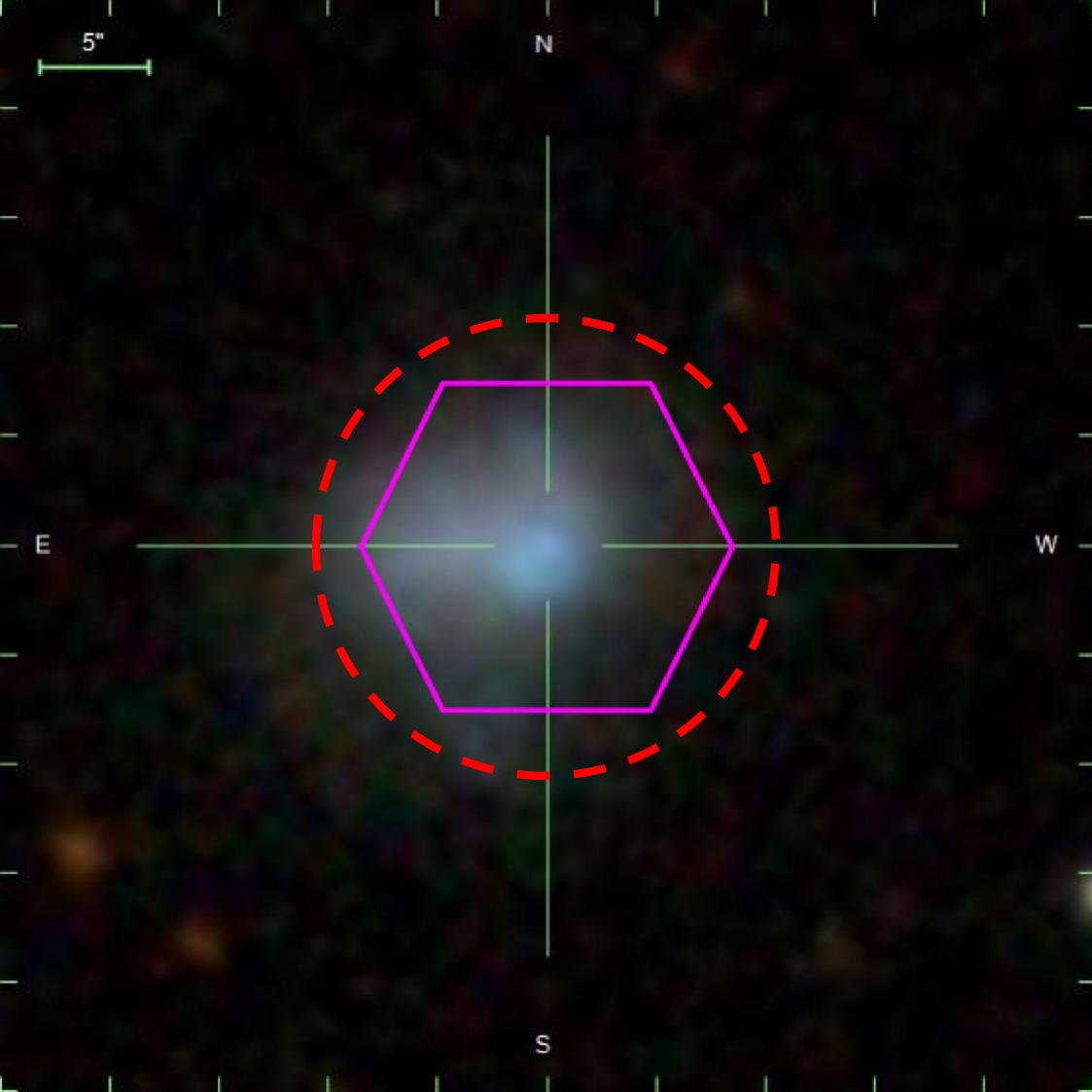}
\includegraphics[bb=5 5 495 405, scale=0.27]{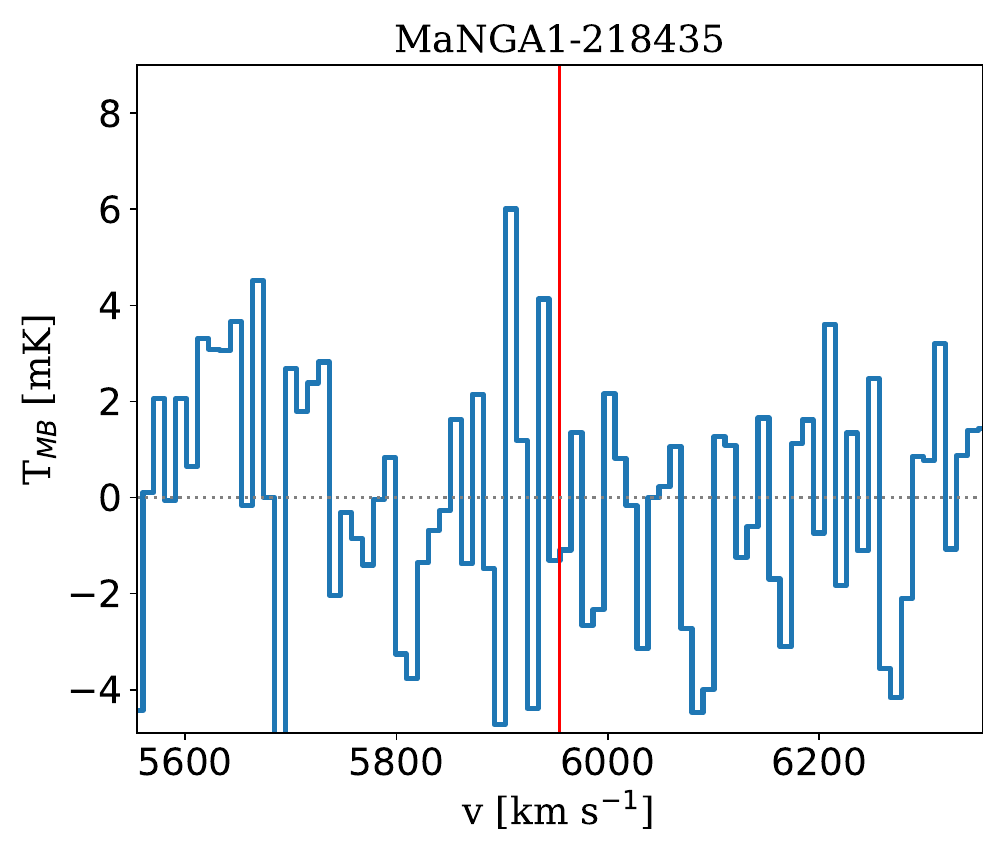}
\end{center}
\caption{continued.}
\end{figure*}

\begin{figure*}
  \ContinuedFloat
\begin{center}
\includegraphics[scale=0.19]{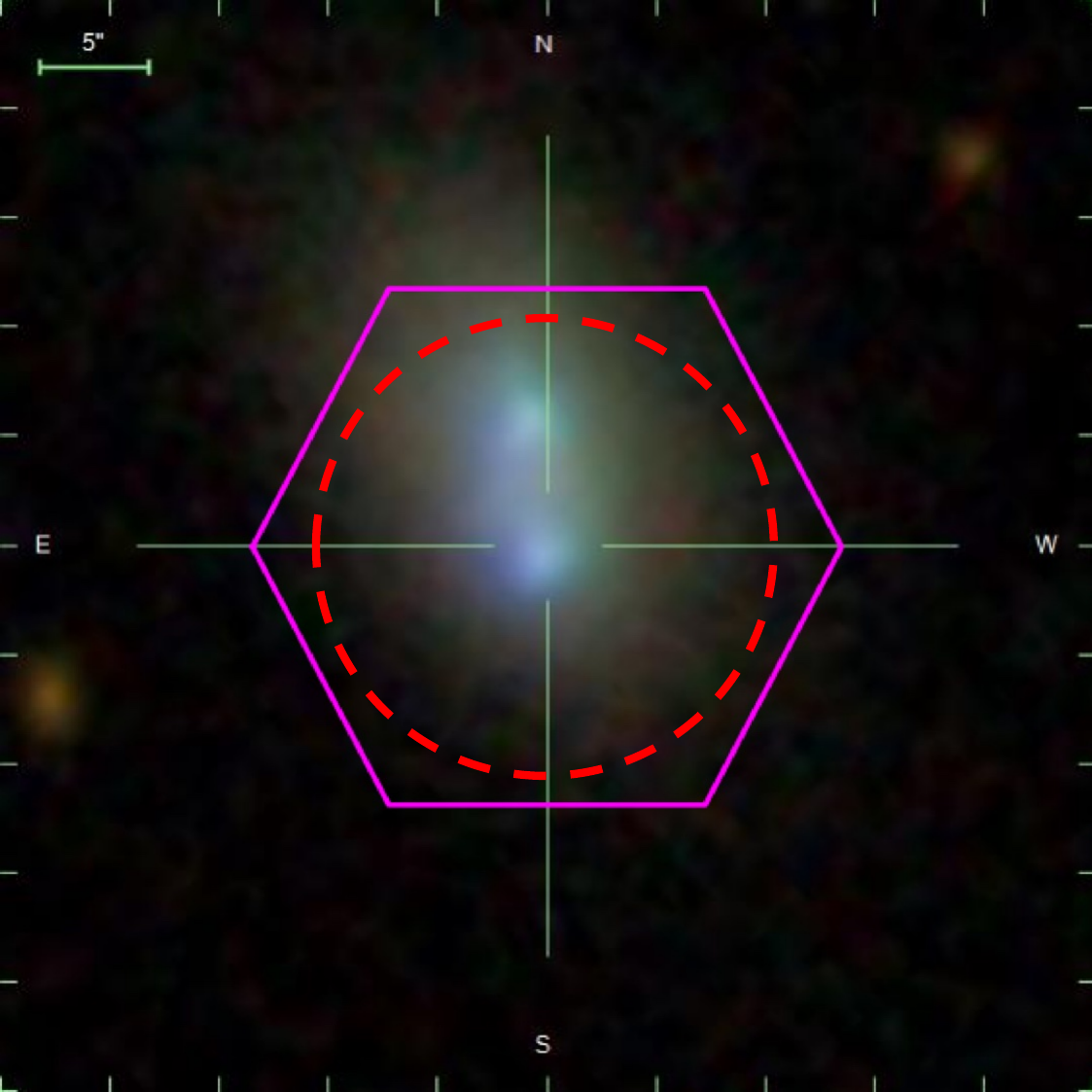}
\includegraphics[bb=5 5 495 405, scale=0.27]{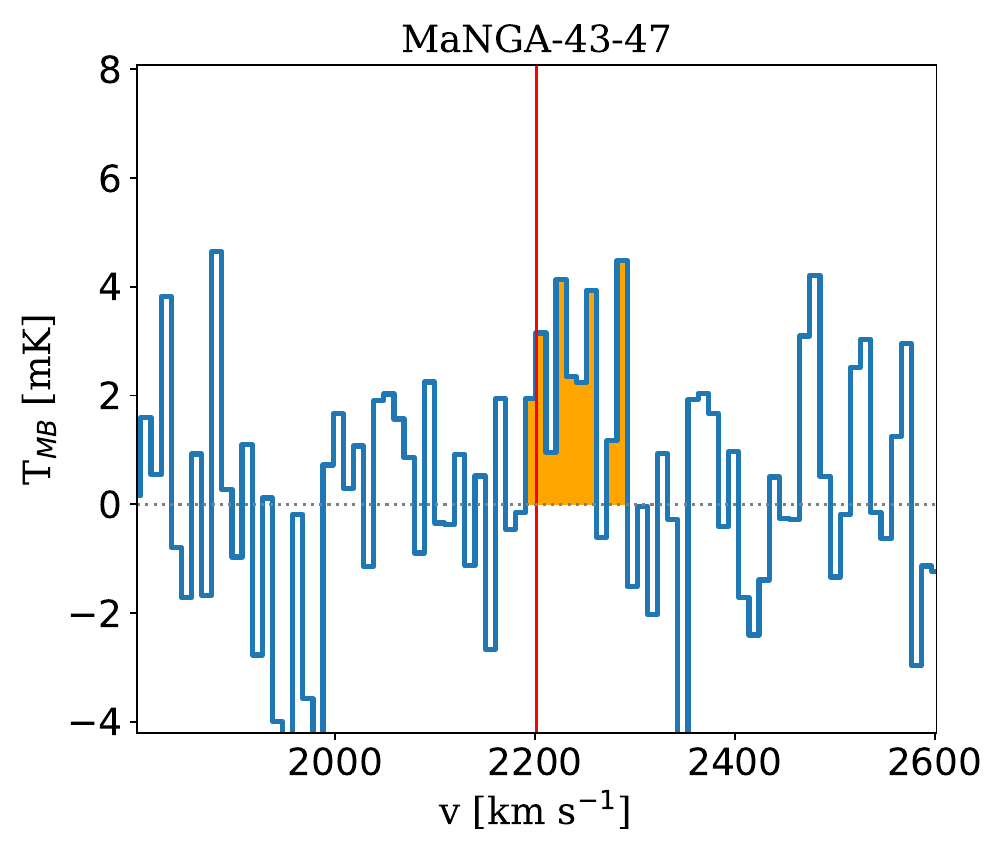}
\includegraphics[scale=0.19]{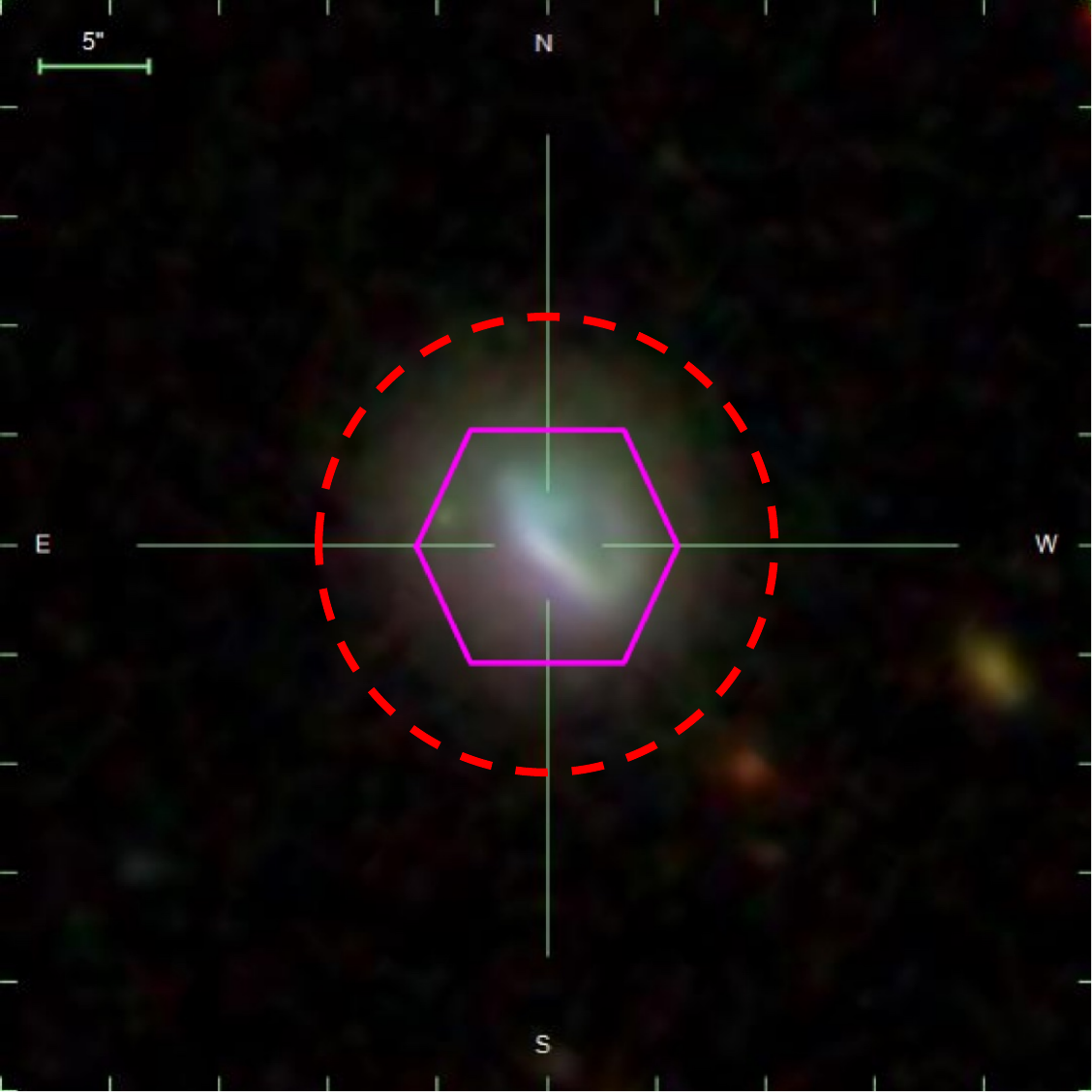}
\includegraphics[bb=5 5 495 405, scale=0.27]{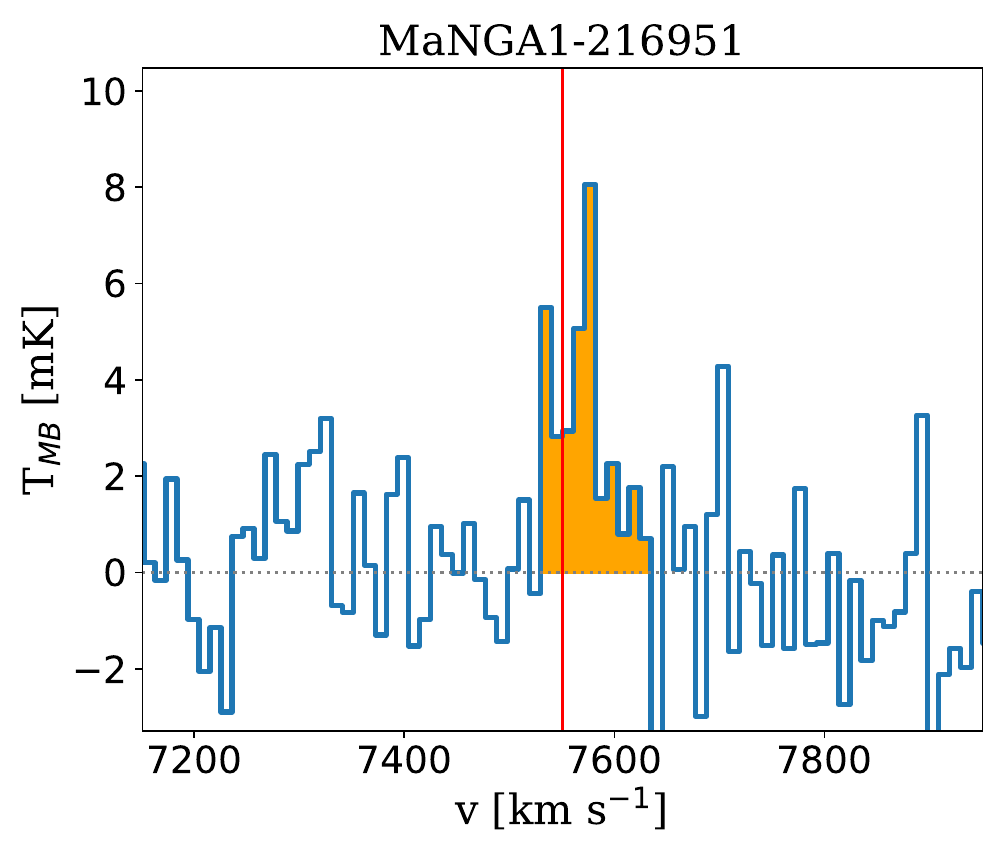}
\includegraphics[scale=0.19]{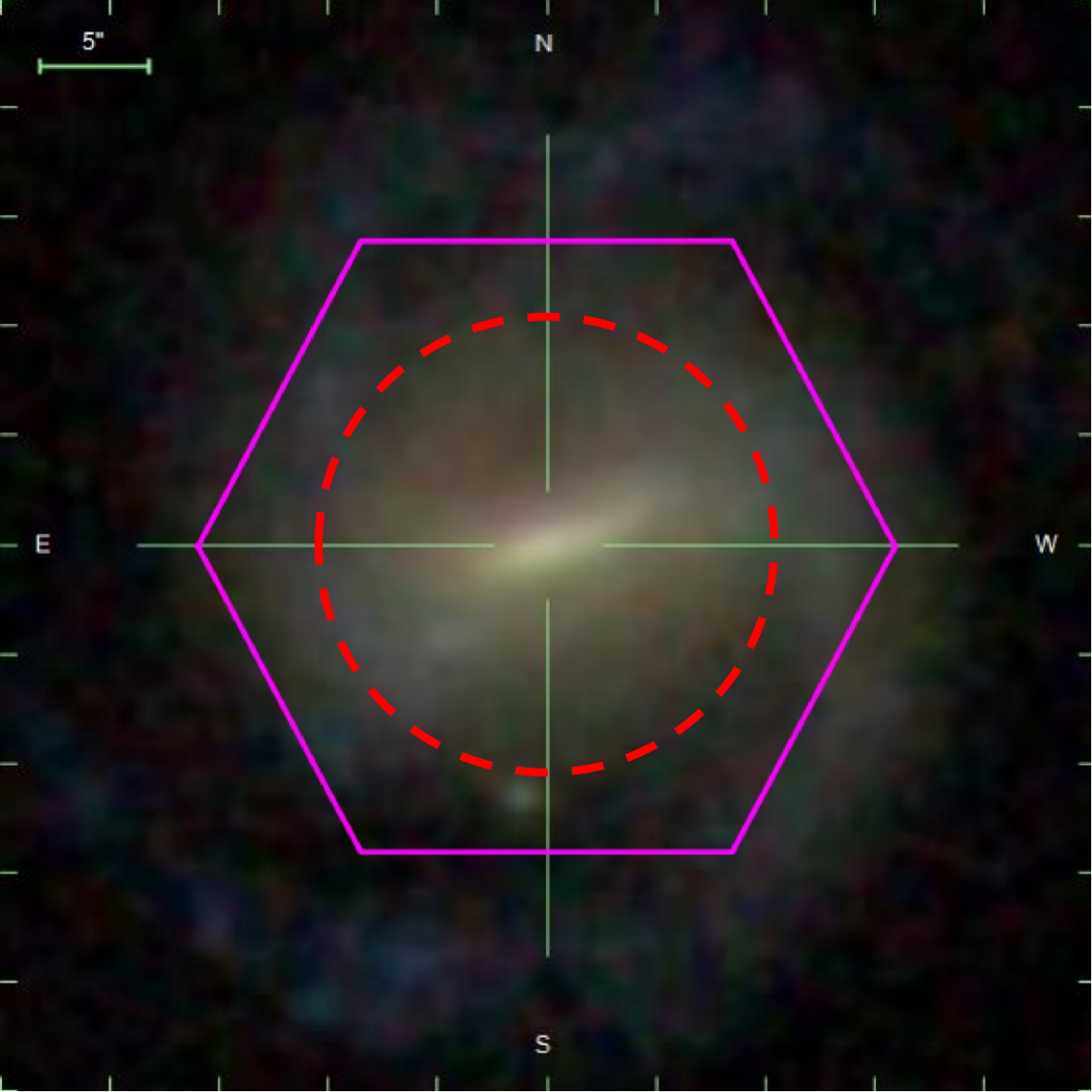}
\includegraphics[bb=5 5 495 405, scale=0.27]{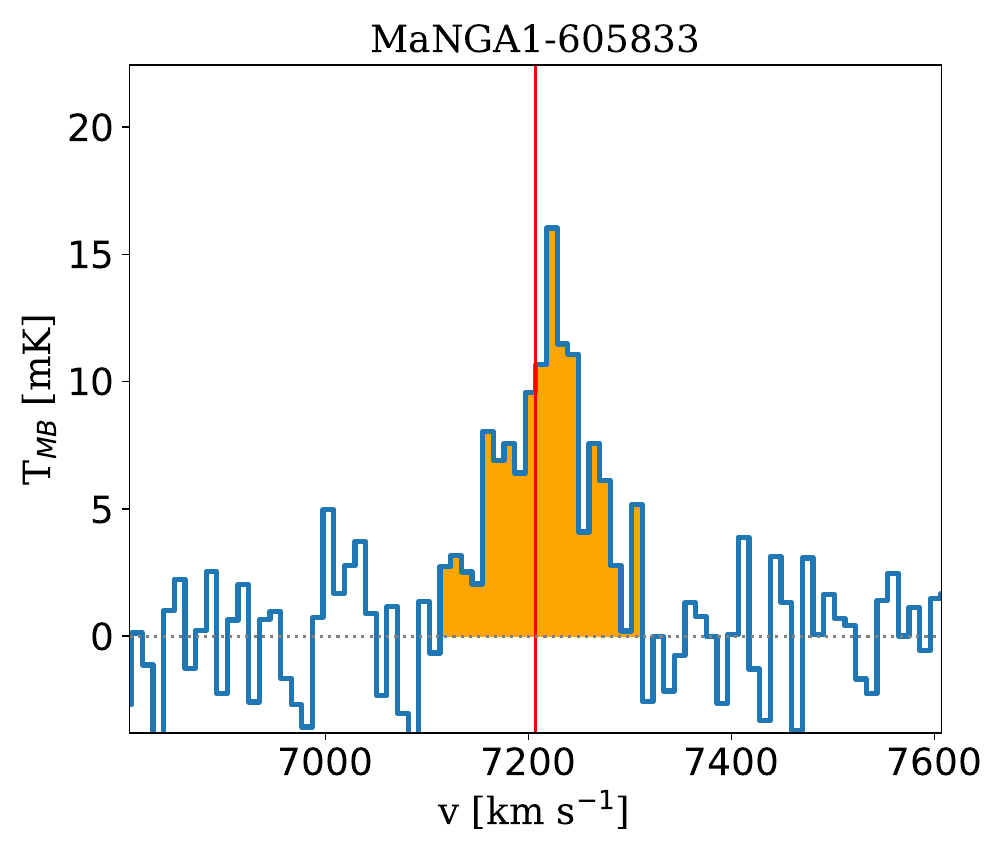}
\includegraphics[scale=0.19]{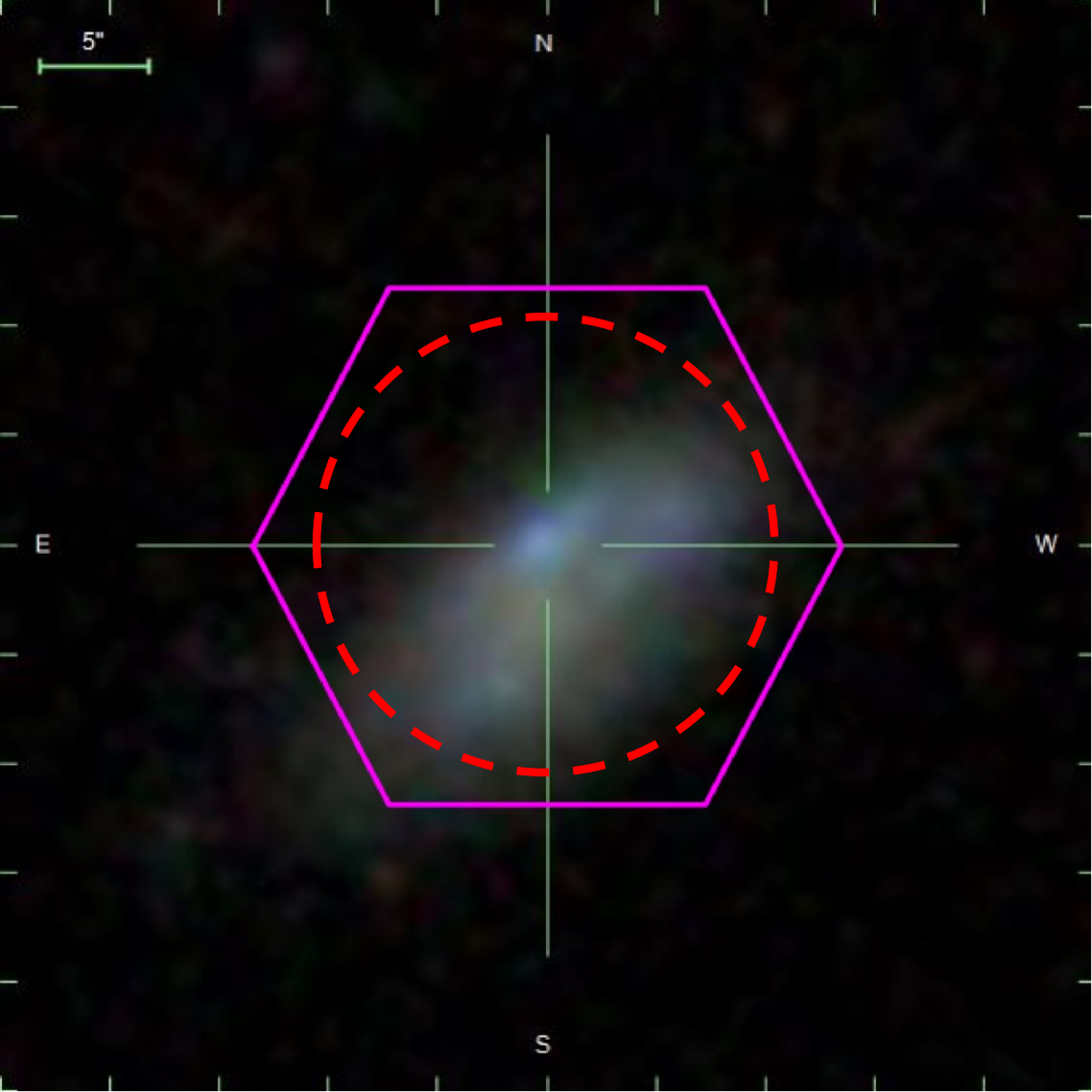}
\includegraphics[bb=5 5 495 405, scale=0.27]{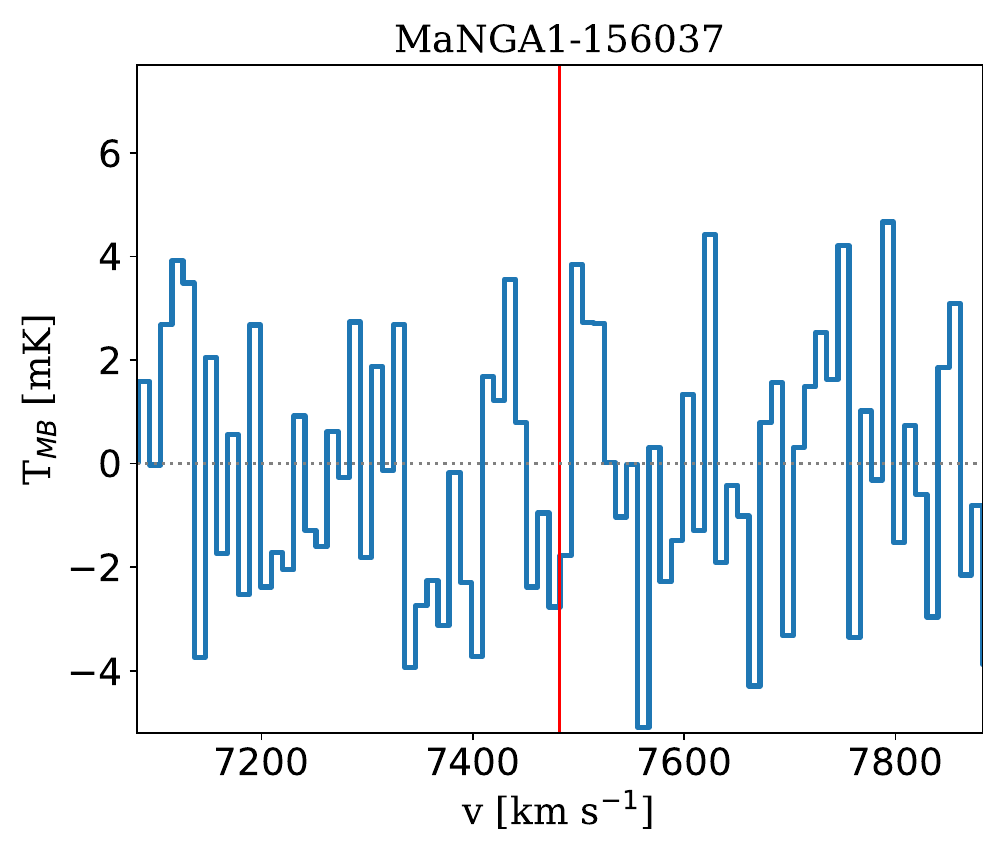}
\includegraphics[scale=0.19]{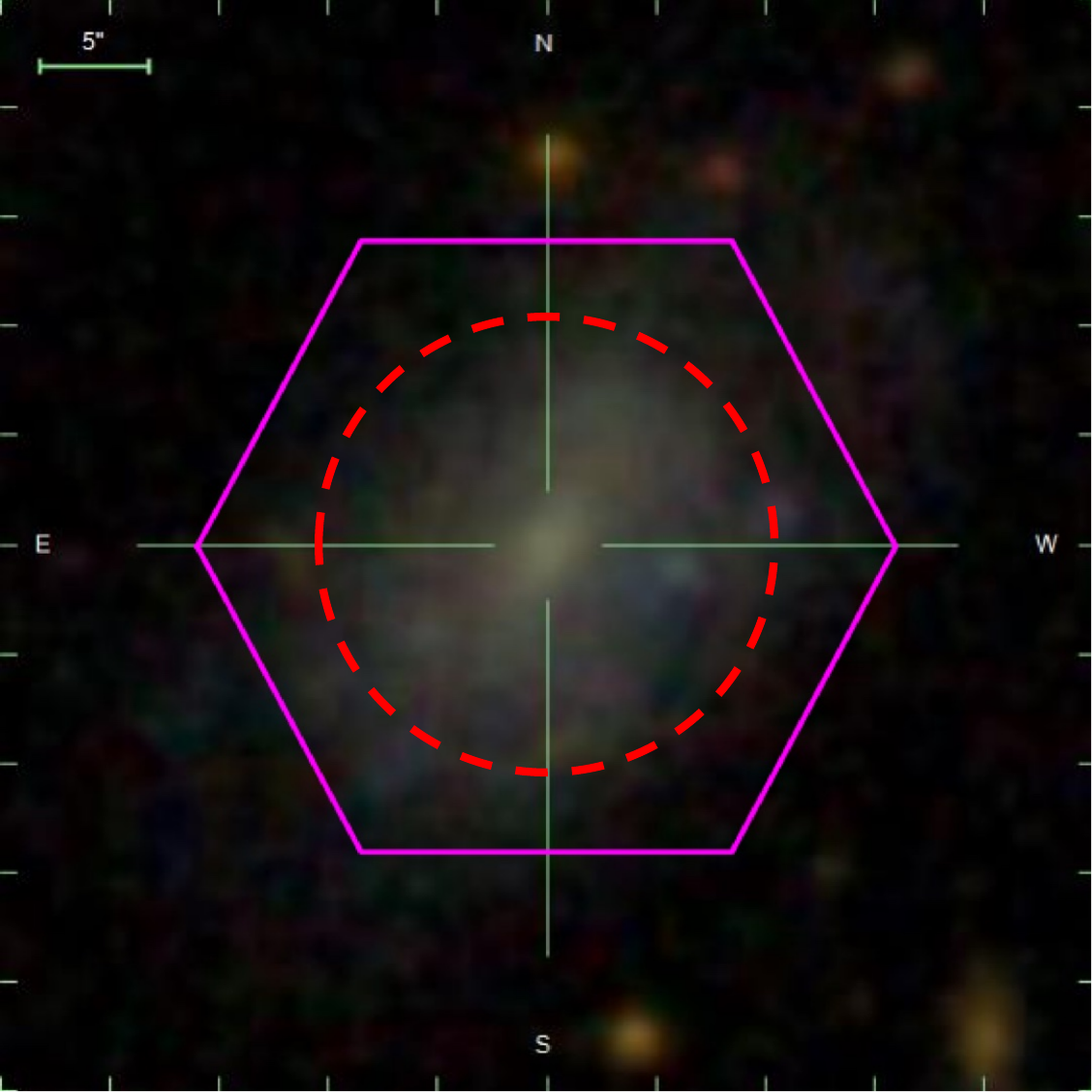}
\includegraphics[bb=5 5 495 405, scale=0.27]{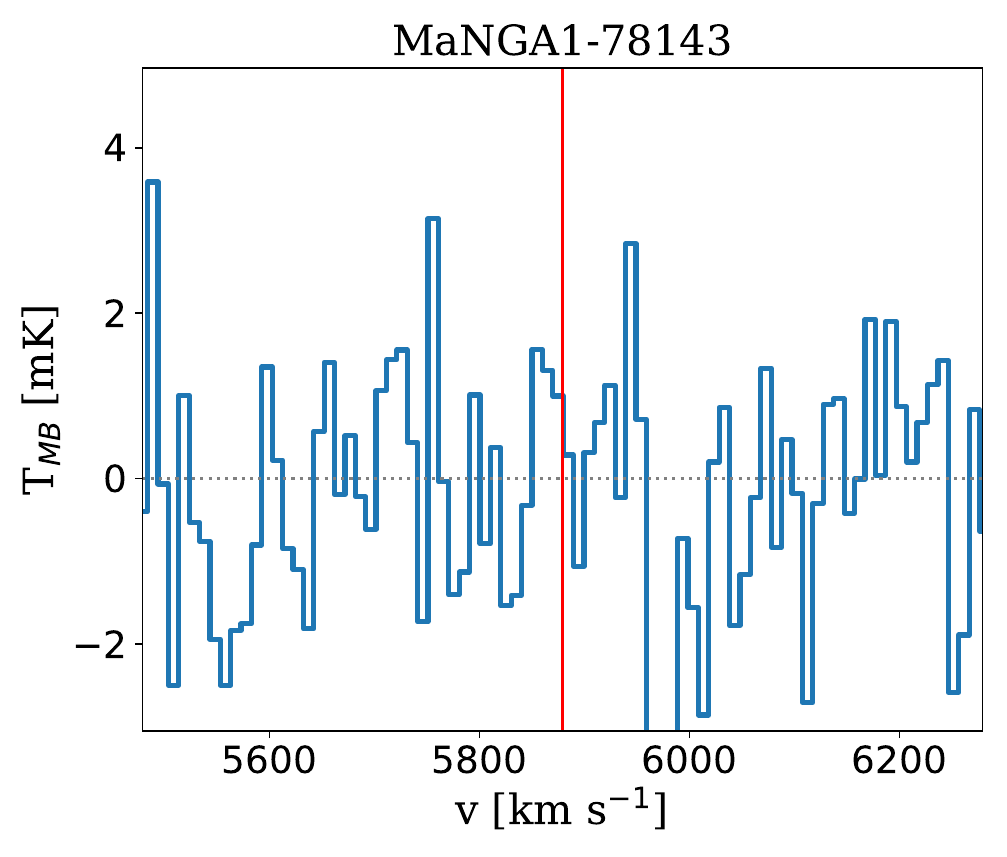}
\includegraphics[scale=0.19]{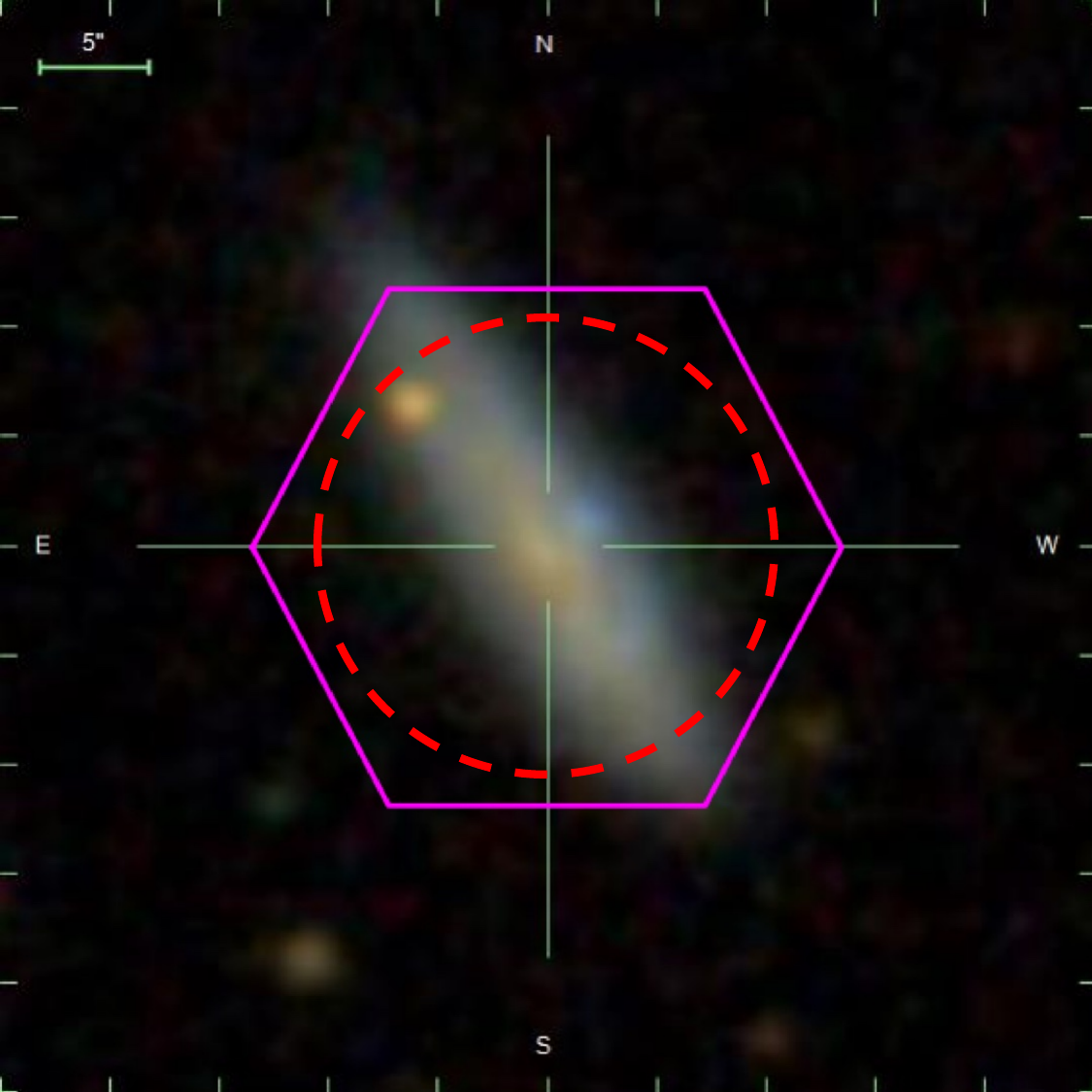}
\includegraphics[bb=5 5 495 405, scale=0.27]{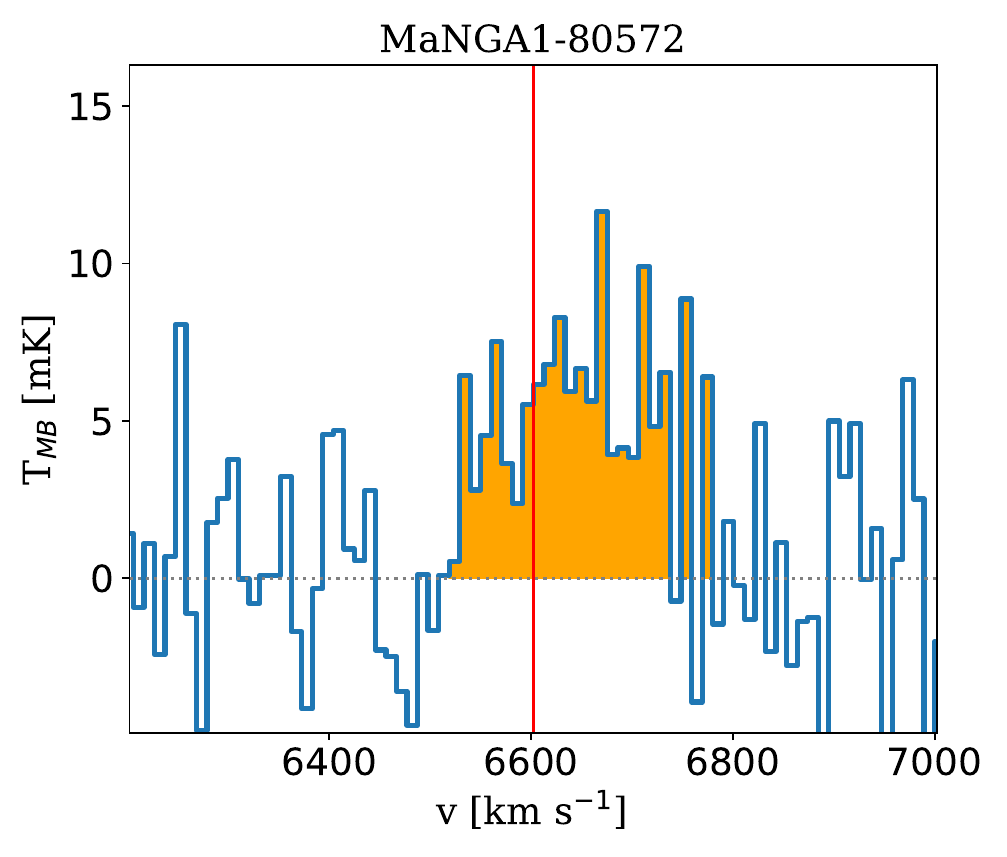}
\includegraphics[scale=0.19]{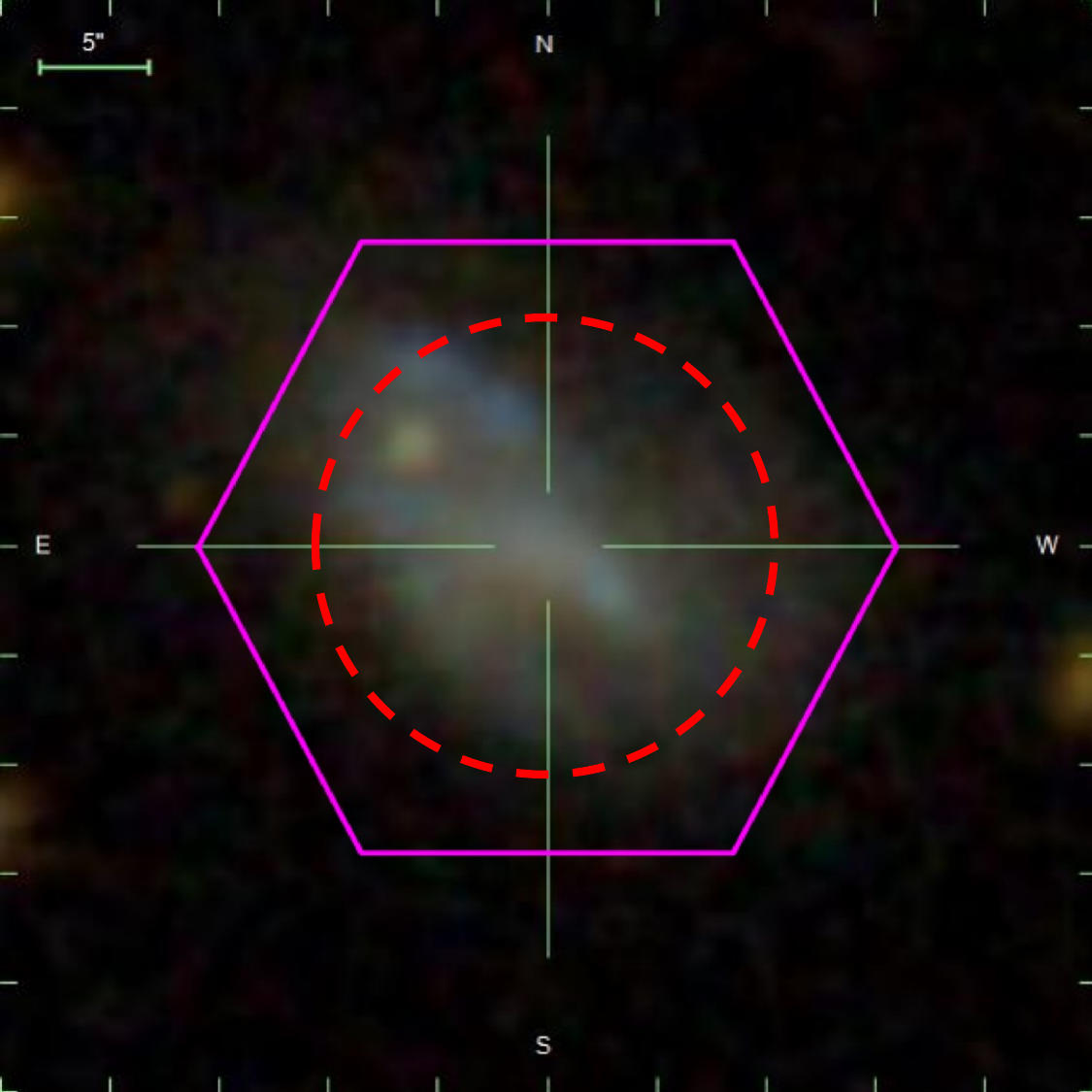}
\includegraphics[bb=5 5 495 405, scale=0.27]{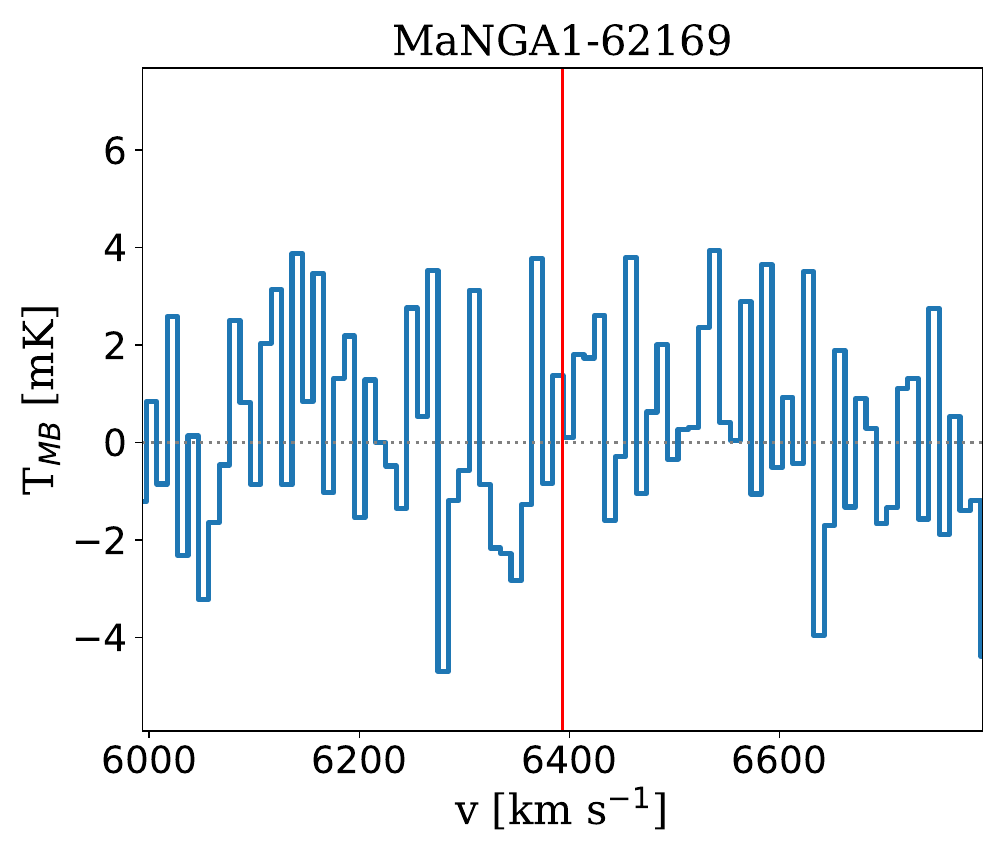}
\includegraphics[scale=0.19]{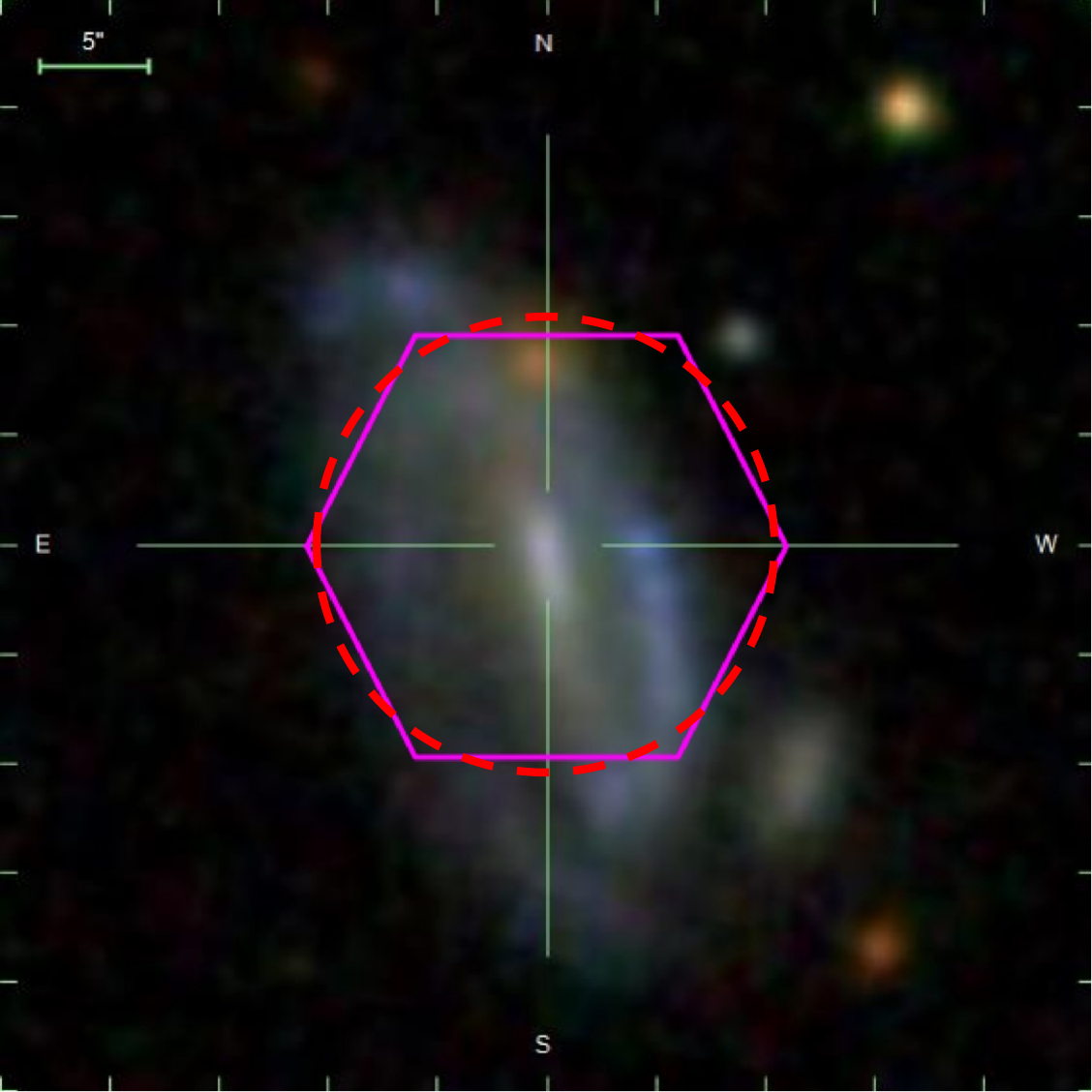}
\includegraphics[bb=5 5 495 405, scale=0.27]{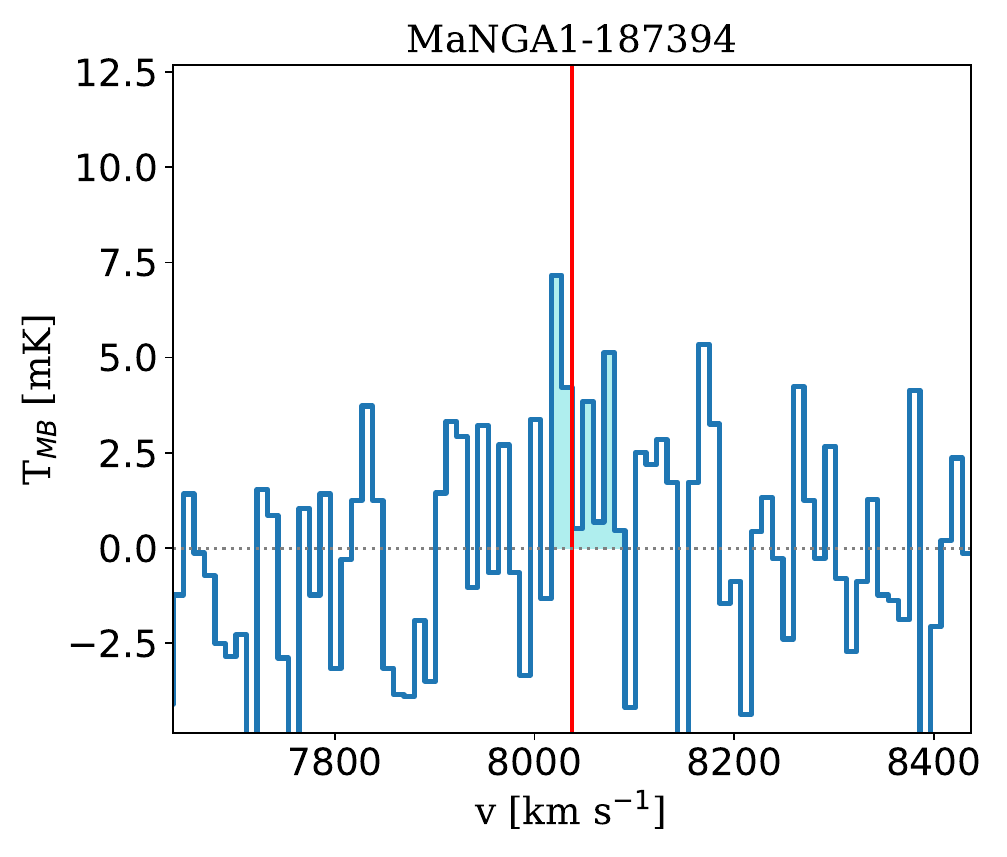}
\includegraphics[scale=0.19]{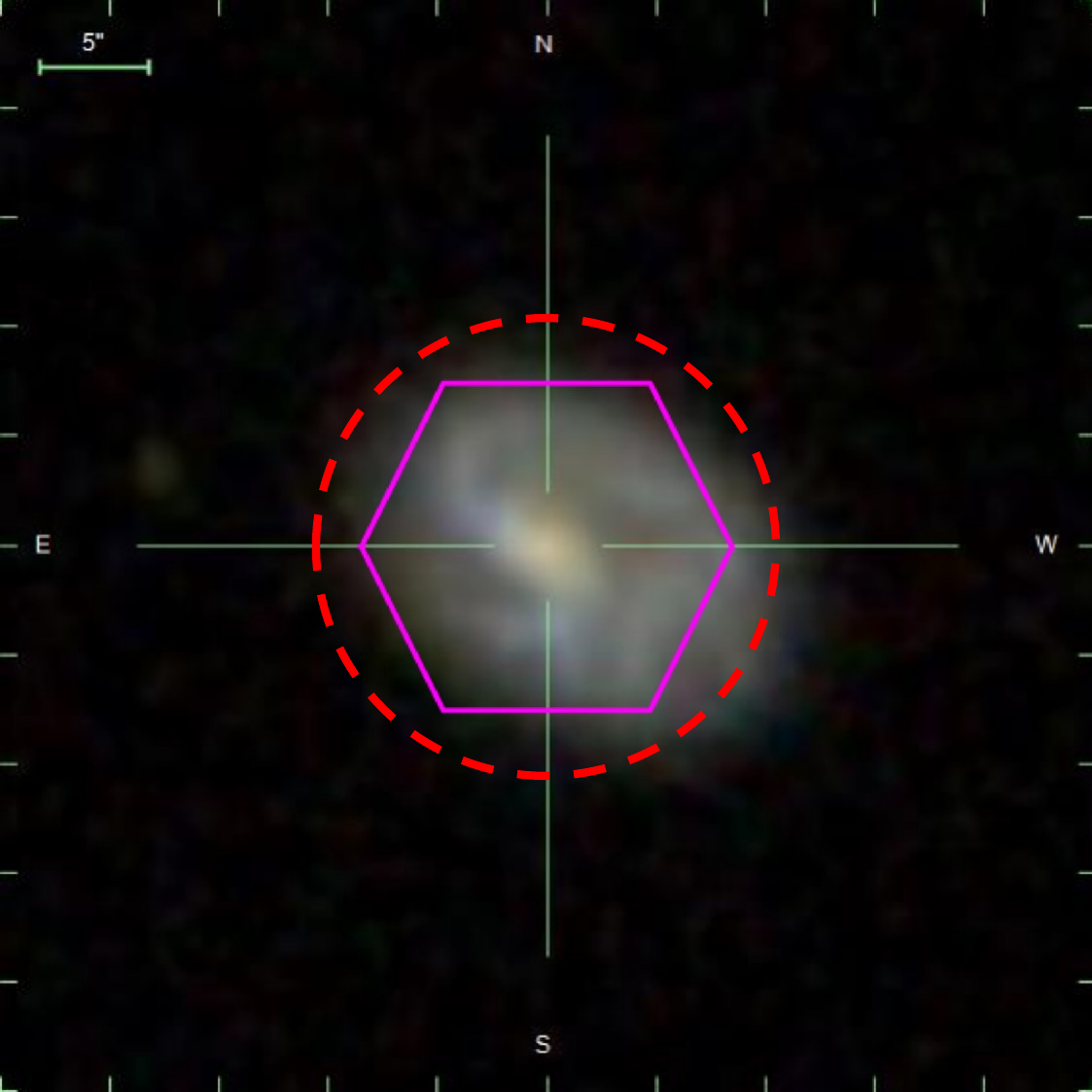}
\includegraphics[bb=5 5 495 405, scale=0.27]{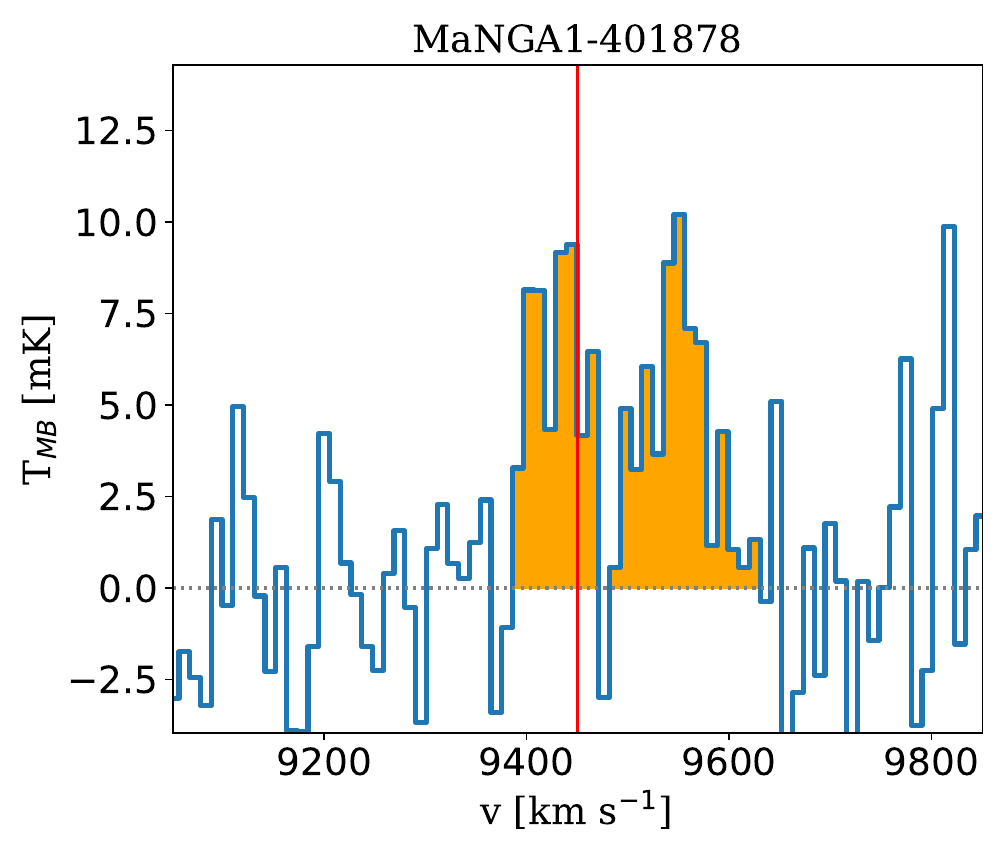}
\includegraphics[scale=0.19]{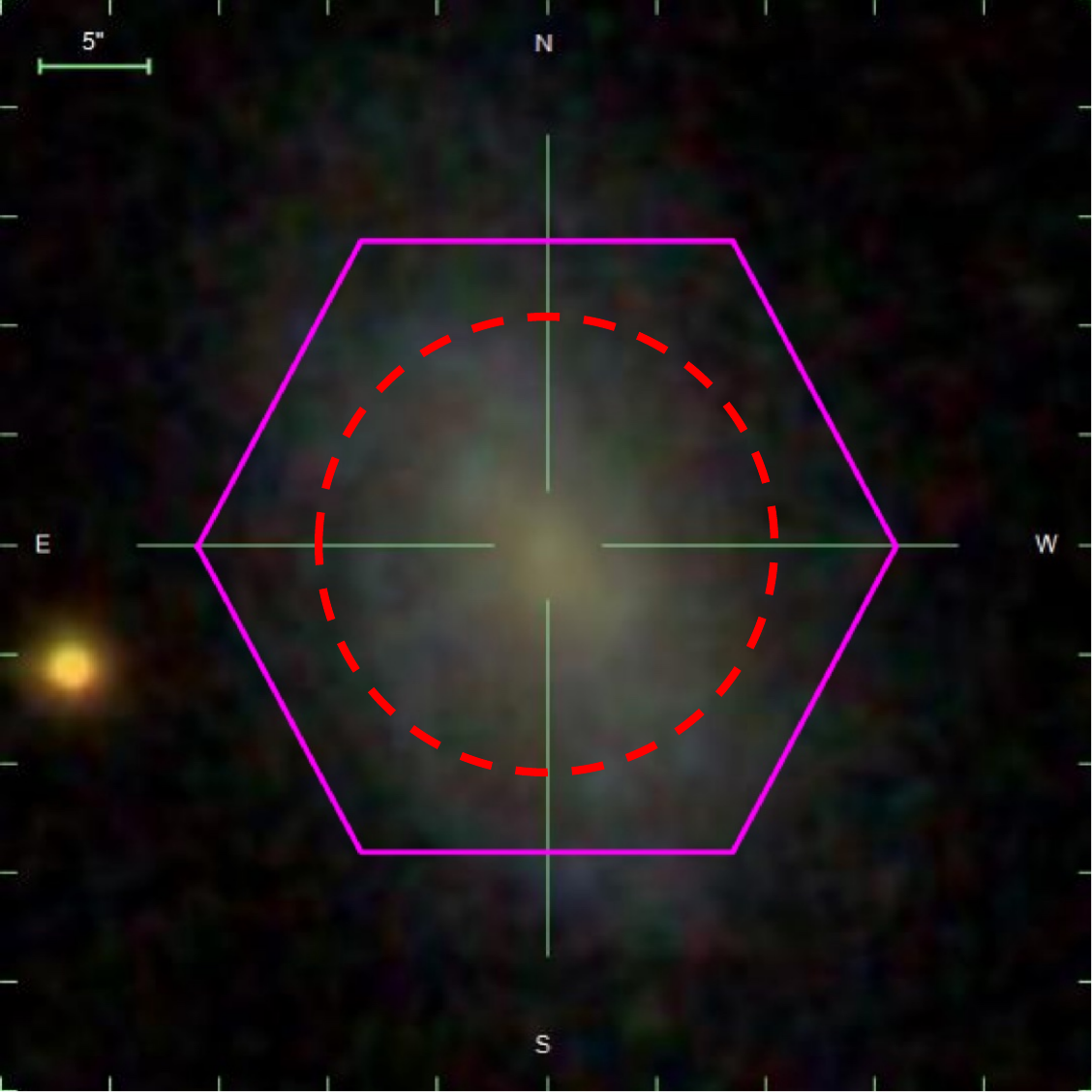}
\includegraphics[bb=5 5 495 405, scale=0.27]{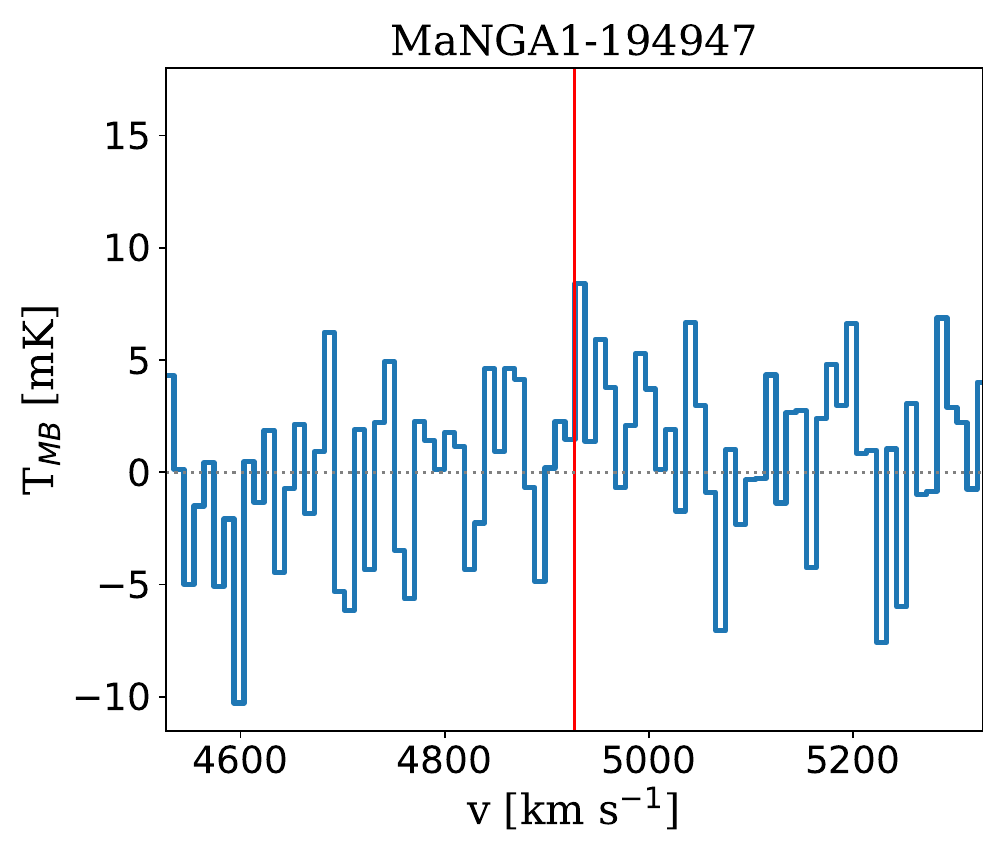}
\includegraphics[scale=0.19]{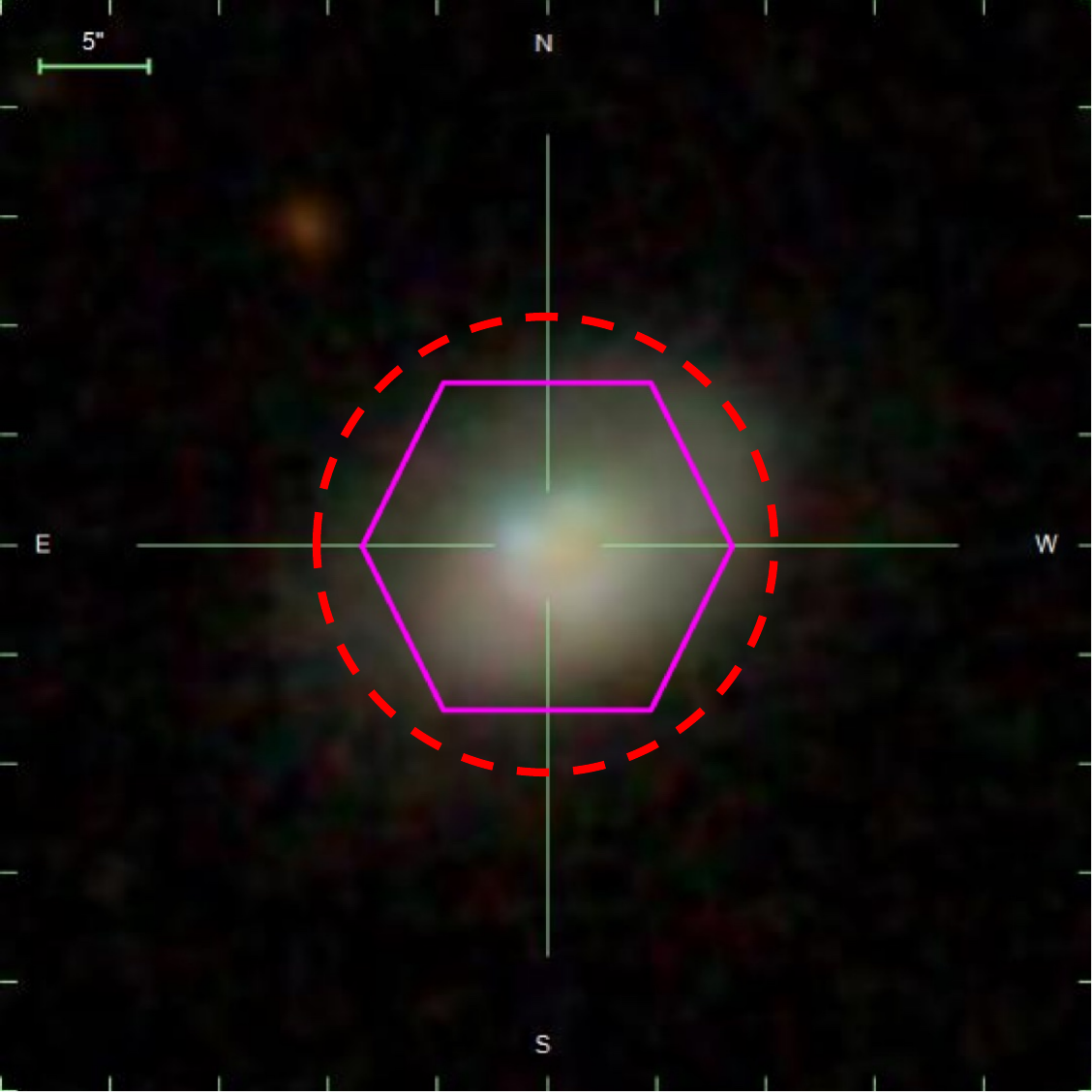}
\includegraphics[bb=5 5 495 405, scale=0.27]{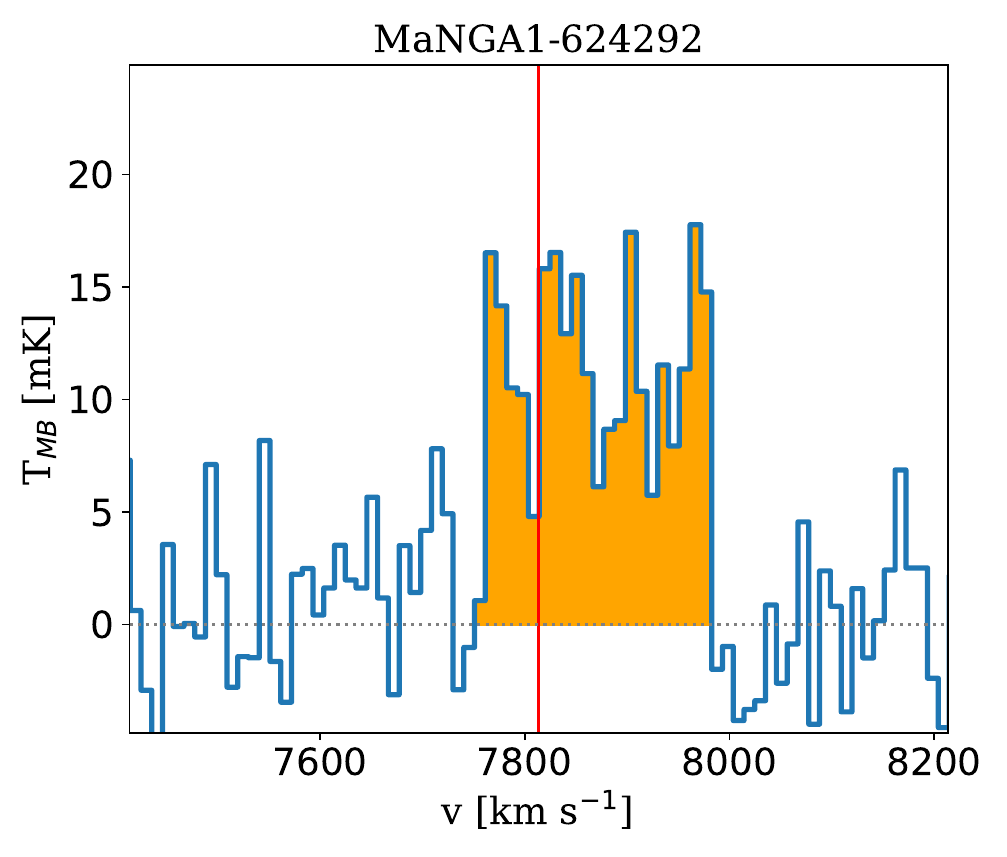}
\includegraphics[scale=0.19]{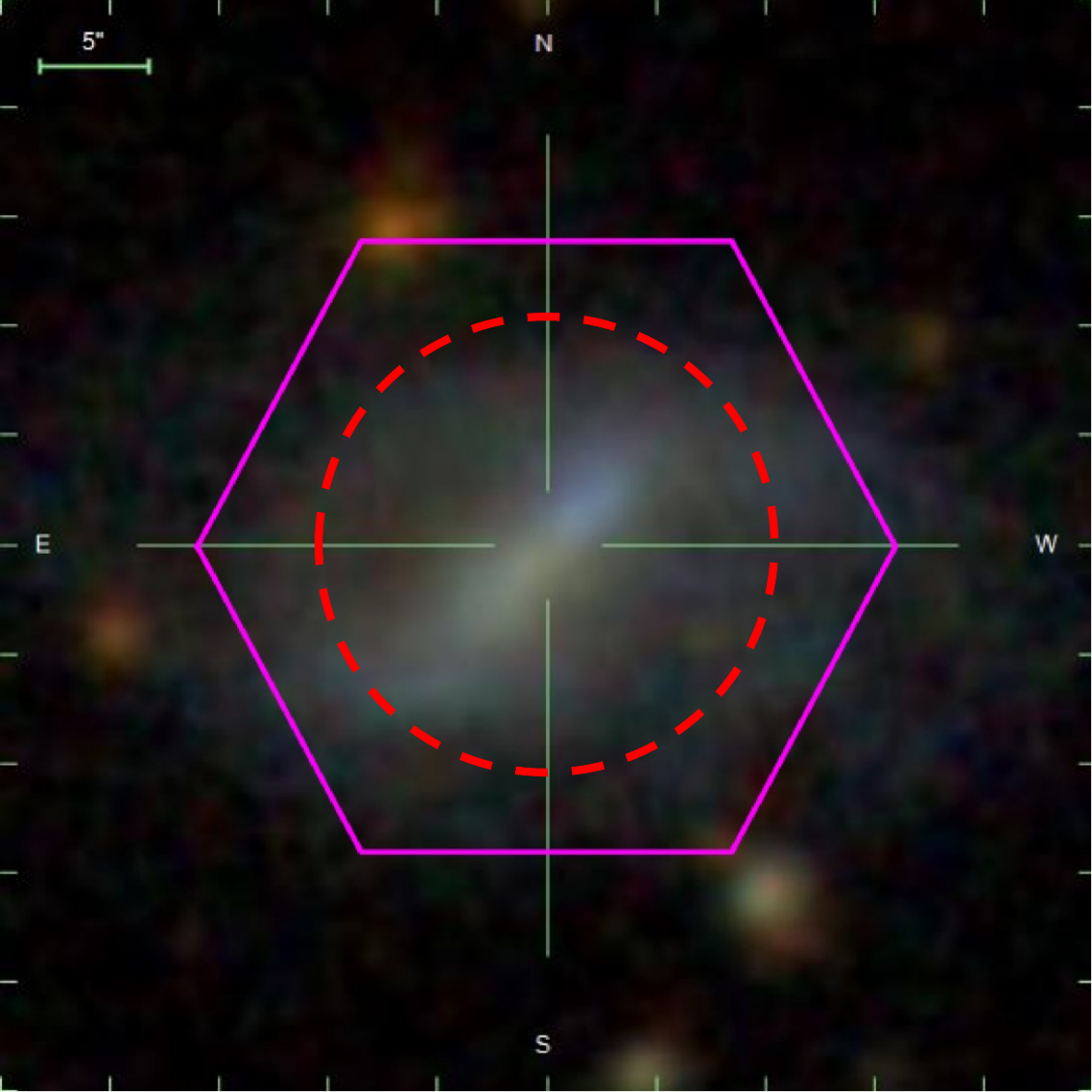}
\includegraphics[bb=5 5 495 405, scale=0.27]{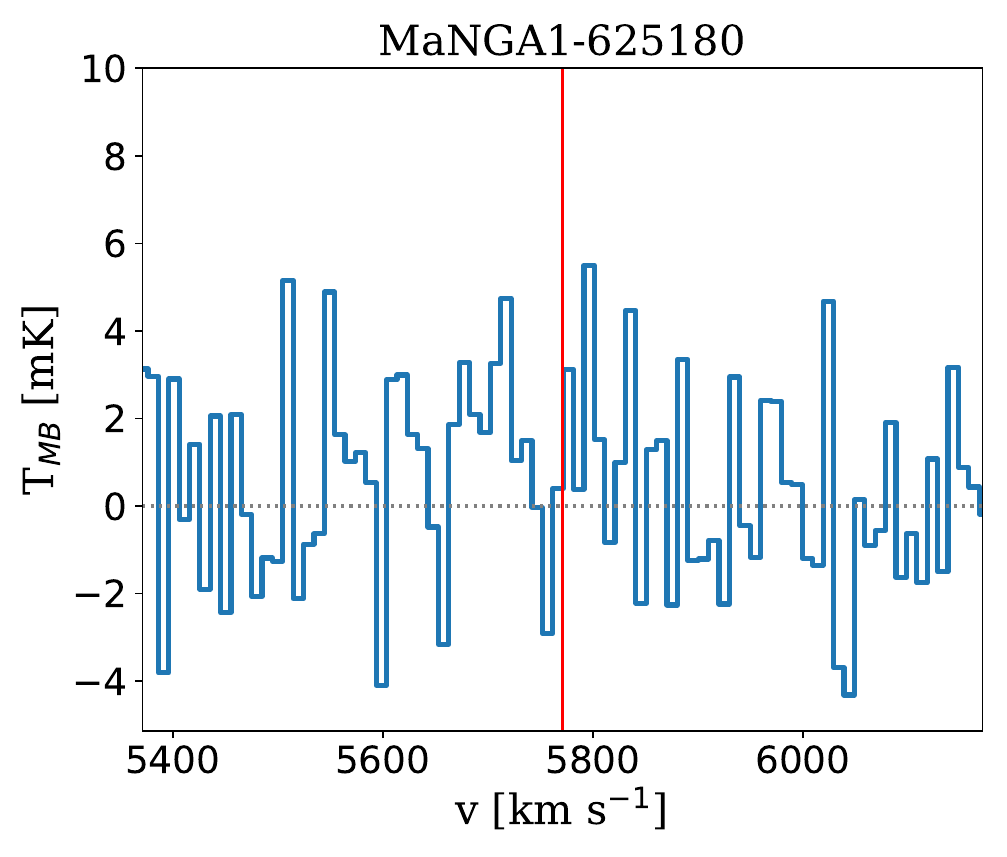}
\caption{continued.}
\end{center}
\end{figure*}

\begin{figure*}
  \ContinuedFloat
\begin{center}
\includegraphics[scale=0.19]{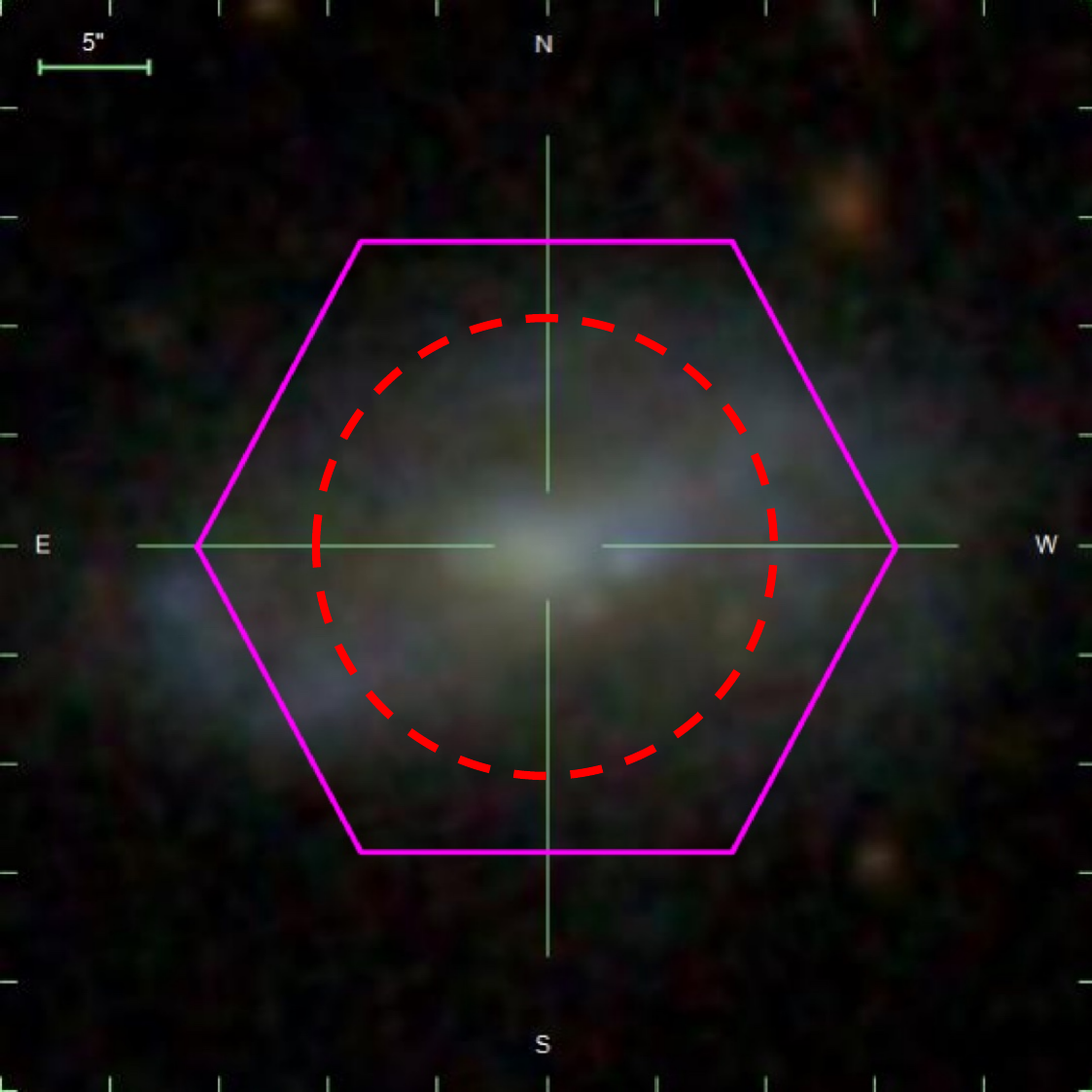}
\includegraphics[bb=5 5 495 405, scale=0.27]{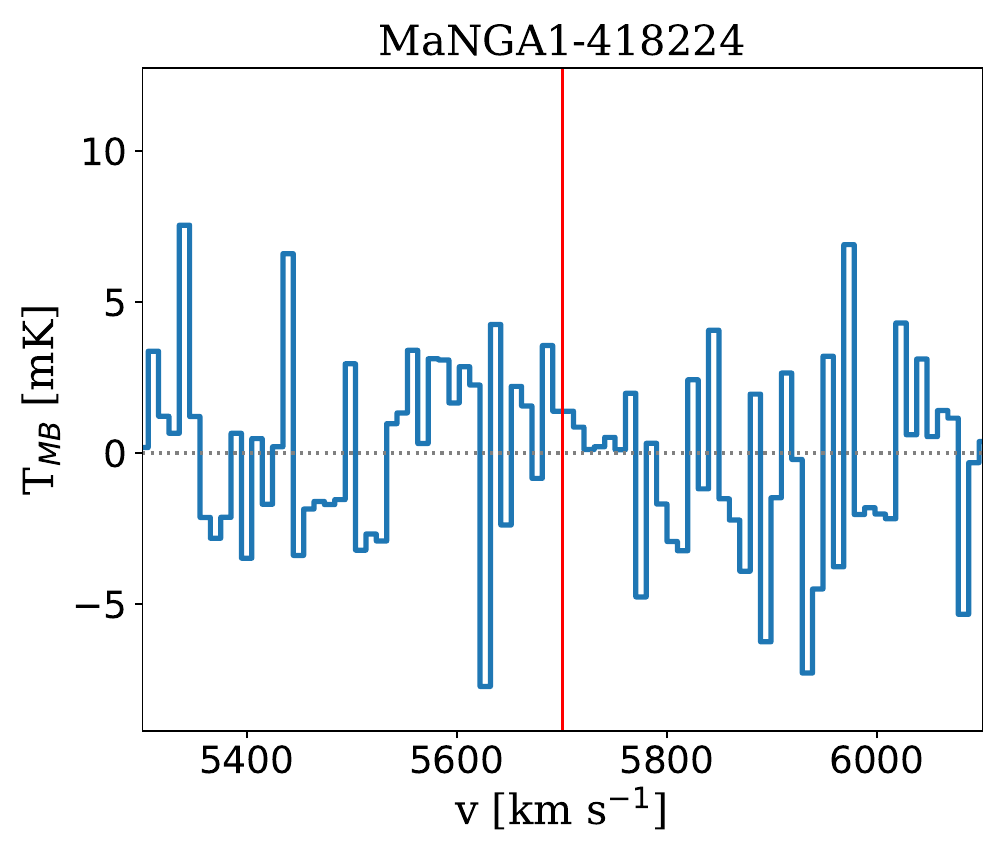}
\includegraphics[scale=0.19]{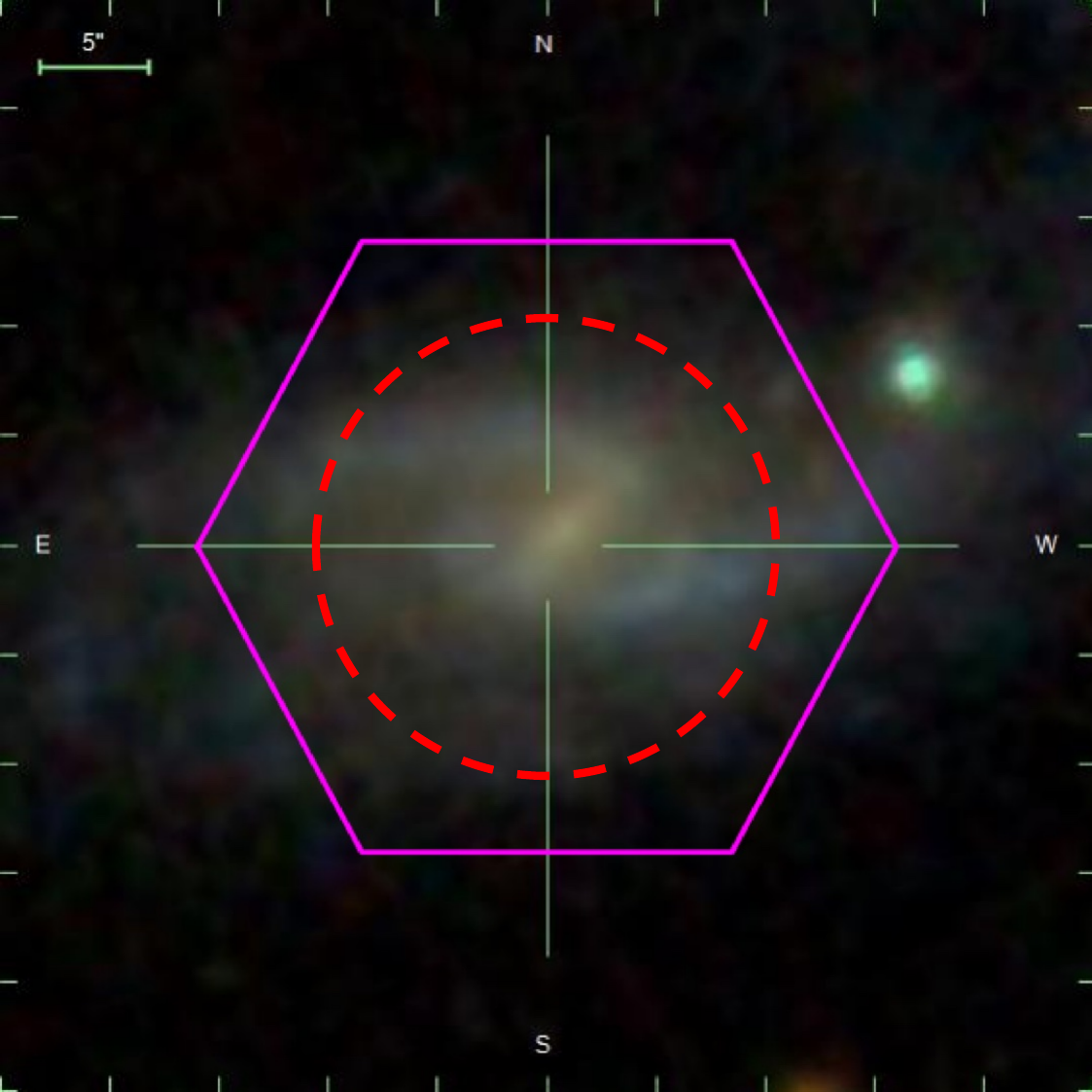}
\includegraphics[bb=5 5 495 405, scale=0.27]{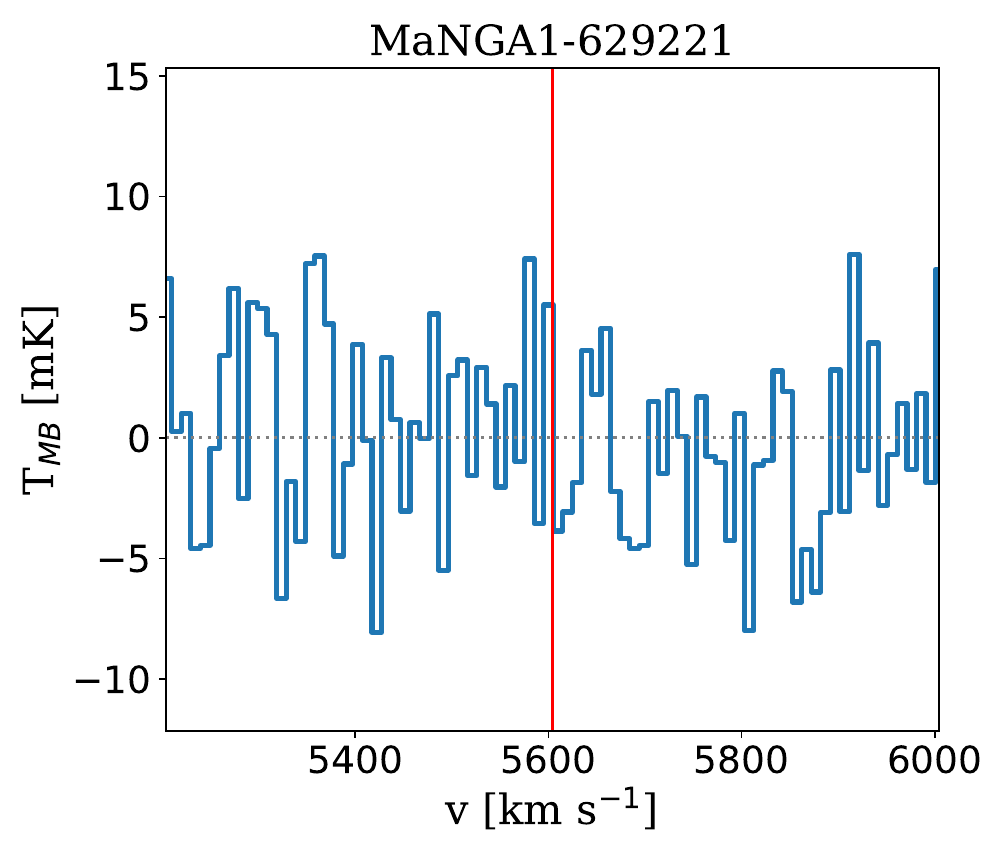}
\includegraphics[scale=0.19]{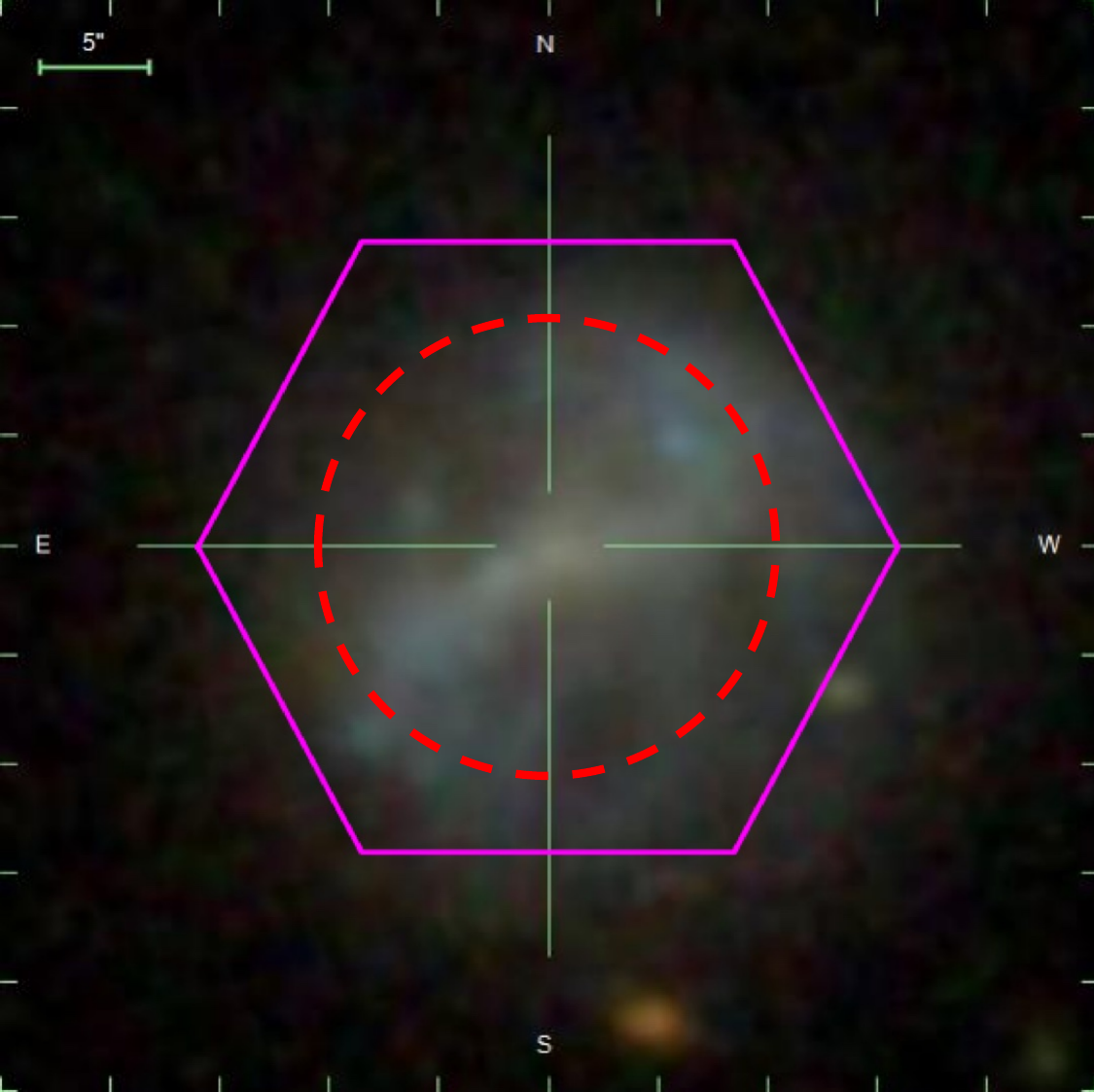}
\includegraphics[bb=5 5 495 405, scale=0.27]{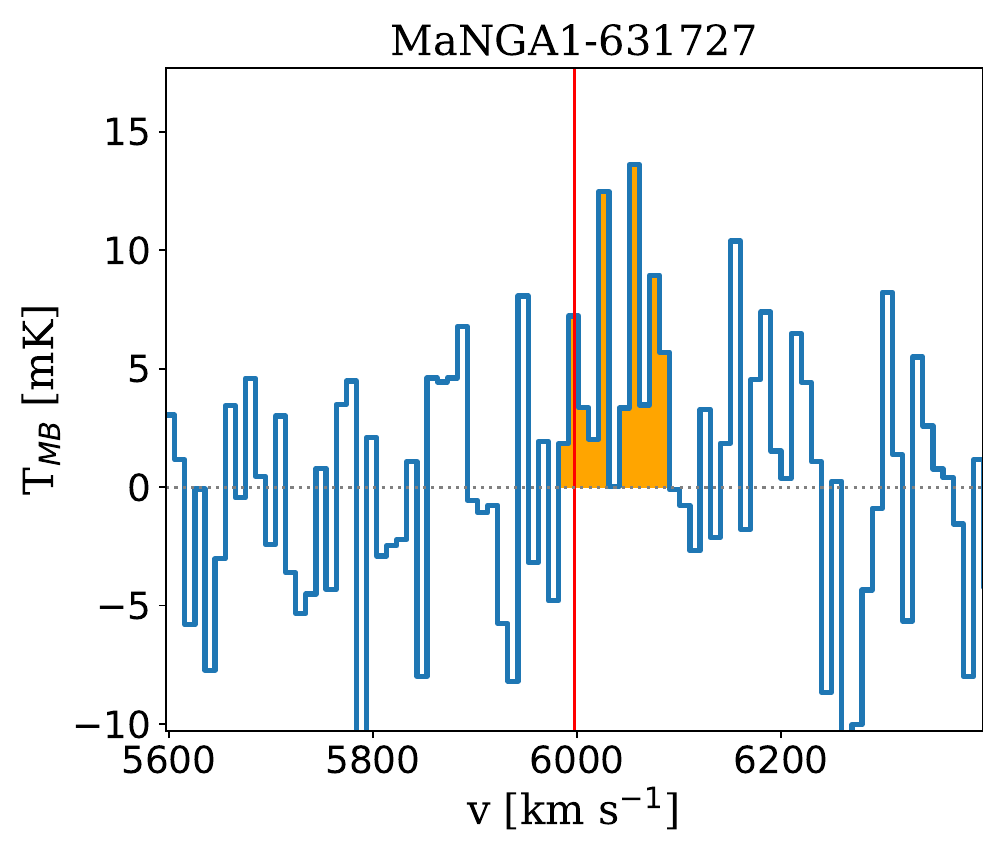}
\includegraphics[scale=0.19]{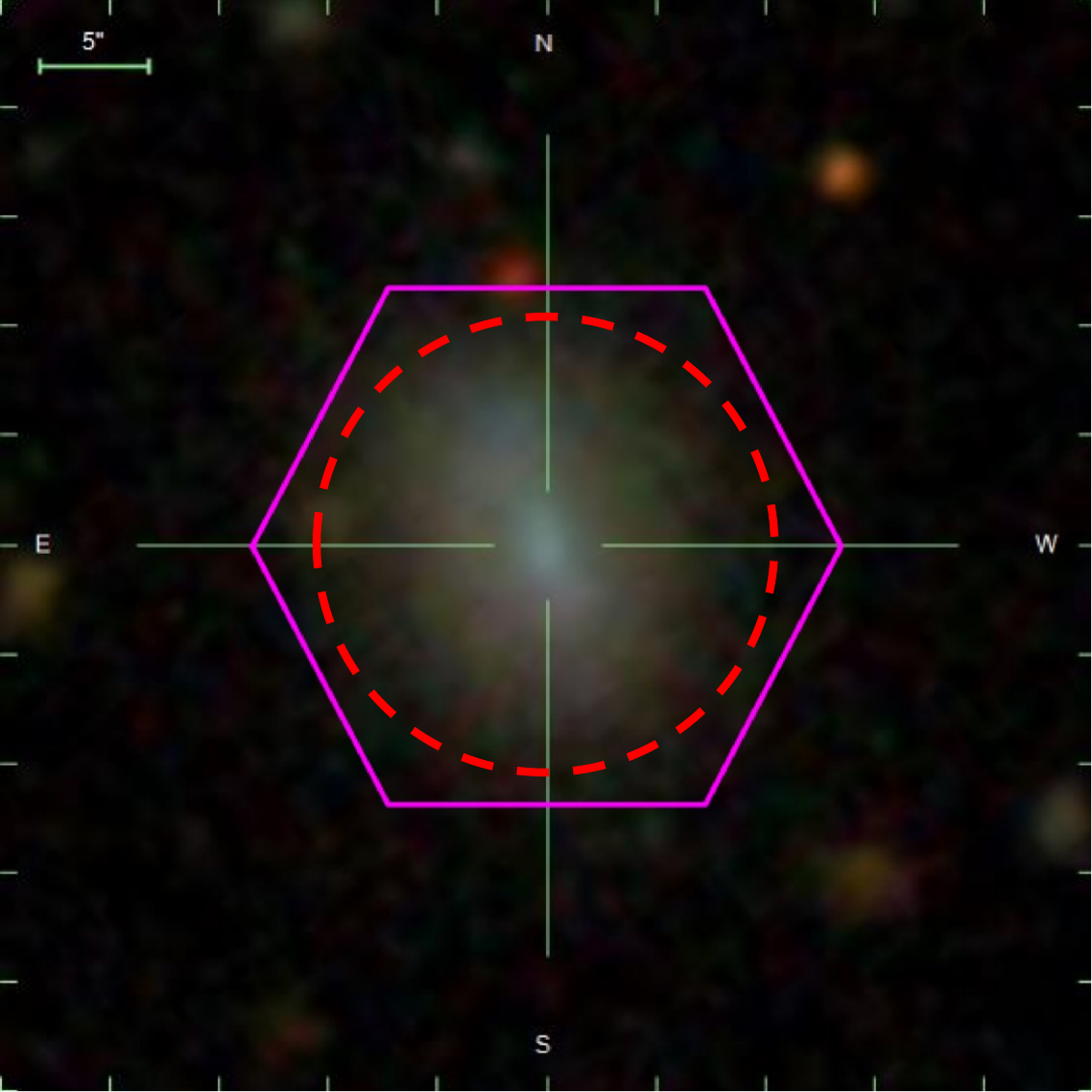}
\includegraphics[bb=5 5 495 405, scale=0.27]{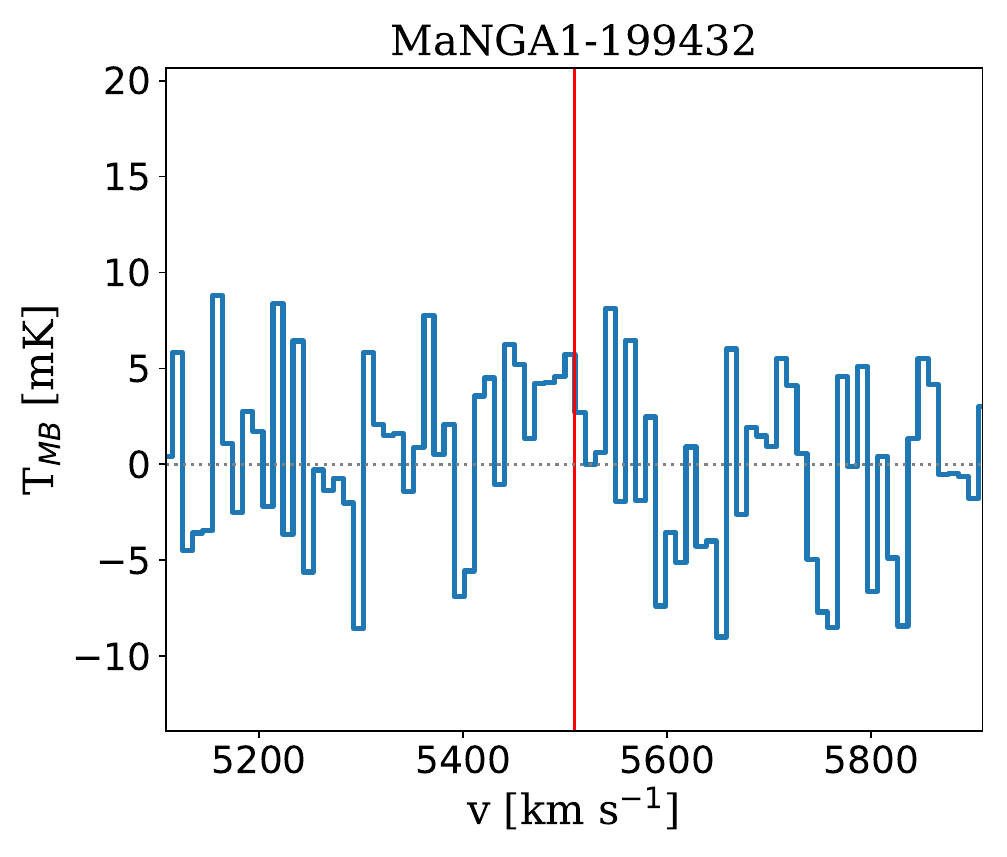}
\includegraphics[scale=0.19]{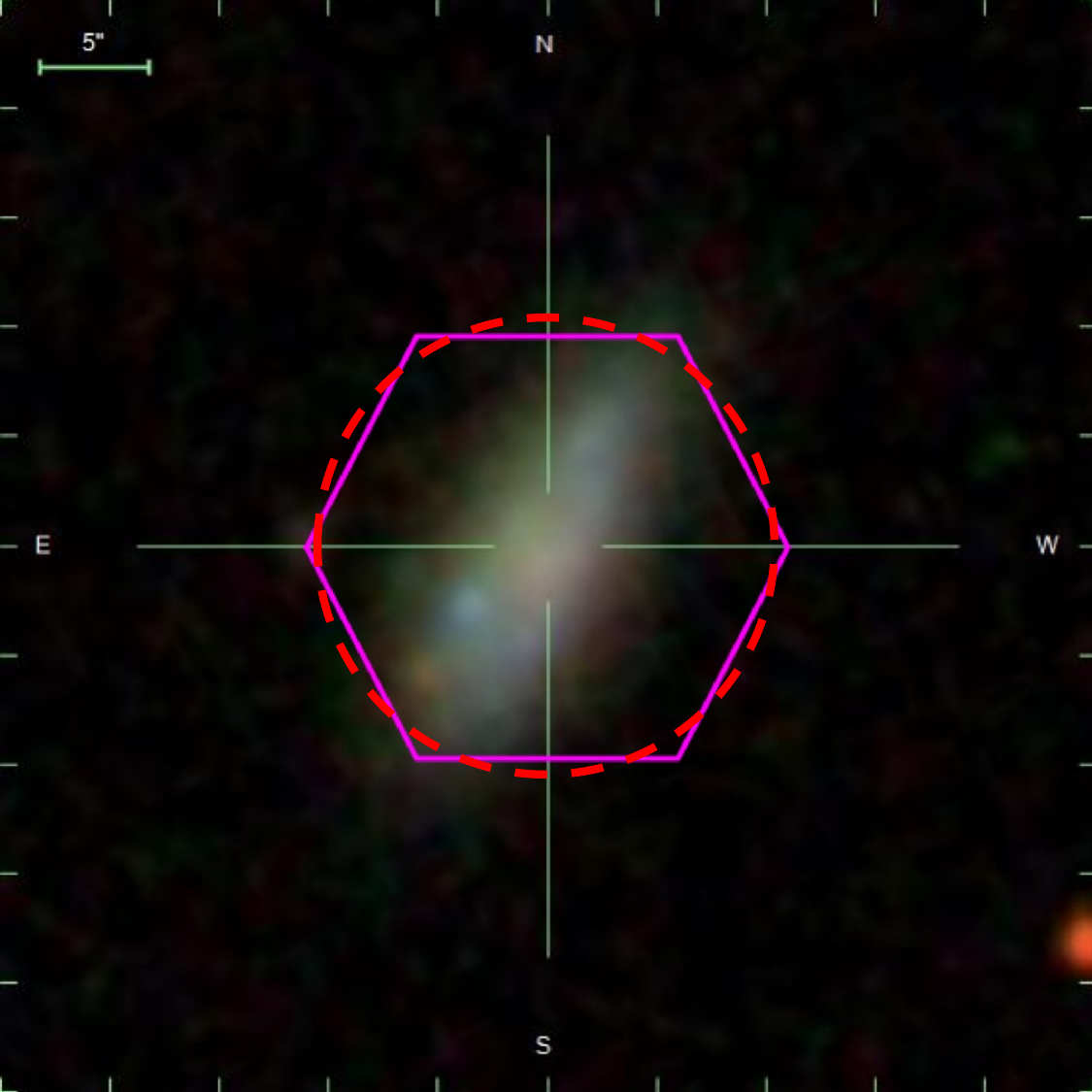}
\includegraphics[bb=5 5 495 405, scale=0.27]{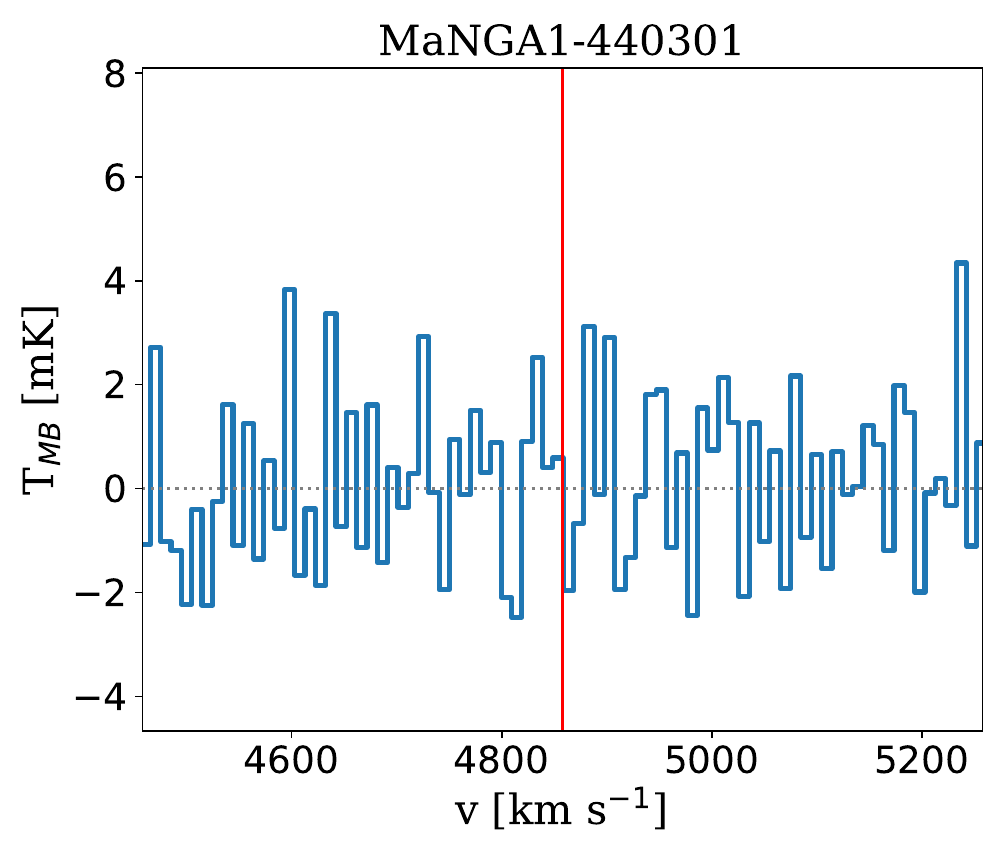}
\includegraphics[scale=0.19]{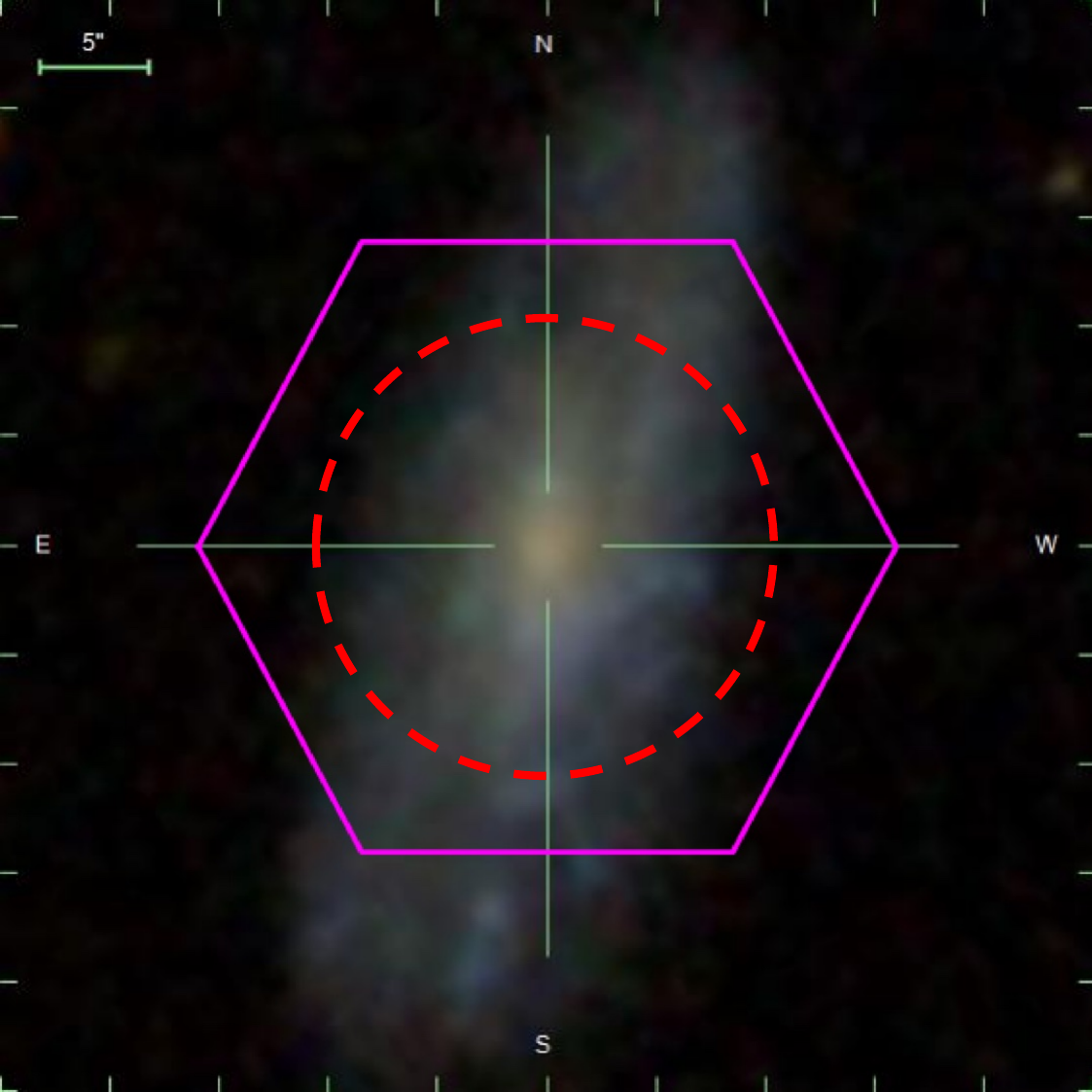}
\includegraphics[bb=5 5 495 405, scale=0.27]{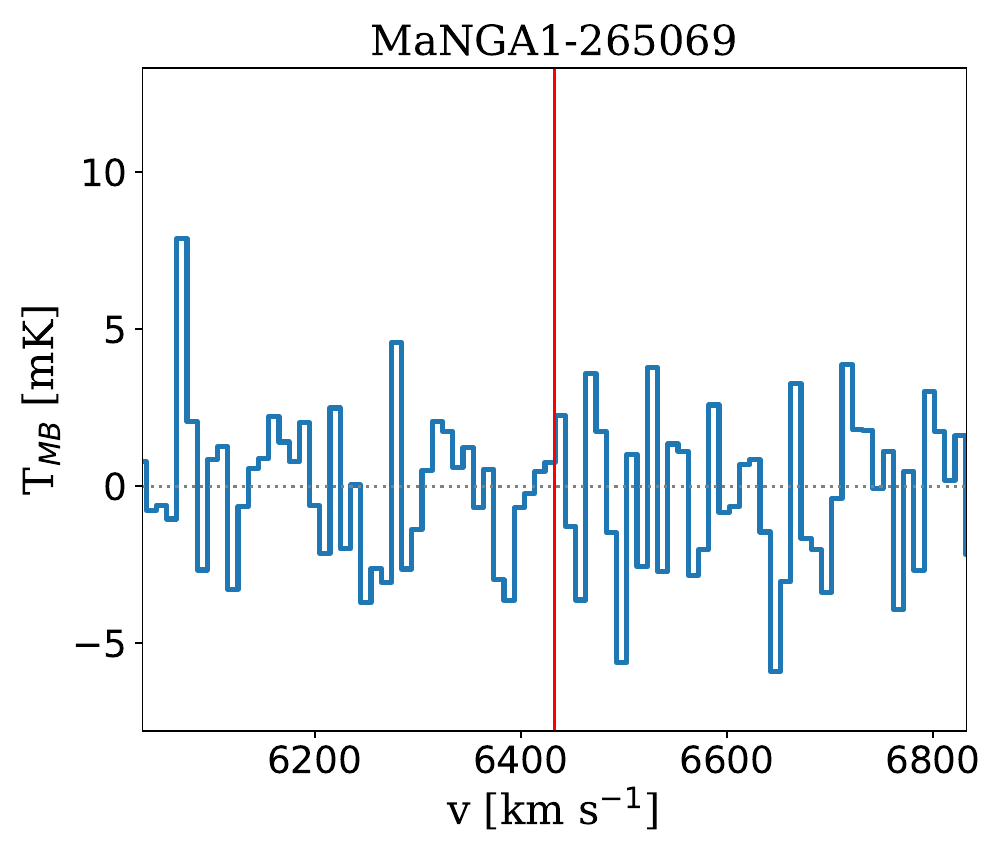}
\includegraphics[scale=0.19]{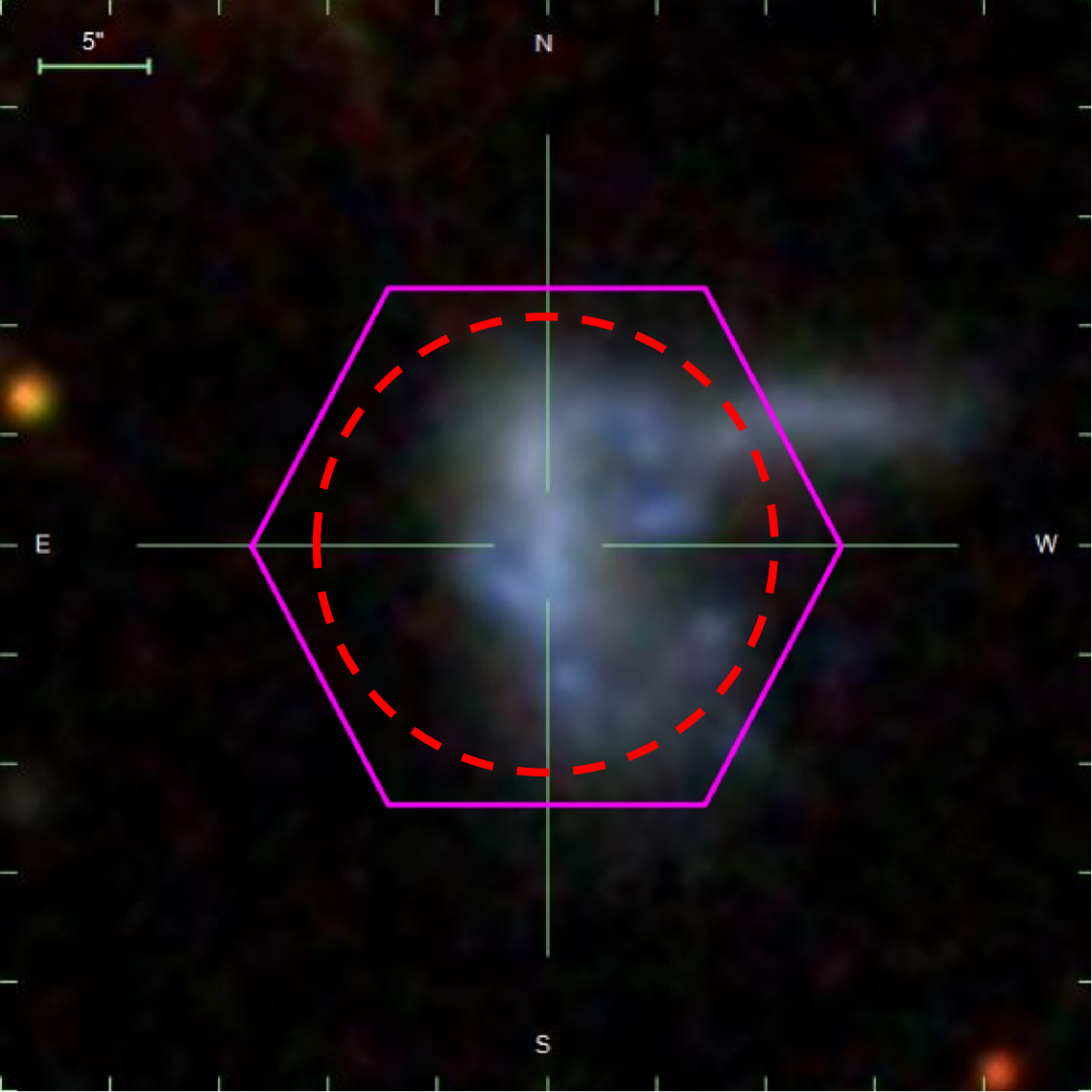}
\includegraphics[bb=5 5 495 405, scale=0.27]{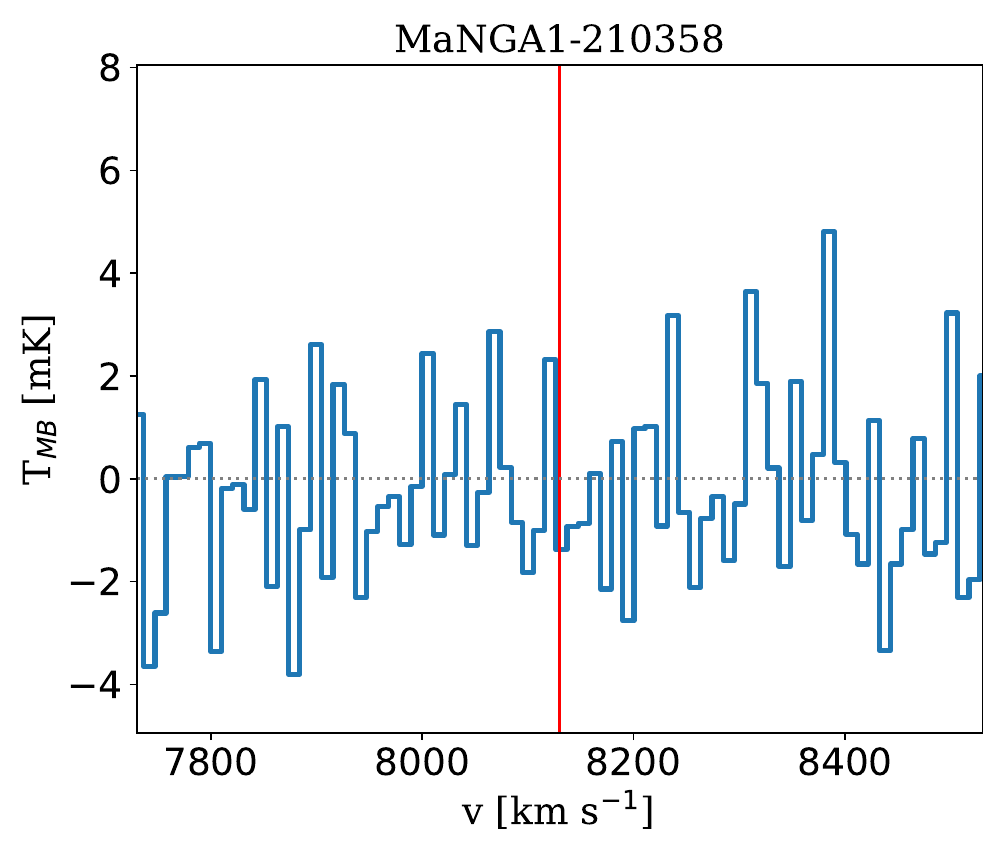}
\includegraphics[scale=0.19]{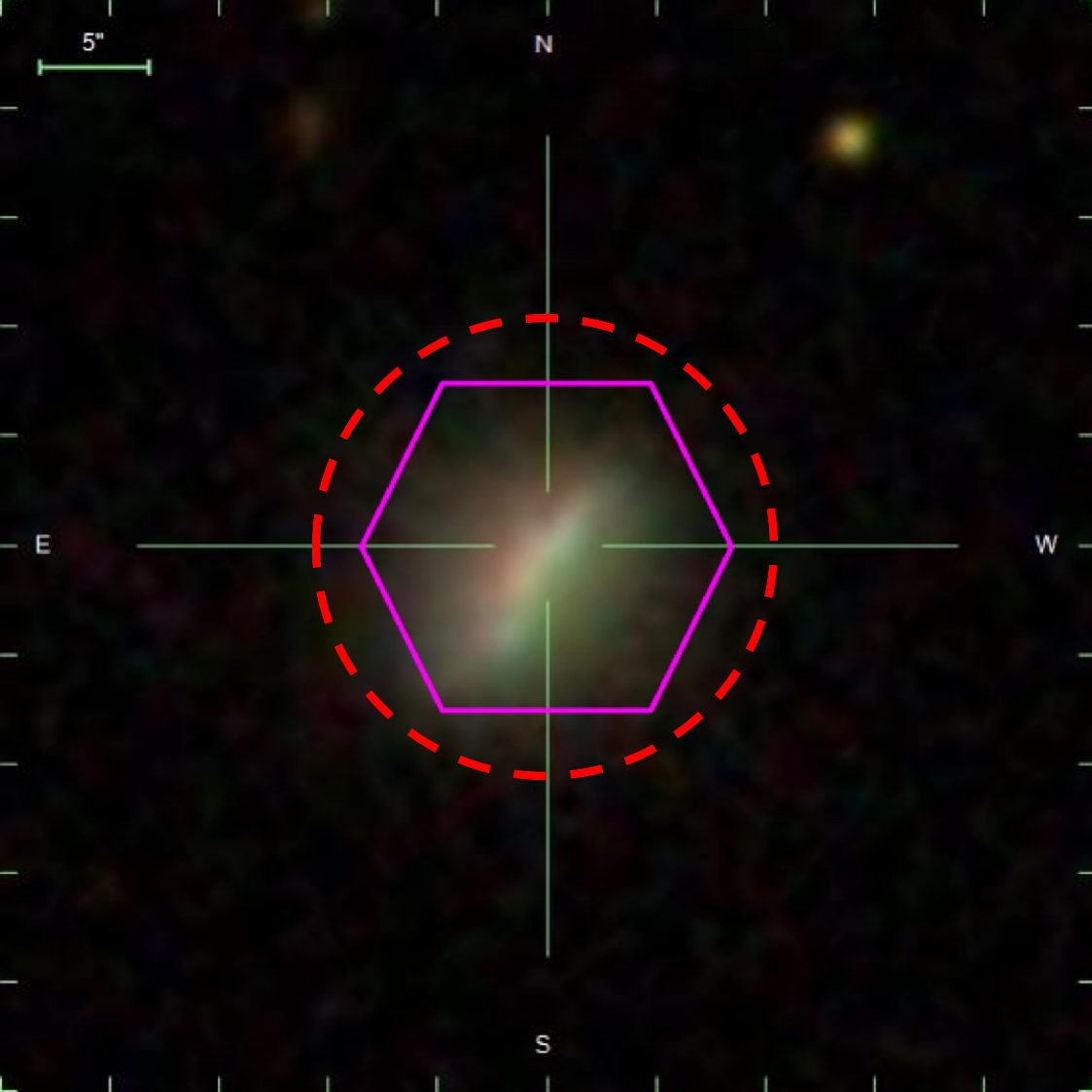}
\includegraphics[bb=5 5 495 405, scale=0.27]{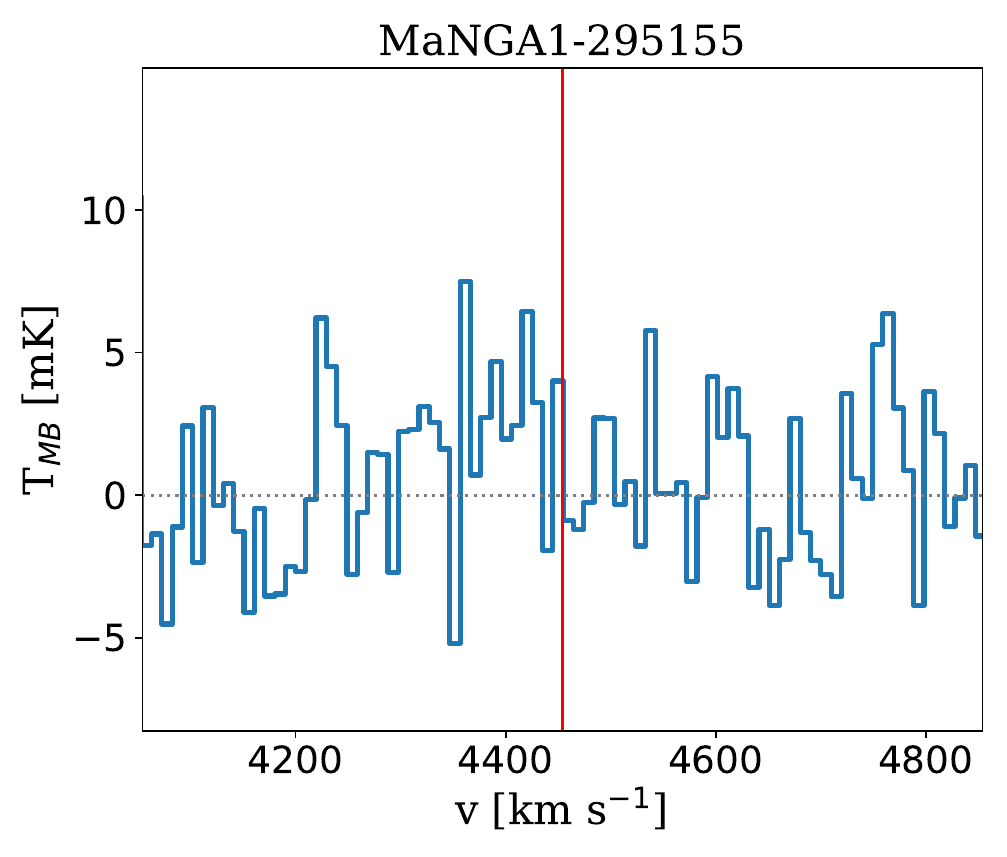}
\includegraphics[scale=0.19]{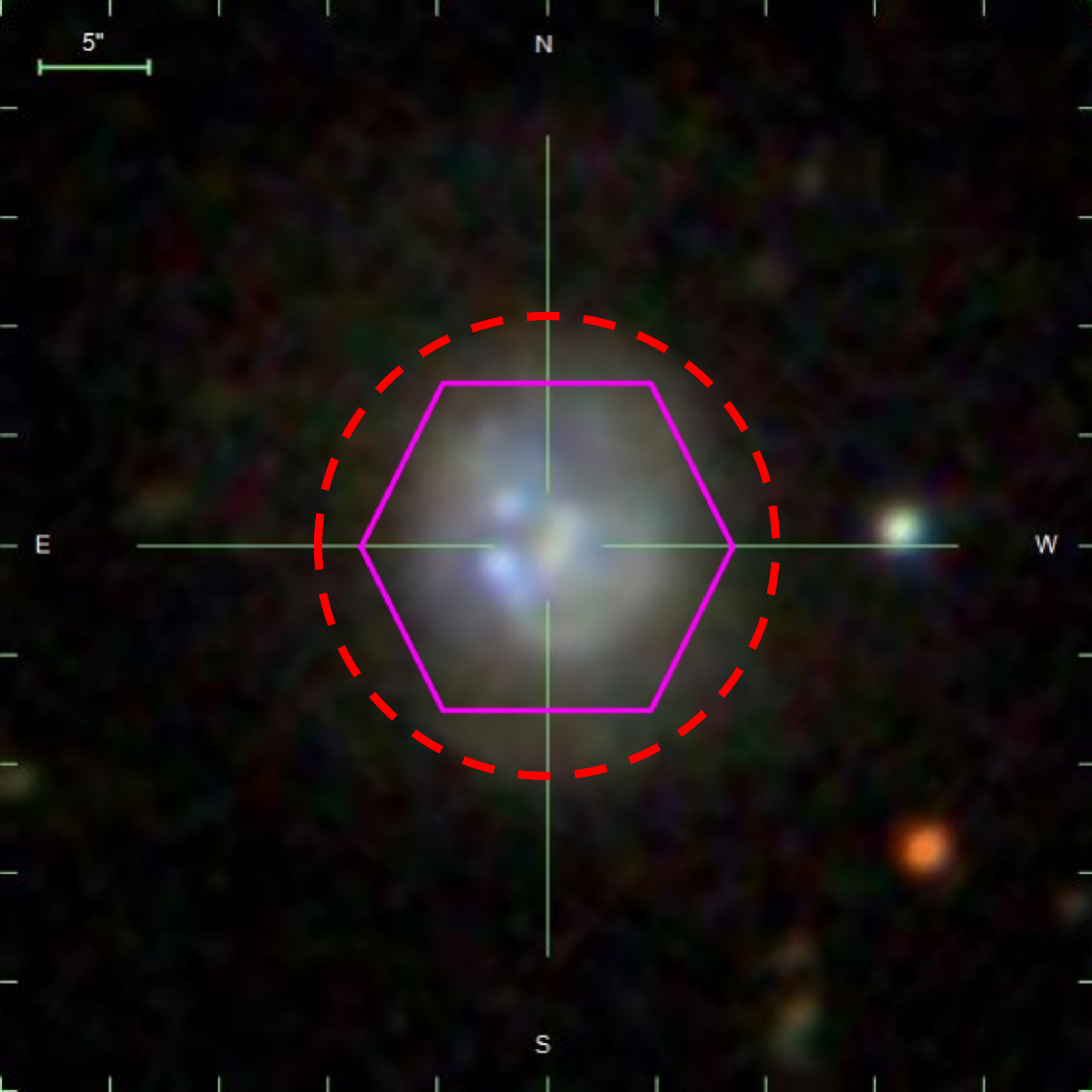}
\includegraphics[bb=5 5 495 405, scale=0.27]{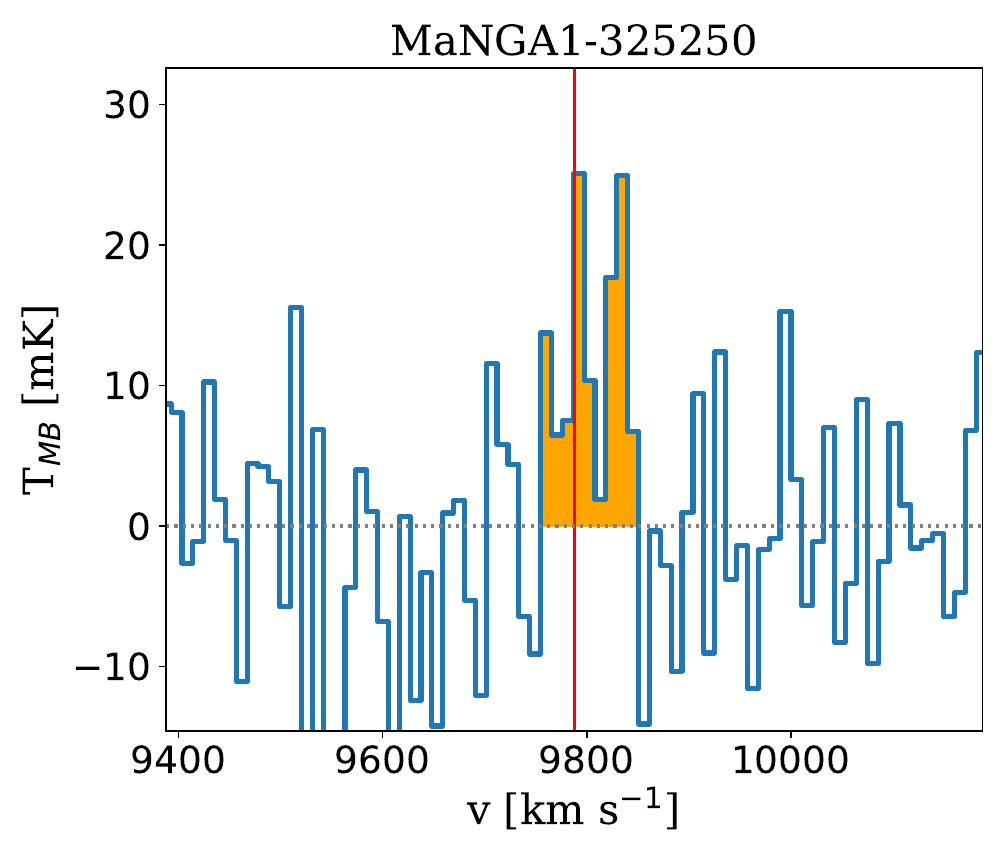}
\includegraphics[scale=0.19]{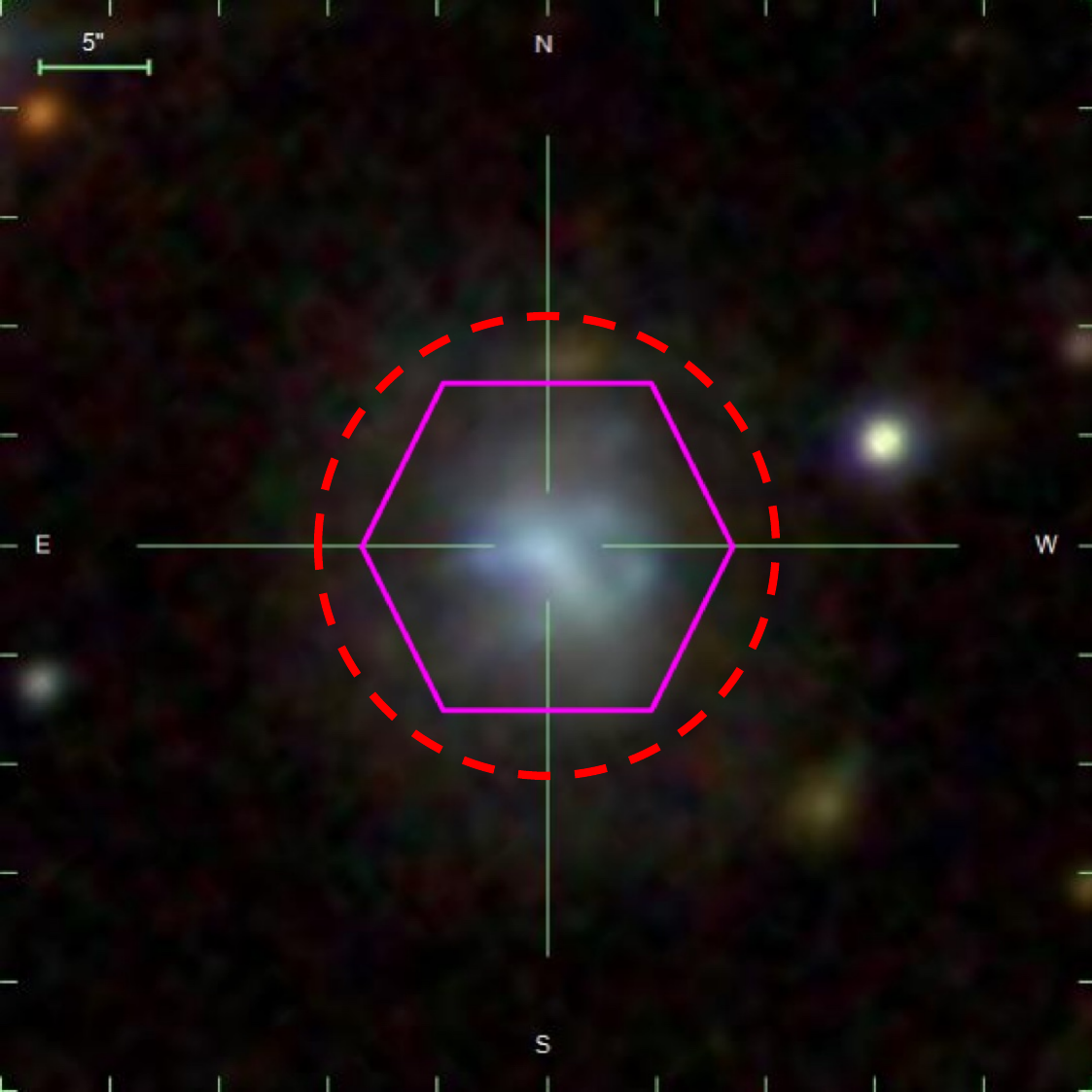}
\includegraphics[bb=5 5 495 405, scale=0.27]{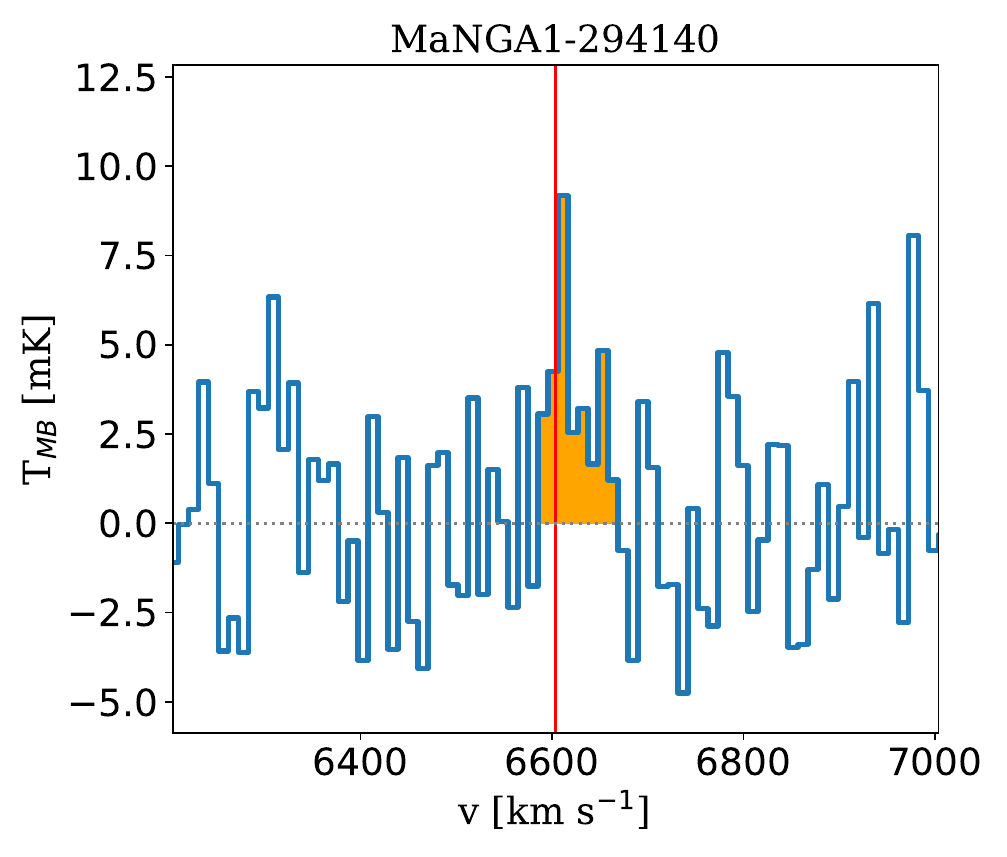}
\includegraphics[scale=0.19]{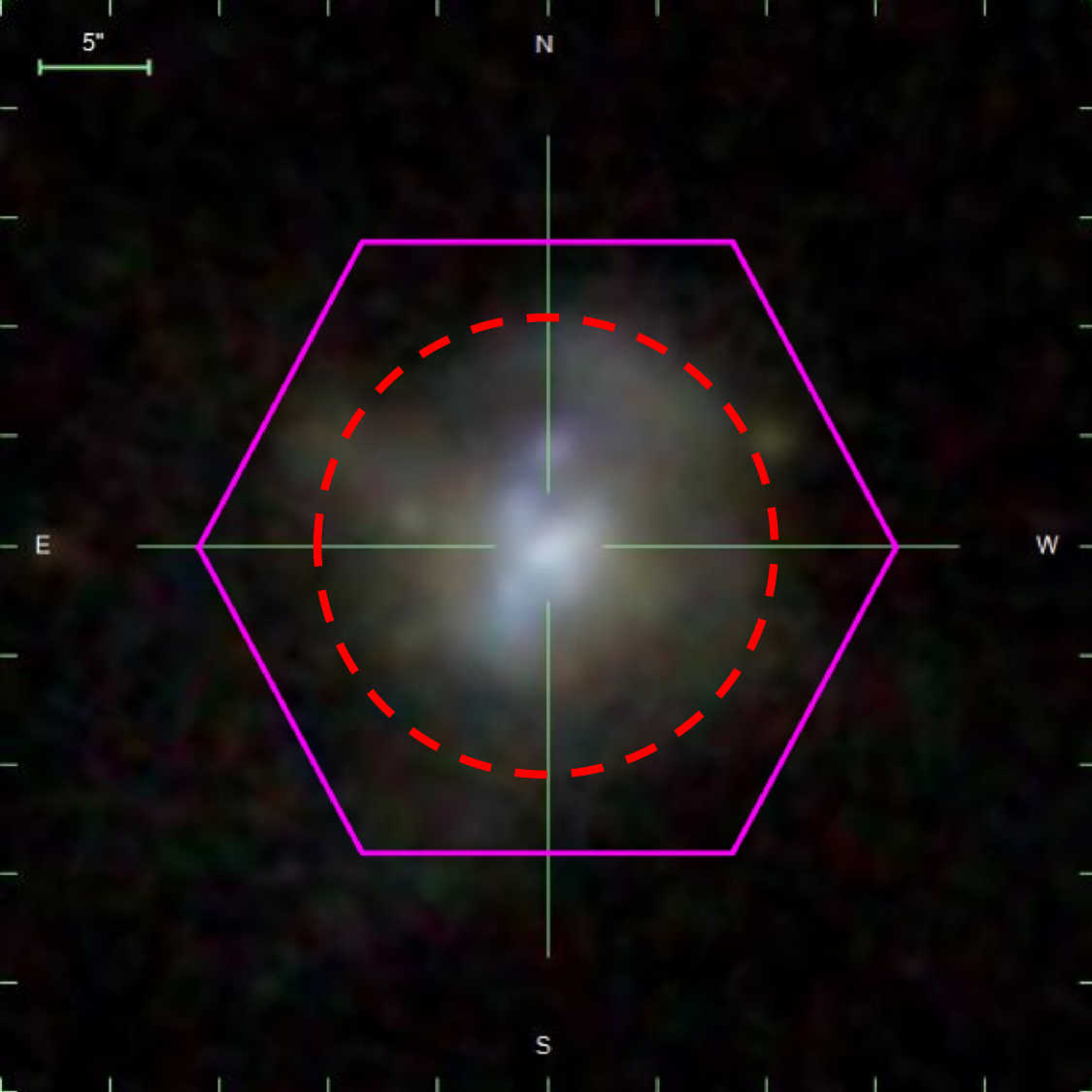}
\includegraphics[bb=5 5 495 405, scale=0.27]{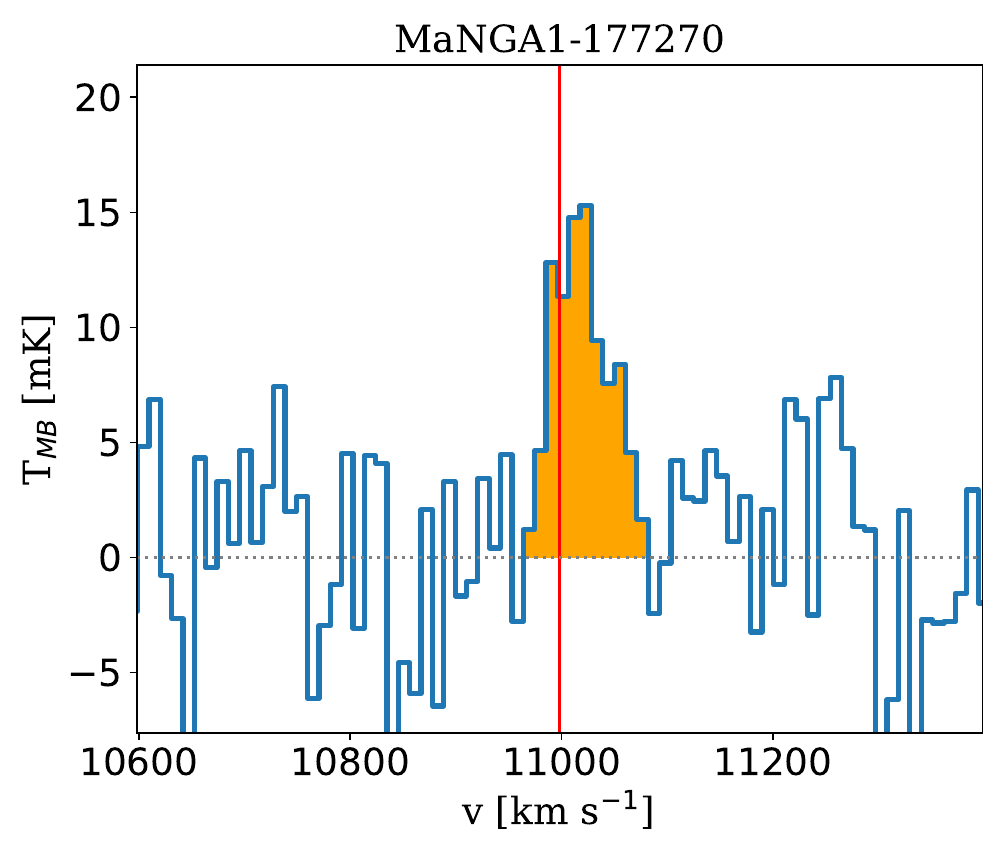}
\includegraphics[scale=0.19]{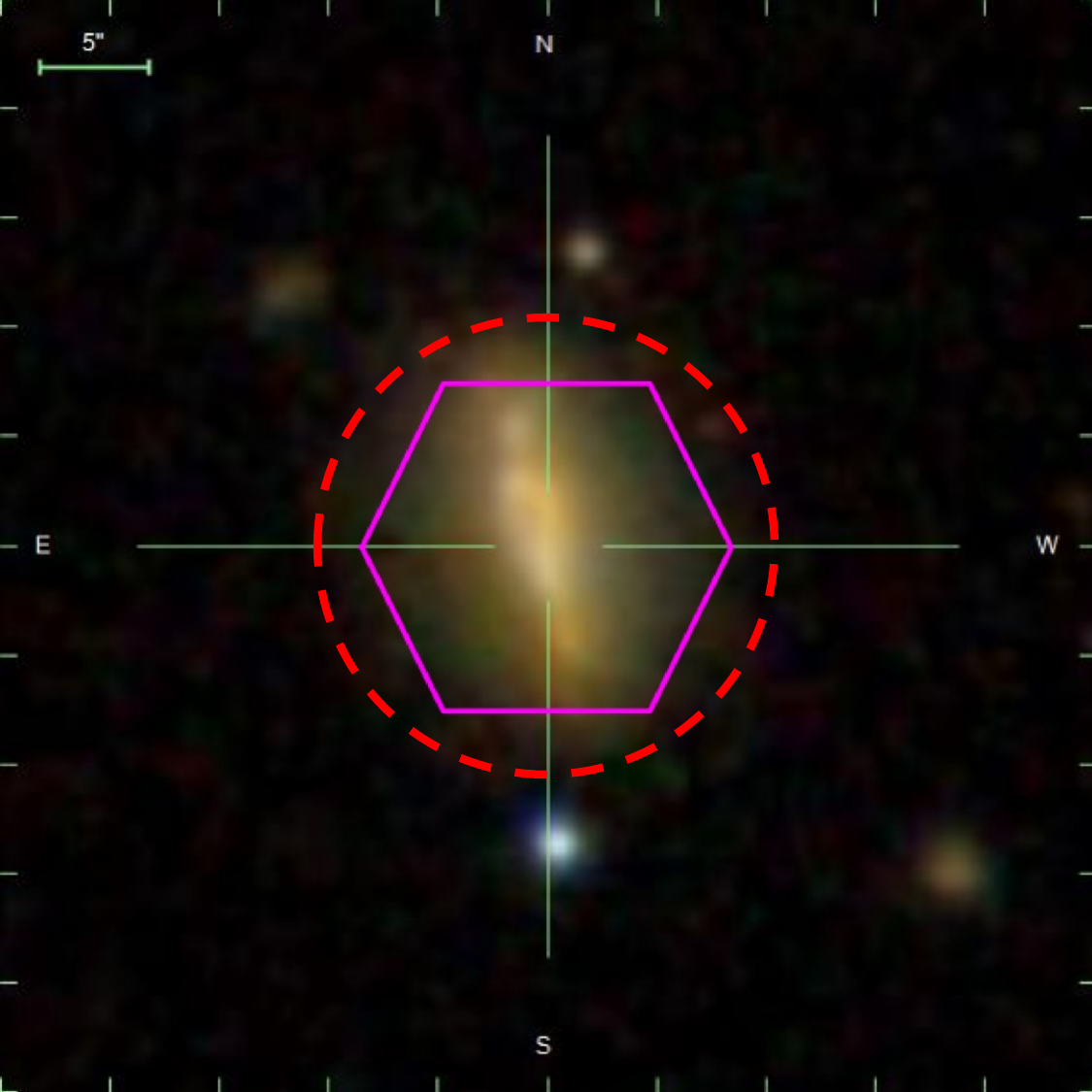}
\includegraphics[bb=5 5 495 405, scale=0.27]{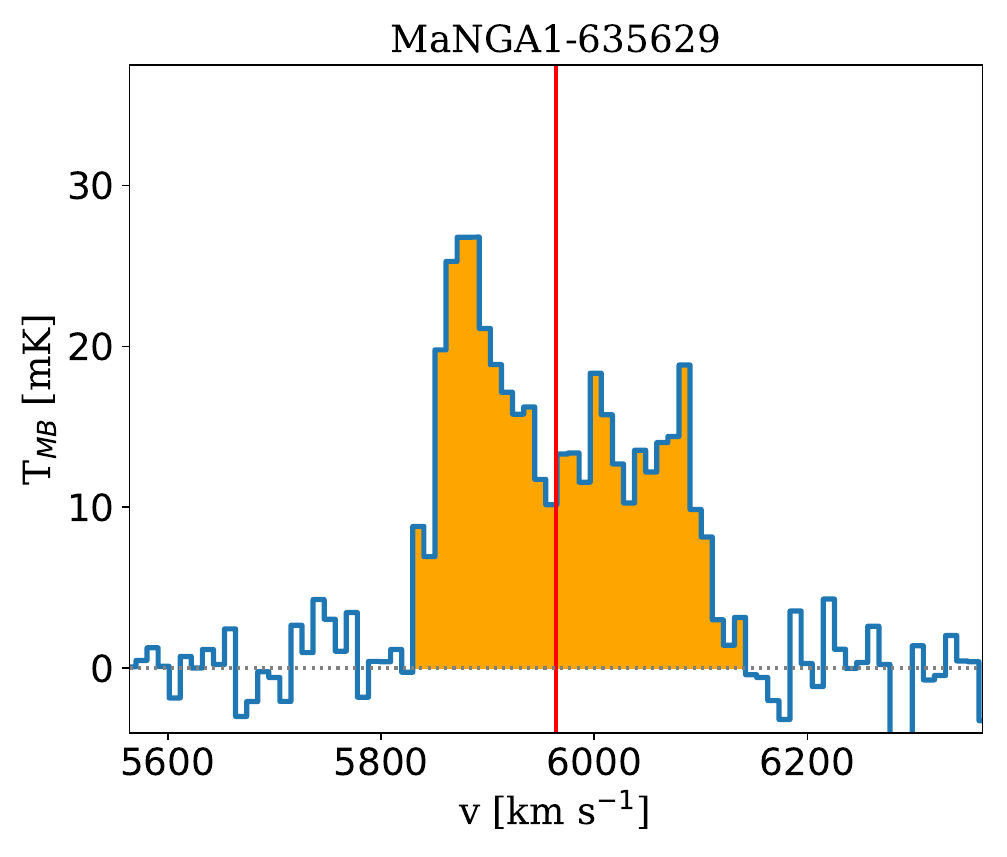}
\caption{continued.}
\label{fig:CO_spectra}
\end{center}
\end{figure*}

\end{appendix}

\end{document}